\documentclass[preprint]{elsarticle}

\usepackage{lineno,hyperref}
\modulolinenumbers[5]
\usepackage[margin=1.5cm]{geometry}

\journal{ ... }

\usepackage{moreverb}

\usepackage{color}
\definecolor{hotcolor}{rgb}{1,0,0}

\usepackage{algorithm}
\usepackage[noend]{algpseudocode}
\usepackage{amsmath}
\usepackage{amsfonts}
\usepackage{amsthm}
\usepackage{url}
\usepackage{bm}
\usepackage{subfigure}
\usepackage{varwidth,xcolor}
\usepackage{stmaryrd}
\usepackage{float}

\newcommand\BibTeX{{\rmfamily B\kern-.05em \textsc{i\kern-.025em b}\kern-.08em
T\kern-.1667em\lower.7ex\hbox{E}\kern-.125emX}}

\newtheorem{mylem}{Lemma}
\newtheorem{myprop}{Proposition}

\begin{document}

\begin{frontmatter}

\title{Comparison of reduced models for blood flow using \\ Runge--Kutta discontinuous Galerkin methods}





\author{Charles Puelz\fnref{affil1}}
\ead{cpuelz@rice.edu}

\author{Sun\v{c}ica \v{C}ani\'c\fnref{affil2}}

\author{B\'eatrice Rivi\`ere\fnref{affil1}}

\author{Craig G. Rusin\fnref{affil3,affil4}}

\fntext[affil1]{Rice University, Department of Computational and Applied Mathematics}
\fntext[affil2]{University of Houston, Department of Mathematics}
\fntext[affil3]{Baylor College of Medicine, Department of Pediatric Cardiology}
\fntext[affil4]{Texas Children's Hospital, Department of Pediatric Medicine--Cardiology}

\begin{abstract}
One--dimensional blood flow models take the general form of nonlinear hyperbolic systems but differ greatly in their formulation.  One class of models considers the physically conserved quantities of mass and momentum, while another class describes mass and velocity.  Further, the averaging process employed in the model derivation requires the specification of the axial velocity profile; this choice differentiates models within each class.  Discrepancies among differing models have yet to be investigated.  In this paper, we systematically compare several reduced models of blood flow for 
physiologically relevant vessel parameters, network topology, and boundary data. The models are discretized by a
class of Runge--Kutta discontinuous Galerkin methods.
\end{abstract}


\end{frontmatter}



\section{Introduction} \label{Section:Intro}

In this paper, we compare several variants of reduced blood flow models expressed as nonlinear hyperbolic systems of conservation laws in one space dimension (the axial dimension of the blood vessel).  We organize the models into two classes: $(1)$ the $(A,Q)$ system and $(2)$ the $(A,U)$ system  modeling vessel cross--sectional area $A$ and average fluid momentum $Q$ or average axial velocity $U$, respectively.  We remark that the $(A,Q)$ system models the physically conserved variables of mass and momentum.  The velocity, however, is never conserved in physical problems, and this is why the $(A,U)$ system does not follow that physical principle. Each class requires the specification of the axial velocity profile as a closure to the averaging process to completely determine the governing equations.  Following the terminology of Hughes, we consider problems where either a {\em flat} profile (axial velocity equal to its average) or a {\em no--slip} profile (axial velocity at the vessel wall is zero) is chosen \cite{Hughes74}.  Models and terminology will be made more precise in the next section.


Despite popularity of reduced models for blood flow in a variety of research contexts (see e.g. \cite{Politi2016, FGNQ01, MN08, SFPF03}), there is little work presenting a systematic comparison of different models using state--of--the--art numerical techniques. Further, to the best of our knowledge, many papers studying these models, for both theoretical investigation and clinical applications, use a simplifying flat velocity profile in the convective part of the equations; we call these {\em flat--profile models}.  As will be made explicit in the next section, flat--profile models are inconsistent with a first principles derivation of the reduced equations, and we are interested in understanding the limits of this model that is popular in the literature.

The literature on reduced blood flow models is vast, partially due to the fact that simulations are much cheaper than full three--dimensional models of the circulatory system \cite{XAF14} and that they perform relatively well in models of vessel networks when compared to physiological data \cite{OP00}.  Below, we give a brief review of literature on models for the axial velocity profile and usage of the $(A,Q)$ and $(A,U)$ systems. 




Several authors have investigated different models for the  axial velocity profile.  Early work of Hughes and Lubliner provided a presentation of now popular classes of reduced blood flow models \cite{HL73, Hughes74}.  In particular, Hughes derived jump conditions for various models arising from flat and no--slip profiles and exhibited numerical experiments comparing different models in a single vessel using Lax--Wendroff--type discretizations \cite{Hughes74}.  Later work by Bessems et. al., using a variant of the $(A,Q)$ system, extended the work of Hughes and Lubliner by describing a novel velocity profile with time dependent core and outer layer. \cite{HL73, BRV07}.  In certain limiting cases, their profile aligns with the model from Hughes and Lubliner.  These authors performed numerical experiments in a single vessel, comparing their model with a profile derived from Poiseuille flow \cite{BRV07}.  Lastly,  Azer and Peskin constructed a profile from Womersley flow and presented numerical results in a single vessel and vessel network \cite{AP07}.  Using the Lax--Wendroff scheme to approximate a version of the $(A,U)$ system, they compared their profile with different models for viscosity and with either ``pure--resistance'' or ``strutured--tree'' outflow conditions at the terminal vessels \cite{AP07}.

For work utilizing the $(A,U)$ class of systems, see e.g. \cite{Coccarelli2015, APRPR15, Sheng95, SFPF03, MN08, MWL09, Boileau15}. Some examples of research using the $(A,Q)$ system with a flat--profile closure include \cite{Muller2015, Zhang2015, Politi2016, Ster92, OP00, FNQ02, Delestre13, WFL14}. A portion of these papers, including work from Formaggia et al., Sherwin et al. and Delestre and Lagr{\'e}e are focused on careful descriptions of discontinuous Galerkin, Taylor Galerkin, and finite--volume schemes for these models; the flat--profile assumption is appealing in this context since these discretizations rely on the expression of the equations in conservative form \cite{FNQ02, SFPF03, Delestre13}.  Other works employing flat--profile models attempt to answer clinical questions, perform physiologically relevant experiments, and validate the models with measured data \cite{OP00, MN08, MWL09}.  Lastly, there has been recent interest in performing systematic comparisons of different numerical schemes applied to flat--profile models \cite{WFL14, Boileau15}.


We remark that systems with a flat profile closure contain mathematical simplifications that lend to their appeal.  In particular, the Riemann invariants can be analytically computed and the $(A,U)$ system can be expressed in conservative form.  Important theoretical work regarding these models include existence of smooth solutions, estimates for shock formation, and analysis of coupling with the three--dimensional Navier--Stokes equations or ordinary differential equation models of the heart and organ beds \cite{CK03, FGNQ01, FMQ05}.    

The first novel contribution of this paper is to present thorough numerical experiments for the comparison reduced blood flow models derived from flat profile {\em and} no--slip profiles, in which the shape of the axial velocity profile is allowed to vary.  In particular, we investigate the effect of profile shape on flow and pressure waveforms by considering both the $(A,Q)$ and $(A,U)$ systems; to the best of our knowledge, a comparison of these systems has not been done before.  We use a discontinuous Galerkin scheme with Runge--Kutta method in time in a large vessel network and compare simple reflection terminal boundary conditions and more physiological three element windkessel terminal boundary conditions.  In our simulations we compare two different numerical fluxes: $(1)$ an upwinding flux in the Riemann invariants and $(2)$ the classical local Lax--Friedrichs flux.  Further, we run experiments in long vessels to study the formation of shocks in both the $(A,Q)$ and $(A,U)$ systems; this phenomena is of interest in modeling physiological conditions where sharp transitions may occur, like aortic regurgitation \cite{ANL71, Rem1956, keener98}.

The second important contribution of this paper is to present theoretical convergence results, stemming from the work of Zhang and Shu, for Runge--Kutta discontinuous Galerkin (DG) discretizations of equations for inviscid blood flow \cite{ZS06}.  These results provide a first step towards analyzing fully discrete approximations schemes for viscous blood flow, vessel networks, and dimensionally heterogenous models of the hemodynamic system (see e.g. \cite{FMQ05} for PDE level analysis for 1d/0d coupled models).  

The outline of the paper is as follows.  The $(A,Q)$ and $(A,U)$ systems are
described in the next two sections. Section~\ref{sec:disc} presents the numerical scheme and states the convergence results.  The treatment of boundary conditions
for vessel networks is given in Section~\ref{sec:bc}.  Numerical simulations and conclusions are given in the last two sections.

\section{Details of the (A,Q) and (A,U) systems}
An asymptotic reduction argument from the three--dimensional incompressible axially symmetric Navier--Stokes equations results in the following set of equations (\v{C}ani\'c and Kim \cite{CK03}):
\begin{align}
&\frac{\partial A}{\partial t} + \frac{\partial Q}{\partial x} = 0 \label{eq:syst1}\\
&\frac{\partial Q}{\partial t} + \frac{\partial}{\partial x}\left(\alpha\frac{Q^2}{A}\right) + \frac{A}{\rho}\frac{\partial p}{\partial x} = 2 \pi \nu R_v \left[\frac{\partial u_x}{\partial r}\right]_{r = R_v}.\label{eq:syst2}
\end{align}
Here, vessels are assumed to be long and slender, $R_v$ is the radius and  $A=\pi R_v^2$ the cross-sectional area.
In addition, $u_x = u_x(x,r,t)$ is the velocity in the axial direction, $\nu := \mu/\rho$ is the kinematic viscosity (with $\mu$ the dynamic viscosity) and $Q:=AU$, where $U = U(x,t)$ is the axial velocity $u_x$ averaged over a cross-section, namely
\begin{equation}
U(x,t) = \frac{1}{\pi R_v^2}\int_0^{2\pi}\int_{0}^{R_v} u_x r dr d\theta = \frac{2}{R_v^2}\int_{0}^{R_v} u_x r dr.
\end{equation}
The function $\alpha = \alpha(x,t)$ is called the \textit{Coriolis coefficient} and is defined
\begin{equation}
\label{eq:alpha}
\alpha = \frac{2}{R_v^2 U^2}\int_0^{R_v} u_x^2 r dr = \frac{\text{average of $u_x^2$}}{(\text{average of $u_x$})^2}.
\end{equation}
To close the system, the axial velocity $u_x$ needs to be specified. As is typical in most of the literature, we assume the following \textit{ad hoc closure}:
\begin{equation}
\label{eq:adhoc}
u_x(x,r,t) = \frac{\gamma + 2}{\gamma}U(x,t)\left[1 - \left(\frac{r}{R_v}\right)^{\gamma}\right], \hspace{0.5cm} \gamma>0.
\end{equation}
Other choices for this closure can also be made, as described in the introduction.  It follows from this equation that $u_x$ over a cross-section is equal to $U(x,t)$ and the no-slip boundary condition is satisfied, i.e. $u_x(x,R,t) = 0$.  In this light, some authors refer to this selection as corresponding to the {\em no--slip theory} for reduced blood flow equations \cite{Hughes74}. Plugging (\ref{eq:adhoc}) in (\ref{eq:alpha}), we obtain the formula
\begin{equation}
\label{eq:gamma}
\gamma = \frac{2-\alpha}{\alpha-1} \quad \text{for} \quad \alpha>1.
\end{equation}
In particular, $\alpha$ is a constant, greater than $1$,  determined by the profile we select.  The system (\ref{eq:syst1})--(\ref{eq:syst2}) simplifies to

\begin{equation}
\label{eq:AQ}
\boxed{\text{$(A,Q)$ system}} \hspace{1cm} \left\{\begin{aligned}
  &\frac{\partial A}{\partial t} + \frac{\partial Q}{\partial x} = 0 \\
  &\frac{\partial Q}{\partial t} + \frac{\partial}{\partial x}\left(\alpha\frac{Q^2}{A}\right) + \frac{A}{\rho}\frac{\partial p}{\partial x} = -2 \pi \nu \frac{\alpha}{\alpha-1} \frac{Q}{A}.
  \end{aligned} \right.
\end{equation}

The $(A,U)$ system follows from (\ref{eq:AQ}) by substituting $Q = AU$, setting $\alpha = 1$ {\em only} in the convective part of the equations, and performing some manipulation. Note that to interchange derivatives, we must assume some smoothness on $A$ and $U$.  One obtains:

\begin{equation}
\label{eq:AU}
\boxed{\text{$(A,U)$ system}} \hspace{1cm} \left\{\begin{aligned}
&\frac{\partial A}{\partial t} + \frac{\partial (AU)}{\partial x} = 0 \\
&\frac{\partial U}{\partial t} + \frac{1}{2}\frac{\partial U^2}{\partial x}+ \frac{1}{\rho}\frac{\partial p}{\partial x} = -2 \pi \nu \frac{\alpha}{\alpha-1} \frac{U}{A}.
\end{aligned} \right.
\end{equation}
One can also make the selection of a {\em flat profile closure}: $u_x(x,r,t) = U(x,t)$, i.e. the axial component of the velocity is constant in the radial direction.  In this case, one obtains $\alpha = 1$.  By definition, this corresponds to inviscid fluid flow, so we assume the kinematic viscosity is equal to zero (see e.g \cite{CK03}).  The inviscid $(A,Q)$ system is given:

\begin{equation}
\label{eq:AQinviscid}
\boxed{\text{inviscid $(A,Q)$ system}} \hspace{1cm} \left\{\begin{aligned}
  &\frac{\partial A}{\partial t} + \frac{\partial Q}{\partial x} = 0 \\
  &\frac{\partial Q}{\partial t} + \frac{\partial}{\partial x}\left(\frac{Q^2}{A}\right) + \frac{A}{\rho}\frac{\partial p}{\partial x} = 0
  \end{aligned} \right.
\end{equation}
and the inviscid $(A,U)$ system is:
\begin{equation}
\label{eq:AUinviscid}
\boxed{\text{inviscid $(A,U)$ system}} \hspace{1cm} \left\{\begin{aligned}
&\frac{\partial A}{\partial t} + \frac{\partial (AU)}{\partial x} = 0 \\
&\frac{\partial U}{\partial t} + \frac{1}{2}\frac{\partial U^2}{\partial x}+ \frac{1}{\rho}\frac{\partial p}{\partial x} = 0.
\end{aligned} \right.
\end{equation}
The flat profile closure allows for the explicit calculation of the Riemann invariants for both the $(A,U)$ and $(A,Q)$ systems, described in the next section.  In our numerical computations, we consider the inviscid systems in addition to two different values for the parameter $\alpha$: $1.1$ and $4/3$.  The choice $\alpha = 4/3$ corresponds to a parabolic profile, while the choice $\alpha = 1.1$ yields a flatter, and some argue, more physiological profile.  These two values are used in the current study since both appear throughout the literature.

\section{Conservative and quasilinear forms}

\subsection{Derivation of different forms}

In this section we derive conservative and quasilinear equations given the following form of the state equation 
\begin{equation}
p = p_0 + \psi(A;A_0,\beta),
\end{equation}
where $A_0$ and $\beta$ are specified parameters, and $p_0$ is the pressure when $A = A_0$.   The function $\psi$ is a monotone increasing function of $A$. For the derivations below, it is assumed that $\beta$ and $A_0$ are constants, but in reality they may depend on $x$. For convenience, define the following notation for the physical variables:
\begin{equation} 
\hat{{\bf U}} := [A,Q]^T \quad \text{and} \quad \tilde{{\bf U}}:=[A,U]^T.
\end{equation}
Further, let:
\begin{equation}
\psi' := \frac{d \psi}{dA} \quad \text{and} \quad \Psi: = \int_{A_0}^A \psi(\xi;A_0,\beta) d \xi.
\end{equation}
For some function ${\bf F}:\mathbb{R}^2 \rightarrow \mathbb{R}^2$ depending on variables ${\bf U}$, we use the following notation for the Jacobian:
\begin{equation}
{\bf F}_{\bf U}:=\nabla_{\bf U}{\bf F}.
\end{equation}
Through simple differentiation, the $(A,Q)$ system in conservative form with flux function $\hat{{\bf F}}$ and source function $\hat{{\bf S}}$ is
\begin{equation}
\label{eq:AQcons}
\frac{\partial}{\partial t}\underbrace{\begin{bmatrix}
A \\
Q
\end{bmatrix}}_{:=\hat{{\bf U}}}
+
\frac{\partial}{\partial x}
\underbrace{\begin{bmatrix}
Q \\
\alpha \frac{Q^2}{A} + \frac{1}{\rho}(A\psi - \Psi)
\end{bmatrix}}_{:=\hat{{\bf F}}({\bf \hat{U}})}
=
\underbrace{\begin{bmatrix}
0 \\
-2 \pi \nu \frac{\alpha}{\alpha-1}\frac{Q}{A}
\end{bmatrix}.}_{:=\hat{{\bf S}}({\bf \hat{U}})}
\end{equation}
Further manipulation reveals the quasilinear form:
\begin{equation}
\label{eq:AQquas}
\frac{\partial}{\partial t}\begin{bmatrix}
A \\
Q
\end{bmatrix}
+
\underbrace{\begin{bmatrix}
0 & 1 \\
A\frac{\psi'}{\rho} - \alpha \frac{Q^2}{A^2} & 2\alpha \frac{Q}{A}
\end{bmatrix}}_{:=\hat{{\bf F}}_{\hat{\bf U}}}
\frac{\partial}{\partial x}\begin{bmatrix}
A \\
Q
\end{bmatrix}
=
\begin{bmatrix}
0 \\
-2 \pi \nu \frac{\alpha}{\alpha-1}\frac{Q}{A}
\end{bmatrix}.
\end{equation}
The conservative form for the $(A,U)$ system is:
\begin{equation}
\label{eq:AUcons}
\frac{\partial}{\partial t}\underbrace{\begin{bmatrix}
A \\
U
\end{bmatrix}}_{:=\tilde{{\bf U}}}
+
\frac{\partial}{\partial x}
\underbrace{\begin{bmatrix}
AU \\
\frac{U^2}{2} + \frac{\psi}{\rho}
\end{bmatrix}}_{:=\tilde{{\bf F}}({\bf \tilde{U}})}
=
\underbrace{\begin{bmatrix}
0 \\
-2 \pi \nu \frac{\alpha}{\alpha-1}\frac{U}{A}
\end{bmatrix}}_{:=\tilde{{\bf S}}({\bf \tilde{U}})},
\end{equation}
and the quasilinear form is
\begin{equation}
\label{eq:AUquas}
\frac{\partial}{\partial t}\begin{bmatrix}
A \\
U
\end{bmatrix}
+
\underbrace{\begin{bmatrix}
U & A \\
\frac{\psi'}{\rho} & U
\end{bmatrix}}_{:=\tilde{{\bf F}}_{\tilde{{\bf U}}}}
\frac{\partial}{\partial x}\begin{bmatrix}
A \\
U
\end{bmatrix}
=
\begin{bmatrix}
0 \\
-2 \pi \nu \frac{\alpha}{\alpha-1}\frac{U}{A}
\end{bmatrix}.
\end{equation}

\subsection{Riemann Invariants for inviscid models}
\label{sec:riemann}

To impose boundary conditions and mathematically model vessel junctions, we use the Riemann invariants of the inviscid $(A,Q)$ and $(A,U)$ systems. From this point forward, to calculate explicit formulas for the invariants, we fix the following form for the state equation:
\begin{equation}
\psi(A; A_0, \beta):= \beta (A^{1/2} - A_0^{1/2}).
\end{equation}
Here $\beta$ is a constant strictly greater than zero. This equation models the elastic properties of arterial walls as a linear elastic membrane \cite{MGC07}. Further, define the function:
\begin{equation}
c(A) := \left( \frac{A}{\rho} \psi' \right)^{1/2}.
\end{equation}
For the $(A,Q)$ system, one computes the eigenvalues $\hat{\lambda}_i$ ($i = 1,2$) and left eigenvectors $\hat{\bf l}_{i}$ ($i = 1,2$) of the Jacobian
\begin{equation}
{\bf \hat{F}}_{\bf \hat{U}}=
\begin{bmatrix}
0 & 1 \\
c^2 - \frac{Q^2}{A^2} & 2\frac{Q}{A}
\end{bmatrix},
\end{equation}
as
\begin{equation}  
\hat{\lambda}_{1} = \frac{Q}{A} + c \quad \text{and} \quad \hat{\lambda}_{2} = \frac{Q}{A} - c,
\end{equation}
\begin{equation}
\hat{{\bf l}}_{1}^T = 
\begin{bmatrix}
-\frac{Q}{A^2} + \frac{c}{A}, & \frac{1}{A}
\end{bmatrix}^T \quad \text{and} \quad \hat{{\bf l}}_{2}^T = 
\begin{bmatrix}
-\frac{Q}{A^2} - \frac{c}{A}, & \frac{1}{A}
\end{bmatrix}^T.
\end{equation}
By definition, the Riemann invariants $\hat{W}_{i}$ ($i = 1,2$) are functions whose gradients $\nabla_{\hat{\bf U}}\hat{W}_{i}$ are parallel to the left eigenvectors.  With the formula for $\psi$, by integration and setting constants to zero, one obtains:
\begin{equation}
\label{eq:AQinv}
\hat{W}_{1} = \frac{Q}{A} + 4c \quad \text{and} \quad \hat{W}_{2} = \frac{Q}{A} - 4c.
\end{equation}
The Riemann invariants can be shifted by a constant; in implementing reflecting boundary conditions, we employ invariants shifted by the constant $4c(A_0)$ so that they vanish when $(A,Q) = (A_0, 0)$.  Later on, for the definition of some numerical fluxes in the DG scheme, we will need the formula for the eigenvalues of the Jacobian of the $(A,Q)$ system with arbitrary $\alpha$.  These are given by
\begin{equation}
\hat{\lambda}_1(\alpha) = \alpha\frac{Q}{A} + \left(\frac{Q^2}{A^2}(\alpha^2-\alpha) + c^2\right)^{1/2} \quad \text{and} \quad \hat{\lambda}_2(\alpha) = \alpha\frac{Q}{A} - \left(\frac{Q^2}{A^2}(\alpha^2-\alpha) + c^2\right)^{1/2}.
\end{equation} 
Similar calculations reveal the eigenvalues and Riemann invariants for the inviscid $(A,U)$ system:
\begin{equation}
\tilde{\lambda}_{1} = U + c \quad \text{and} \quad \tilde{\lambda}_{2} = U - c,
\end{equation}
\begin{equation}
\label{eq:AUinv}
\tilde{W}_{1} = U \pm 4c \quad \text{and} \quad \tilde{W}_{2} = U - 4c.
\end{equation}

\section{Numerical discretization}
\label{sec:disc}

\subsection{DG formulation and time discretization}

Let us describe the discontinuous Galerkin formulation we implement for a general hyperbolic system in conservative form, namely:
\begin{align}
&\frac{\partial {\bf U}}{\partial t} + \frac{\partial}{\partial x}{\bf F}({\bf U}) = {\bf S}({\bf U}), \quad \text{in } [0,L] \times (0,T],\\
&{\bf U}(\cdot,0) = {\bf U}_0(\cdot) \quad \text{in } [0,L]. 
\end{align}
Let the collection of intervals $\{I_e\}_{e=0}^{N}$ be a uniform partition of the interval $[0,L]$, with $I_e = [x_e,x_{e+1}]$ of  size $h$. Let $\mathbb{P}^k(I_e)$ be the space of polynomials of degree $k$ on the interval $I_e$.  The approximation space is
\begin{equation}
\mathbb{V}_h^k:=\{ \phi:[0,L] \rightarrow \mathbb{R} \text{ such that } \phi|_{I_e} \in \mathbb{P}^k(I_e) \text{ for all $e = 0, \ldots, N$} \}.
\end{equation}
Define the notation for traces of a function $\phi:[0,L]\rightarrow \mathbb{R}$ to the interior boundaries of the intervals:
\begin{align}
\phi^\pm|_{x_e} &:= \lim_{\varepsilon \rightarrow 0 \text{ and } \varepsilon > 0} \phi(x_e \pm \varepsilon), \quad e = 1, \ldots, N, \\
\phi^+|_{x_0} &:= \lim_{\varepsilon \rightarrow 0 \text{ and } \varepsilon > 0} \phi(x_0 + \varepsilon), \\
\phi^-|_{x_{N+1}} &:= \lim_{\varepsilon \rightarrow 0 \text{ and } \varepsilon > 0} \phi(x_{N+1} -\varepsilon).
\end{align}

Lastly, define
\begin{align}
\mathcal{H}_e({\bf U}_h, {\bf \Phi}_h) &:= \int_{I_e}{\bf F}({\bf U}_h)\cdot \frac{d {\bf \Phi}_h}{dx} 
+ \int_{I_e}{\bf S}({\bf U}_h)\cdot {\bf \Phi}_h \\
&-{\bf F}^{nf}({\bf U}_h)|_{x_{e+1}} \cdot {\bf \Phi}^-|_{x_{e+1}} 
+ {\bf F}^{nf}({\bf U}_h)|_{x_e} \cdot {\bf \Phi}^+|_{x_{e}} \quad e = 0, \ldots, N.
\end{align}
The semi--discrete formulation is given as follows:
seek ${\bf U}_h \in \mathbb{V}_h^k \times \mathbb{V}_h^k$ satisfying
\begin{equation}
 \int_{I_e}\frac{\partial {\bf U}_h}{\partial t} \cdot {\bf \Phi}_h  = \mathcal{H}_e({\bf U}_h, {\bf \Phi}_h)
\end{equation}
for all ${\bf \Phi}_h \in \mathbb{V}_h^k \times \mathbb{V}_h^k$ and $e = 0, \ldots, N$.
\vspace{0.2cm}

The function ${\bf F}^{nf}({\bf U}_h)$ yet to be defined is the \textit{numerical flux}.  The following definitions are used in all cases except at a junction of vessels in a network (below we provide the definition of the numerical flux in this case).  For boundary conditions, we denote:
\begin{align}
{\bf U}_{\text{inlet}}&:=\text{ boundary data at inlet ($x = 0$)}, \\
{\bf U}_{\text{outlet}}&:=\text{ boundary data at outlet ($x = L$).}
\end{align}
The average of the flux function at the interface between elements is:
\begin{align}
\{{\bf F}({\bf U}_h)\}|_{x_e}&:=\frac{1}{2}\left({\bf F}\left({\bf U}_h^+|_{x_e}\right) + {\bf F}\left({\bf U}_h^-|_{x_e}\right)\right), \quad e = 1, \ldots, N,\\
\{{\bf F}({\bf U}_h)\}|_{x_0}&:=\frac{1}{2}\left({\bf F}\left({\bf U}_h^+|_{x_0}\right) + {\bf F}\left({\bf U}_{\text{inlet}}\right)\right), \\
\{{\bf F}({\bf U}_h)\}|_{x_{N+1}}&:=\frac{1}{2}\left({\bf F}\left({\bf U}_{\text{outlet}}\right) + {\bf F}\left({\bf U}_h^-|_{x_{N+1}}\right)\right)
\end{align}
The jump of the discrete function ${\bf U}_h$ at an interface is:
\begin{align}
\llbracket {\bf U}_h \rrbracket|_{x_e}&:= {\bf U}_h^+|_{x_e} - {\bf U}_h^-|_{x_e}, \quad e = 1, \ldots, N, \\
\llbracket {\bf U}_h \rrbracket|_{x_0}&:= {\bf U}_h^+|_{x_0} - {\bf U}_{\text{inlet}}, \\
\llbracket {\bf U}_h \rrbracket|_{x_{N+1}}&:= {\bf U}_{\text{outlet}} - {\bf U}_h^-|_{x_{N+1}}.
\end{align}
Lastly, let the symbols $\lambda_i$ ($i = 1,2$) be the eigenvalues of the Jacobian of ${\bf F}$ and denote:
\begin{align}
\lambda_i^+|_{x_e} &:= \lambda_i \left( {\bf U}_h^+|_{x_e}\right), \quad e = 0, \ldots, N, \\ 
\lambda_i^-|_{x_e} &:= \lambda_i \left( {\bf U}_h^+|_{x_e}\right), \quad e = 1, \ldots, N+1, \\ 
\lambda_i^+|_{x_{N+1}} &:= \lambda_i \left( {\bf U}_{\text{outlet}} \right), \\ 
\lambda_i^-|_{x_0} &:= \lambda_i \left( {\bf U}_{\text{inlet}}\right). 
\end{align}
We will compare two different fluxes.  The first one is the {\bf local Lax--Friedrichs flux} (LLF) given by:
\begin{equation}
{\bf F}^{nf}({\bf U}_h)|_{x_e}=\{{\bf F}({\bf U}_h)\}|_{x_e} - \frac{1}{2}\max \left(\left|\lambda_1^\pm|_{x_e}\right|, \left|\lambda_2^\pm|_{x_e}\right|\right)\llbracket{\bf U}_h\rrbracket|_{x_e}, \quad e = 0, \ldots N 
\end{equation}
For a description of this numerical flux and others used for hyperbolic systems, see the work of Cockburn et al. \cite{CS_3}.

The second flux is the {\bf upwinding flux} (UP), determined by ``upwinding'' in the Riemann invariants given by the inviscid systems, i.e. equations (\ref{eq:AQinv}) and (\ref{eq:AUinv}).  This flux was defined and used in the work of Sherwin et al. \cite{SFPF03}.  In order to define this flux, we must make some general assumptions that are satisfied by the reduced blood flow models considered in this paper.  First, assume that we can view the Riemann invariants as functions of physical variables $W_i = W_i({\bf U})$ ($i = 1,2$) and vice versa ${\bf U} = {\bf U}(W_1, W_2)$.  Further, assume $\lambda_1>0$ and $\lambda_2<0$. 

This numerical flux is computed in several steps.  First, compute the invariants
\begin{align}
W_{1}^{up}|_{x_e} &:= W_1({\bf U}_h^-|_{x_e}), \quad e = 1, \ldots, N+1, \\
W_{2}^{up}|_{x_e} &:= W_2({\bf U}_h^+|_{x_e}), \quad e = 0, \ldots, N, \\
W_{1}^{up}|_{x_0} &:= W_1({\bf U}_{\text{inlet}}),\\
W_{2}^{up}|_{x_{N+1}} &:= W_2({\bf U}_{\text{outlet}}),
\end{align}
and then the physical variables
\begin{align}
{\bf U}^{up}|_{x_e} := {\bf U}(W_1^{up}|_{x_e},W_2^{up}|_{x_e}), \quad e=0, \ldots, N+1
\end{align}
The flux is defined by evaluating the flux function at these upwinded values of the physical variables:
\begin{equation}
{\bf F}^{nf}({\bf U}_h)|_{x_e}:={\bf F}\left( {\bf U}^{up}|_{x_e}\right), \quad e = 0, \ldots, N+1.
\end{equation}

For the time discretization, we employ a second order Runge--Kutta scheme \cite{ZS06}.  Let the time step be denoted by $\Delta t$ and the final time be $T$.  Define $M = T/\Delta t$ and assume for simplicitly it is an integer.  The fully discrete form is as follows:  

\vspace{0.2cm}
\noindent
For $n = 1, \ldots, M-1$, given ${\bf U}_h^{n-1}$, compute ${\bf V}_h^n$ and ${\bf U}_h^{n+1}$ in $\mathbb{V}_h^k \times \mathbb{V}_h^k$ satisfying
\begin{align}
\label{eq:RKDG1}
\int_{I_e}{\bf V}^n_h \cdot {\bf \Phi_h}  &= \int_{I_e}{\bf U}_h^n \cdot {\bf \Phi}_h  + \Delta t \mathcal{H}_e({\bf U}_h^n, {\bf \Phi}_h) \\
\label{eq:RKDG2}
\int_{I_e}{\bf U}^{n+1}_h \cdot {\bf \Phi}_h  &= \frac{1}{2}\int_{I_e}{\bf V}_h^n \cdot {\bf \Phi}_h  + \frac{1}{2}\int_{I_e}{\bf U}_h^n \cdot {\bf \Phi}_h  + \frac{\Delta t}{2} \mathcal{H}_e({\bf V}_h^n, {\bf \Phi}_h).
\end{align}
for all ${\bf \Phi}_h \in \mathbb{V}_h^k \times \mathbb{V}_h^k$ and $e = 0, \ldots, N$.
\vspace{0.2cm}

\subsection{Symmetrizability and analysis}

The notion of {\em symmetrizability} for systems of conservation laws often simplifies numerical analysis for discretizations of the equations.  Further, it is a reasonable assumption for systems modeling physical phenomena, and has been shown to be equivalent to the existence of a so--called {\em convex entropy function} \cite{Har83}.  More concretely, to demonstrate symmetrizability, one seeks a transformation from the physical variables ${\bf U}$ to {\em entropy variables} ${\bf V}$ so the Jacobian ${\bf U}_{\bf V}$ is symmetric positive definite (SPD) and ${\bf F}_{\bf V} = {\bf F}_{\bf U}{\bf U}_{\bf V}$ is symmetric.  Then, with zero source term, the conservation law takes the form:
\begin{equation}
{\bf U}_{\bf V}\frac{\partial {\bf U}}{\partial t} + {\bf F}_{\bf U}{\bf U}_{\bf V}\frac{\partial {\bf V}}{\partial x} = 0.
\end{equation}
One can show under reasonable assumptions that an entropy function derived from the energy of the system symmetrizes the equations for reduced blood flow.  With symmetrizability, we can leverage numerical analysis performed by Zhang and Shu to arrive at convergence estimates for discontinuous Galerkin discretizations of the equations \cite{ZS06}.

The entropy for the inviscid models was derived by Formaggia et al. \cite{FGNQ01}:
\begin{align}
\boxed{(A,Q)} \quad \hat{\mathcal{E}}(x,t) &= \frac{1}{2}\rho \frac{Q^2}{A} + \int_{A_0}^A \psi(\xi;A_0,\beta)d\xi, \\ 
\boxed{(A,U)} \quad \tilde{\mathcal{E}}(x,t) &= \frac{1}{2}\rho A U^2 + \int_{A_0}^A \psi(\xi;A_0,\beta)d\xi.
\end{align}

\begin{mylem}
If $A(x,t) >\delta > 0$ for $(x,t) \in [0,L]\times[0,T]$, the inviscid $(A,Q)$ system is symmetrizable.
\begin{proof}
With the physical variables ${\bf U} = [A,Q]^T$, following Harten's work \cite{Har83}, 
the entropy variables ${\bf V} = [V_1, V_2]^T$ are:
\begin{align}
V_1({\bf U}) &= \frac{\partial \hat{\mathcal{E}}}{\partial A} = -\frac{\rho}{2}\frac{Q^2}{A^2} + \psi(A), \\
V_2({\bf U}) &= \frac{\partial \hat{\mathcal{E}}}{\partial Q} = \rho \frac{Q}{A}.
\end{align}
The definition of the entropy variables defines a transformation ${\bf U} \rightarrow {\bf V}({\bf U})$.  The inverse of this transformation, denoted ${\bf U}({\bf V})$, will symmetrize the equations.  Recall the notation and assumption on $\psi$:
\begin{equation}
\psi' := \frac{d\psi}{dA} > 0 \text{ for $A>0$}.
\end{equation}
So the function $\psi'$ is positive since $A>\delta>0$.  Here is the Jacobian of ${\bf V}({\bf U})$:
\begin{equation}
{\bf U}_{\bf V}^{-1} = {\bf V}_{\bf U} = 
\begin{bmatrix}
\rho \frac{Q^2}{A^3} + \psi' & -\rho \frac{Q}{A^2} \\
-\rho \frac{Q}{A^2} & \frac{\rho}{A}
\end{bmatrix}.
\end{equation}
Its determinant, eigenvalues, denoted as the set $\sigma({\bf V}_{\bf U})$, and trace  are:
\begin{equation}
\det {\bf V}_{\bf U} = \frac{\rho}{A} \psi',
\end{equation}
\begin{equation}
\sigma({\bf V}_{\bf U}) = \left\{\frac{1}{2}\left( \frac{\rho}{A} + \rho\frac{Q^2}{A^3} + \psi' \pm \sqrt{ \left(\frac{\rho}{A} - \psi' \right)^2 + 2\rho^2 \frac{Q^2}{A^2} + 2\rho \psi' \frac{Q^2}{A^3} + \rho^2 \frac{Q^4}{A^6} }\right) \right\},
\end{equation}
\begin{equation}
\text{tr } {\bf V}_{\bf U} = \rho \frac{Q^2}{A^3} + \psi' + \frac{\rho}{A}.
\end{equation}
It is clear that the determinant and trace are both positive, proving that ${\bf V}_{\bf U}$ is SPD.  Thus, the inverse ${\bf U}_{\bf V}$ is also SPD.  Lastly, using the form of ${\bf F}_{\bf U}$ given in (\ref{eq:AQquas}) with $\alpha = 1$, one sees that
\begin{align}
{\bf F}_{\bf U} {\bf U}_{\bf V} &= 
\frac{A}{\rho \psi'}\begin{bmatrix}
0 & 1 \\
\frac{A}{\rho}\psi' - \frac{Q^2}{A^2} & 2\frac{Q}{A} 
\end{bmatrix}
\begin{bmatrix}
\frac{\rho}{A} & \rho \frac{Q}{A^2} \\
\rho \frac{Q}{A^2} & \rho \frac{Q^2}{A^3} + \psi'
\end{bmatrix} \\
&=\frac{A}{\rho \psi'}
\begin{bmatrix}
\rho \frac{Q}{A^2} & \rho \frac{Q^2}{A^3} + \psi' \\
\rho \frac{Q^2}{A^3} + \psi' & 3 \frac{Q}{A}\psi' + \rho \frac{Q^3}{A^4} 
\end{bmatrix},
\end{align} 
verifying the matrix is symmetric.
\end{proof}
\end{mylem}

\begin{mylem}
If $A(x,t)>\delta>0$ and $|U(x,t)| < c(A(x,t))$ for $(x,t) \in [0,L]\times[0,T]$, the inviscid $(A,U)$ system is symmetrizable.  
\begin{proof}
The proof follows closely the proof of the previous lemma. Let 
${\bf U} = [A,U]^T$.
From the entropy function given above, define the transformation as
\begin{align}
V_1({\bf U}) &= \frac{\partial \tilde{\mathcal{E}}}{\partial A} = \frac{\rho}{2}U^2 + \psi(A),\\
V_2({\bf U}) &= \frac{\partial \tilde{\mathcal{E}}}{\partial U} = \rho AU.
\end{align}
The Jacobian is 
\begin{equation}
{\bf V}_{\bf U} = 
\begin{bmatrix}
\psi' & \rho U \\
\rho U & \rho A
\end{bmatrix}.
\end{equation}
Its determinant and trace are
\begin{equation}
\det {\bf V}_{\bf U} = \rho^2 \left(\frac{A}{\rho} \psi' - U^2 \right) = \rho^2(c^2 - U^2), 
\end{equation}
\begin{equation}
\text{tr } {\bf V}_{\bf U} = \psi' + \rho A,
\end{equation}
where $c = c(A)$ is the speed appearing in the formula for the convective velocity of the Riemann invariants.  The additional assumption that $|U| < c$ implies the determinant is positive.  Since the trace is positive, ${\bf V}_{\bf U}$ is SPD, so its inverse ${\bf U}_{\bf V}$ is SPD.  Lastly, symmetry is verified using ${\bf F}_{\bf U}$ from (\ref{eq:AUquas}):
\begin{align}
{\bf F}_{\bf U} {\bf U}_{\bf V} &=
\frac{1}{\rho^2 (c^2 - U^2)}
\begin{bmatrix}
U & A \\
\frac{\psi'}{\rho} & U
\end{bmatrix}
\begin{bmatrix}
\rho A & -\rho U \\
-\rho U & \psi'
\end{bmatrix} \\
&=\frac{1}{\rho^2 (c^2 - U^2)}
\begin{bmatrix}
0 & \rho U^2 + A \psi' \\
-\rho U^2 + A \psi' & 0
\end{bmatrix}
\end{align}
\end{proof}
\end{mylem} 
Now we may combine symmetrizability with several other assumptions to conclude convergence of a discontinuous Galerkin scheme with a second order Runge-Kutta time discretization \cite{ZS06}.  In particular, the numerical flux must be in the class of {\em generalized E--fluxes} (the local--Lax Friedrich's flux falls in this class).  The result is summarized for the inviscid $(A,Q)$ and $(A,U)$ models in the following propositions.  The statements closely follow \cite{ZS06}.

\begin{myprop}[Convergence for the inviscid $(A,Q)$ system.]
Assume the area satisfies
\begin{equation}
A(x,t) > \delta > 0 \text{ for $(x,t) \in [0,L]\times[0,T]$},
\end{equation}
the exact solution ${\bf U} = [A,Q]^T$ is smooth enough with bounded derivatives, and the function ${\bf F}({\bf U})$ is $C^3(\mathbb{R}^2)$ with bounded derivatives.  Further, assume the solution is compactly supported or the boundary conditions are periodic and the numerical flux is locally Lipschitz continuous.  Let $B$ be a constant that does not depend on $h$ and $\Delta t$, and suppose a CFL condition of the form $\Delta t \leq B h^{4/3}$ is satisfied. Then for some constant $C>0$ that does not depend on $\left({\bf U}_h^n\right)_{n=0}^M$, $\Delta t$, and $h$, the numerical solution $\left({\bf U}_h^n\right)_{n=0}^M$ from the scheme defined by (\ref{eq:RKDG1})-(\ref{eq:RKDG2}), with the local Lax--Friedrichs numerical flux, satisfies the following bound:
\begin{equation}
\label{eq:est1}
\max_{n = 0, \ldots, M} \|{\bf U}(n\Delta t) - {\bf U}_h^n\|_{L^2(0,L)} \leq C(h^{k + 1/2} + \Delta t^2).
\end{equation} 
\end{myprop}

The same result holds true for the inviscid $(A,U)$ system with an additional assumption on the exact solution:

\begin{myprop}[Convergence for the inviscid $(A,U)$ system.]
Assume the exact solution ${\bf U} = [A,U]^T$ satisfies
\begin{equation}
A(x,t) > \delta > 0 \text{ and } |U(x,t)| < c(A(x,t)) \text{ for $(x,t) \in [0,L] \times [0,T]$},
\end{equation}
is smooth enough with bounded derivatives, and the function ${\bf F}({\bf U})$ is $C^3(\mathbb{R}^2)$ with bounded derivatives.  Further, assume the solution is compactly supported or the boundary conditions are periodic and the numerical flux is locally Lipschitz continuous. Let $B$ be a constant that does not depend on $h$ and $\Delta t$, and suppose a CFL condition of the form $\Delta t \leq B h^{4/3}$ is satisfied. Then for some constant $C>0$ that does not depend on $\left({\bf U}_h^n\right)_{n=0}^M$, $\Delta t$, and $h$, the numerical solution $\left({\bf U}_h^n\right)_{n=0}^M$ from the scheme defined by (\ref{eq:RKDG1})--(\ref{eq:RKDG2}), with the local Lax--Friedrichs numerical flux, satisfies the following bound:
\begin{equation}
\label{eq:est2}
\max_{n = 0, \ldots, M} \|{\bf U}(n\Delta t) - {\bf U}_h^n\|_{L^2(0,L)} \leq C(h^{k + 1/2} + \Delta t^2).
\end{equation} 
\end{myprop}

The symmetrizability assumption is employed in the analysis via the introduction of a norm depending on ${\bf U}_{\bf V}$; smoothness of the exact solutions and the SPD assumption allow one to conclude that this norm is equivalent to the $L^2$ norm used in the estimate.  Also, notice this theoretical perspective differentiates the $(A,Q)$ and $(A,U)$ models, i.e. verification of symmetrizability for the $(A,U)$ model requires the additional assumption $|U| < c(A)$ rendering the $(A,U)$ system strictly hyperbolic.

\section{Boundary Conditions}
\label{sec:bc}

We employ standard approaches for boundary conditions in the form of Dirichlet data, at vessel junctions in a network, and at the terminal vessels of a network.  These conditions are summarized below.

\subsection{Dirichlet data}

We describe the process for imposing Dirichlet boundary data ${\bf U}_{\text{inlet}}$ at the inlet $x_0$ for the $(A,Q)$ class of systems.  An analogous approach may be used for outlet data and for the $(A,U)$ class of systems.

This process relies on the Riemann invariants $W_i$ ($i = 1,2$) derived in Section~\ref{sec:riemann}.  At time step $n$, suppose we presribe the area $A_{\text{inlet}}^n$.  The corresponding value for the fluid momentum at the inlet $Q_{\text{inlet}}^n$ is determined by first extrapolating the right--to--left moving Riemann invariant to the boundary using the solution at the previous time step $W_2^{n-1}$:
\begin{equation}
\label{eq:extrap}
W_{2,{\text{approx}}}^n := W_2^{n-1}(x_0 - \Delta t \lambda_{2,\text{approx}}^{n-1}),
\end{equation} 
with
\begin{equation}
\lambda_{2,\text{approx}}^{n-1} := \lambda_2^{n-1}(A^{+,n-1}|_{x_0}, Q^{+,n-1}|_{x_0}).
\end{equation}
We use the value of the approximated Riemann invariant at the boundary and rearrange the formula to solve for the fluid momentum.
\begin{equation}
Q^n_{\text{inlet}} := A^n_{\text{inlet}} \left( W_{2,{\text{approx}}}^n + 4c(A^n_{\text{inlet}})\right).
\end{equation}
This process determines the Dirichlet boundary conditions at the inlet, ${\bf U}_\text{inlet}$, that are then built into the numerical flux function as described above.  A similar approach may be used for prescribing the fluid momentum $Q_\text{inlet}^n$ or the left--to--right moving invariant $W_{1,\text{inlet}}^n$ at the inlet of the vessel.

\subsection{Vessel junctions}

Boundary conditions at vessel junctions are determined by holding constant the values of the Riemann invariants and enforcing continuity of total pressure and conservation of fluid momentum.  More precisely, suppose at a junction there are $N_{\text{in}}$ incoming vessels and $N_{\text{out}}$ outgoing vessels.  We need to determine the values of the physical variables at the incoming vessels $A_{\text{in}}^{(k)}, Q_{\text{in}}^{(k)}$ ($k = 1, \ldots, N_{\text{in}}$) and at the outgoing vessels $A_{\text{out}}^{(j)}, Q_{\text{out}}^{(j)}$ ($j = 1, \ldots, N_{\text{out}}$).  For simplicitly of presentation, we diverge from our previous notation and let $W_{1,\text{in}}^{(k)}$ and $W_{2,\text{out}}^{(j)}$ denote the traces of the Riemann invariants at the incoming and outgoing vessels respectively.  The requirements at the vessel junction may be specified mathematically in the following nonlinear system of algebraic equations:
\begin{align}
W_{1,\text{in}}^{(k)} &= \frac{Q_{\text{in}}^{(k)}}{A_{\text{in}}^{(k)}} + 4 c\left(A_{\text{in}}^{(k)}\right) \quad \text{ for $k = 1, \ldots, N_{\text{in}}$}, \\
W_{2,\text{out}}^{(j)} &= \frac{Q_{\text{out}}^{(j)}}{A_{\text{out}}^{(j)}} - 4 c\left(A_{\text{out}}^{(j)}\right) \quad \text{ for $j = 1, \ldots, N_{\text{out}}$}, \\
\sum_{k = 1}^{N_{\text{in}}} Q_{\text{in}}^{(k)} &= \sum_{j = 1}^{N_{\text{out}}} Q_{\text{out}}^{(j)}, \\
\frac{\rho}{2}\left(\frac{Q_{\text{in}}^{(1)}}{A_{\text{in}}^{(1)}}\right)^2 + p\left(A_{\text{in}}^{(1)}\right) &= \frac{\rho}{2}\left(\frac{Q_{\text{in}}^{(k)}}{A_{\text{in}}^{(k)}}\right)^2 + p\left(A_{\text{in}}^{(k)}\right) \quad \text{ for $k = 2, \ldots, N_{\text{in}}$}, \\
\frac{\rho}{2}\left(\frac{Q_{\text{in}}^{(1)}}{A_{\text{in}}^{(1)}}\right)^2 + p\left(A_{\text{in}}^{(1)}\right) &= \frac{\rho}{2}\left(\frac{Q_{\text{out}}^{(j)}}{A_{\text{out}}^{(j)}}\right)^2 + p\left(A_{\text{out}}^{(j)}\right) \quad \text{ for $j = 1, \ldots, N_{\text{out}}$}.
\end{align}

These equations are solved with Newton's method.  The definition of the numerical flux at a vessel junction is different from above and is redefined as follows: at the outlet of the incoming vessels it is defined as
\begin{equation}
{\bf F}^{nf}({\bf U})|_{x_{N+1}}:={\bf F}(A_{\text{in}}^{(k)}, Q_{\text{in}}^{(k)}).
\end{equation}
Similarly, at the inlet of the outgoing vessels we have:
\begin{equation}
{\bf F}^{nf}({\bf U})|_{x_0}:={\bf F}(A_{\text{out}}^{(j)}, Q_{\text{out}}^{(j)}).
\end{equation}

\subsection{Reflection boundary conditions for terminal vessels}

For code validation with the fifty--vessel network given in \cite{SFPF03}, we employ reflection boundary conditions used by these authors at terminal vessels in the network.  More specifically, at the outlets of the  terminal vessels in a given network, we expect reflections due to the resistive nature of organ beds.  Described below is a simple approach for resistance boundary conditions using the Riemann invariants of the inviscid systems \cite{SFPF03}.  The Riemann invariants for the inviscid $(A,Q)$ system satisfy the following system:
\begin{align}
\frac{\partial \hat{W}_1}{\partial t} &+ \hat{\lambda}_1 \frac{\partial \hat{W}_1}{\partial x} = 0, \\
\frac{\partial \hat{W}_2}{\partial t} &+ \hat{\lambda}_2 \frac{\partial \hat{W}_2}{\partial x} = 0.
\end{align}
Assuming that $\lambda_1$ remains positive and $\lambda_2$ remains negative, this system is well-defined when $\hat{W}_1$ is specified at the inlet of the vessel ($x = 0$) and $\hat{W}_2$ is specified at the outlet ($x = L$) of the vessel.  Prescribing $\hat{W}_2(x = L,t) = 0$ on the outlet yields no effect on the characteristic variables in the interior of the domain.  Alternatively, one may specify an outlet boundary condition depending on the incoming characteristic, i.e. for some $0\leq R \leq 1$, let 
\begin{align}
\label{eq:res1}
\hat{W}_2(L, t) &= -R \hat{W}_1(L,t), \\
\label{eq:res2}
A^+|_L &= A^-|_L.
\end{align} 
The above equations allow us to specify the values of the physical variables to the right of the outlet, $A^+|_L$, $Q^+|_L$, given the known values to the left of the outlet, $A^-|_L$, $Q^-|_L$.  Consider the following definition of the Riemann invariants now shifted by the constant $c_0 = c(A_0)$.
\begin{equation}
\hat{W}_{1} = \frac{Q}{A} + 4(c-c_0) \quad \text{and} \quad \hat{W}_{2} = \frac{Q}{A} - 4(c - c_0).
\end{equation}
With this definition, (\ref{eq:res1}) and (\ref{eq:res2}) become the following:
\begin{align}
A^+|_L &= A^-|_L, \\ 
Q^+|_L &= (1-R)A^-|_L \left[ \frac{Q^-|_L}{A^-|_L} + 4\left(c(A^-|_L) - c_0|_L\right)\right] - Q^-|_L. 
\end{align}
The numerical flux at $x = L$ is then evaluated at these values.  Similarly for the $(A,U)$ system, one has
\begin{align}
A^+|_L &= A^-|_L, \\ 
U^+|_L &= (1-R)\left[ U^-|_L + 4\left(c(A^-|_L) - c_0|_L\right)\right] - U^-|_L. 
\end{align}

\subsection{Three element windkessel boundary conditions for terminal vessels}

\begin{figure}[h]
\begin{center}
\includegraphics[scale=0.5]{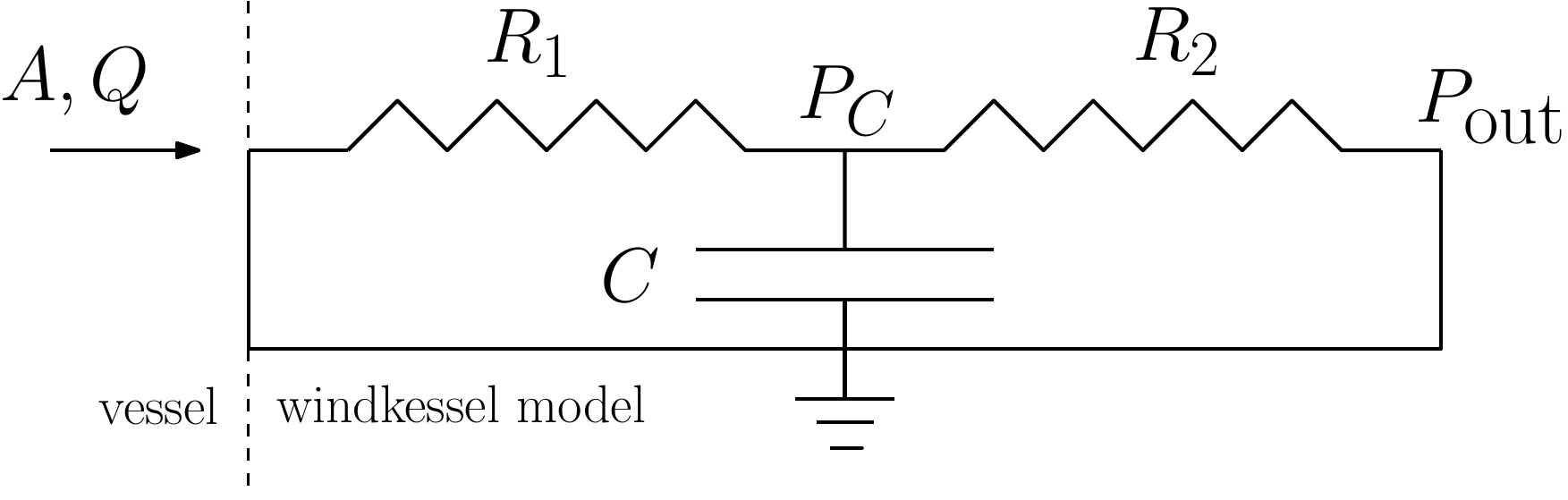}
\caption{A schematic of the three element windkessel model used for boundary conditions at the outlets of the terminal vessels in the network.}
\label{fig:schem_wind}
\end{center}
\end{figure}

For a physiologically meaningful comparison of the $(A,Q)$ and $(A,U)$ systems, we employ a three element windkessel boundary condition at the end of each terminal vessel.  This terminal model, mathematically described by an RC circuit, accounts for both organ bed resistance and compliance. A schematic, adapted from \cite{Boileau15}, is given in Figure \ref{fig:schem_wind}.
Ohm's and Kirhchoff's laws for this model are given respectively:
\begin{align}
\label{eq:wind1}
Q &= \frac{p(A) - P_C}{R_1} \\
\label{eq:wind2}
C \frac{dP_C}{dt} &= Q - \frac{P_C - P_{\text{out}}}{R_2}.
\end{align}
At the $n$th timestep, given $P_C^n$ and $A^n, Q^n$ evaluated at the outlet of the terminal vessel, $P_C^{n+1}, A^{n+1}, Q^{n+1}$ are computed as the solution to the following system:
\begin{align}
Q^{n+1} &= \frac{p(A^{n+1}) - P_C^{n+1}}{R_1} \\
P_C^{n+1} &= P_C^n + \frac{\Delta t}{C} \left( Q^{n+1} - \frac{P_C^{n+1} - P_{\text{out}}}{R_2}\right) \\
W_1^n &= \frac{Q^{n+1}}{A^{n+1}} + 4 c(A^{n+1}).
\end{align}
This approximation is the same as the process described in \cite{APS12} except we use backward Euler to discretize the differential equation for $P_C$. The numerical flux at $x=L$ is evaluated at $A^{n+1}, Q^{n+1}$.

\section{Results}

\subsection{Convergence rates for numerical scheme}

In this section we use the method of manufactured solution to obtain numerical convergence rates for the spatial discretization of the scheme. The results presented here are only for the $(A,Q)$ system, but we observe similar results for the $(A,U)$ system.  The domain is the unit interval and the exact solution is chosen as:
\begin{align}
\label{eq:exactsol1}
A(x,t) &= \cos(2 \pi x) \cos(t) + 2,\quad \forall x, t, \\
\label{eq:exactsol2}
Q(x,t) &= \sin(2 \pi x) \cos(t),\quad \forall x, t.
\end{align}
The errors in the $L^2$ norm between the approximate and exact solution are computed on a sequence of uniformly refined meshes (from $h=1/2$ to $h=1/32$).  Numerical convergence rates are derived
from the numerical errors. We also study the effect of the approximation order by varying the polynomial degree ($k=1, 2, 3$).
For these computations, $\Delta t = 2 \times 10^{-5}$, small enough for the temporal error to be negligible, and the scheme is evolved for ten timesteps.  

\begin{table}[!htb]
\begin{center}
\begin{tabular}{ c | c c | c c | c c }
 & $k=1$ & & $k=2$ & & $k = 3$ &   \\ \hline
$h$ & $L^2$ error & rate & $L^2$ error & rate & $L^2$ error & rate  \\ \hline
5.0000e-01 & 8.5047e-02 & -- & 8.5047e-02 & -- & 2.7752e-03 & -- \\
2.5000e-01 & 6.2774e-02 & 4.3810e-01 & 8.3827e-03 & 3.3428e+00 & 8.3376e-04 & 1.7349e+00 \\
1.2500e-01 & 1.6123e-02 & 1.9610e+00 & 1.0714e-03 & 2.9680e+00 & 5.3267e-05 & 3.9683e+00 \\
6.2500e-02 & 4.0652e-03 & 1.9878e+00 & 1.3478e-04 & 2.9907e+00 & 3.3771e-06 & 3.9794e+00 \\
3.1250e-02 & 1.0246e-03 & 1.9883e+00 & 1.6936e-05 & 2.9925e+00 & 2.1726e-07 & 3.9583e+00 \\
\end{tabular}
\caption{Errors and rates for $A$ from the inviscid $(A,Q)$ system with the upwinding flux.}
\label{table:rates1}
\begin{tabular}{c | c c | c c | c c}
 & $k=1$ & & $k=2$ & & $k = 3$ &  \\ \hline
$h$ & $L^2$ error & rate & $L^2$ error & rate & $L^2$ error & rate  \\ \hline
5.0000e-01 & 3.0776e-01 & -- & 1.7275e-02 & -- & 1.7264e-02 & -- \\
2.5000e-01 & 6.2775e-02 & 2.2936e+00 & 8.3839e-03 & 1.0430e+00 & 8.3365e-04 & 4.3722e+00 \\
1.2500e-01 & 1.6123e-02 & 1.9611e+00 & 1.0719e-03 & 2.9675e+00 & 5.3233e-05 & 3.9690e+00 \\
6.2500e-02 & 4.0651e-03 & 1.9877e+00 & 1.3501e-04 & 2.9890e+00 & 3.3745e-06 & 3.9796e+00 \\
3.1250e-02 & 1.0254e-03 & 1.9871e+00 & 1.7032e-05 & 2.9867e+00 & 2.1756e-07 & 3.9552e+00 \\
\end{tabular}
\caption{Errors and rates for $Q$ from the inviscid $(A,Q)$ system with the upwinding flux.}
\label{table:rates2}
\end{center}
\end{table}

\begin{table}[!htb]
\begin{center}
\begin{tabular}{c | c c | c c | c c}
 & $k=1$ & & $k=2$ & & $k = 3$ &   \\ \hline
$h$ & $L^2$ error & rate & $L^2$ error & rate & $L^2$ error & rate  \\ \hline
5.0000e-01 & 8.5047e-02 & -- & 8.5047e-02 & -- & 2.7751e-03 & -- \\
2.5000e-01 & 6.2783e-02 & 4.3789e-01 & 8.3826e-03 & 3.3428e+00 & 8.3452e-04 & 1.7335e+00 \\
1.2500e-01 & 1.6127e-02 & 1.9609e+00 & 1.0719e-03 & 2.9673e+00 & 5.3345e-05 & 3.9675e+00 \\
6.2500e-02 & 4.0669e-03 & 1.9875e+00 & 1.3513e-04 & 2.9877e+00 & 3.3843e-06 & 3.9784e+00 \\
3.1250e-02 & 1.0252e-03 & 1.9880e+00 & 1.7116e-05 & 2.9809e+00 & 2.1754e-07 & 3.9595e+00 \\
\end{tabular}
\caption{Errors and rates for $A$ from the inviscid $(A,Q)$ system with the local Lax--Friedrichs flux.}
\label{table:rates3}
\begin{tabular}{c | c c | c c | c c}
 & $k=1$ & & $k=2$ & & $k = 3$ &  \\ \hline
$h$ & $L^2$ error & rate & $L^2$ error & rate & $L^2$ error & rate  \\ \hline
5.0000e-01 & 3.0777e-01 & -- & 1.7275e-02 & -- & 1.7265e-02 & -- \\
2.5000e-01 & 6.2773e-02 & 2.2936e+00 & 8.3845e-03 & 1.0429e+00 & 8.3348e-04 & 4.3725e+00 \\
1.2500e-01 & 1.6121e-02 & 1.9612e+00 & 1.0722e-03 & 2.9672e+00 & 5.3210e-05 & 3.9694e+00 \\
6.2500e-02 & 4.0645e-03 & 1.9878e+00 & 1.3513e-04 & 2.9881e+00 & 3.3717e-06 & 3.9802e+00 \\
3.1250e-02 & 1.0251e-03 & 1.9873e+00 & 1.7080e-05 & 2.9840e+00 & 2.1733e-07 & 3.9555e+00 \\
\end{tabular}
\caption{Errors and rates for $Q$ from the inviscid $(A,Q)$ system with the local Lax--Friedrichs flux.}
\label{table:rates4}
\end{center}
\end{table}

\begin{table}[!htb]
\begin{center}
\begin{tabular}{c | c c | c c | c c}
 & $k=1$ & & $k=2$ & & $k = 3$ &   \\ \hline
$h$ & $L^2$ error & rate & $L^2$ error & rate & $L^2$ error & rate  \\ \hline
5.0000e-01 & 8.5047e-02 & -- & 8.5047e-02 & -- & 2.7752e-03 & -- \\
2.5000e-01 & 6.2774e-02 & 4.3810e-01 & 8.3827e-03 & 3.3428e+00 & 8.3376e-04 & 1.7349e+00 \\
1.2500e-01 & 1.6123e-02 & 1.9610e+00 & 1.0714e-03 & 2.9680e+00 & 5.3266e-05 & 3.9684e+00 \\
6.2500e-02 & 4.0651e-03 & 1.9878e+00 & 1.3478e-04 & 2.9907e+00 & 3.3767e-06 & 3.9795e+00 \\
3.1250e-02 & 1.0245e-03 & 1.9884e+00 & 1.6935e-05 & 2.9926e+00 & 2.1710e-07 & 3.9591e+00 \\
\end{tabular}
\caption{Errors and rates for $A$ from the $(A,Q)$ system, $\alpha = 1.1$, with the upwinding flux.}
\label{table:rates5}
\begin{tabular}{c | c c | c c | c c}
 & $k=1$ & & $k=2$ & & $k = 3$ &  \\ \hline
$h$ & $L^2$ error & rate & $L^2$ error & rate & $L^2$ error & rate  \\ \hline
5.0000e-01 & 3.0776e-01 & -- & 1.7275e-02 & -- & 1.7265e-02 & -- \\
2.5000e-01 & 6.2777e-02 & 2.2935e+00 & 8.3839e-03 & 1.0430e+00 & 8.3379e-04 & 4.3720e+00 \\
1.2500e-01 & 1.6125e-02 & 1.9609e+00 & 1.0719e-03 & 2.9675e+00 & 5.3273e-05 & 3.9682e+00 \\
6.2500e-02 & 4.0671e-03 & 1.9872e+00 & 1.3503e-04 & 2.9889e+00 & 3.3822e-06 & 3.9774e+00 \\
3.1250e-02 & 1.0270e-03 & 1.9855e+00 & 1.7041e-05 & 2.9861e+00 & 2.1887e-07 & 3.9498e+00 \\
\end{tabular}
\caption{Errors and rates for $Q$ from the $(A,Q)$ system, $\alpha = 1.1$, with the upwinding flux.}
\label{table:rates6}
\end{center}
\end{table}

\begin{table}[!htb]
\begin{center}
\begin{tabular}{c | c c | c c | c c}
 & $k=1$ & & $k=2$ & & $k = 3$ &   \\ \hline
$h$ & $L^2$ error & rate & $L^2$ error & rate & $L^2$ error & rate  \\ \hline
5.0000e-01 & 8.5047e-02 & -- & 8.5047e-02 & -- & 2.7751e-03 & -- \\
2.5000e-01 & 6.2784e-02 & 4.3787e-01 & 8.3827e-03 & 3.3428e+00 & 8.3459e-04 & 1.7334e+00 \\
1.2500e-01 & 1.6128e-02 & 1.9608e+00 & 1.0720e-03 & 2.9672e+00 & 5.3352e-05 & 3.9675e+00 \\
6.2500e-02 & 4.0671e-03 & 1.9875e+00 & 1.3518e-04 & 2.9873e+00 & 3.3847e-06 & 3.9784e+00 \\
3.1250e-02 & 1.0253e-03 & 1.9880e+00 & 1.7142e-05 & 2.9793e+00 & 2.1745e-07 & 3.9603e+00 \\
\end{tabular}
\caption{Errors and rates for $A$ from the $(A,Q)$ system, $\alpha = 1.1$, with the local Lax--Friedrichs flux.}
\label{table:rates7}
\begin{tabular}{c | c c | c c | c c}
 & $k=1$ & & $k=2$ & & $k = 3$ &  \\ \hline
$h$ & $L^2$ error & rate & $L^2$ error & rate & $L^2$ error & rate  \\ \hline
5.0000e-01 & 3.0778e-01 & -- & 1.7275e-02 & -- & 1.7265e-02 & -- \\
2.5000e-01 & 6.2775e-02 & 2.2936e+00 & 8.3848e-03 & 1.0428e+00 & 8.3356e-04 & 4.3724e+00 \\
1.2500e-01 & 1.6123e-02 & 1.9610e+00 & 1.0722e-03 & 2.9671e+00 & 5.3242e-05 & 3.9687e+00 \\
6.2500e-02 & 4.0662e-03 & 1.9874e+00 & 1.3516e-04 & 2.9879e+00 & 3.3785e-06 & 3.9781e+00 \\
3.1250e-02 & 1.0266e-03 & 1.9858e+00 & 1.7093e-05 & 2.9832e+00 & 2.1857e-07 & 3.9502e+00 \\
\end{tabular}
\caption{Errors and rates for $Q$ from the $(A,Q)$ system, $\alpha = 1.1$, with the local Lax--Friedrichs flux.}
\label{table:rates8}
\end{center}
\end{table}

Table~\ref{table:rates1} and Table~\ref{table:rates2} show the errors and rates for the inviscid $(A,Q)$ system for the
DG method with the upwinding flux.  We observe that we recover the convergence rate
 $k+1$ for polynomial degree $k$, which is the optimal convergence rate for a
scalar hyperbolic conservation law in one dimension, as noted in \cite{ZS06}.   For
general systems of conservation laws, the theoretical rates are suboptimal, as described by
(\ref{eq:est1}) and (\ref{eq:est2}).  
For comparison, in Table~\ref{table:rates3} and Table~\ref{table:rates4}, we repeat the experiments for the DG method with the local Lax-Friedrichs flux. Similar conclusions can be made. The choice of the numerical flux does not have
any effect on the errors or rates.

The theoretical rates apply only to the DG approximation of inviscid reduced blood flow models with the local Lax--Friedrichs flux. We next investigate
the numerical rates for the general $(A,Q)$ system with $\alpha=1.1$.  
Table~\ref{table:rates5} and Table~\ref{table:rates6} show the errors and rates, with the
same set-up as the previous experiments. The upwinding flux is used.  The numerical rates
are optimal.  Theoretical error estimates remain an open question. Table~\ref{table:rates7}
and Table~\ref{table:rates8} show optimal rates for the case of the local Lax-Friedrichs flux.
Results are comparable to those obtained with the upwinding flux.

\clearpage
\subsection{Validation of fifty--five vessel network with reflection boundary conditions}

\begin{figure}[!htb]
\begin{center}
\includegraphics[scale=0.35]{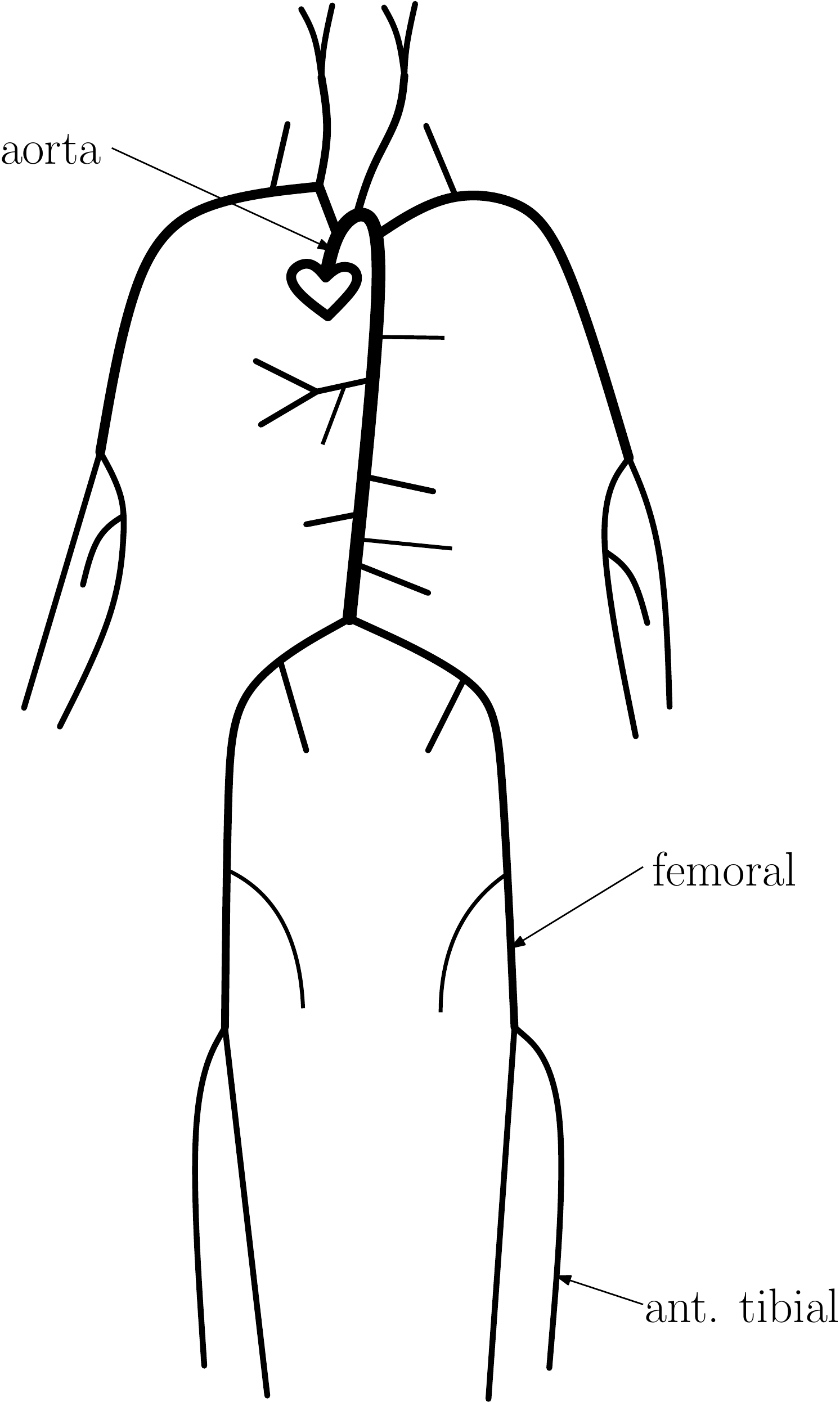}
\includegraphics[scale=0.45]{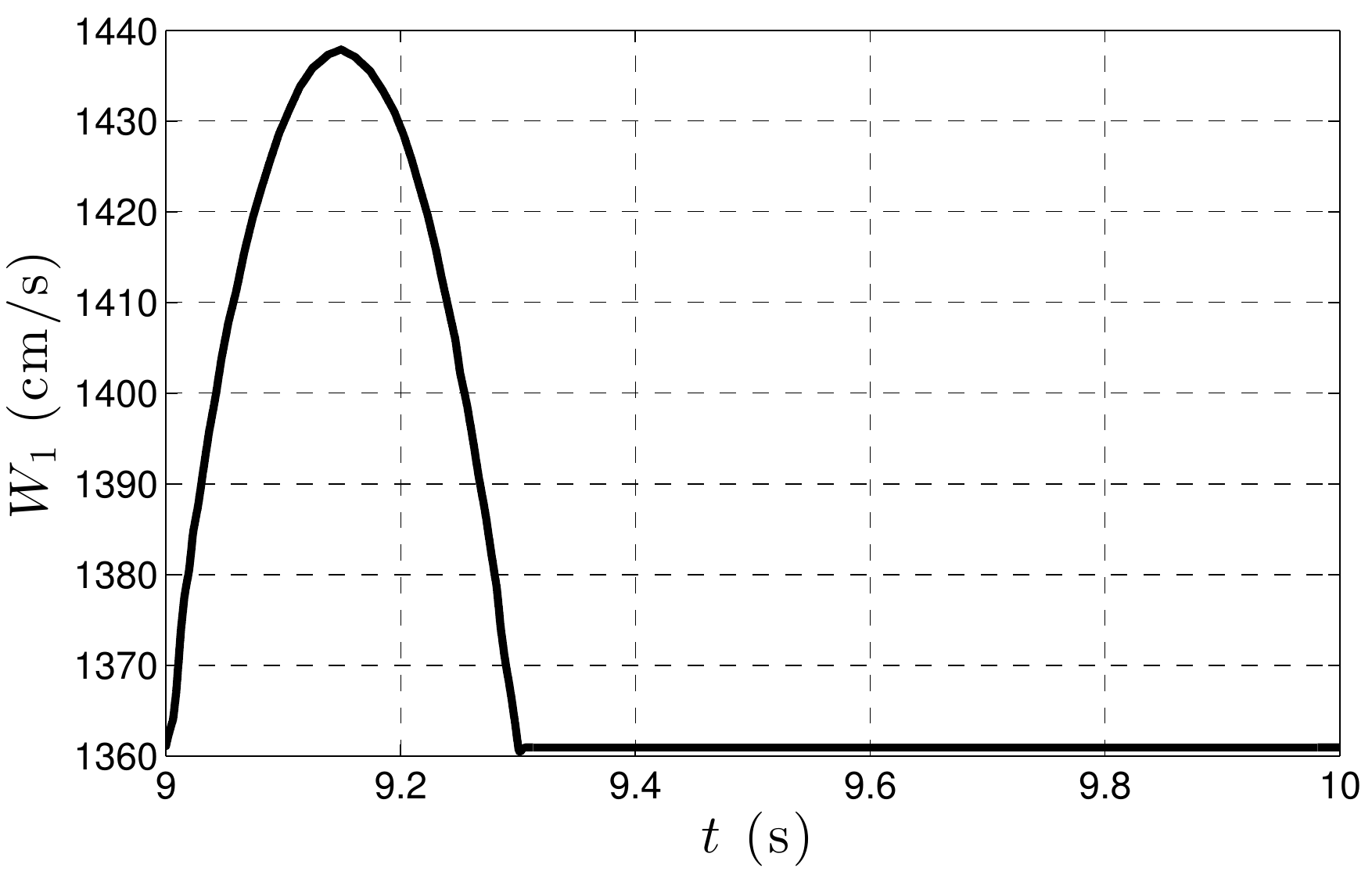}
\caption{The schematic on the left depicts the topology of the vessel network used throughout this paper, from Sherwin et al. \cite{SFPF03}.  The labels indicate the ascending aorta, where the inlet boundary condition is specified, and the femoral and anterior tibial arteries, where waveforms are measured.  The figure on the right is the inlet boundary condition at the ascending aorta, also from \cite{SFPF03}.}
\label{fig:sher1}
\end{center}
\end{figure}

In this section, we verify our numerical scheme by simulating blood flow in a fifty--five vessel network and comparing with results from \cite{SFPF03}.  We provide results for both the inviscid  $(A,Q)$ and inviscid $(A,U)$ systems in this section.  The incoming Riemann invariant $W_1$ is prescribed at the inlet of the ascending aorta, and the reflection boundary conditons (\ref{eq:res1}) and (\ref{eq:res2}) are used at the outlets of the terminal vessels.  Figure \ref{fig:sher1} displays the vessel network and the inlet boundary condition.  Vessel  parameters are taken from \cite{SFPF03} and the numerical and physical parameters are given in Tables \ref{table:Sherwin1} and \ref{table:Sherwin2}.

\begin{table}[!htb]
\begin{center}
\begin{tabular}{| c | c |}
\hline
 $\nu$ $(\text{cm}^2\text{/s})$ & $p_0$ (mmHg) \\ \hline
0 & 75 \\ \hline
\end{tabular}
\caption{Physical parameters for the fifty--five vessel network simulation to compare with results from Sherwin et al. \cite{SFPF03}.}
\label{table:Sherwin1}
\end{center}
\end{table}

\begin{table}[!htb]
\begin{center}
\begin{tabular}{| c | c | c | c |}
\hline
$\Delta t$ (s) & $h$ (cm) & $k$ (polynomial degree) & numerical flux  \\ \hline
$10^{-4}$ & 1 & 1 & UP\\ \hline
\end{tabular}
\caption{Numerical parameters for the validation experiment with the network and boundary conditions from \cite{SFPF03}.}
\label{table:Sherwin2}
\end{center}
\end{table}

\begin{figure}[!htb]
\begin{center}
\includegraphics[scale=0.35]{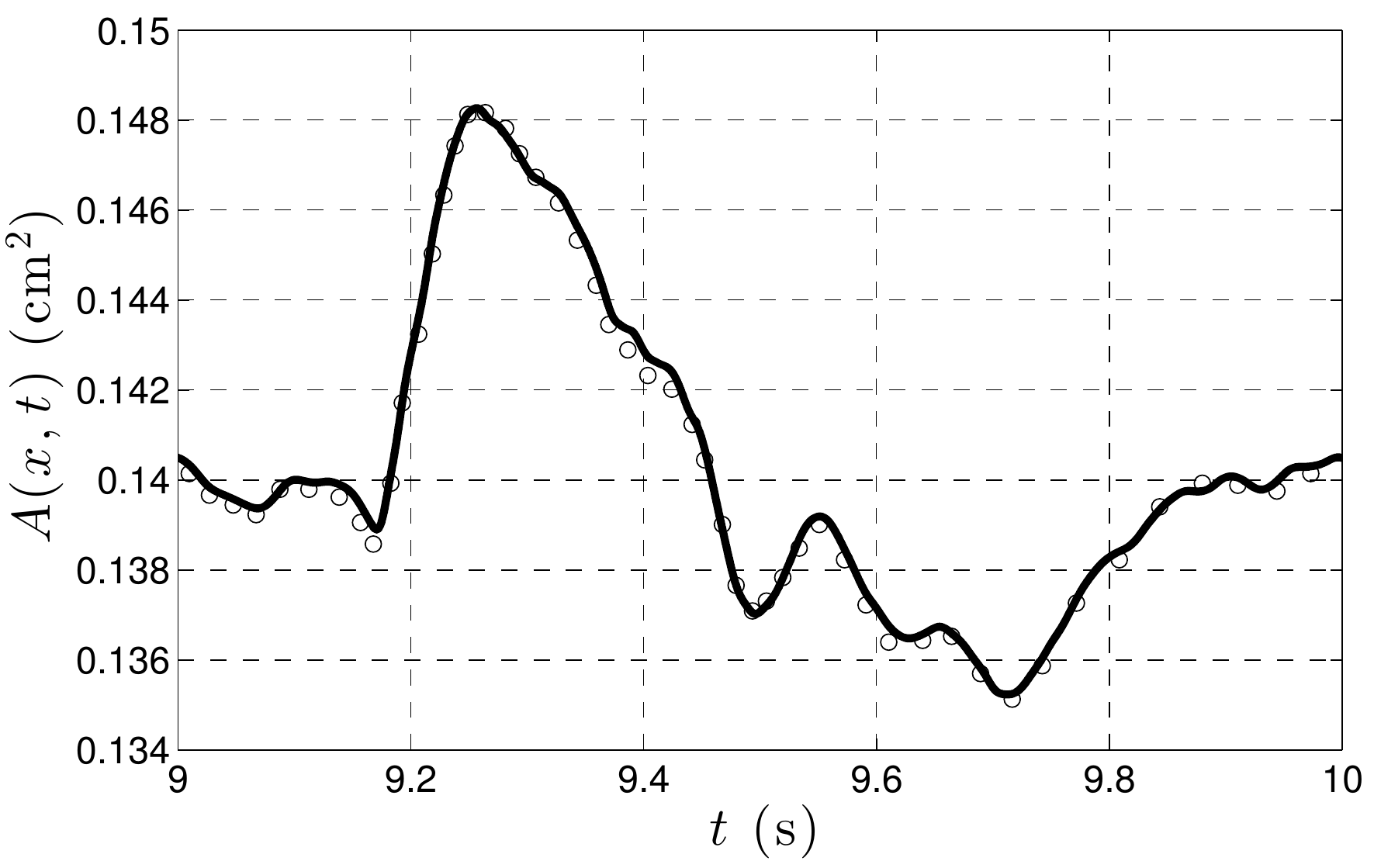}
\includegraphics[scale=0.35]{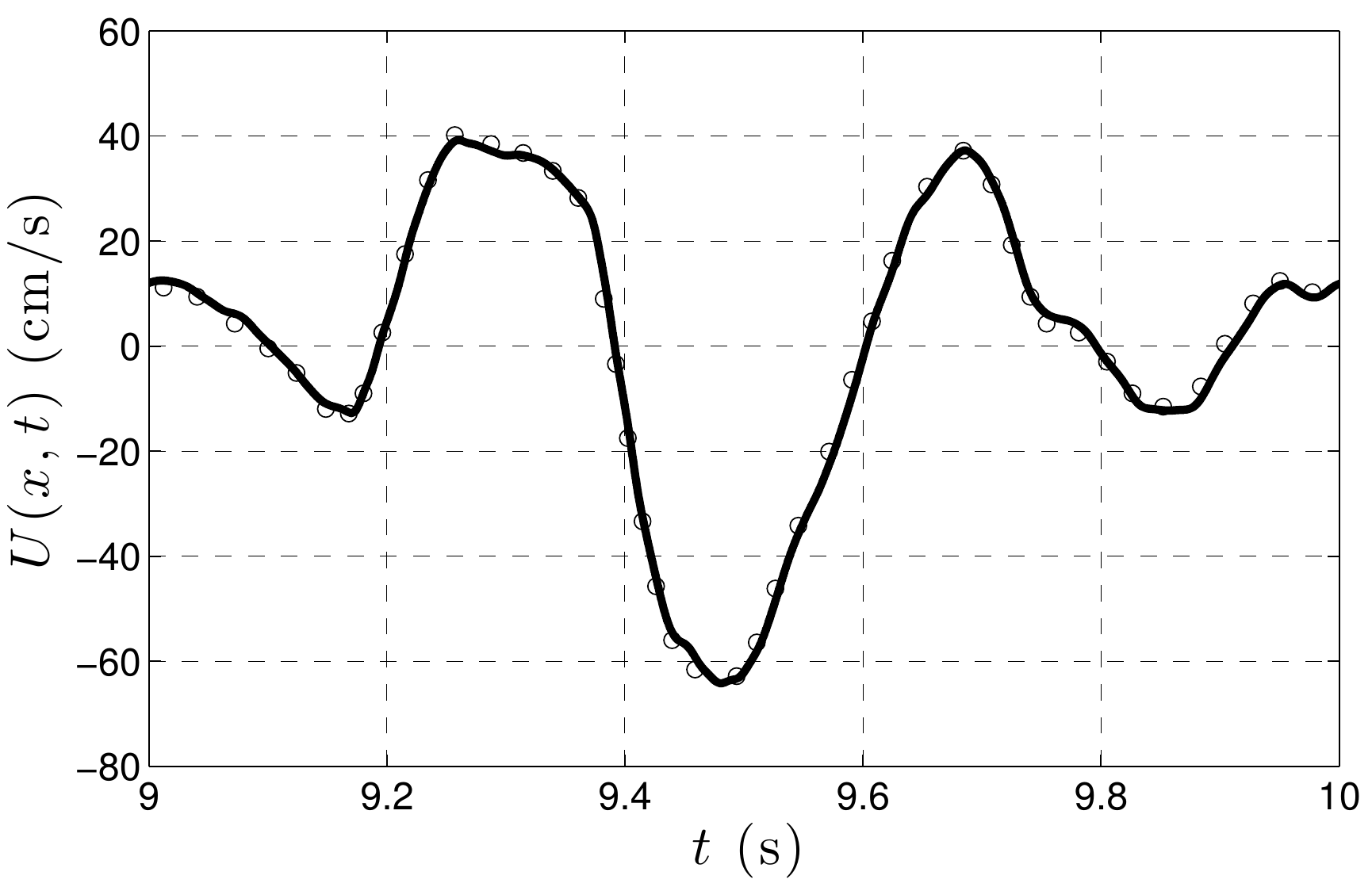} \\
\caption{Waveforms from the $(A,U)$ system obtained at the inlet of the left femoral artery.  Our numerical results are plotted with the solid line  and the circles are data taken from Sherwin et al. \cite{SFPF03}.}
\label{fig:sher2}
\end{center}
\end{figure}

\begin{figure}[!htb]
\begin{center}
\includegraphics[scale=0.35]{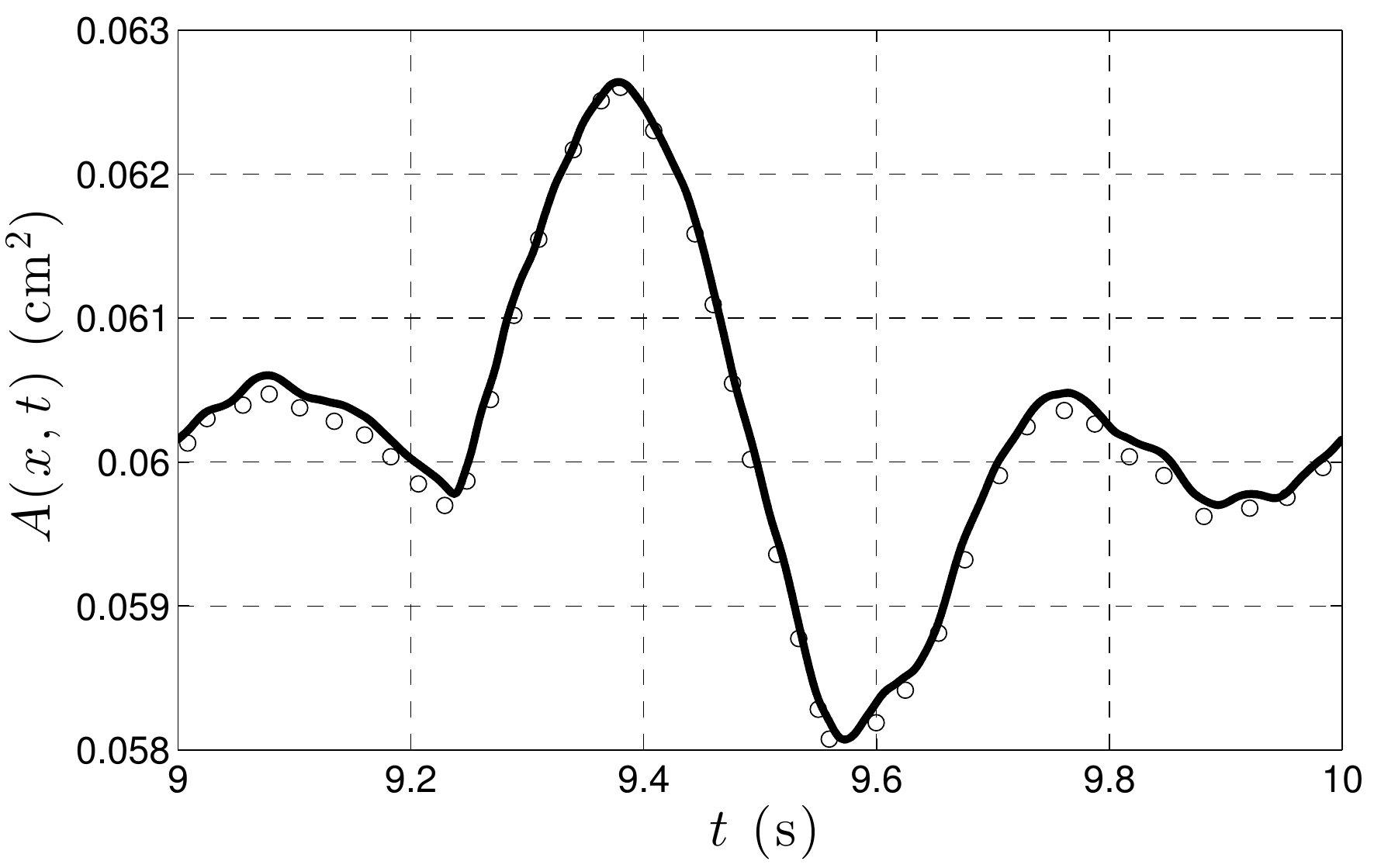}
\includegraphics[scale=0.35]{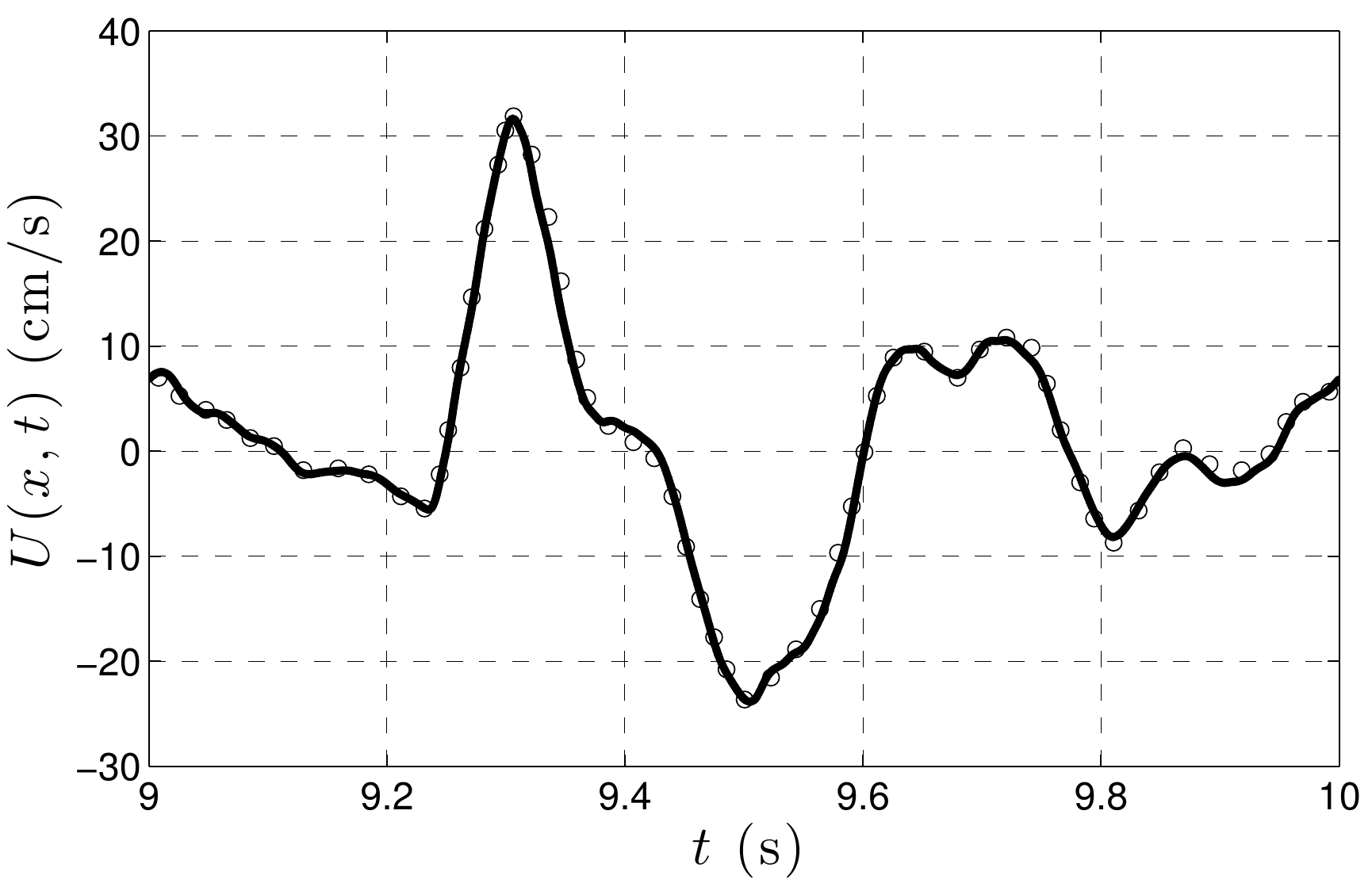} \\
\caption{Waveforms from the $(A,U)$ system obtained at the inlet of the left anterior tibial artery.  Our numerical results are plotted with the solid line  and the circles are data taken from Sherwin et al. \cite{SFPF03}.}
\label{fig:sher3}
\end{center}
\end{figure}

\begin{figure}[!htb]
\begin{center}
\includegraphics[scale=0.35]{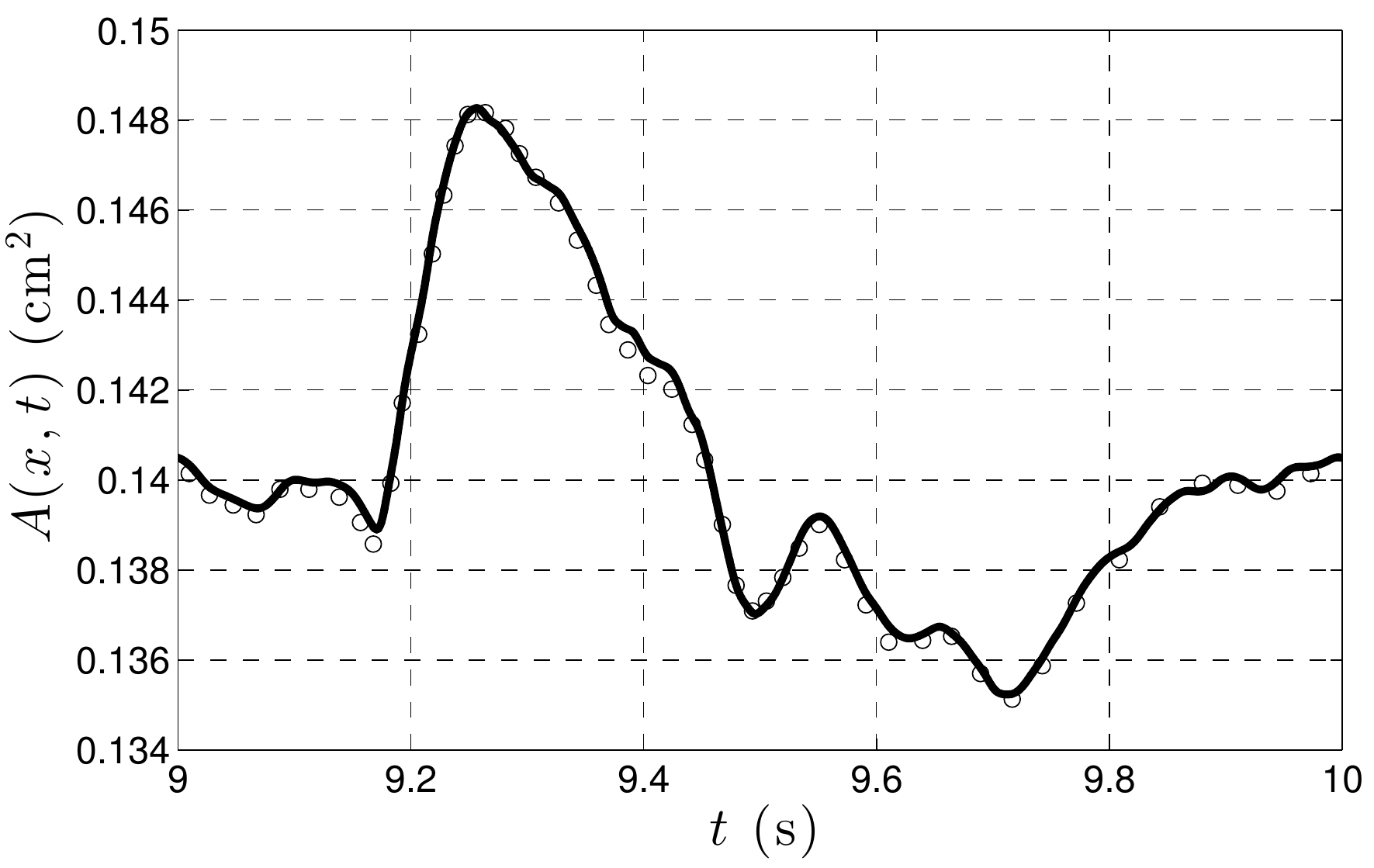}
\includegraphics[scale=0.35]{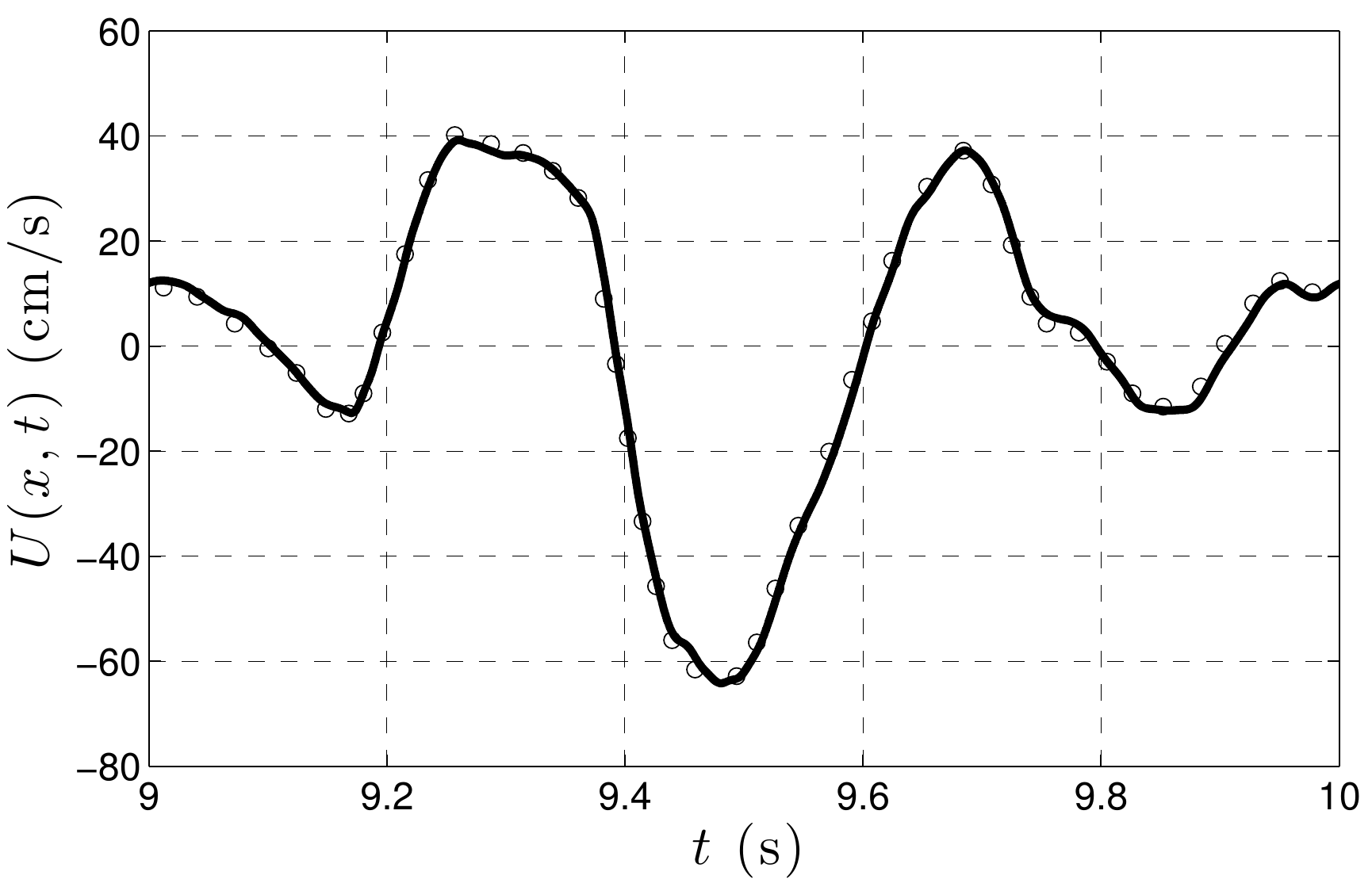} \\
\caption{Waveforms from the $(A,Q)$ system obtained at the inlet of the left femoral artery.  Our numerical results are plotted with the solid line  and the circles are data taken from Sherwin et al. \cite{SFPF03}.}
\label{fig:sher4}
\end{center}
\end{figure}

\begin{figure}[!htb]
\begin{center}
\includegraphics[scale=0.35]{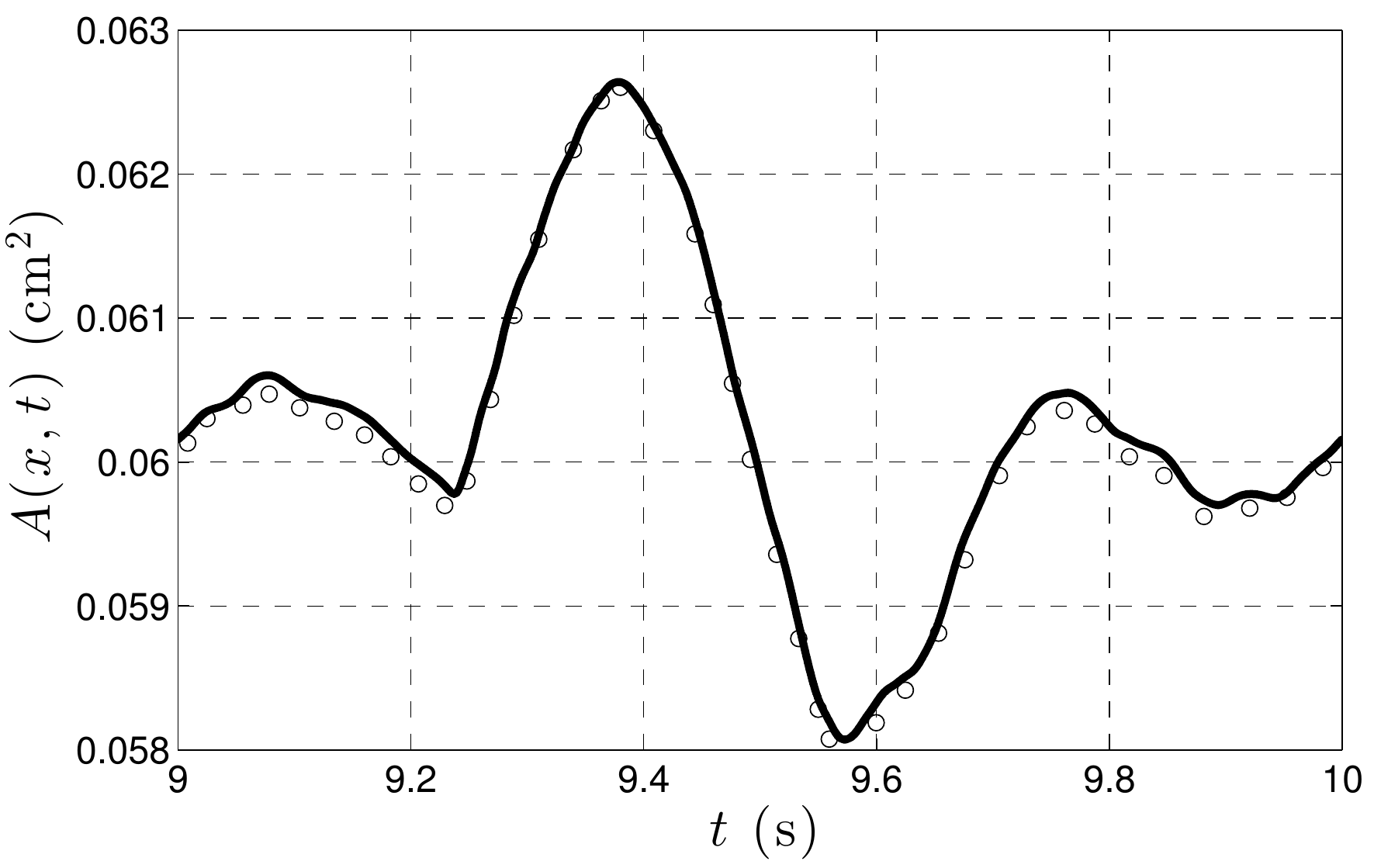}
\includegraphics[scale=0.35]{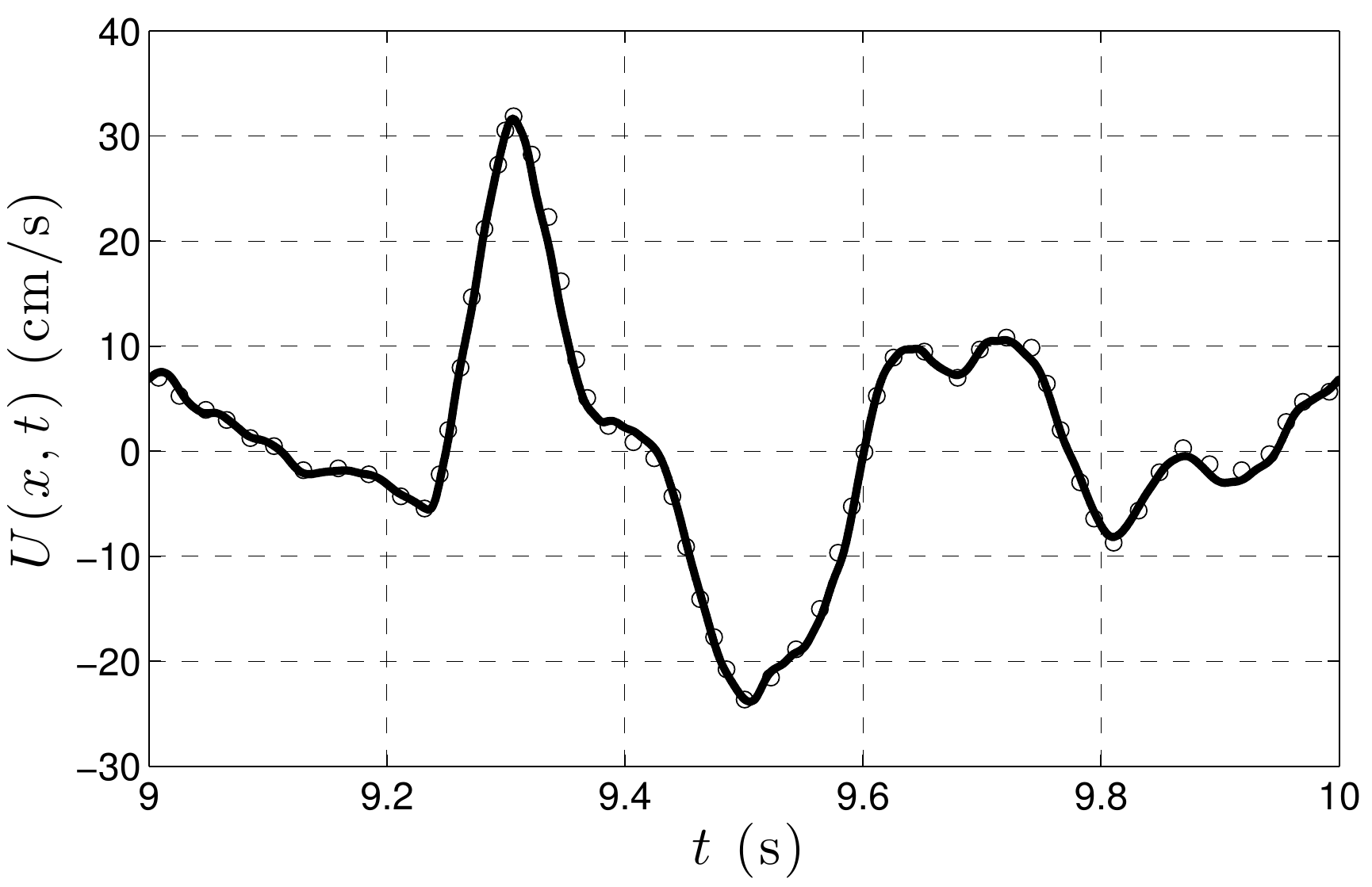} \\
\caption{Waveforms from the $(A,Q)$ system obtained at the inlet of the left anterior tibial artery.  Our numerical results are plotted with the solid line  and the circles are data taken from Sherwin et al. \cite{SFPF03}.}
\label{fig:sher5}
\end{center}
\end{figure}

The waveforms displayed are obtained during the tenth cardiac cycle at the inlet of the left femoral and left anterior tibial vessels.  We plot results for the $(A,U)$ system in Figures \ref{fig:sher2} and \ref{fig:sher3} and the $(A,Q)$ system in Figures \ref{fig:sher4}  and \ref{fig:sher5}. Also plotted in circles are data taken from the waveforms in \cite{SFPF03} obtained with different polynomial degree and timestep.  We observe excellent agreement between our results and those from \cite{SFPF03}.



\clearpage
\subsection{Comparison of waveforms obtained with windkessel and reflection boundary conditions}

In this section, we study the effect of different terminal boundary conditions on the waveforms.  We compare the reflection boundary conditions given in equations (\ref{eq:res1}) and (\ref{eq:res2}), with three element windkessel boundary conditions specified by equations (\ref{eq:wind1}) and (\ref{eq:wind2}).  The parameters for the windkessel boundary conditions are taken from \cite{Boileau15} and for the reflection boundary condition from \cite{SFPF03}.  The vessel network and inlet boundary data are the same as in the previous section, and the numerical and physical parameters are given in Tables \ref{table:bccomp1} and \ref{table:bccomp2}.

\begin{table}[!htb]
\begin{center}
\begin{tabular}{| c | c | c |}
\hline
 $\nu$ $(\text{cm}^2\text{/s})$ & $p_0$ (mmHg) & $P_\text{out}$ (mmHg)  \\ \hline
0 & 75 & 0 \\ \hline
\end{tabular}
\caption{Physical parameters for experiment studying the effects of different terminal boundary conditions.}
\label{table:bccomp1}
\end{center}
\end{table}

\begin{table}[!htb]
\begin{center}
\begin{tabular}{| c | c | c | c |}
\hline
$\Delta t$ (s) & $h$ (cm) & $k$ (polynomial degree) & numerical flux  \\ \hline
$10^{-4}$ & 1 & 1 & UP\\ \hline
\end{tabular}
\caption{Numerical parameters for experiment studying the effects of different terminal boundary conditions.}
\label{table:bccomp2}
\end{center}
\end{table}

\begin{figure}[!htb]
\begin{center}
\includegraphics[scale=0.35]{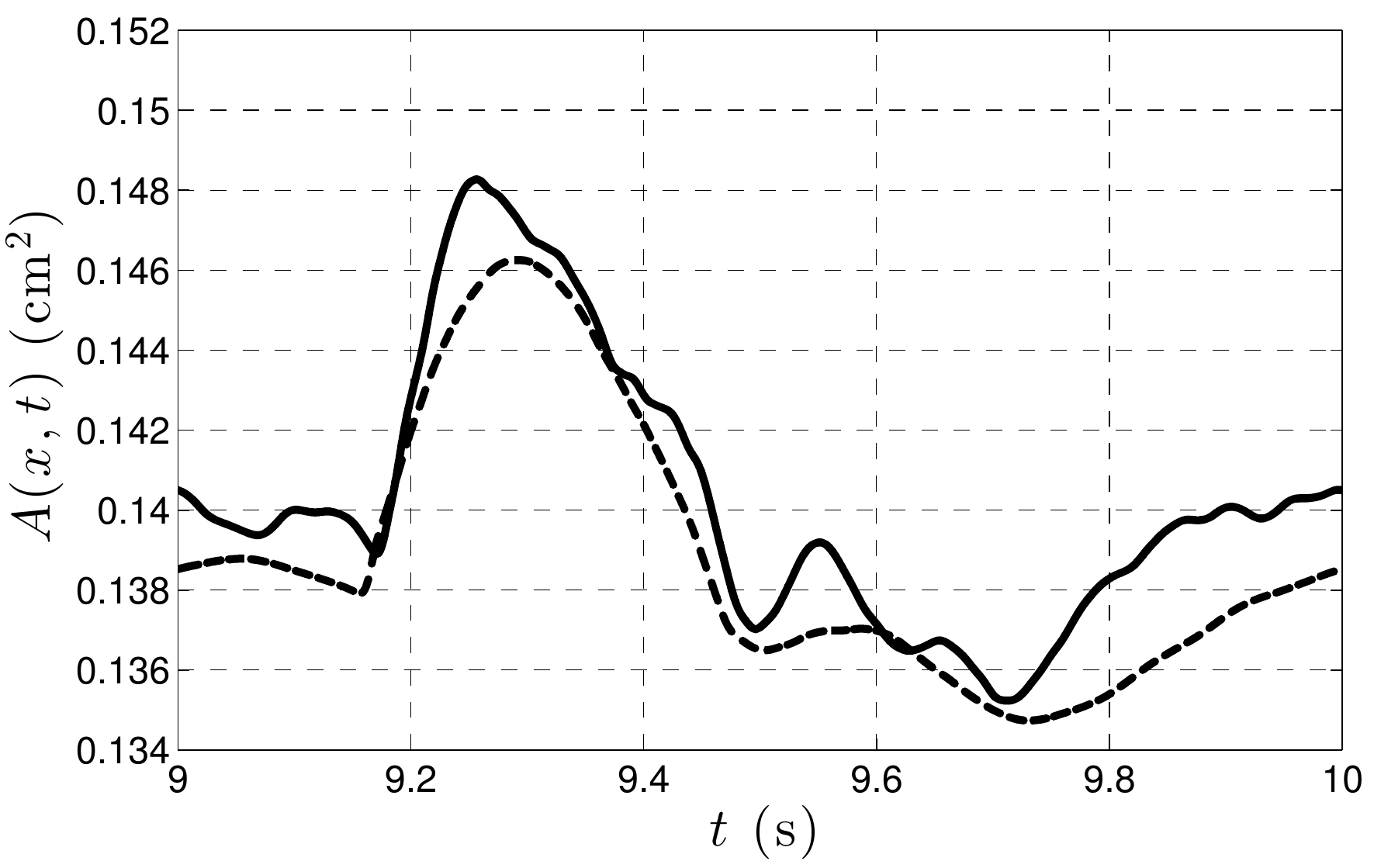}
\includegraphics[scale=0.35]{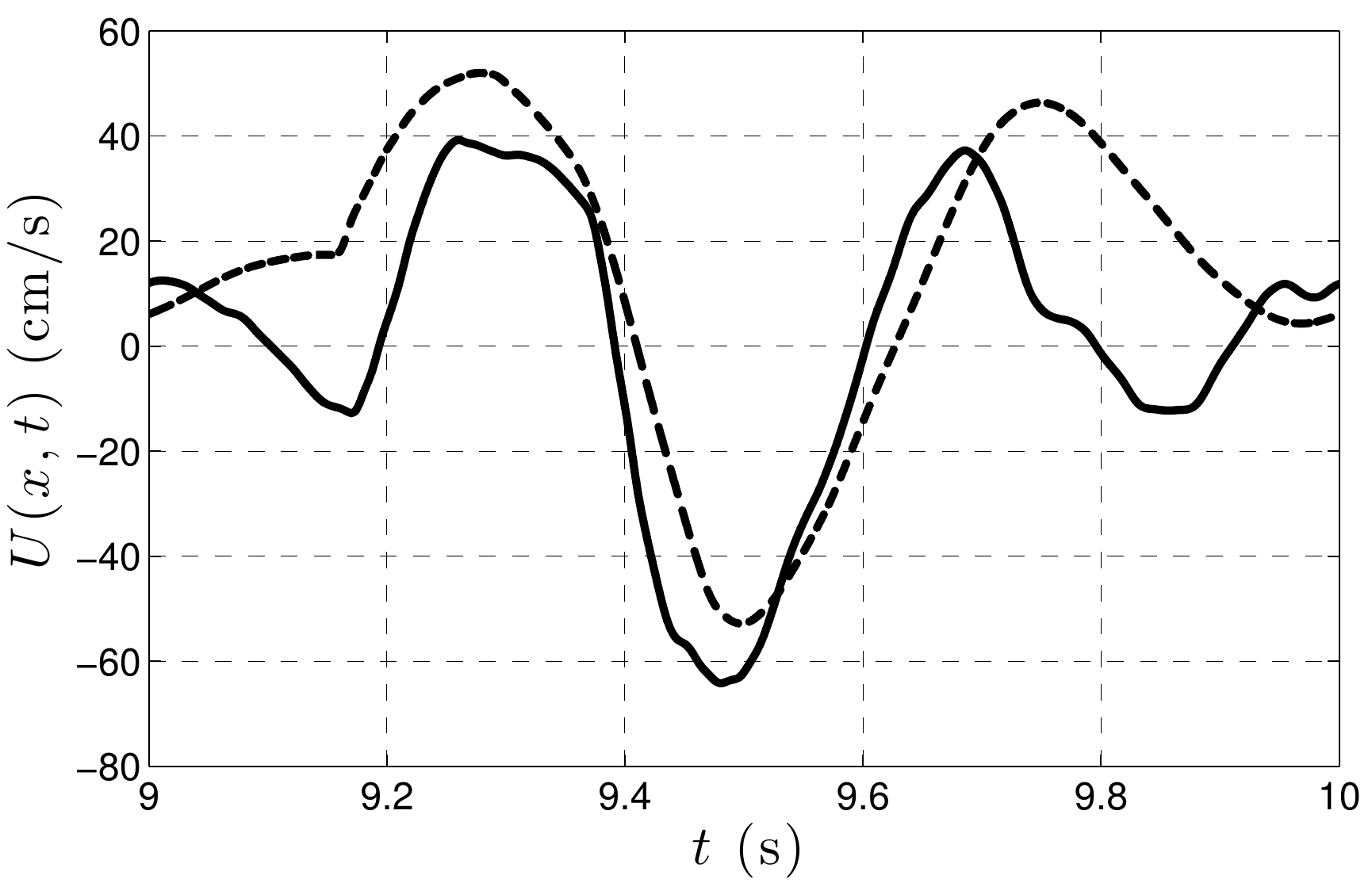} \\
\caption{Waveforms from the $(A,U)$ system obtained at the inlet of the left femoral artery.  Solid line corresponds to results with reflection boundary conditions and dashed line corresponds to results with windkessel boundary conditions.}
\label{fig:bccomp1}
\end{center}
\end{figure}

\begin{figure}[!htb]
\begin{center}
\includegraphics[scale=0.35]{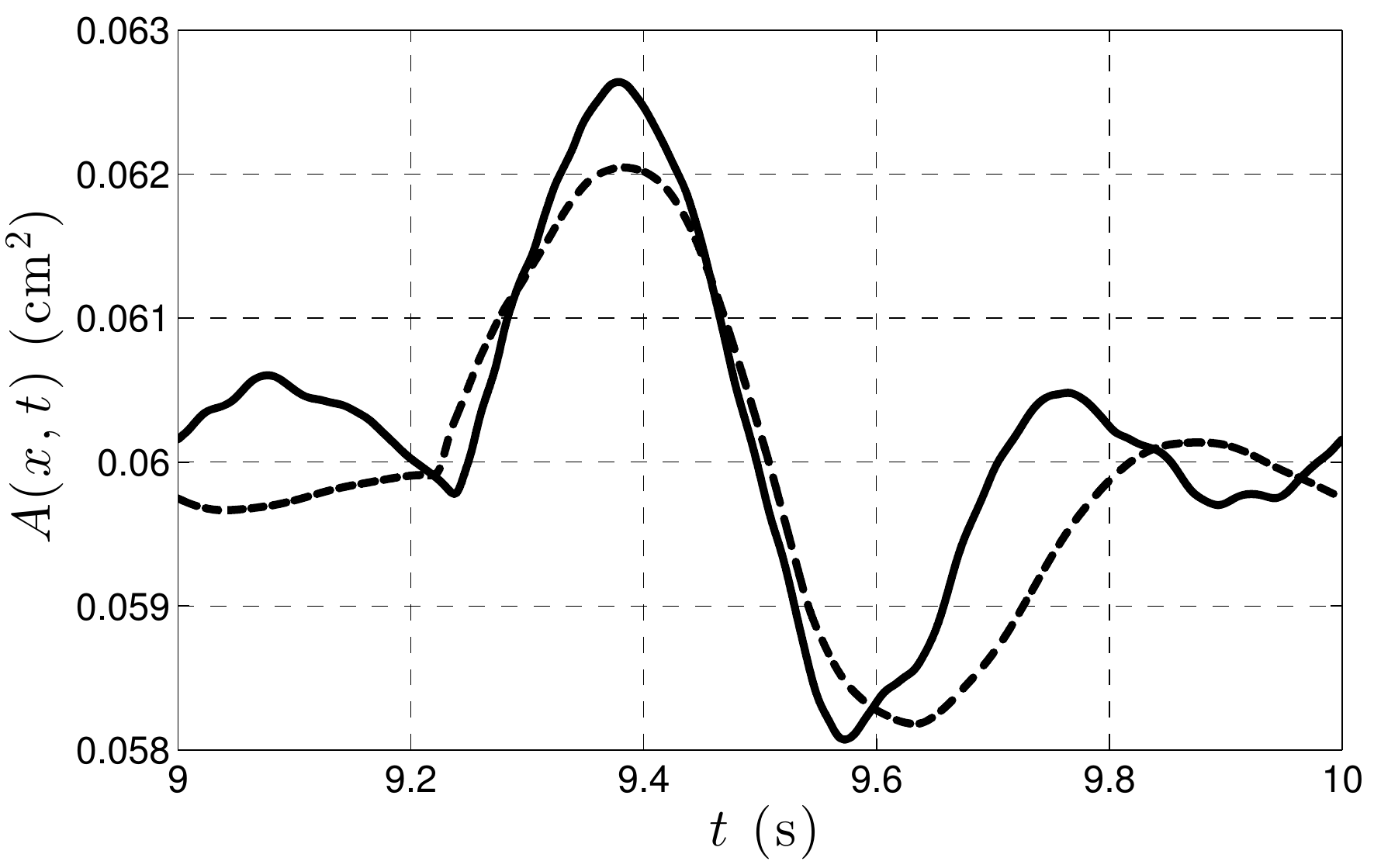}
\includegraphics[scale=0.35]{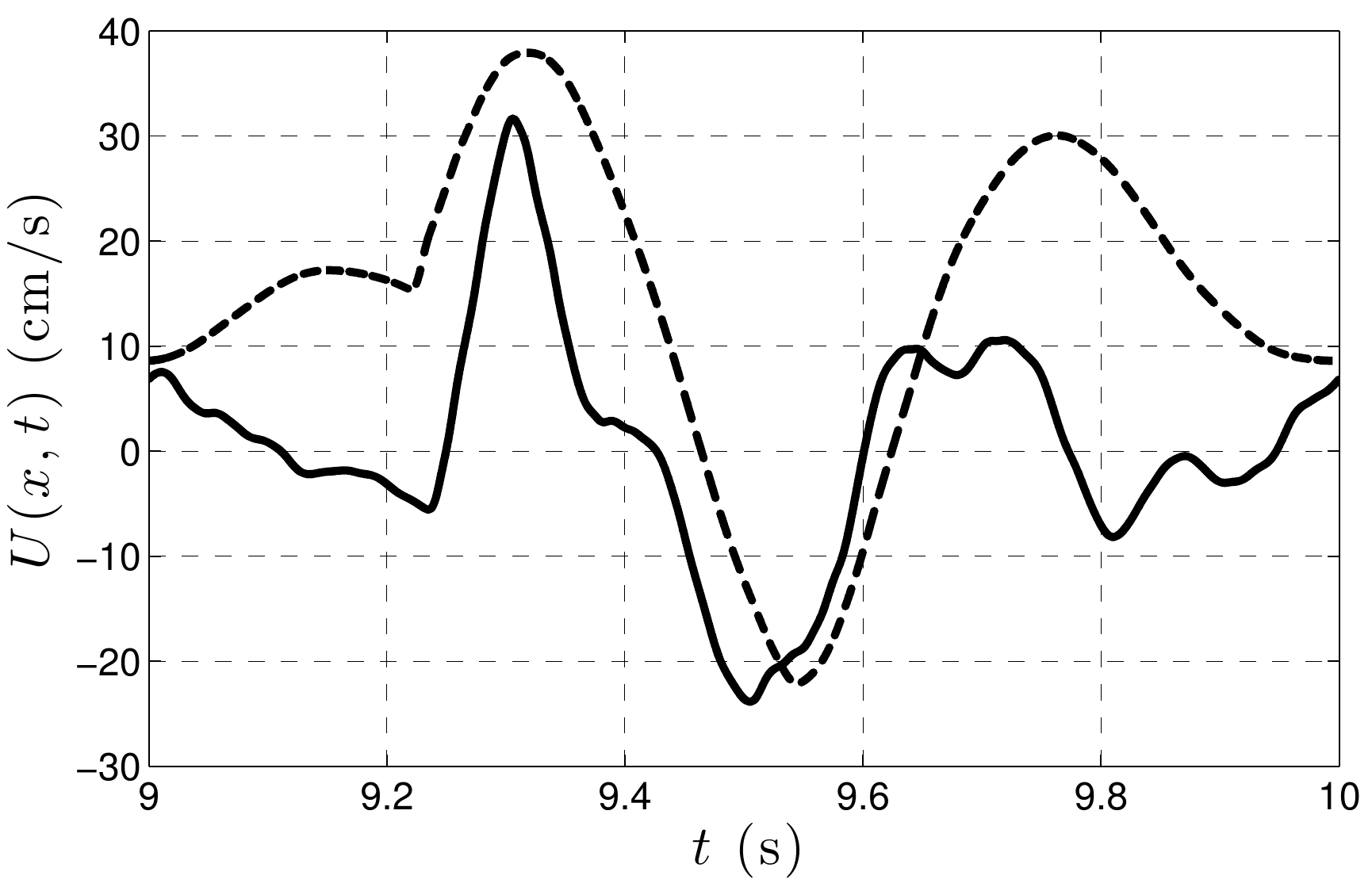} \\
\caption{Waveforms from the $(A,U)$ system obtained at the inlet of the left anterior tibial artery.  Solid line corresponds to results with reflection boundary conditions and dashed line corresponds to results with windkessel boundary conditions.}
\label{fig:bccomp2}
\end{center}
\end{figure}

\begin{figure}[!htb]
\begin{center}
\includegraphics[scale=0.35]{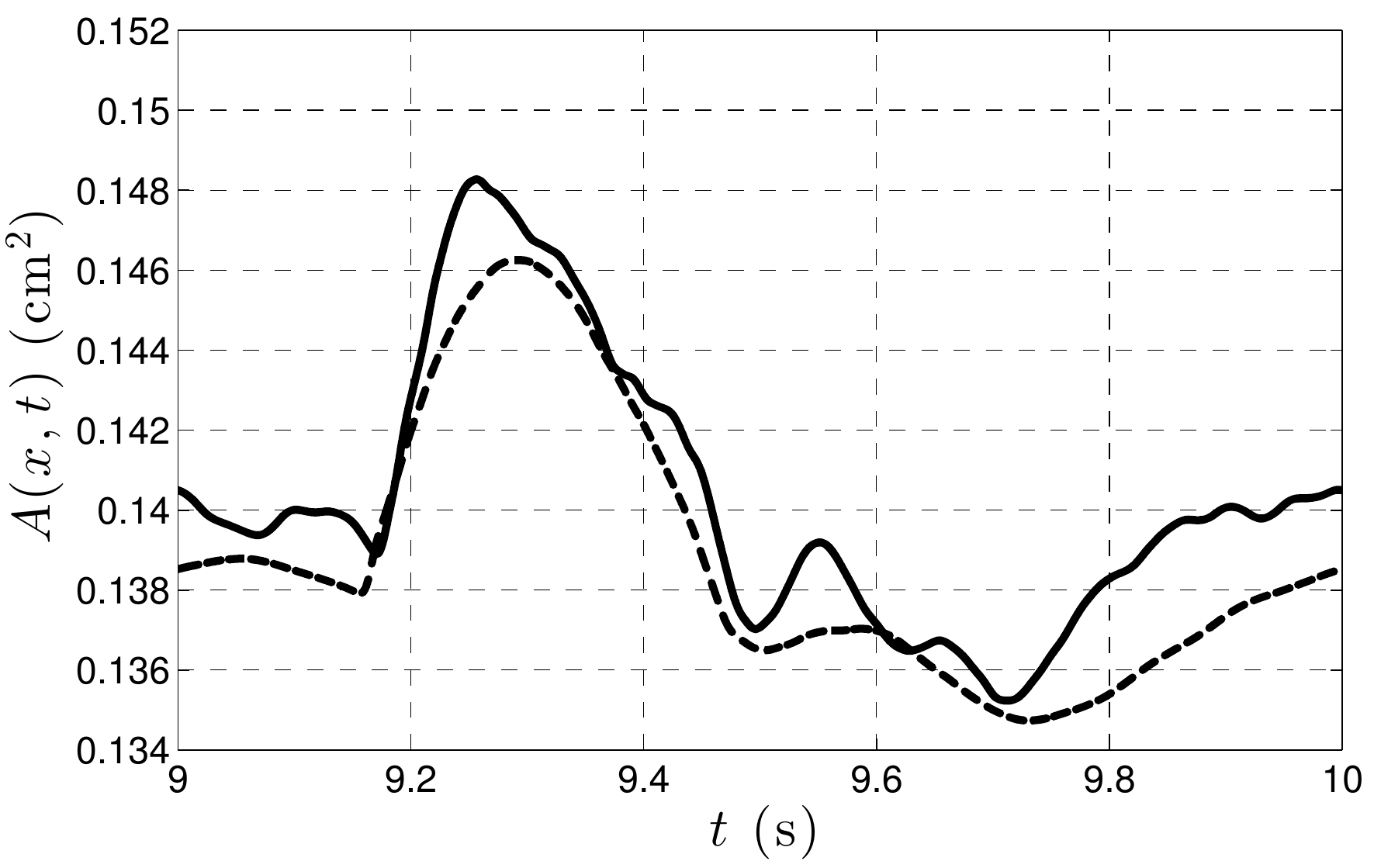}
\includegraphics[scale=0.35]{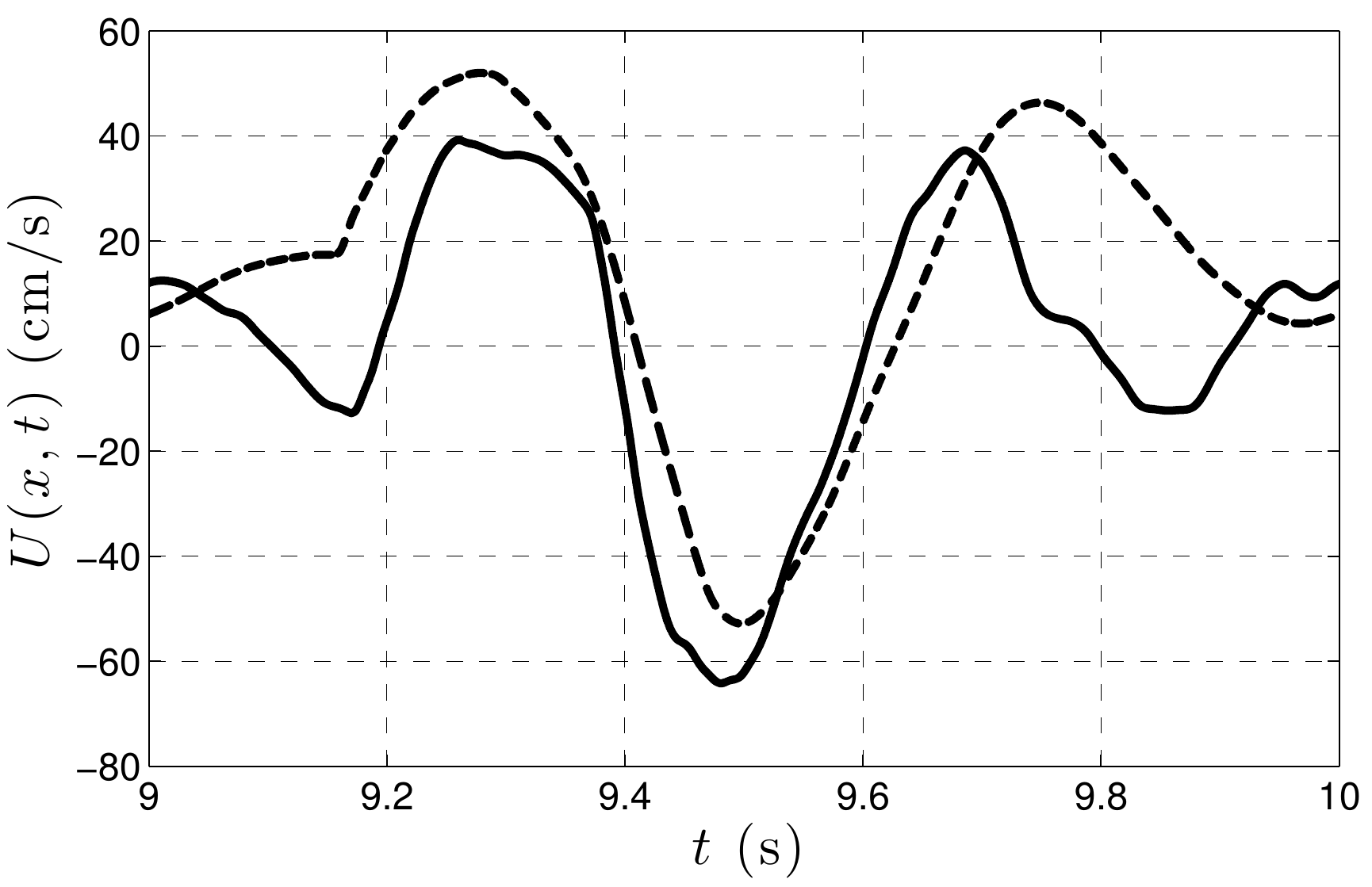} \\
\caption{Waveforms from the $(A,Q)$ system obtained at the inlet of the left femoral artery.  Solid line corresponds to results with reflection boundary conditions and dashed line corresponds to results with windkessel boundary conditions.}
\label{fig:bccomp3}
\end{center}
\end{figure}

\begin{figure}[!htb]
\begin{center}
\includegraphics[scale=0.35]{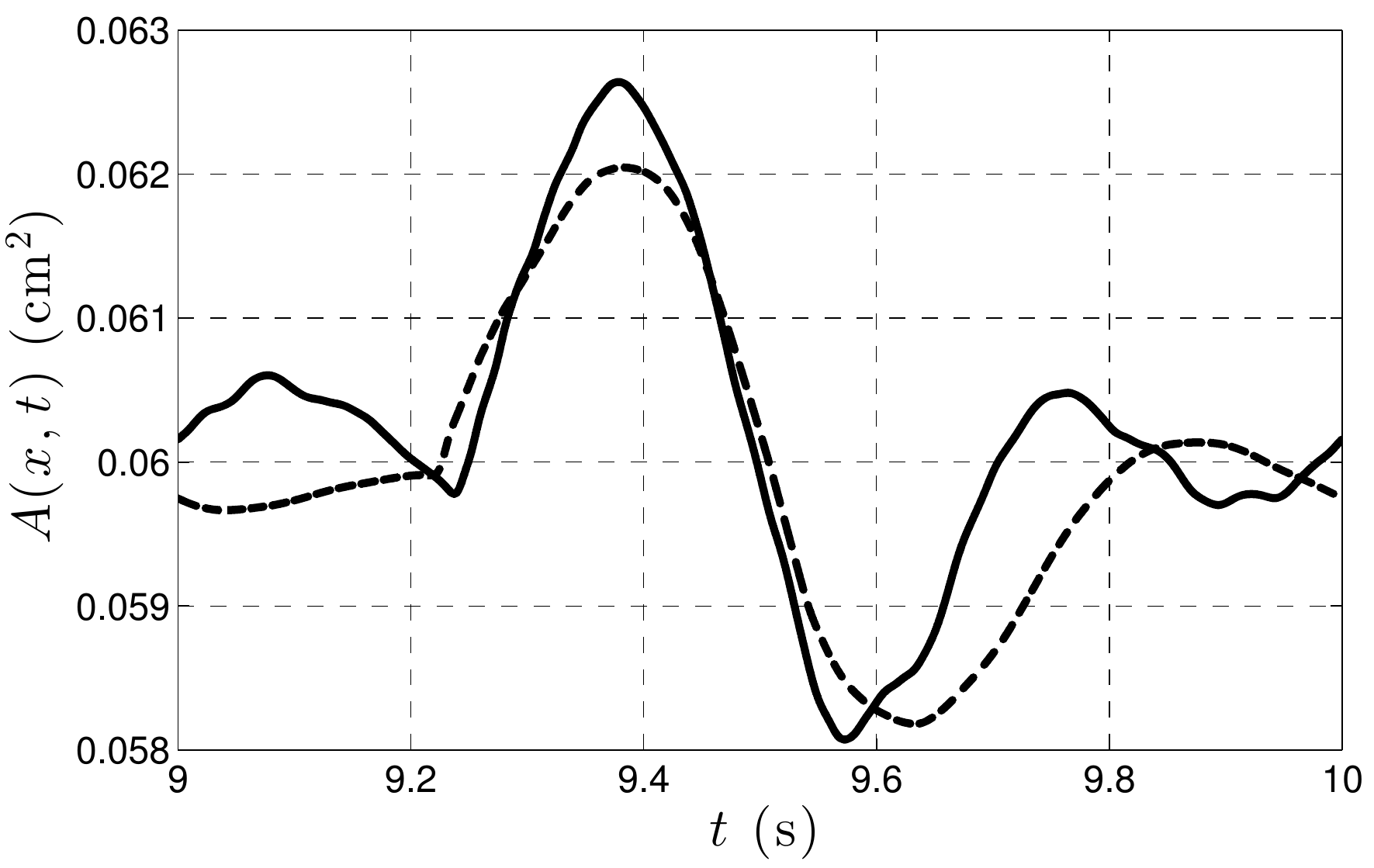}
\includegraphics[scale=0.35]{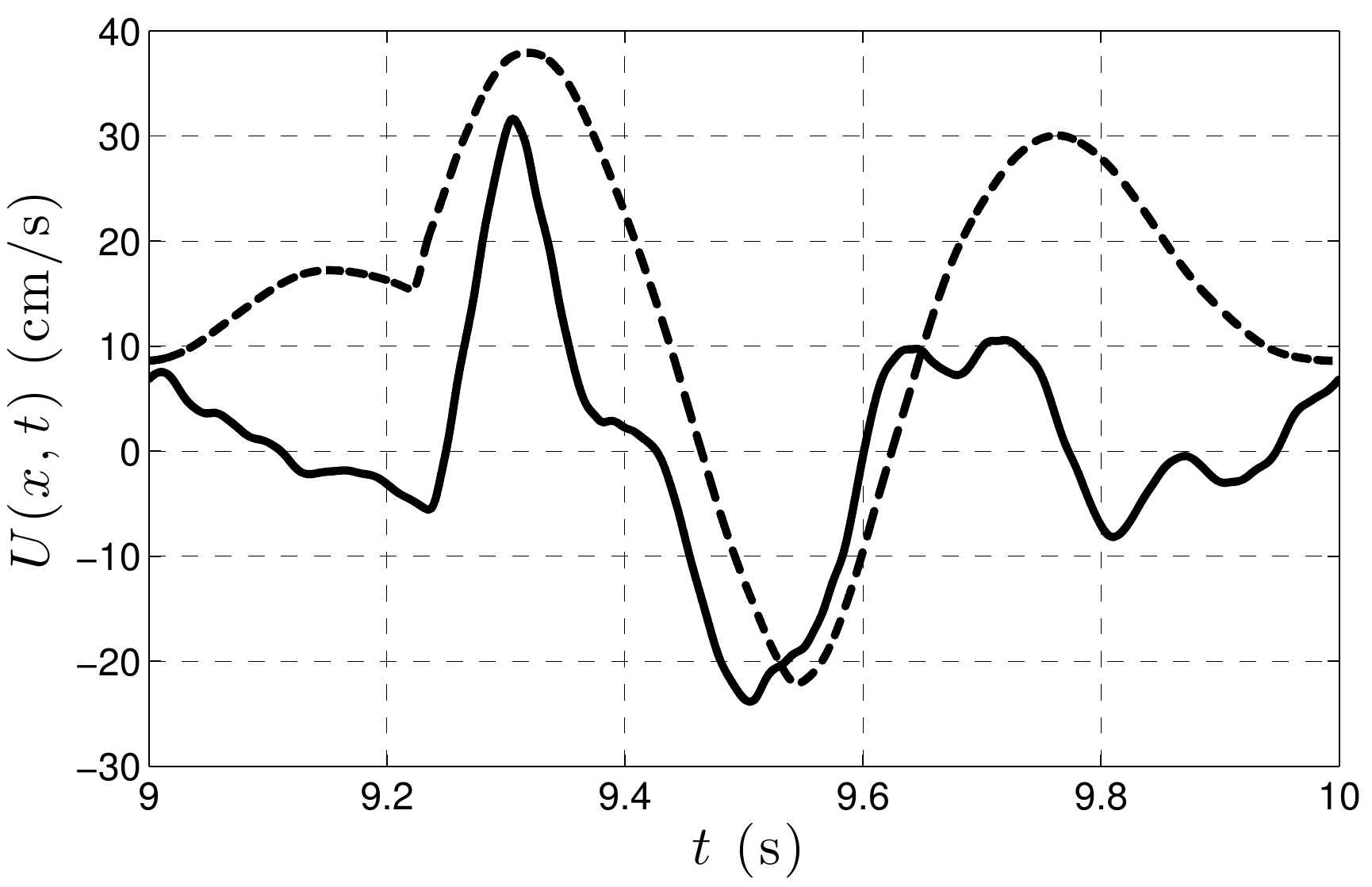} \\
\caption{Waveforms from the $(A,Q)$ system obtained at the inlet of the left anterior tibial artery.  Solid line corresponds to results with reflection boundary conditions and dashed line corresponds to results with windkessel boundary conditions.}
\label{fig:bccomp4}
\end{center}
\end{figure}

In Figures \ref{fig:bccomp1}--\ref{fig:bccomp4}, the solid line waveforms are produced with the reflection boundary condition and the dashed line waveforms are produced with the three element windkessel model.  Figures \ref{fig:bccomp1} and \ref{fig:bccomp2} are from the $(A,U)$ system and Figures \ref{fig:bccomp3} and \ref{fig:bccomp4} are from the $(A,Q)$ system.  First, we note that the waveforms from either $(A,Q)$ or $(A,U)$ are similar for a given choice of boundary conditions. Second, we observe that both terminal boundary conditions produce different waveforms with relative similar shape and magnitude. The reflection conditions create higher frequency oscillations while the windkessel model yields distinctly smoother features.  Since the oscillations from the reflection conditions are arguably less physiological, we use the windkessel conditions in the remainder of the paper.




\clearpage
\subsection{Fifty--five vessel network with physiological inlet data}

\begin{figure}[!htb]
\begin{center}
\includegraphics[scale=0.375, trim=0 0 -100 0]{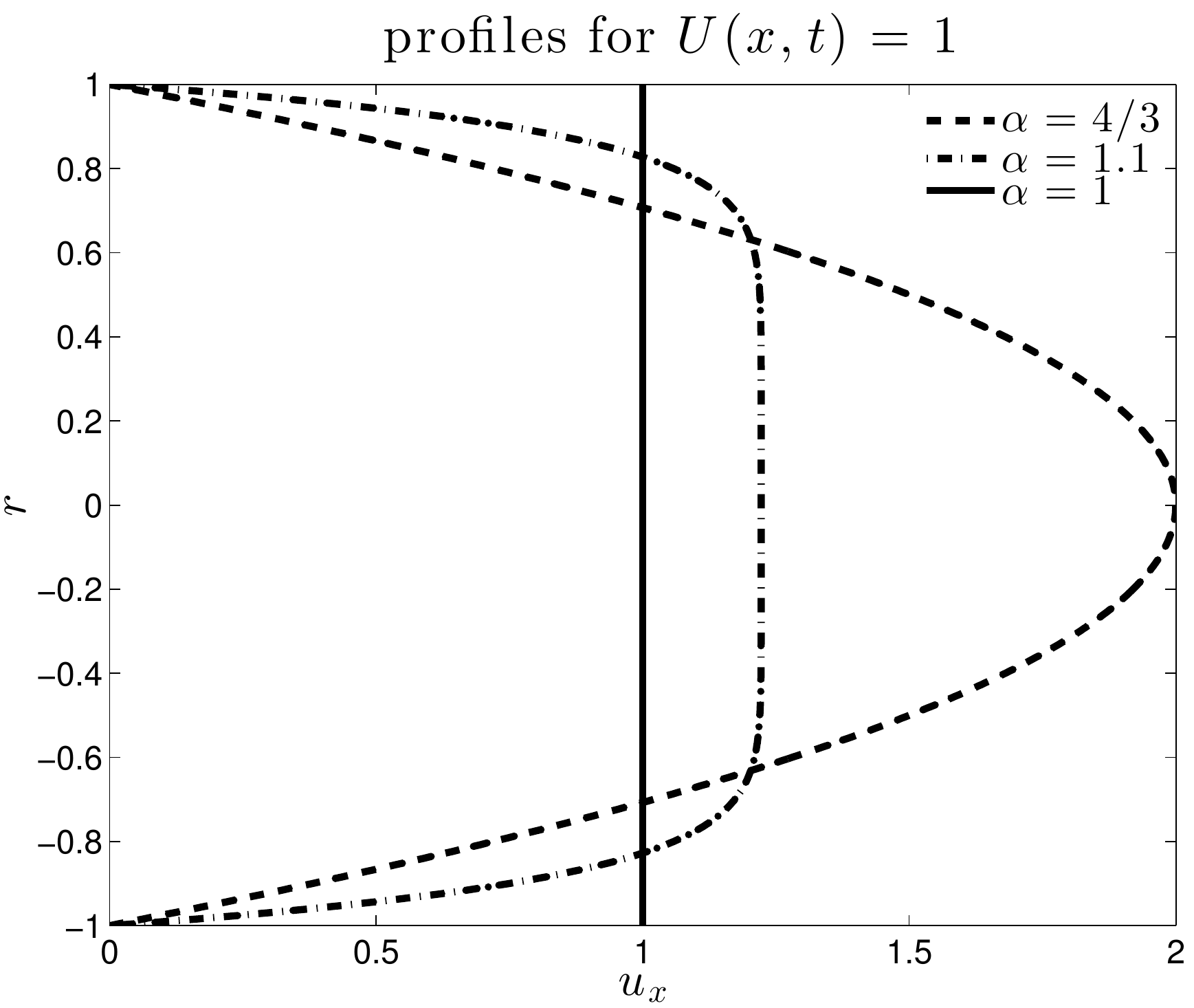}
\includegraphics[scale=0.4]{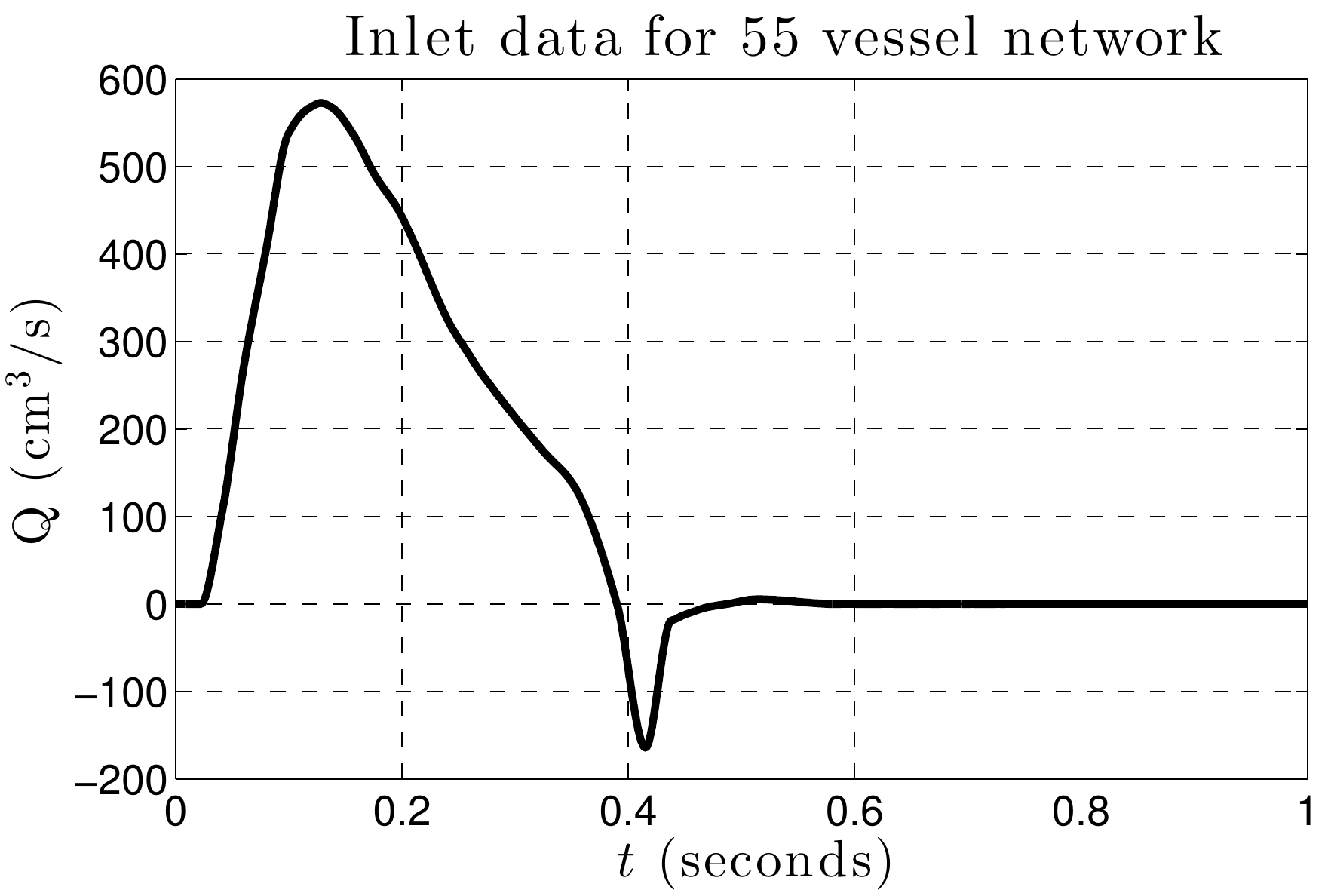}
\caption{The plot on the left shows the different velocity profiles compared in this section (with $U=1$), along with the flat profile corresponding to $\alpha = 1$.  The plot on the right depicts boundary data for $Q$ at the inlet of the ascending aorta.} 
\label{fig:55ves0}
\end{center}
\end{figure}

In this section, we study the effect of changing the shape of the velocity profile (changing $\alpha$) on waveforms produced by the $(A,Q)$ and $(A,U)$ systems with physiological inlet data. For this simulation, we use the fifty--five vessel network and windkessel boundary conditions described in the previous section. The inlet condition at the ascending aorta is the momentum $Q$ given in Figure \ref{fig:55ves0} and is obtained from experimental data \cite{Boileau15, murgo1980}.


\begin{table}[!htb]
\begin{center}
\begin{tabular}{| c | c | c |}
\hline
 $\nu$ $(\text{cm}^2\text{/s})$ & $p_0$ (mmHg) & $P_{\text{out}}$ (mmHg) \\ \hline
$3.302\times 10^{-2}$ & 75 & 0 \\ \hline
\end{tabular}
\caption{Physical parameters for fifty--five vessel network with physiological momentum inlet data.}
\label{table:windkessel1}
\end{center}
\end{table}

\begin{table}[!htb]
\begin{center}
\begin{tabular}{| c | c | c | c |}
\hline
$\Delta t$ (s) & $h$ (cm) & $k$ (polynomial degree) & numerical flux  \\ \hline
$10^{-4}$ & 1 & 1 & UP \text{ and } LLF \\ \hline
\end{tabular}
\caption{Numerical parameters for fifty--five vessel network with windkessel terminal boundary conditions.}
\label{table:windkessel2}
\end{center}
\end{table}

\begin{figure}[!htb]
\begin{center}
\includegraphics[scale=0.35]{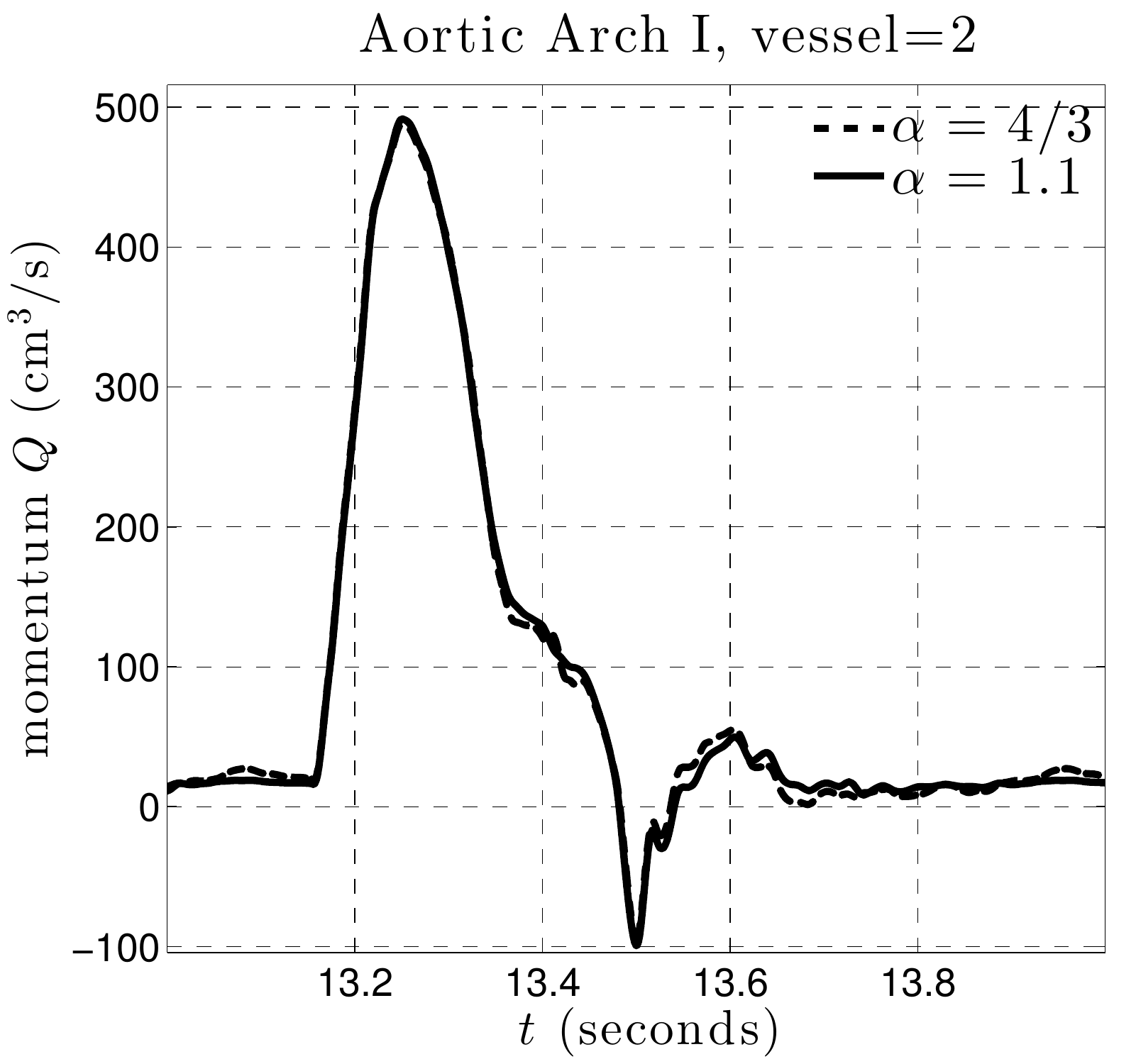} \includegraphics[scale=0.35]{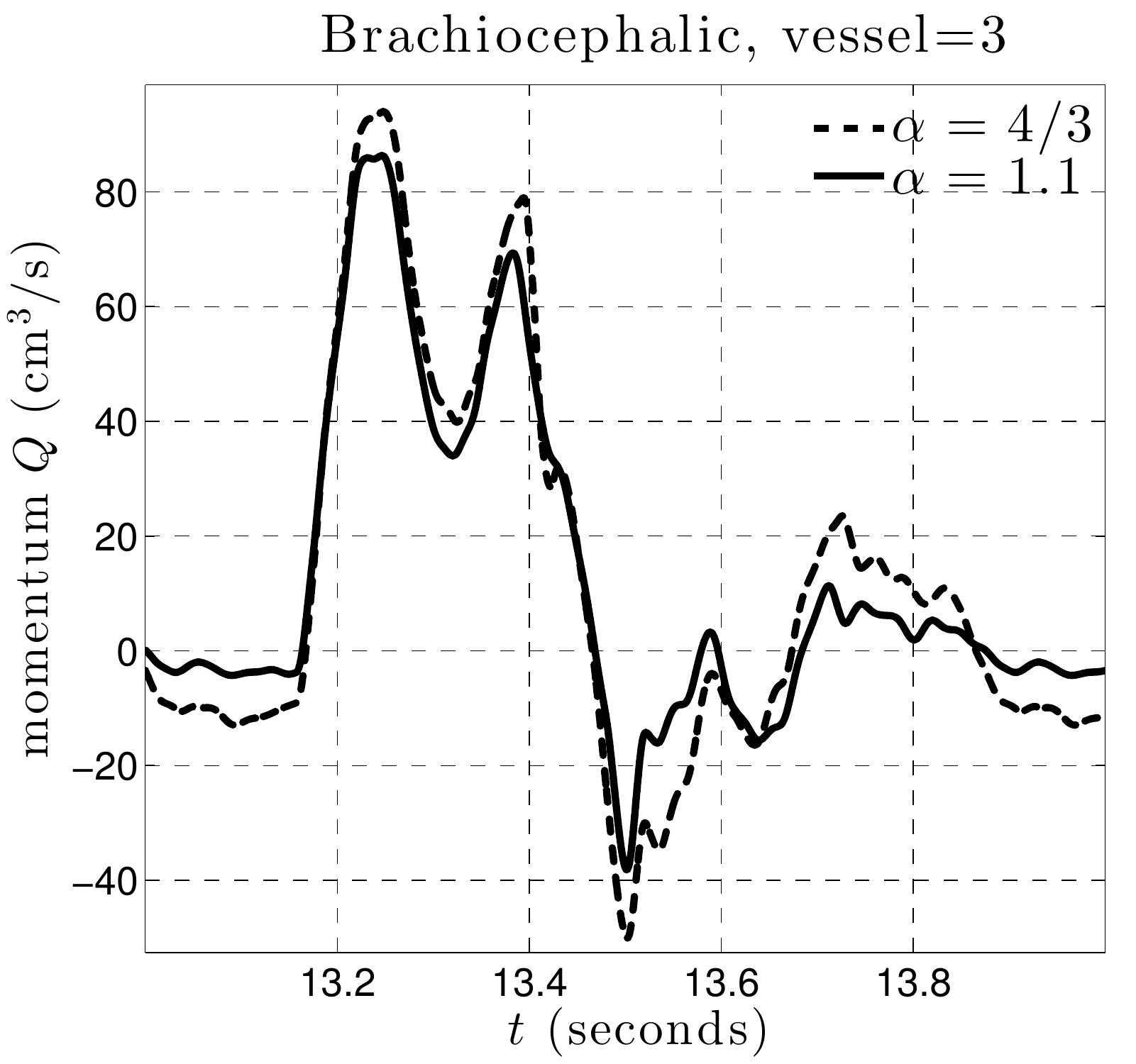} \includegraphics[scale=0.35]{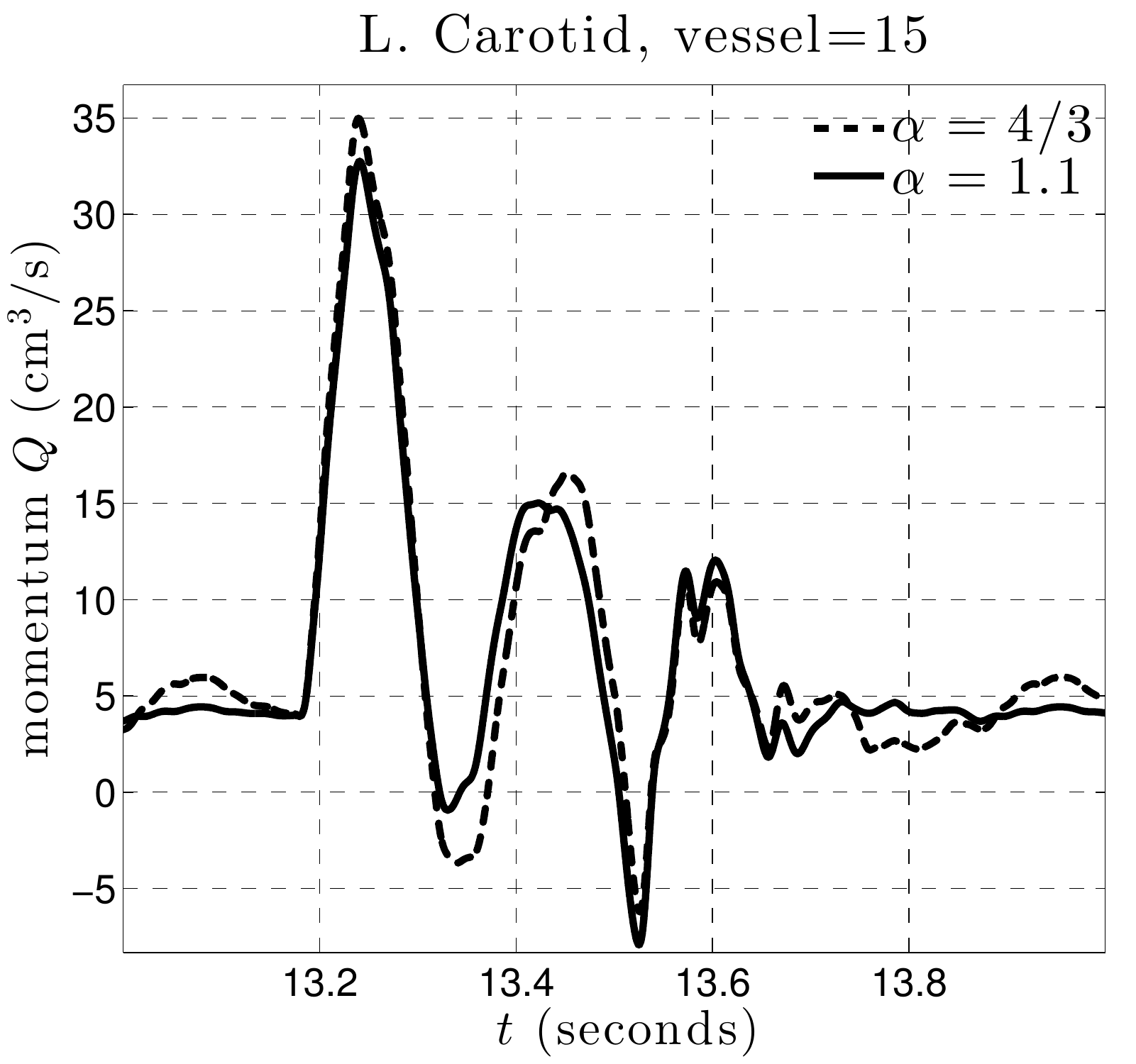} \\
\includegraphics[scale=0.35]{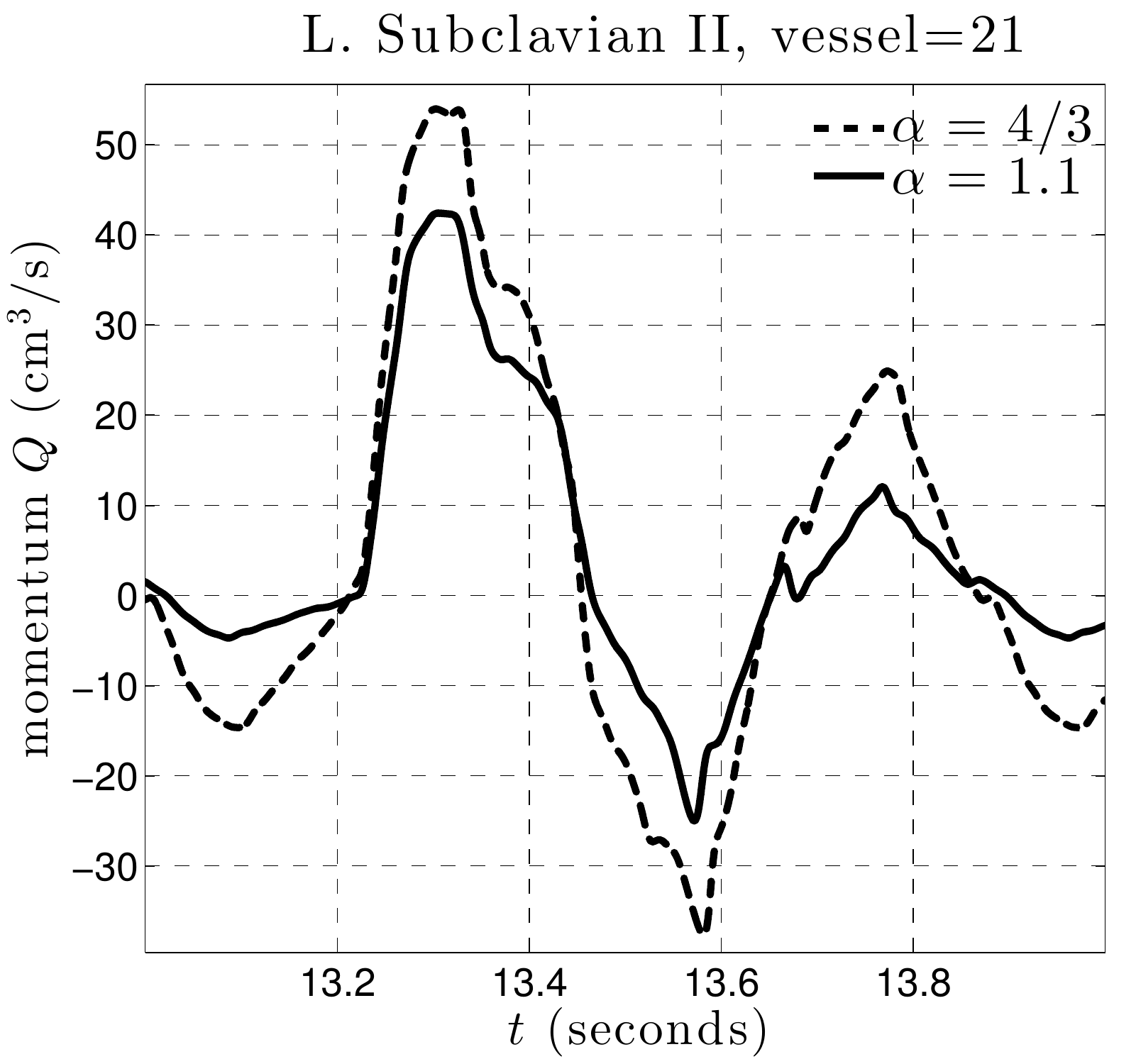} \includegraphics[scale=0.35]{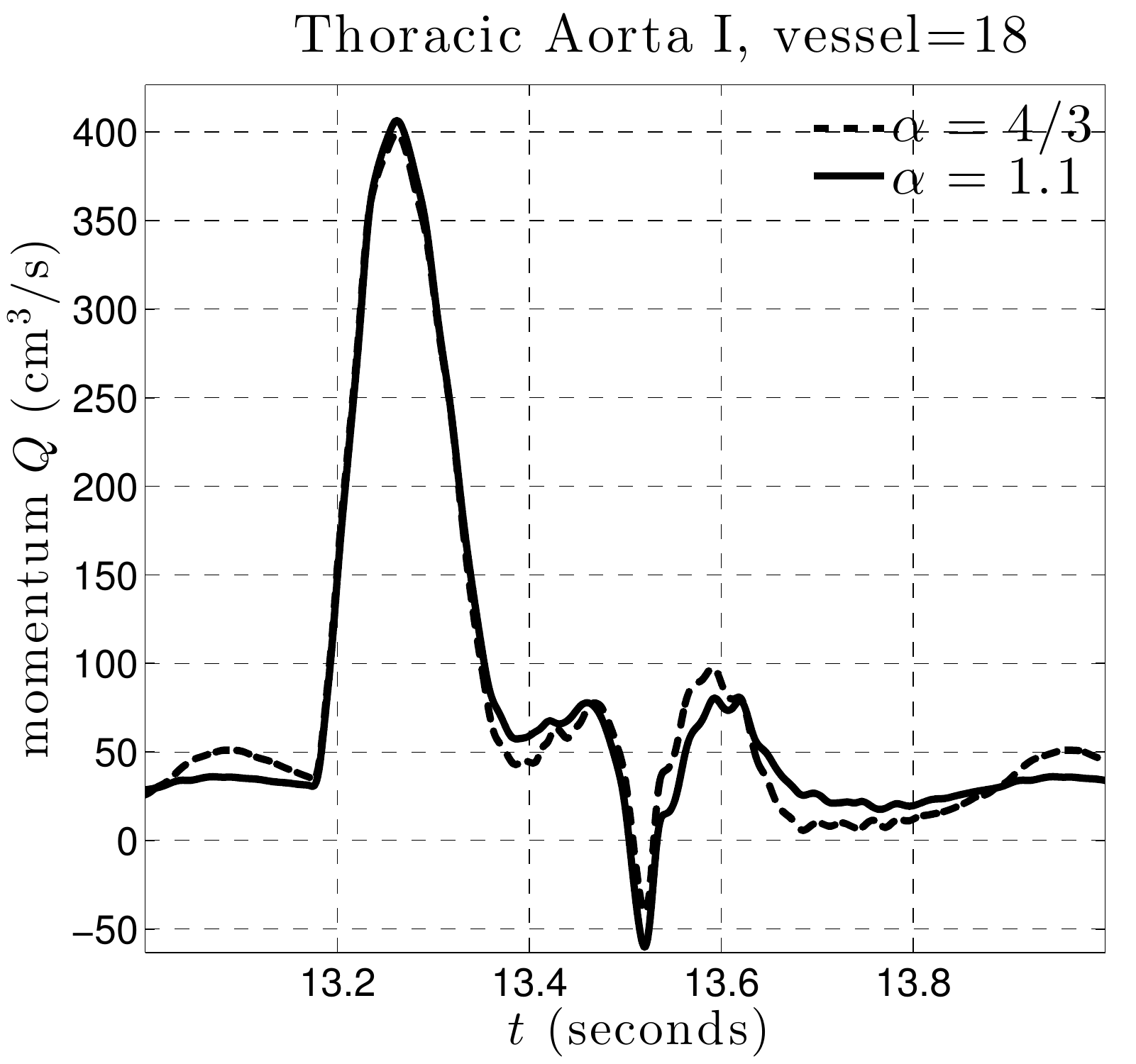} \includegraphics[scale=0.35]{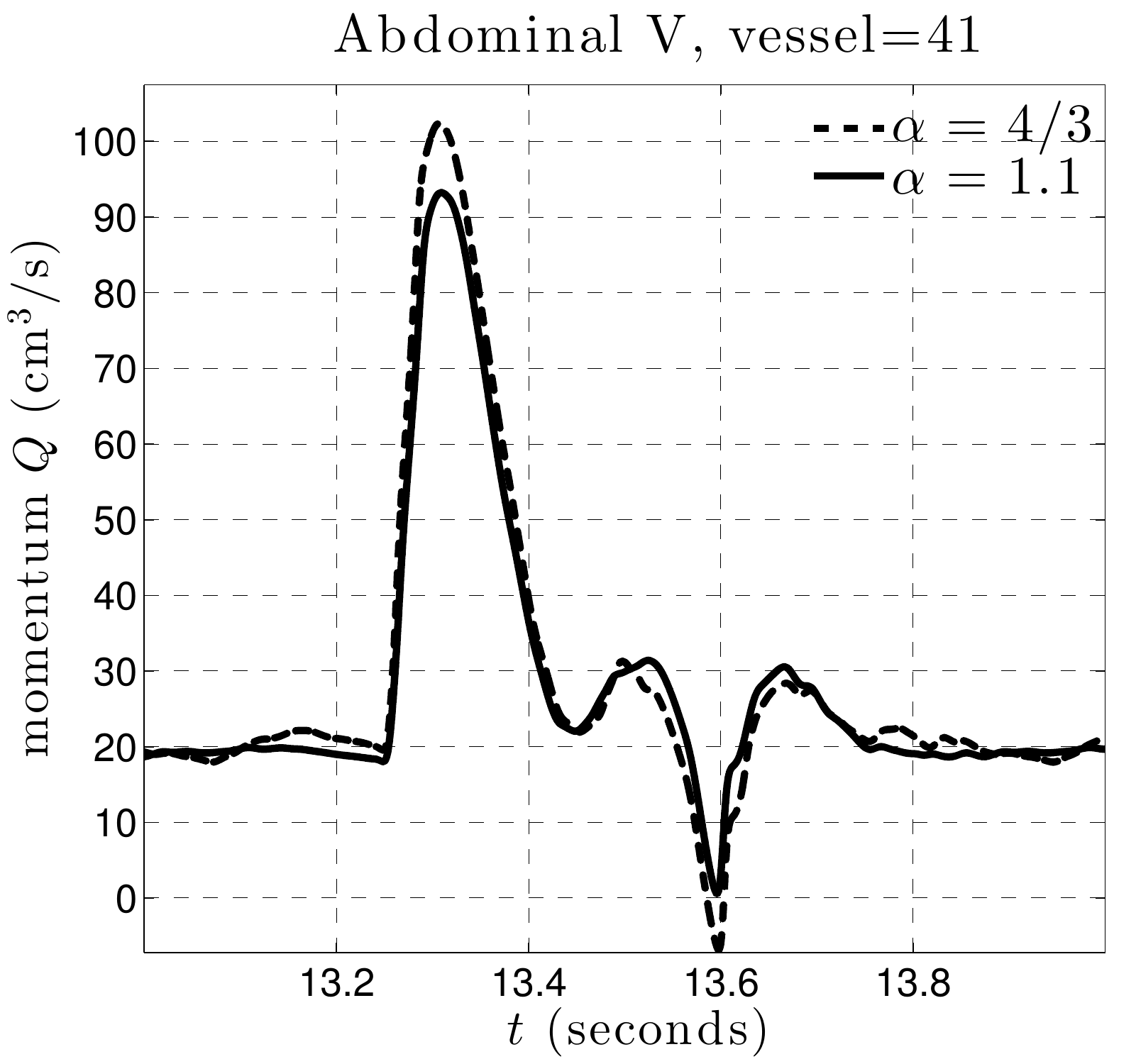} \\
\includegraphics[scale=0.35]{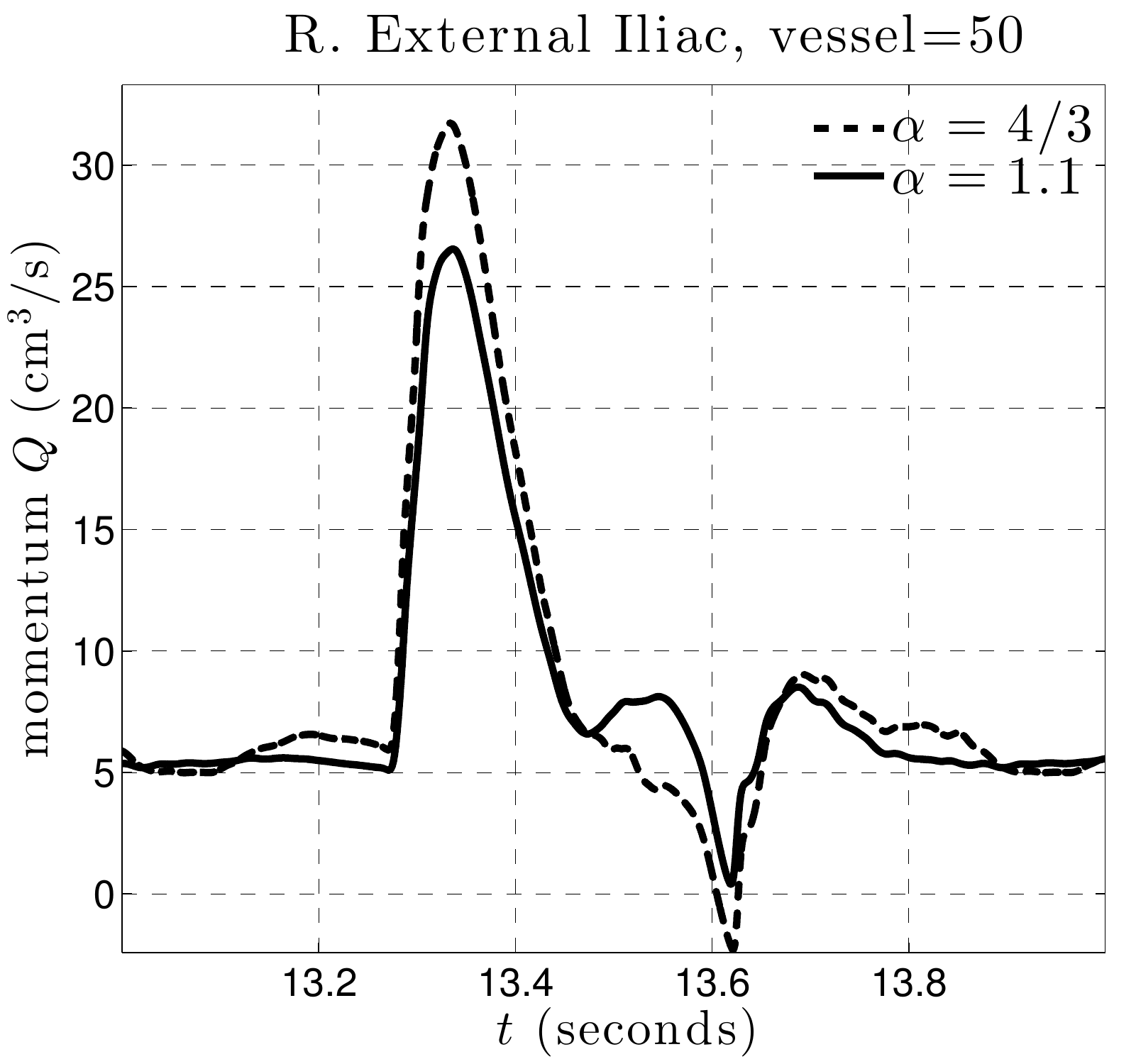} \includegraphics[scale=0.35]{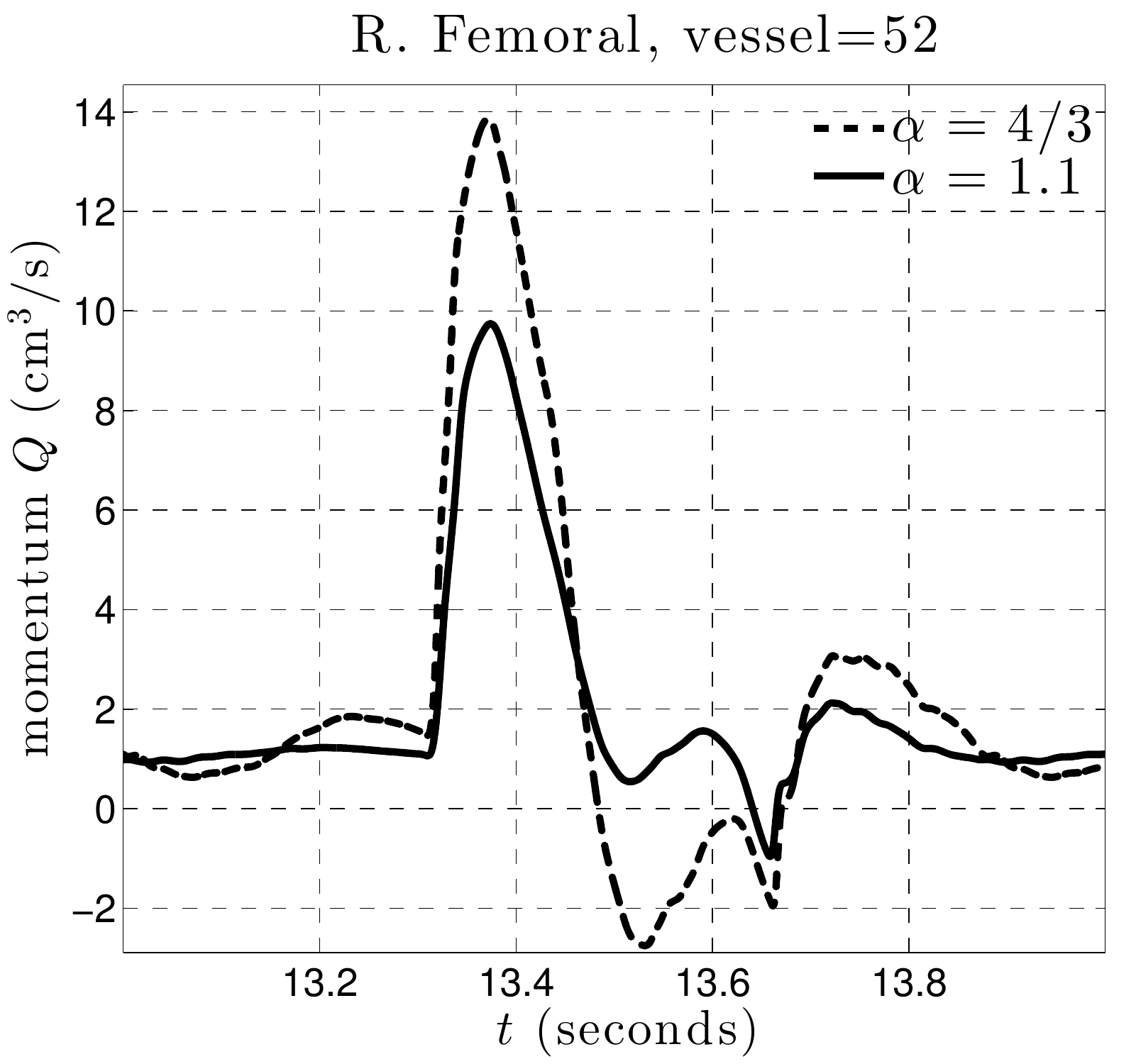} \includegraphics[scale=0.35]{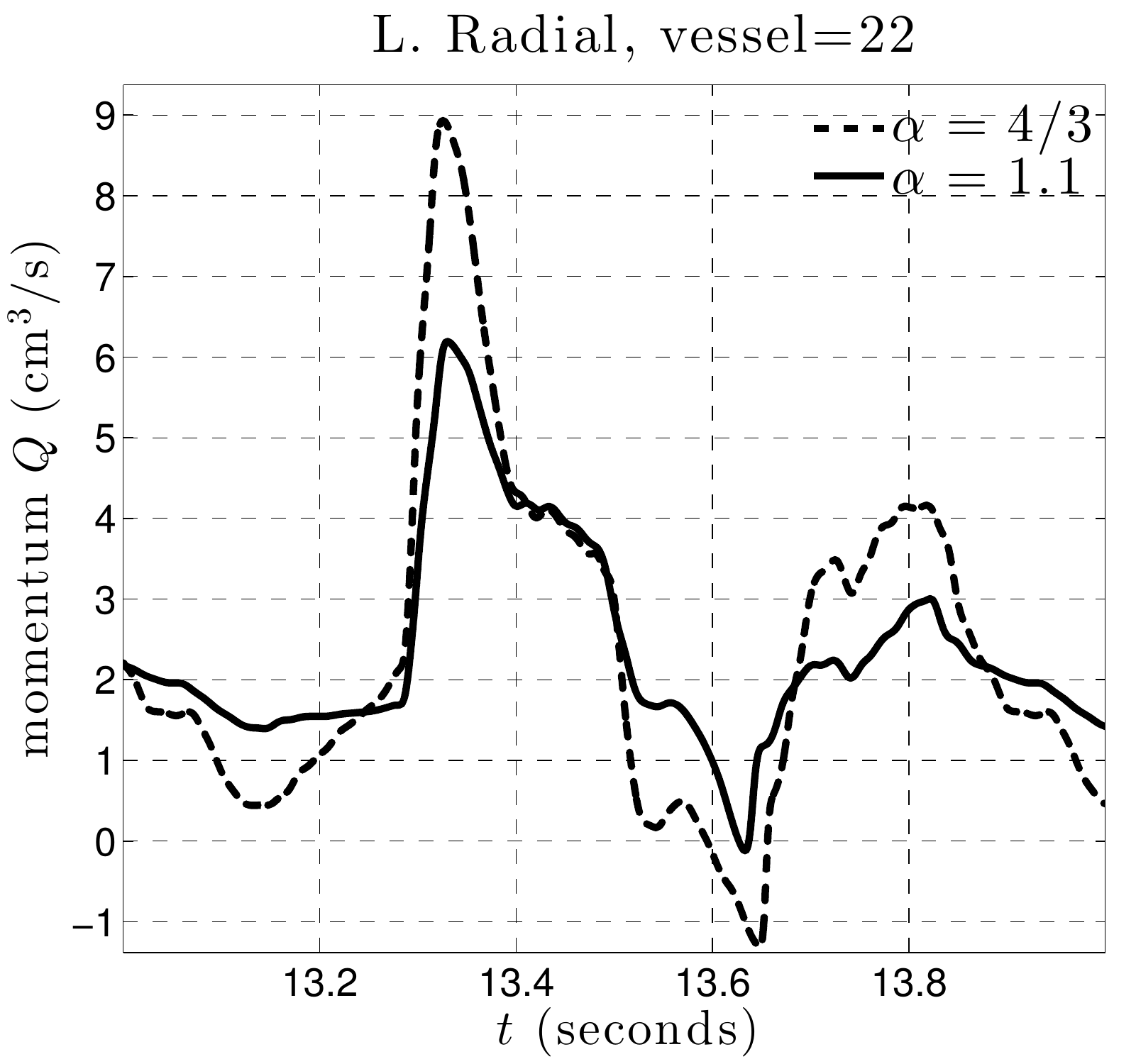} \\
\caption{A comparison of the momentum waveforms from the $(A,Q)$ system with $\alpha = 1.1$ and $\alpha=4/3$.}
\label{fig:55vesmom1}
\end{center}
\end{figure}

\begin{figure}[!htb]
\begin{center}
\includegraphics[scale=0.35]{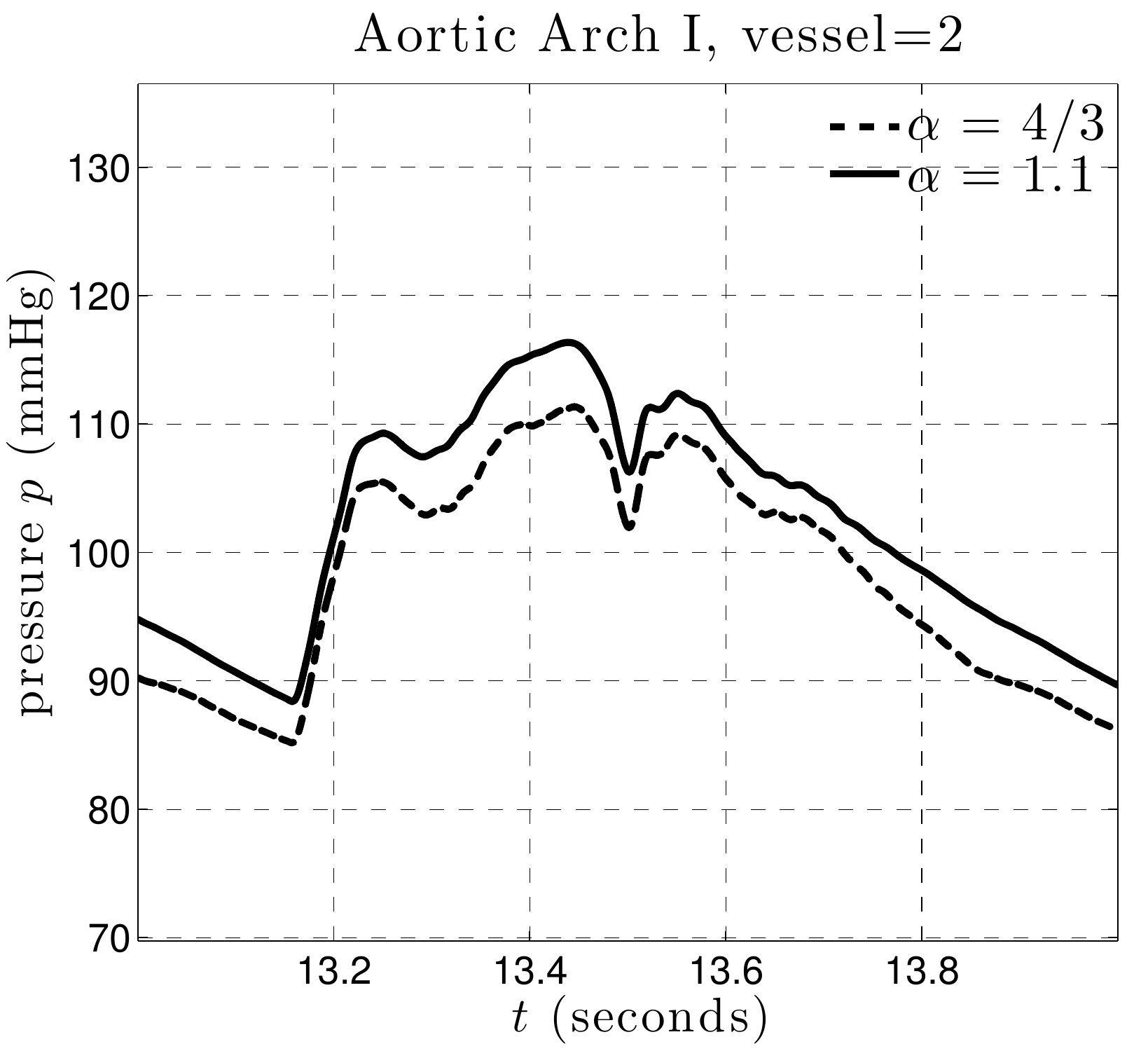} \includegraphics[scale=0.35]{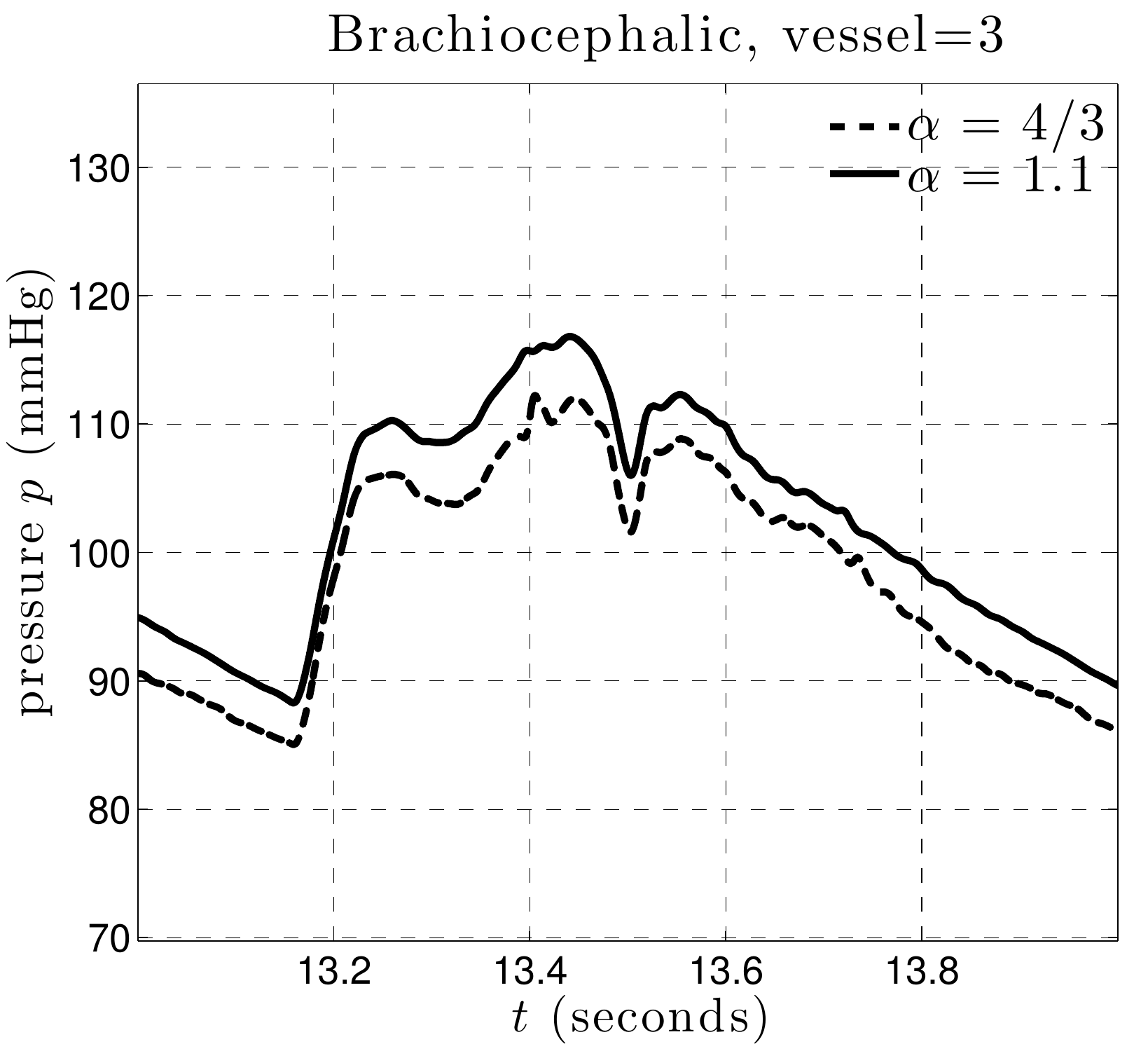} \includegraphics[scale=0.35]{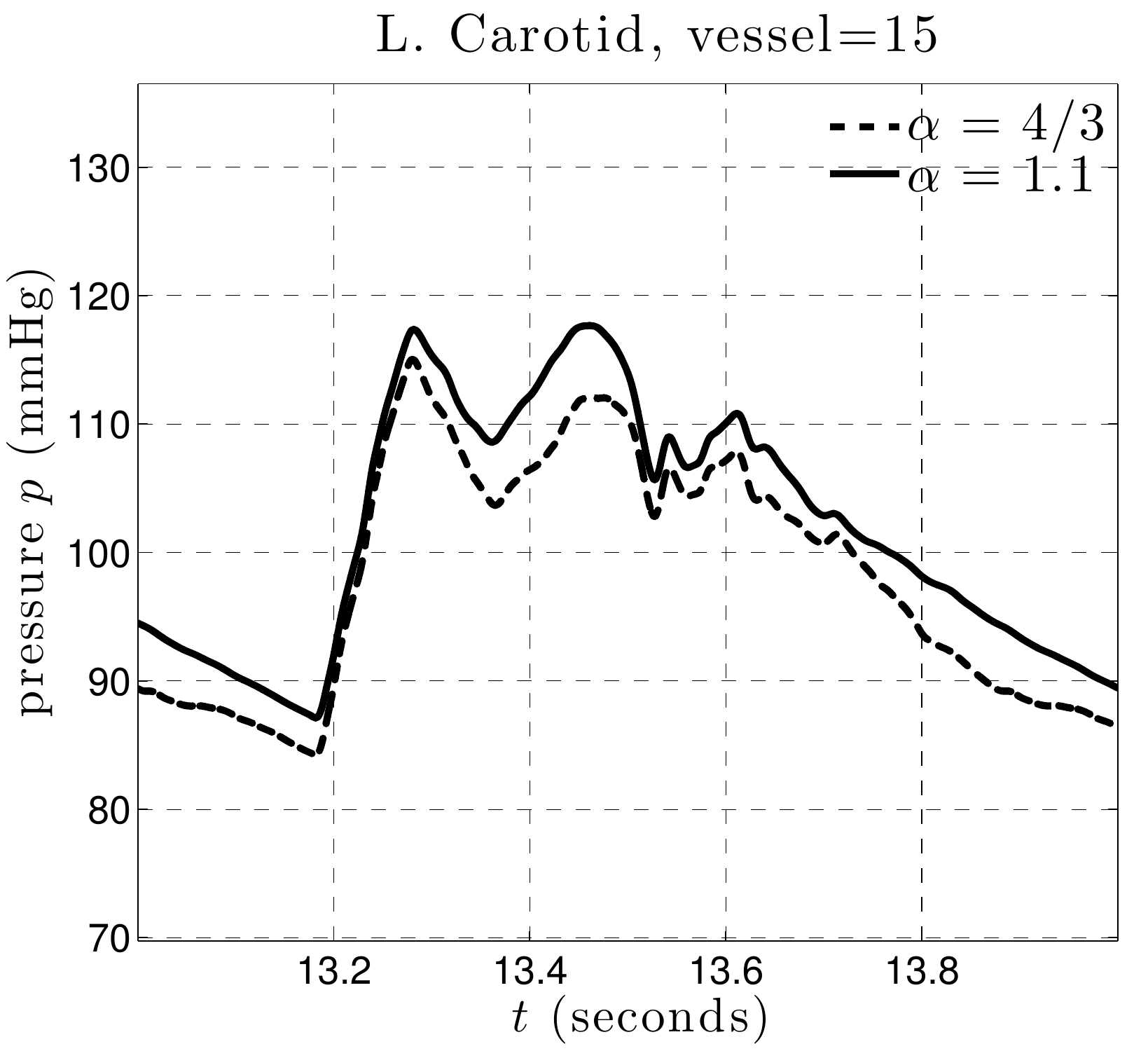} \\
\includegraphics[scale=0.35]{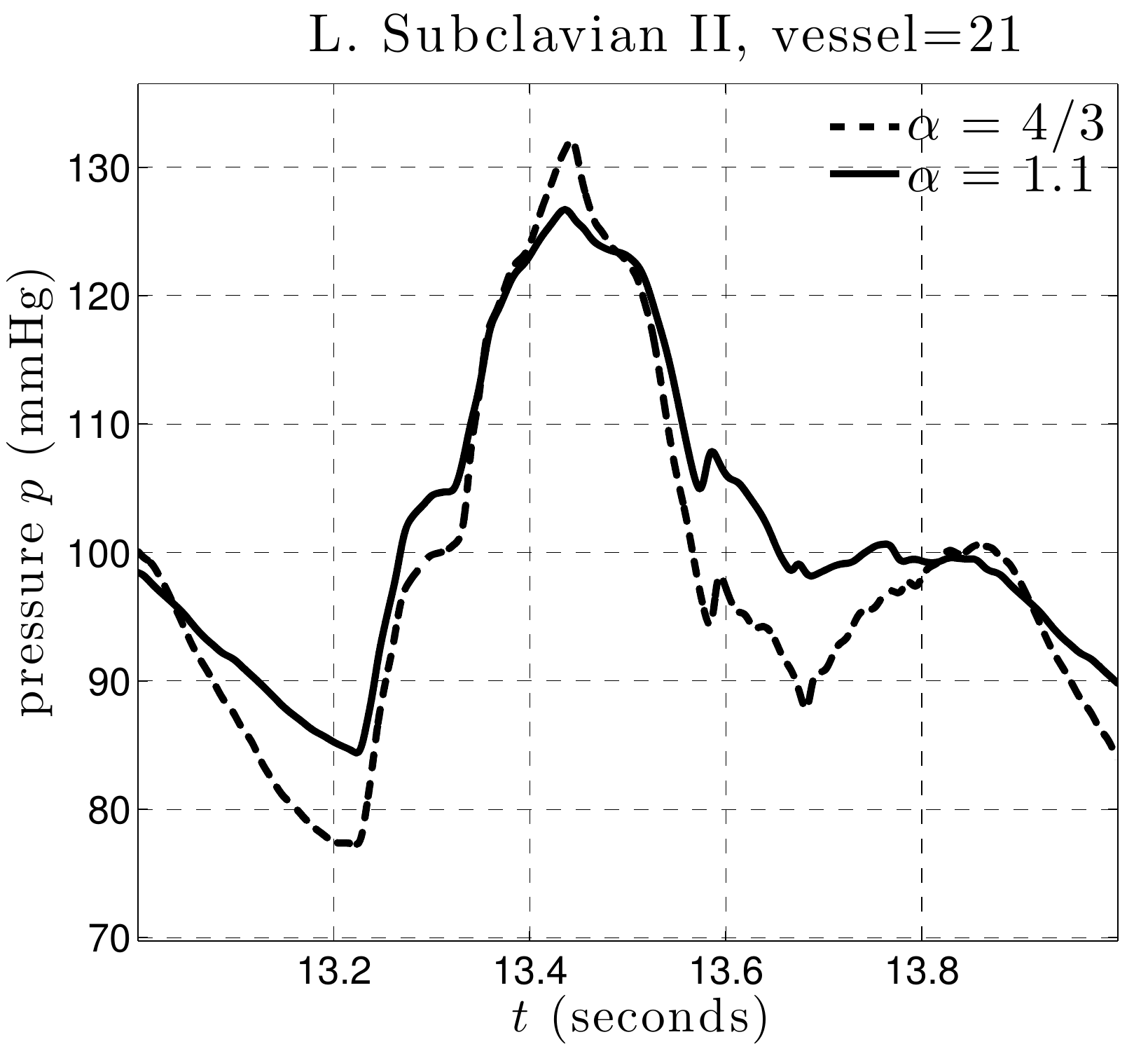} \includegraphics[scale=0.35]{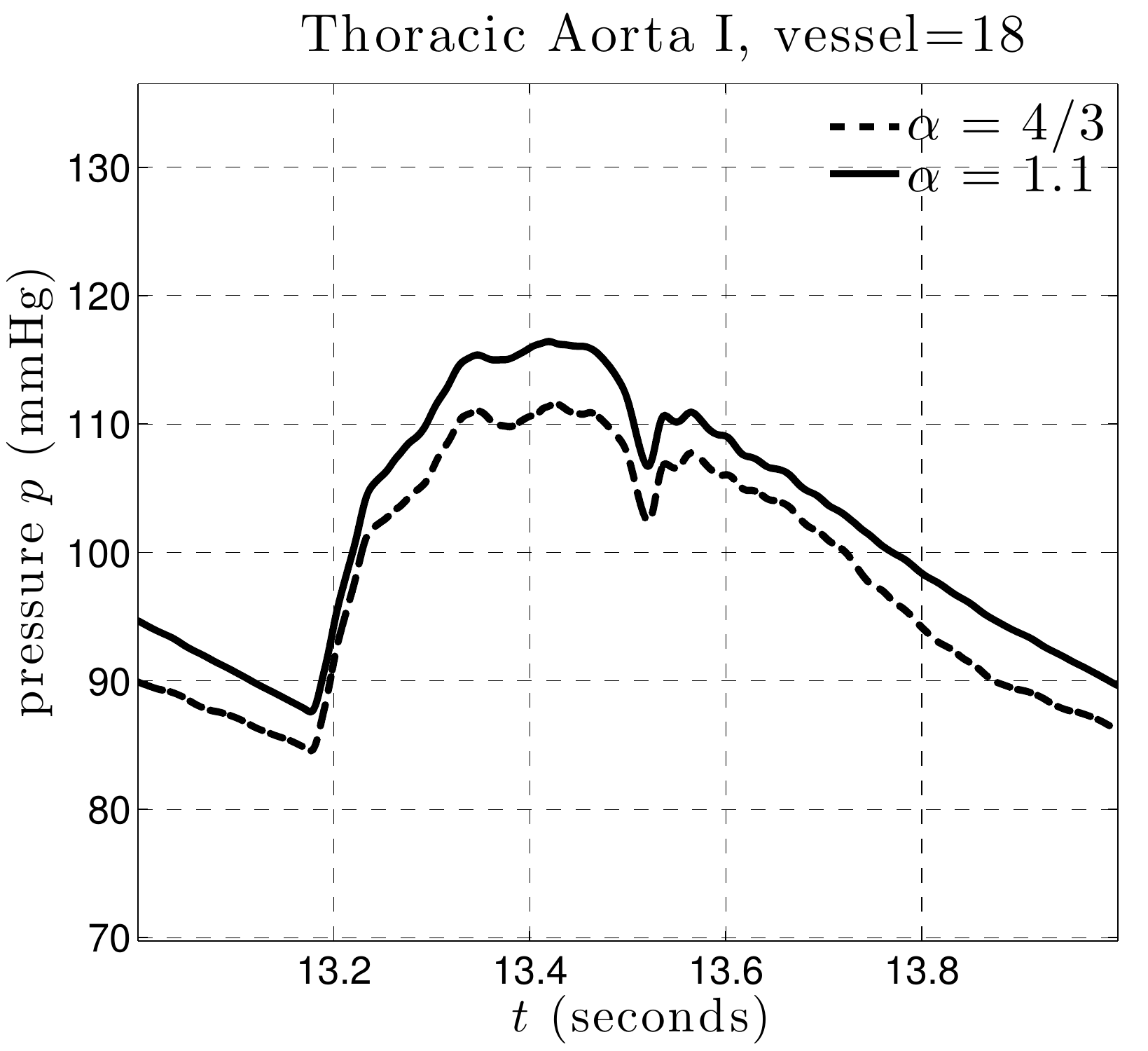} \includegraphics[scale=0.35]{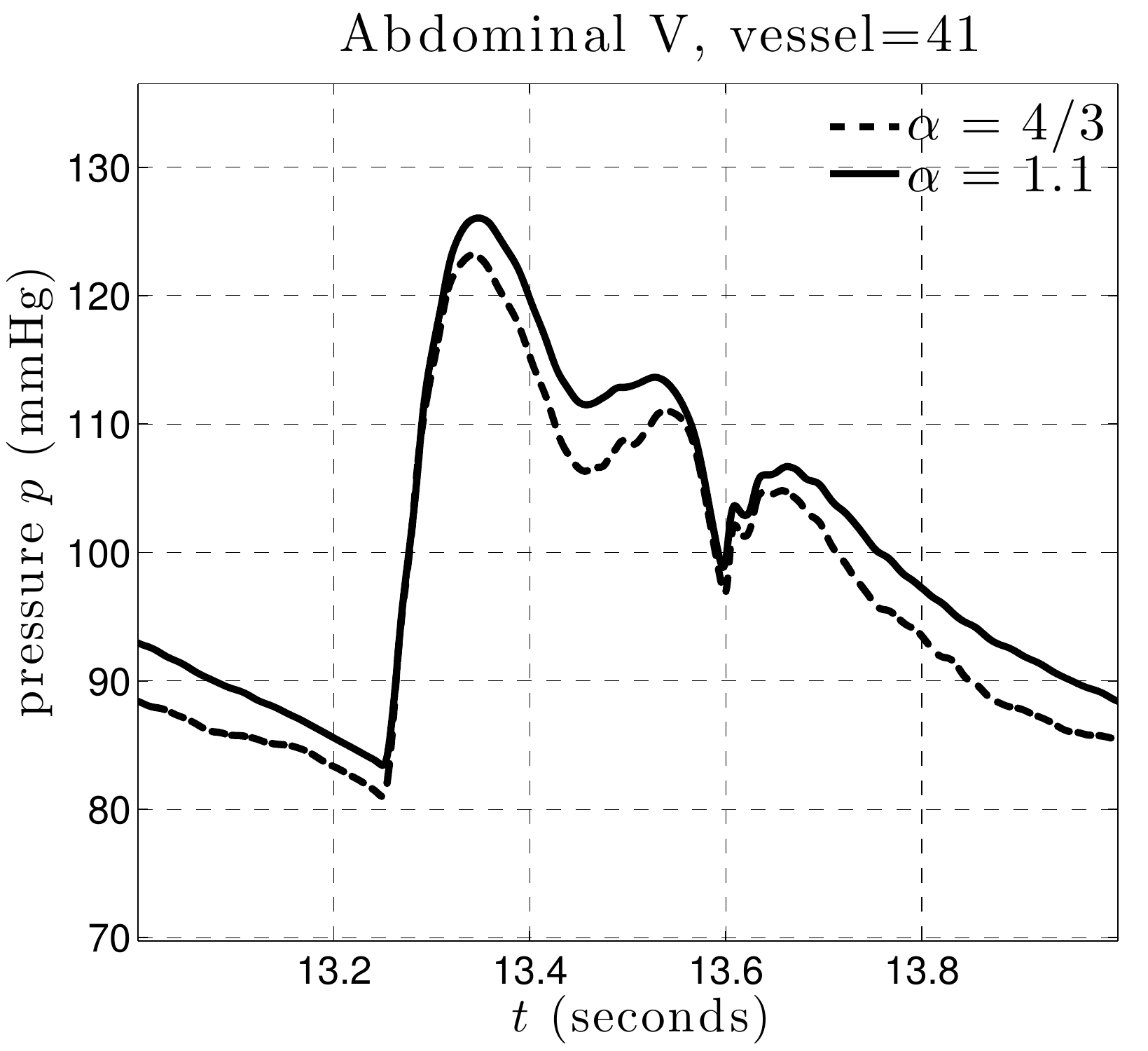} \\
\includegraphics[scale=0.35]{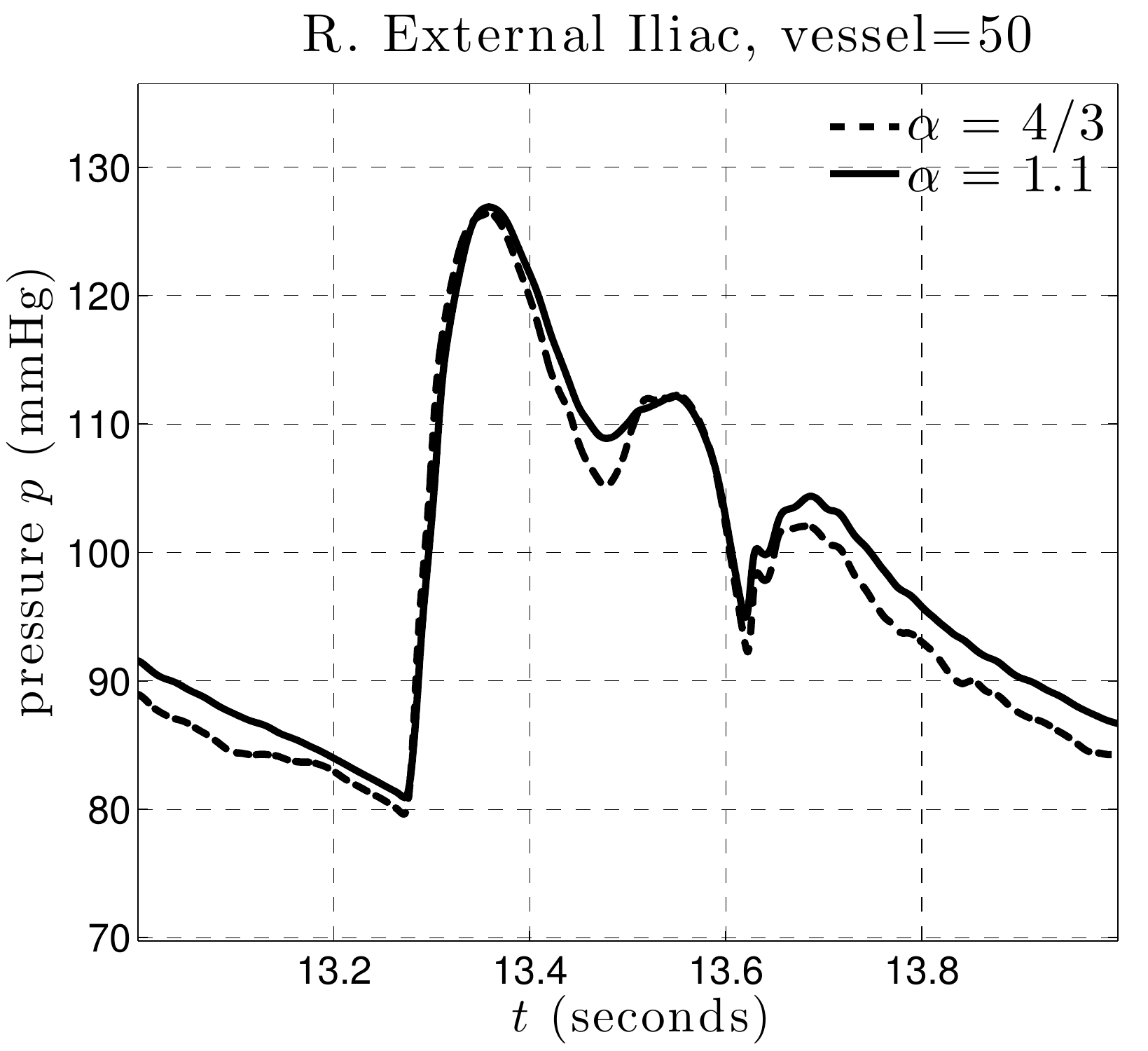} \includegraphics[scale=0.35]{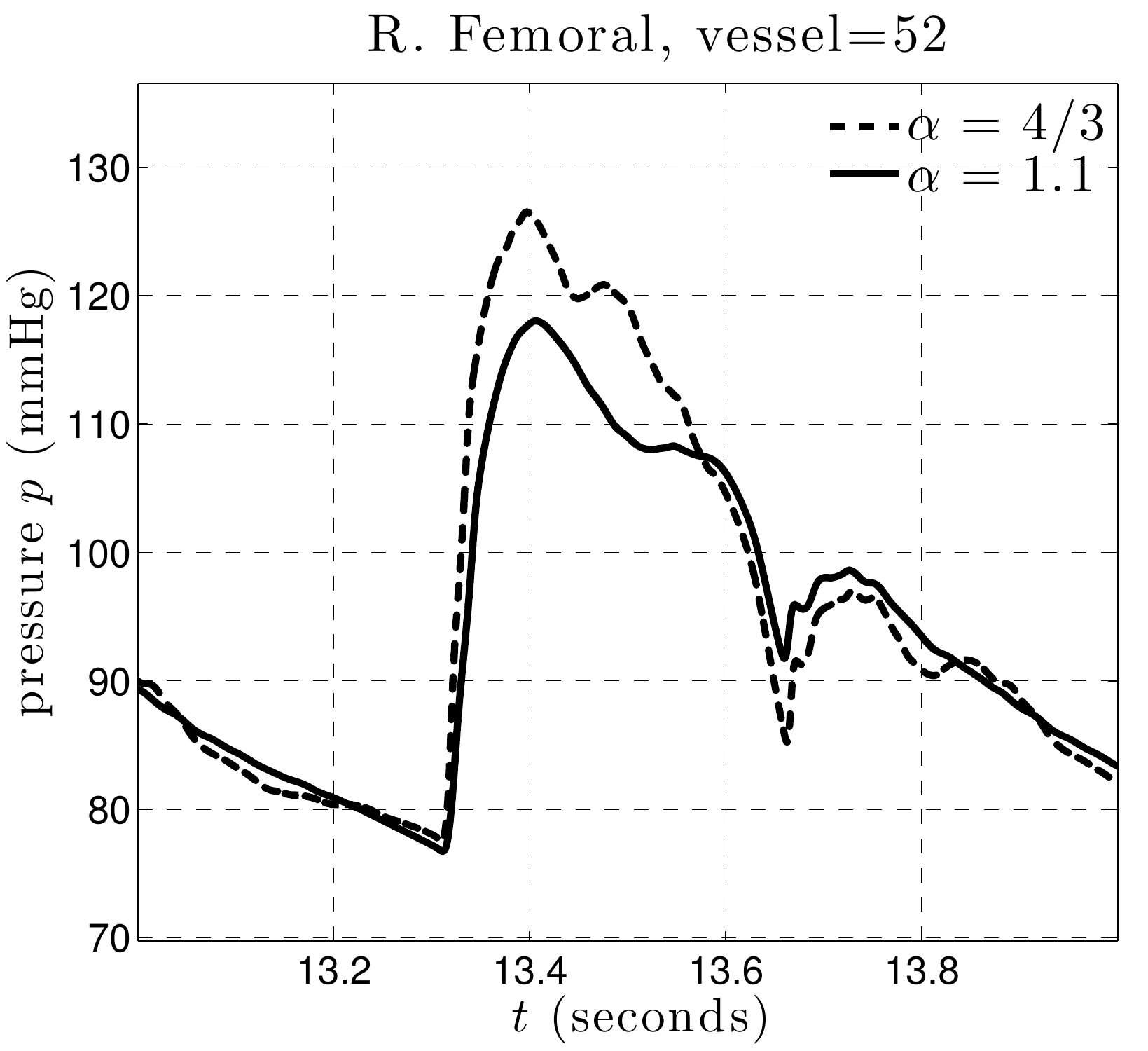} \includegraphics[scale=0.35]{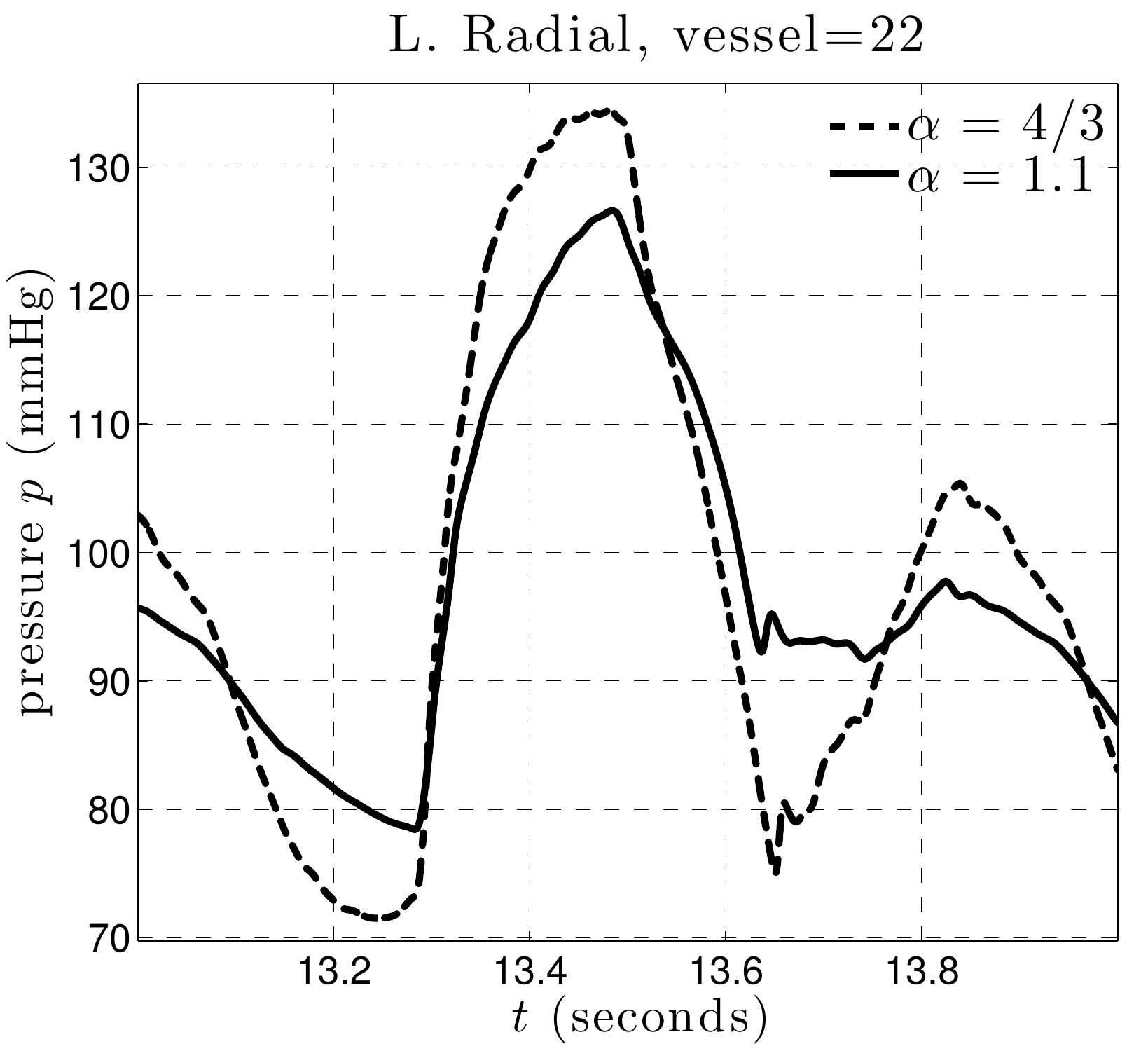} \\
\caption{A comparison of the pressure waveforms from the $(A,Q)$ system with $\alpha = 1.1$ and $4/3$.}
\label{fig:55vespress1}
\end{center}
\end{figure}

\begin{figure}[!htb]
\begin{center}
\includegraphics[scale=0.35]{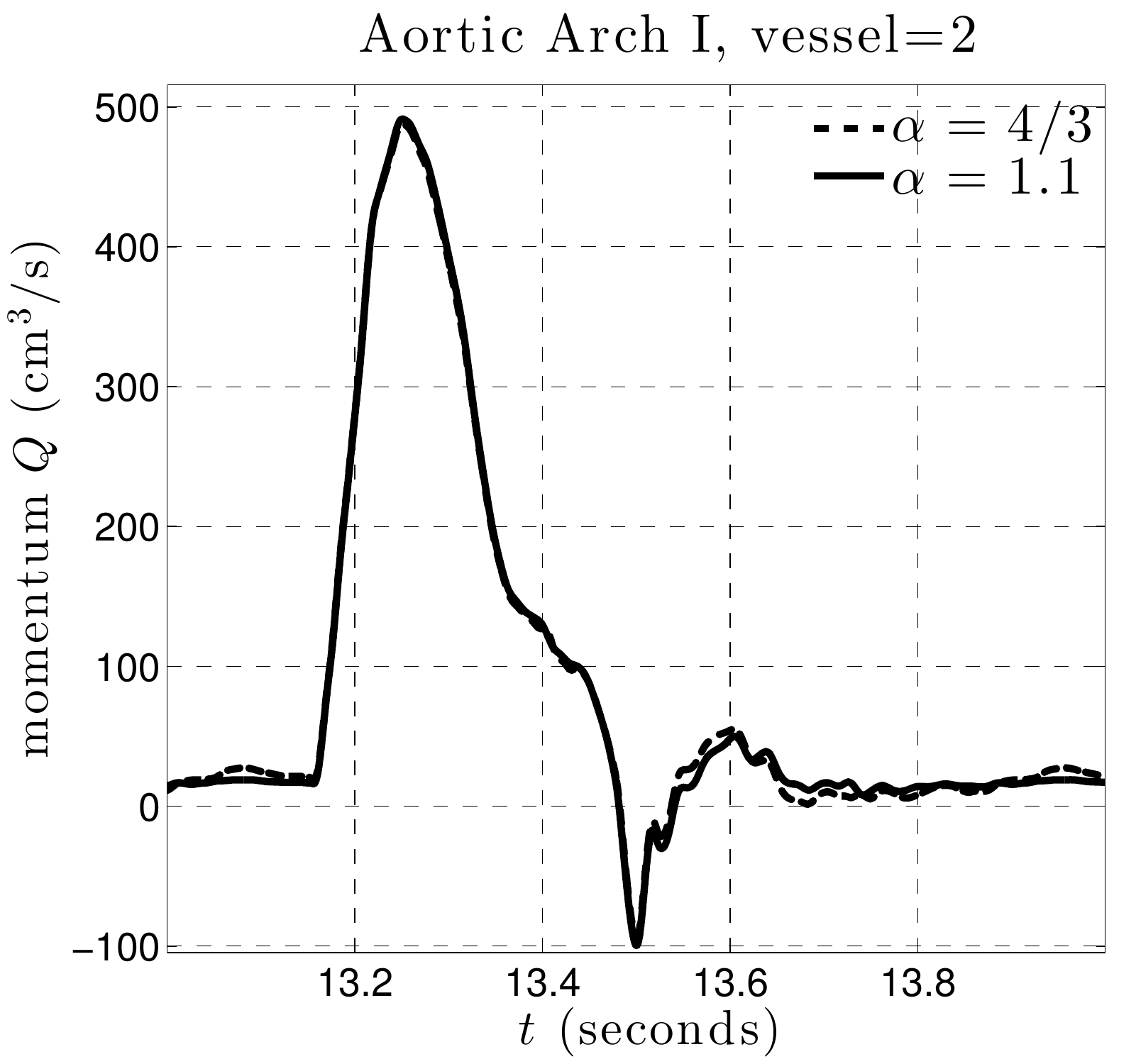} \includegraphics[scale=0.35]{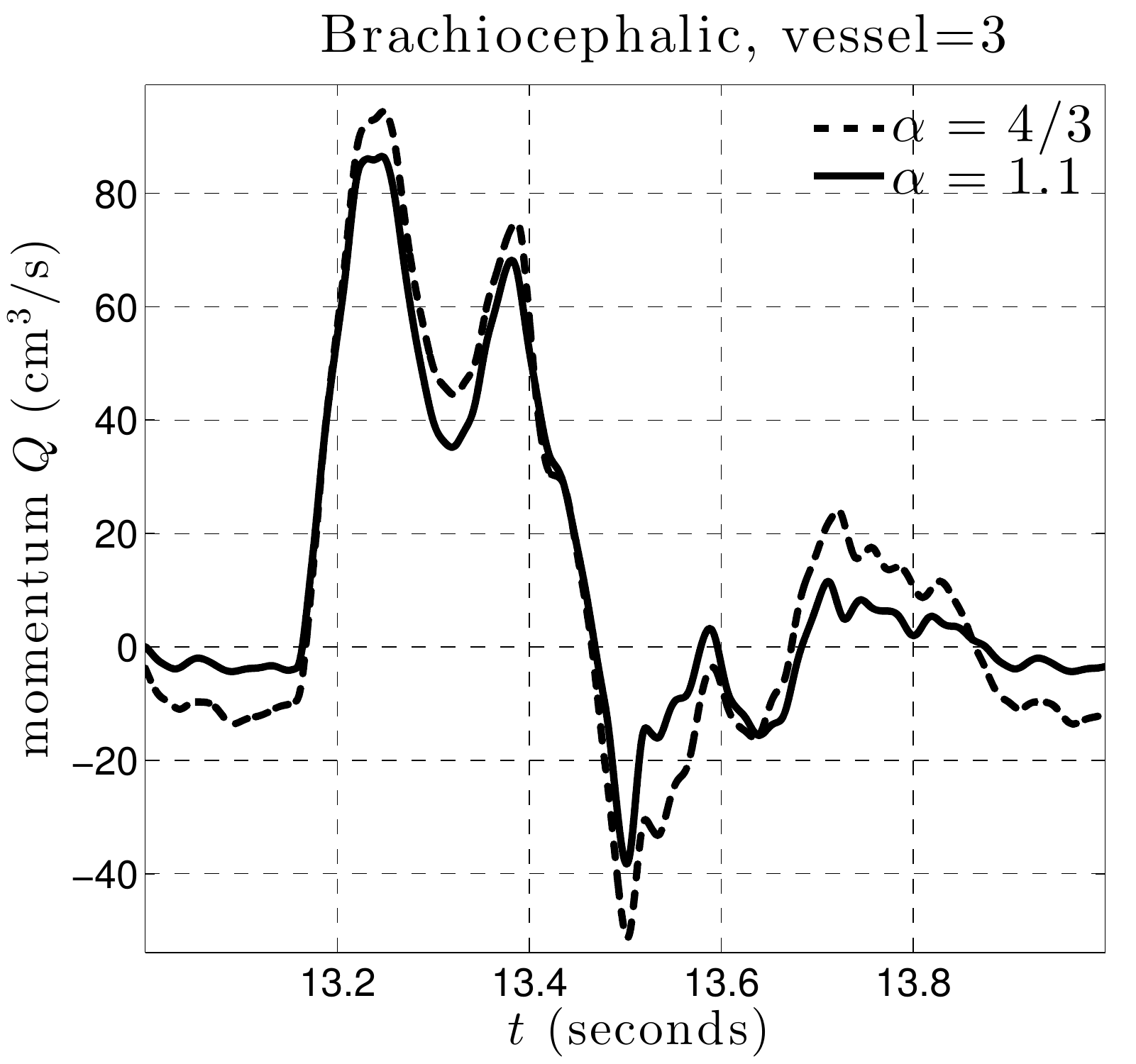} \includegraphics[scale=0.35]{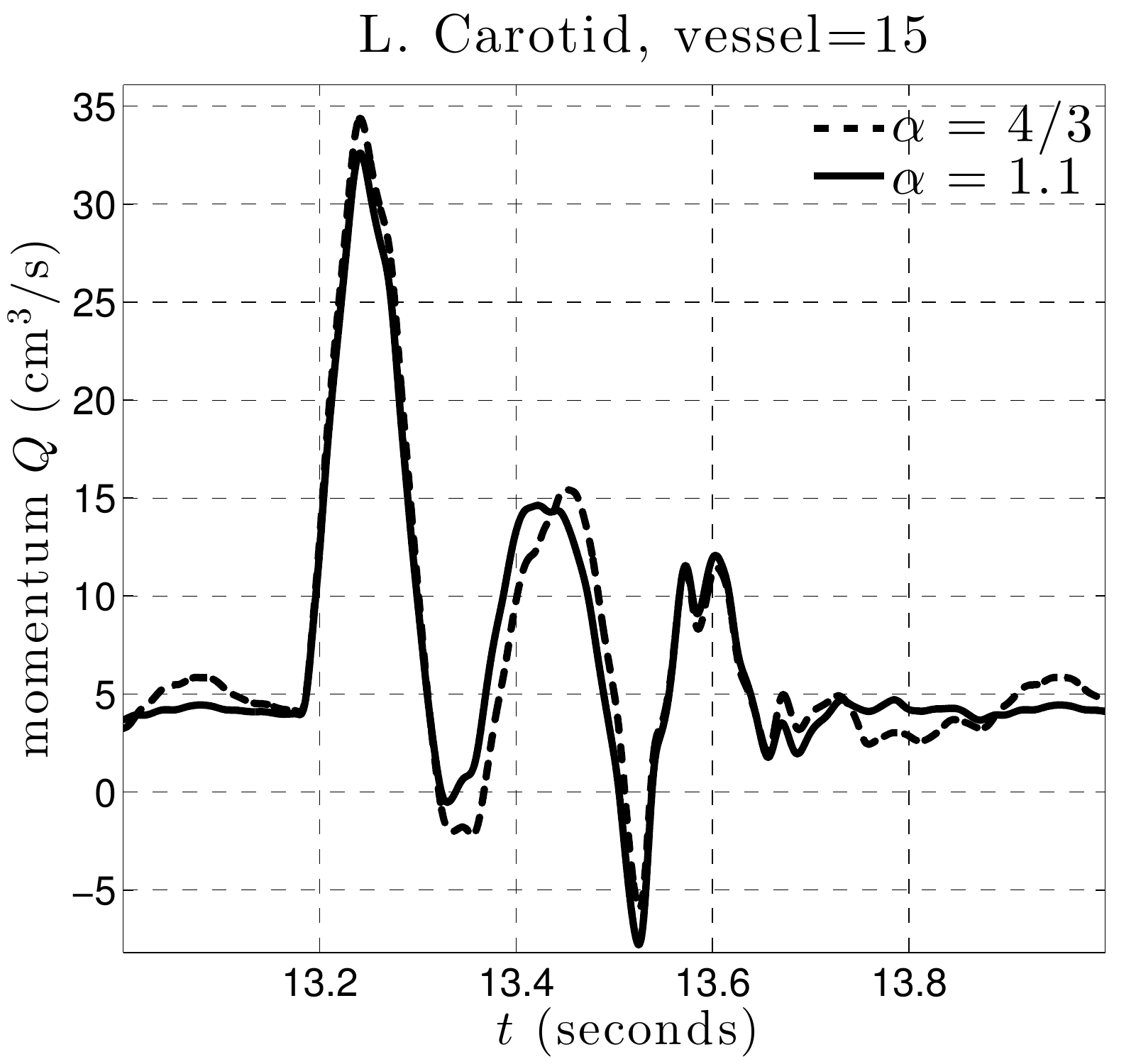} \\
\includegraphics[scale=0.35]{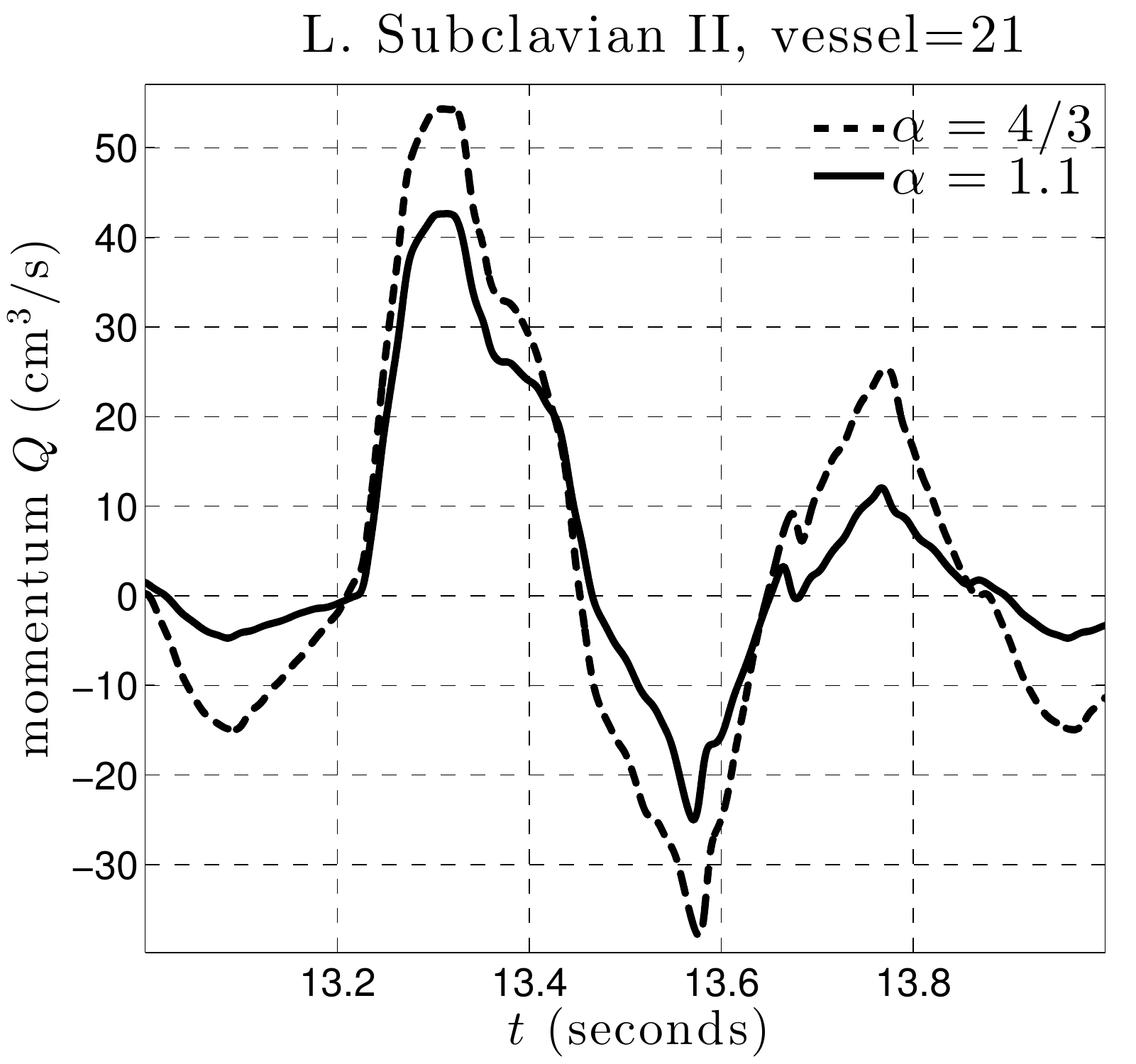} \includegraphics[scale=0.35]{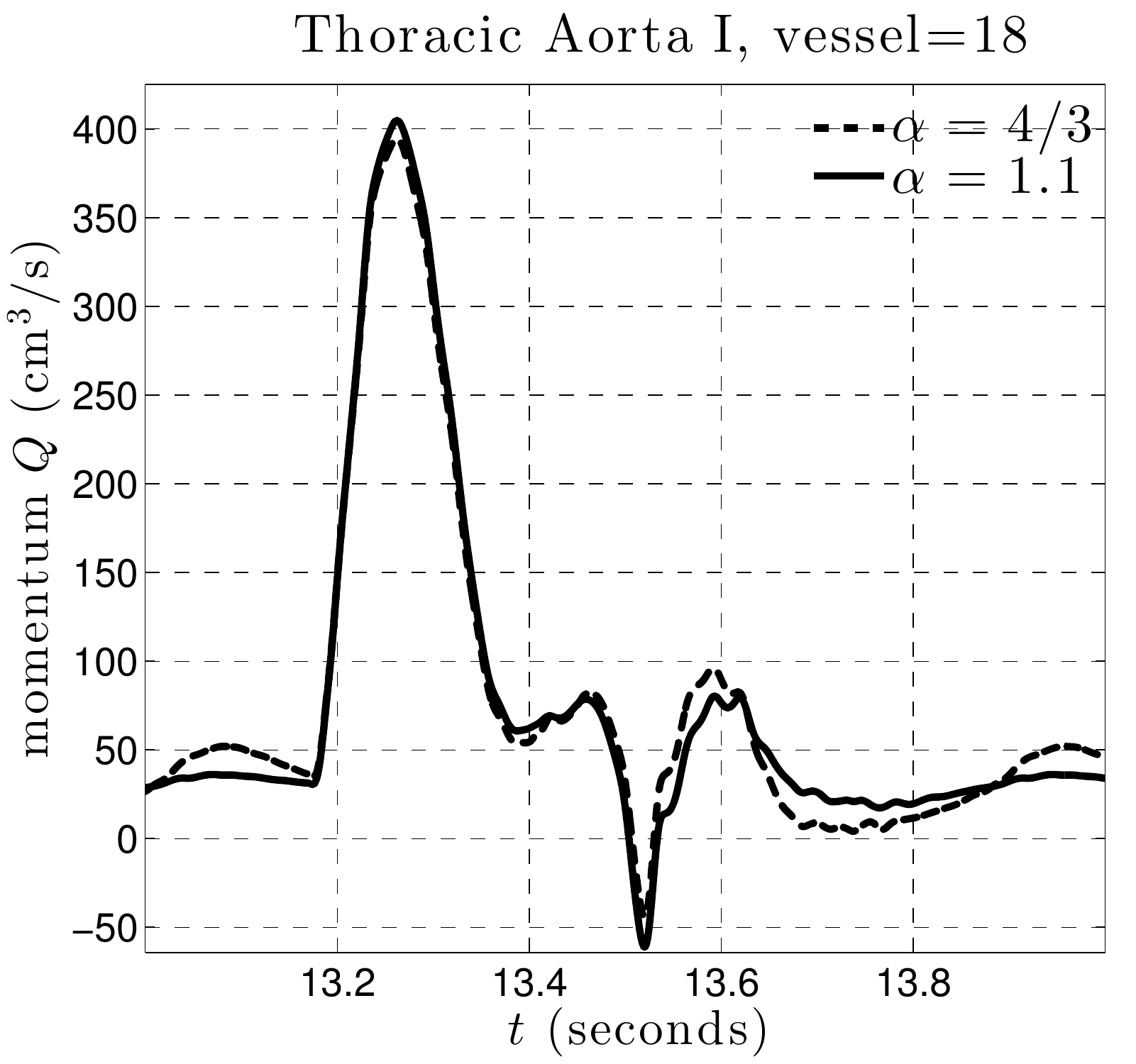} \includegraphics[scale=0.35]{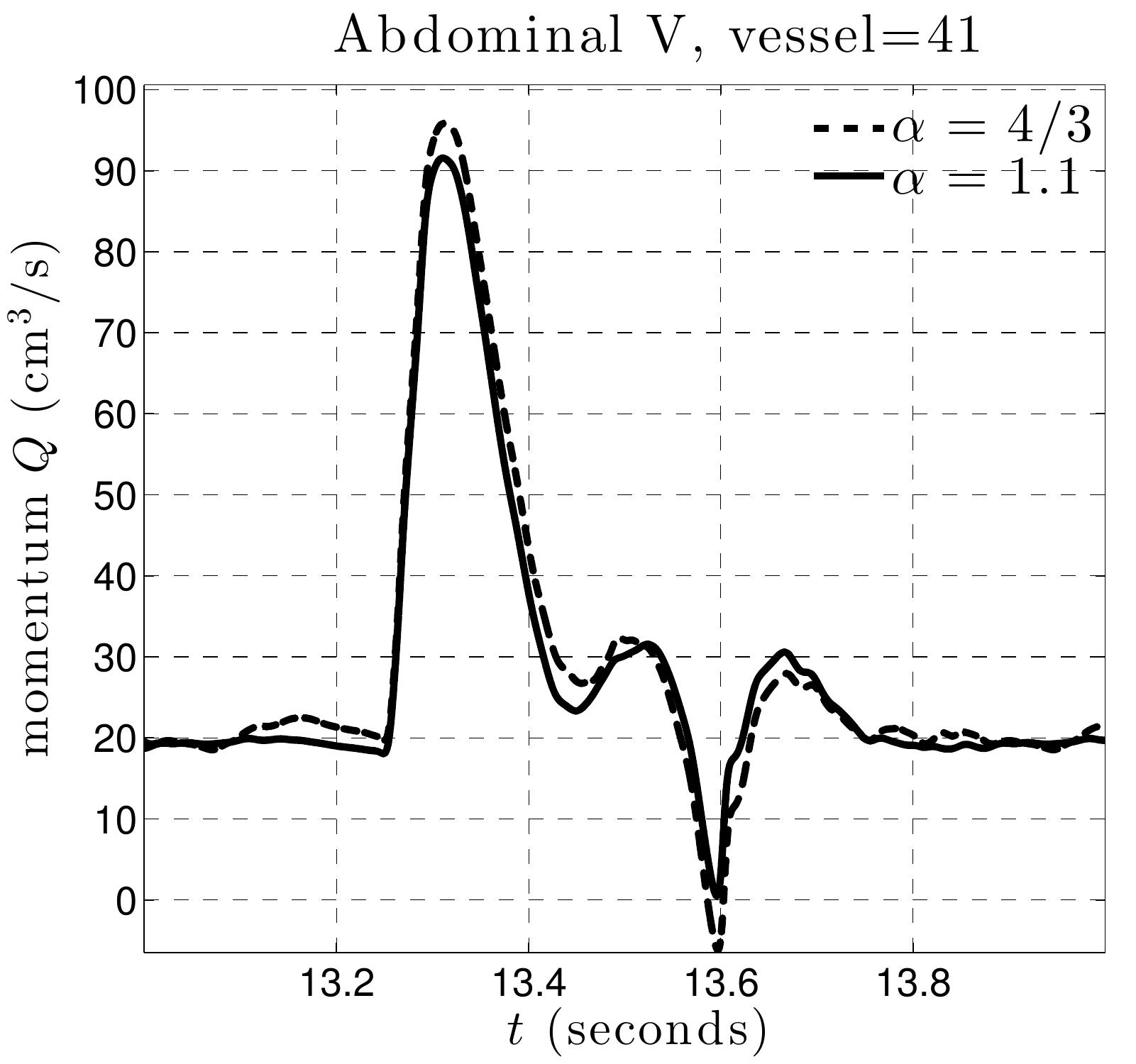} \\
\includegraphics[scale=0.35]{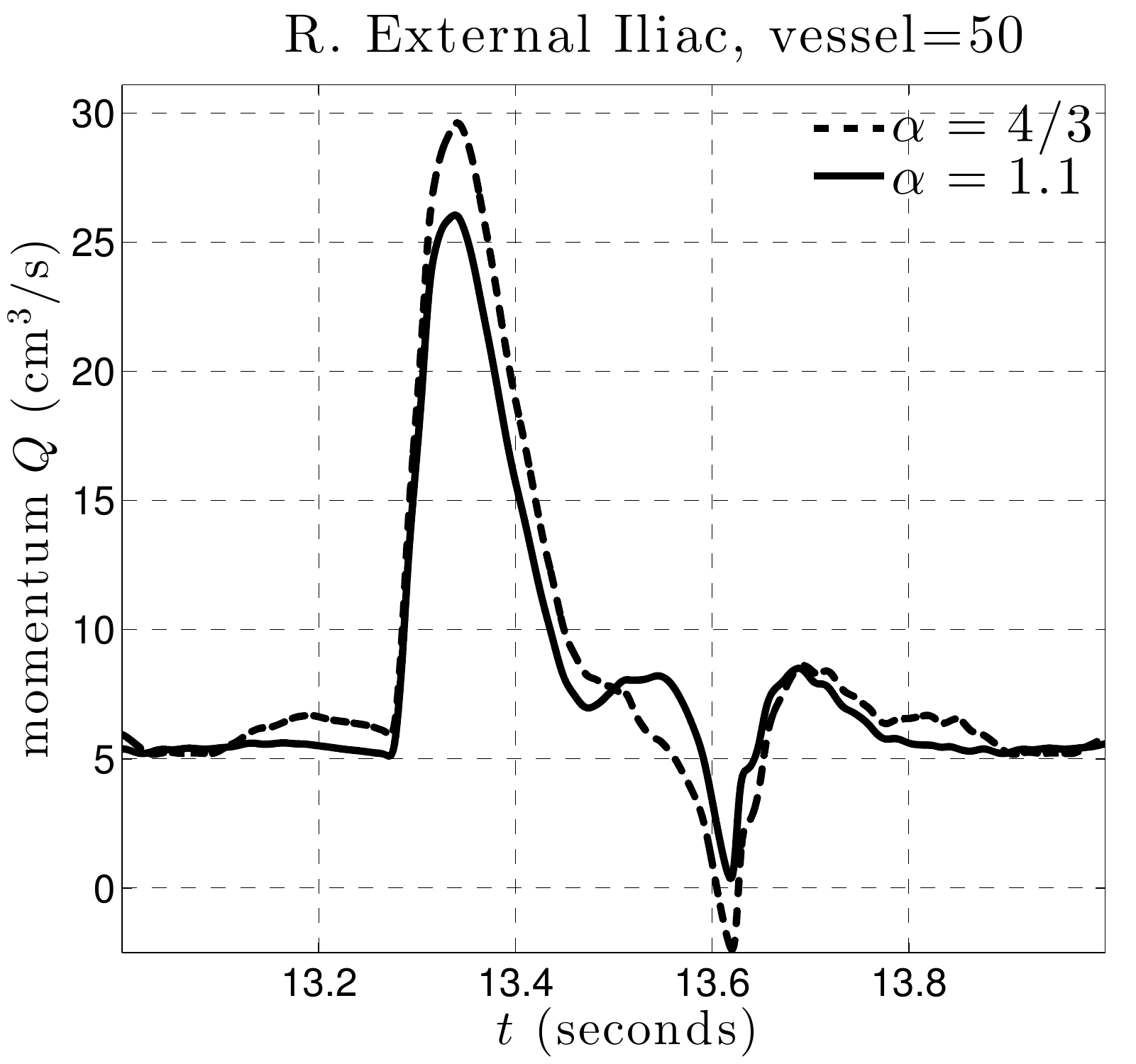} \includegraphics[scale=0.35]{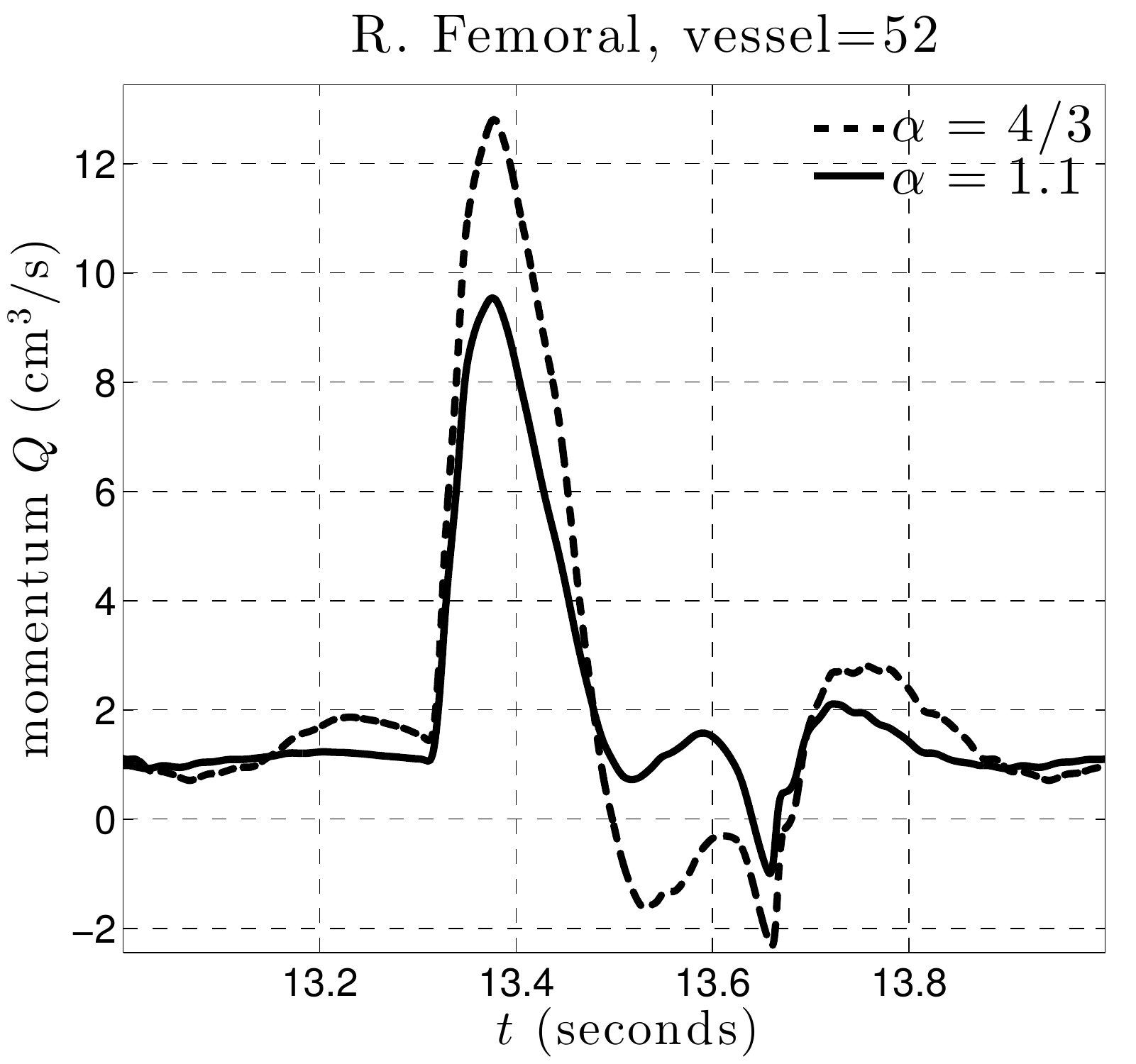} \includegraphics[scale=0.35]{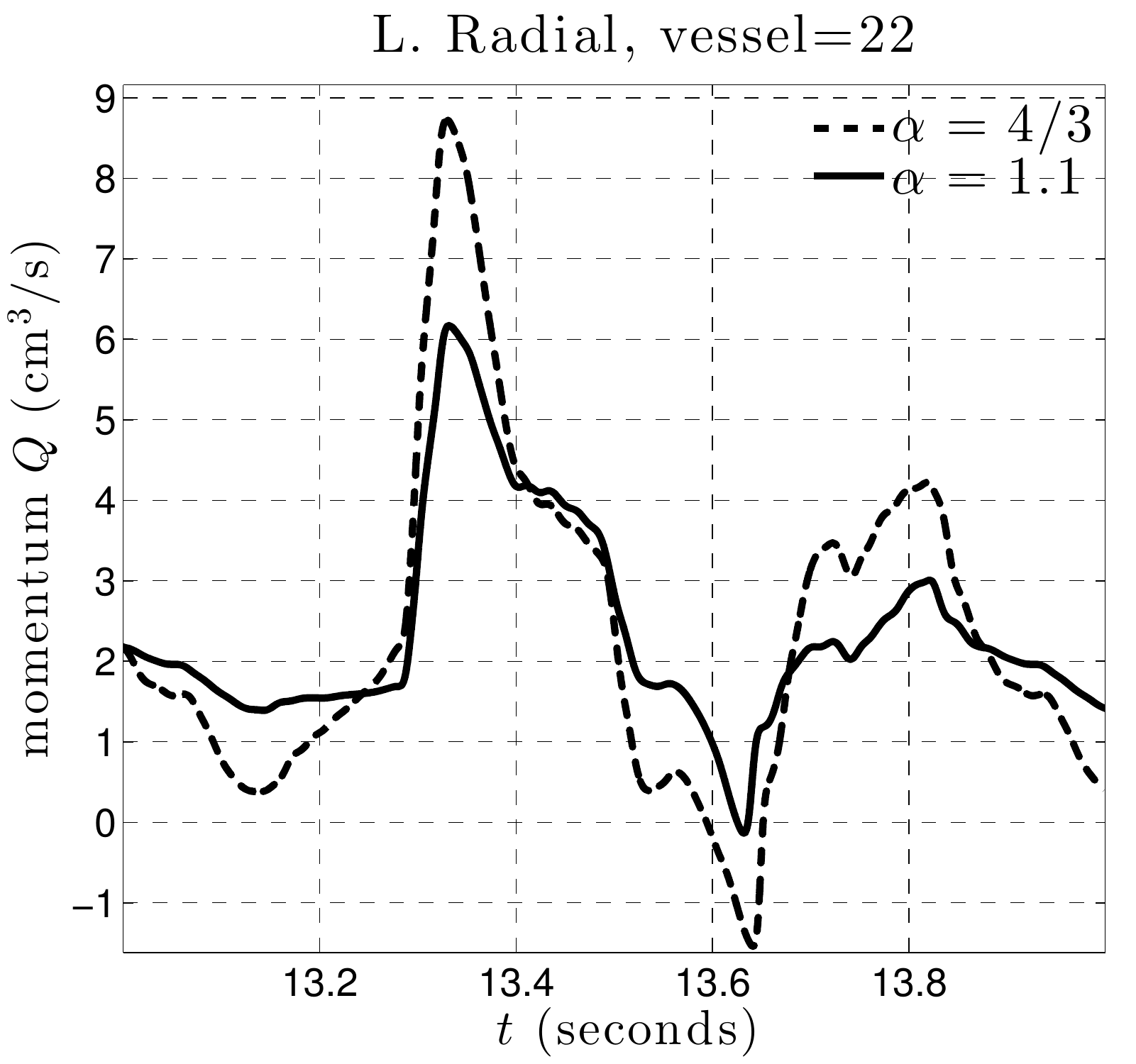} \\
\caption{A comparison of the momentum waveforms from the $(A,U)$ system with $\alpha = 1.1$ and $\alpha=4/3$.}
\label{fig:55vesmom2}
\end{center}
\end{figure}

\begin{figure}[!htb]
\begin{center}
\includegraphics[scale=0.35]{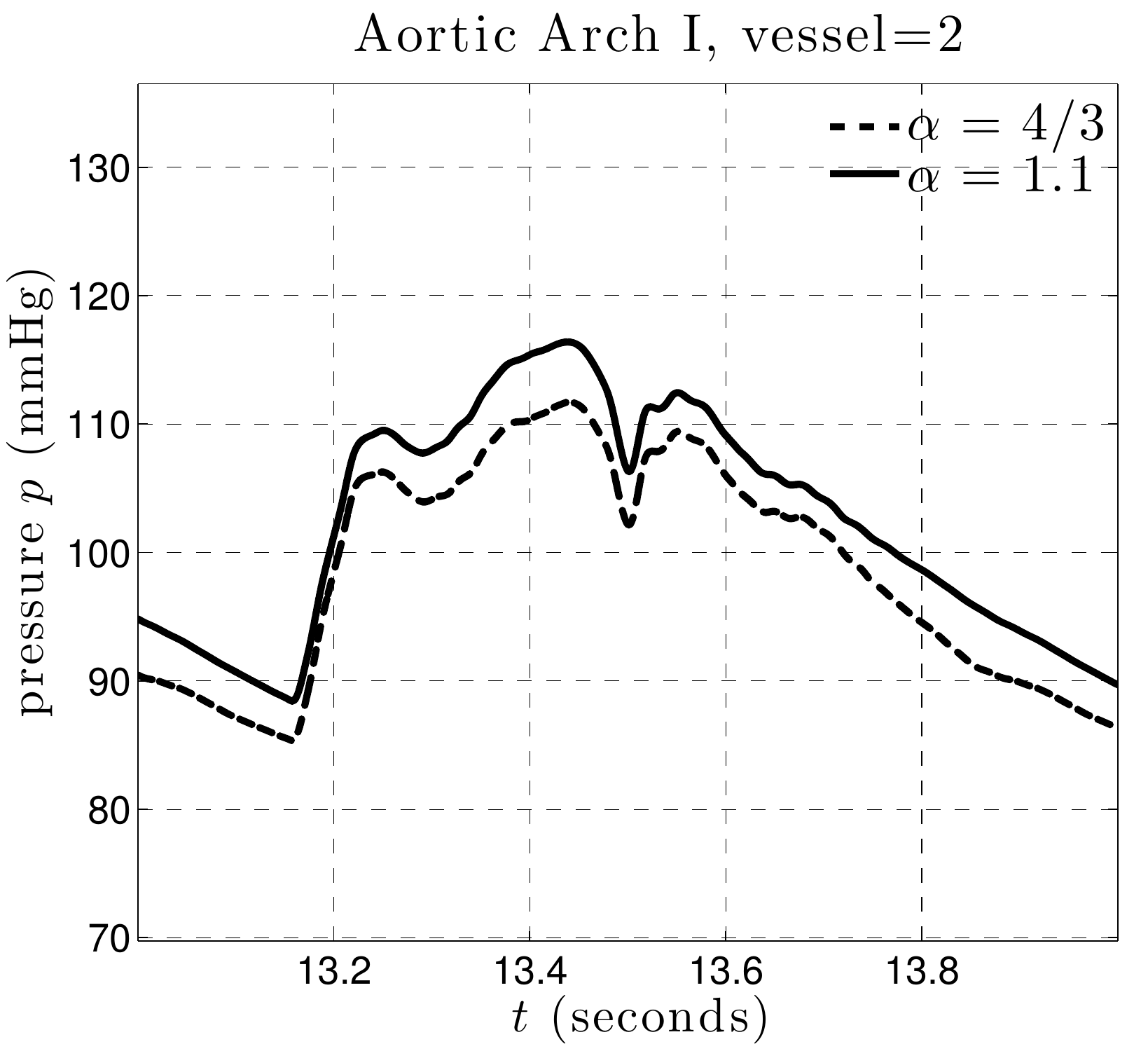} \includegraphics[scale=0.35]{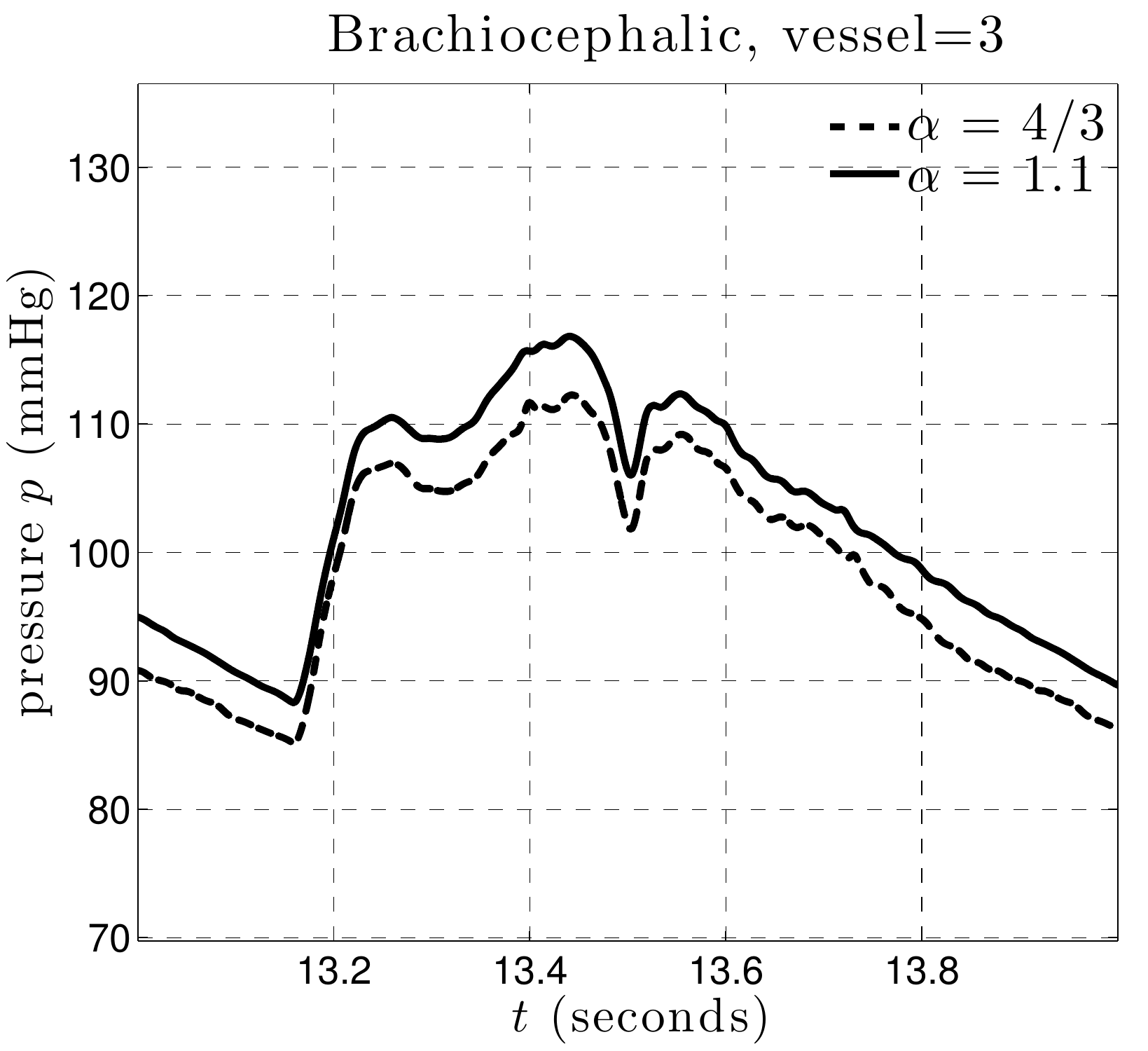} \includegraphics[scale=0.35]{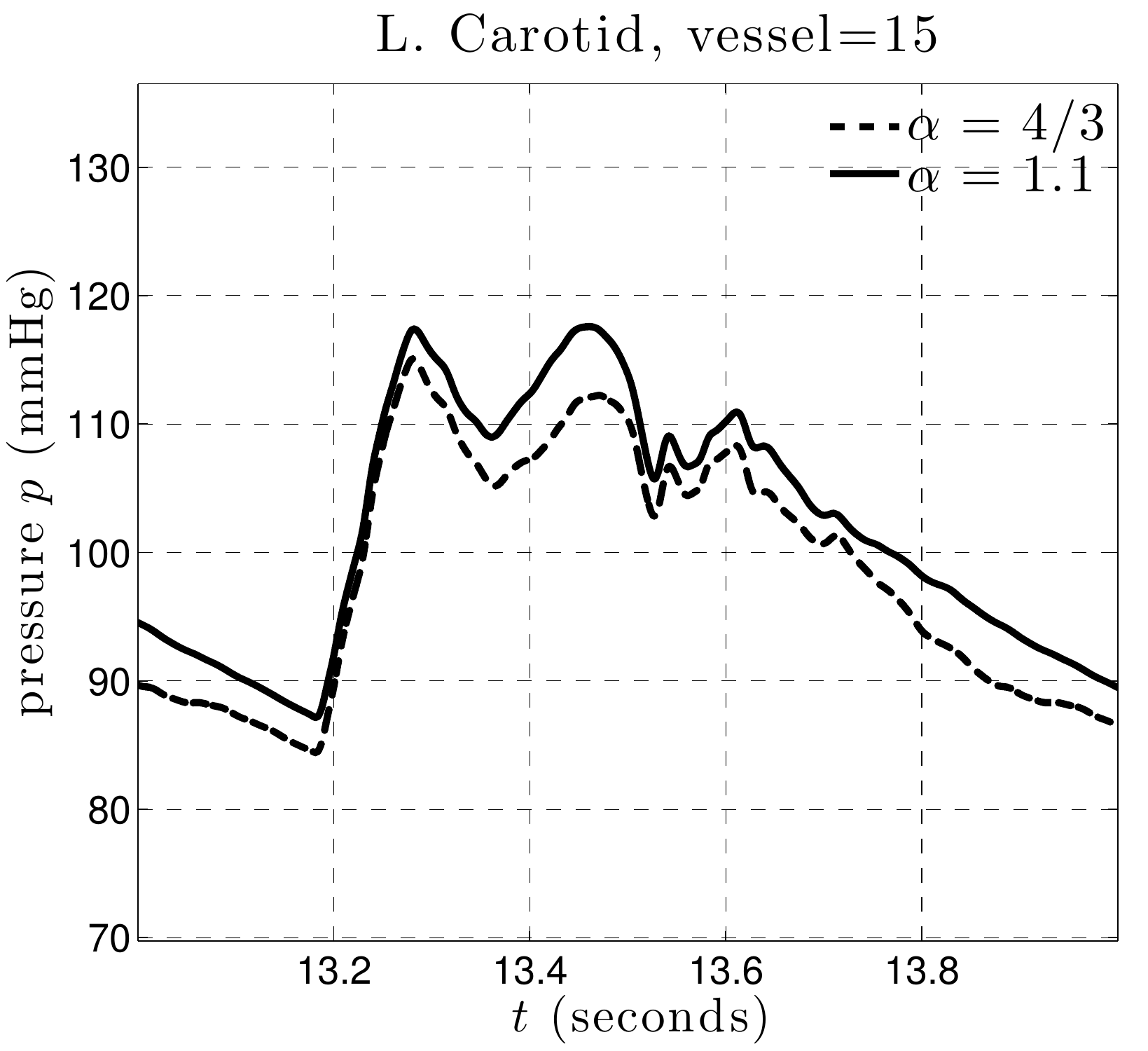} \\
\includegraphics[scale=0.35]{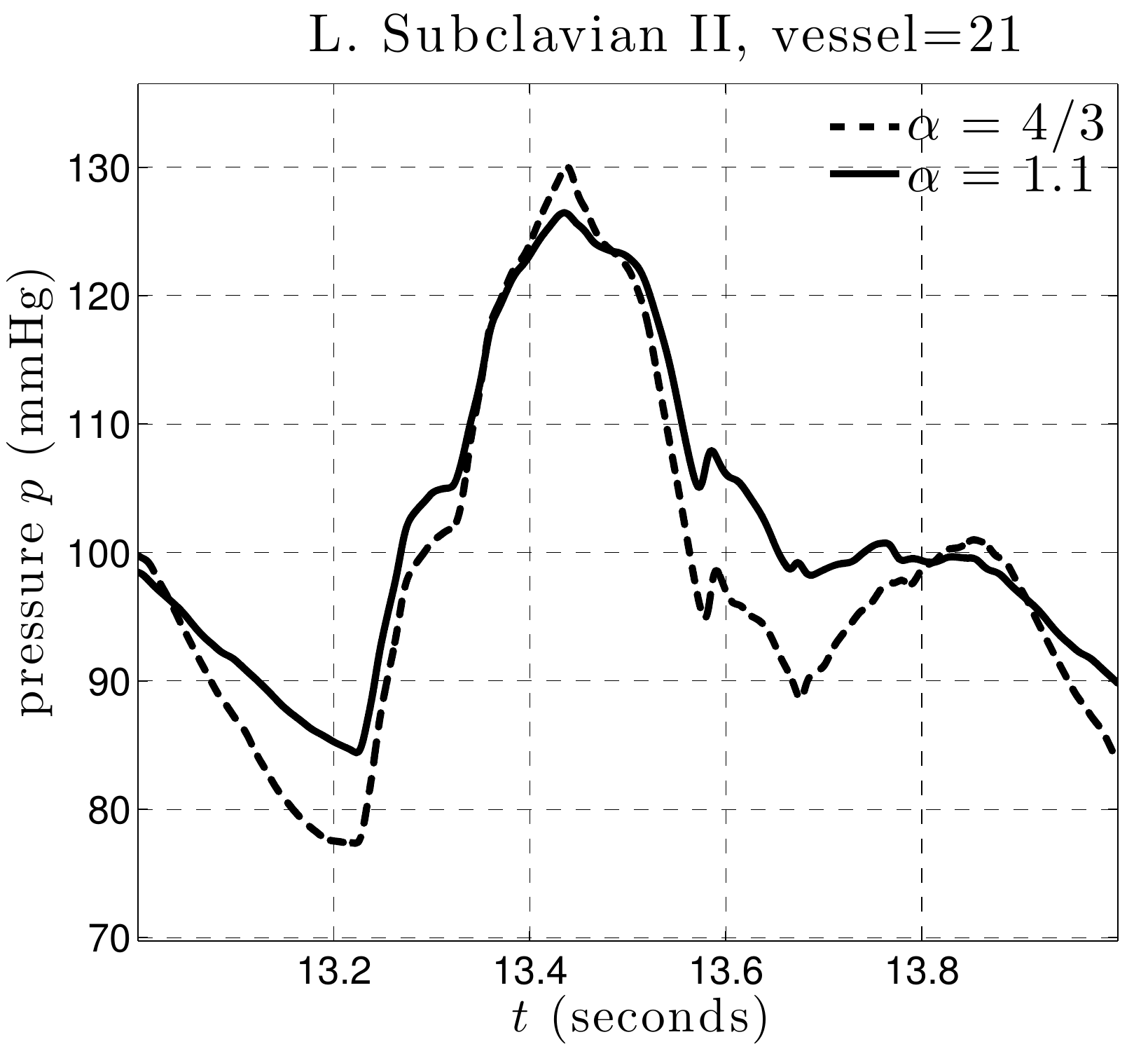} \includegraphics[scale=0.35]{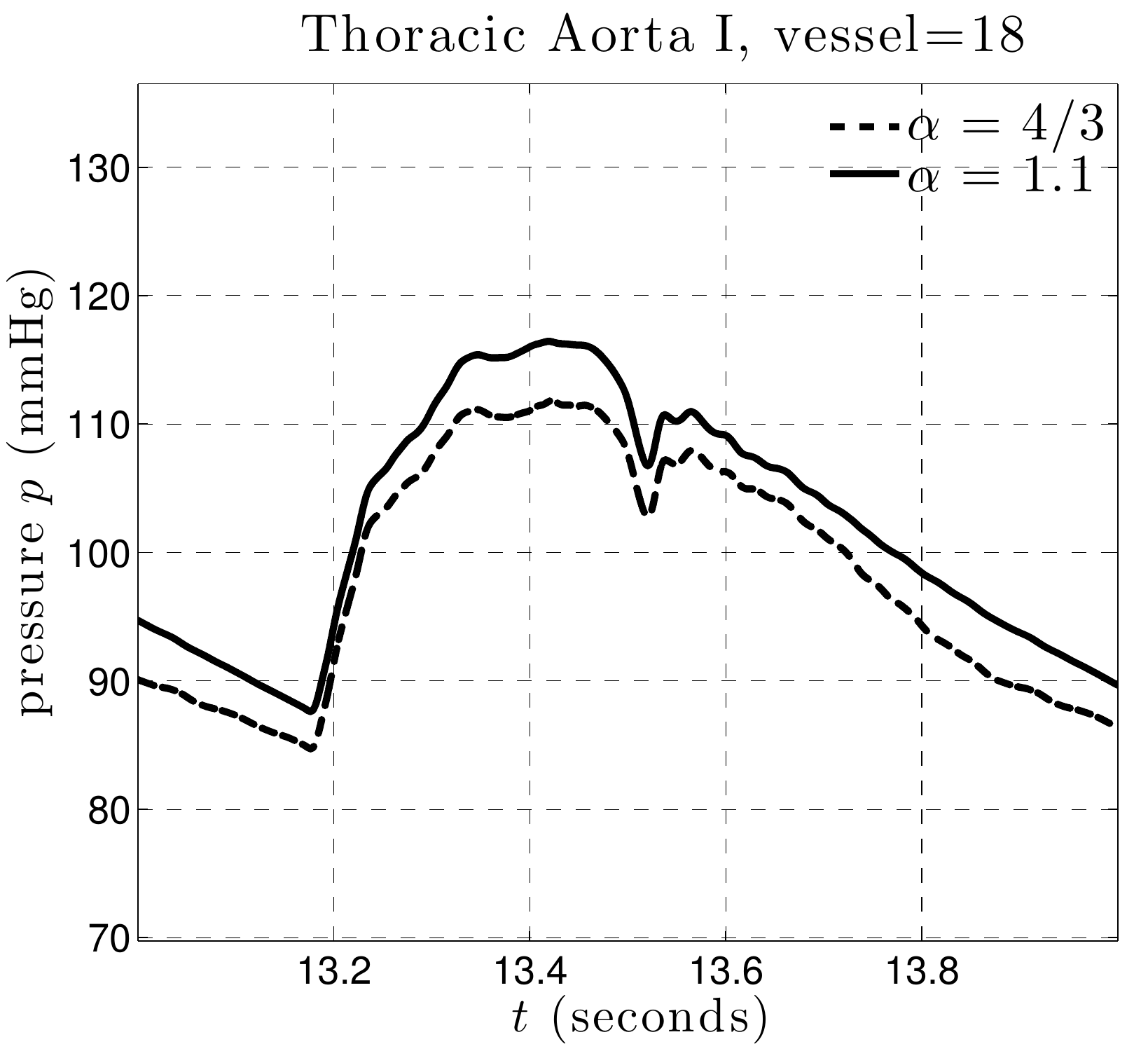} \includegraphics[scale=0.35]{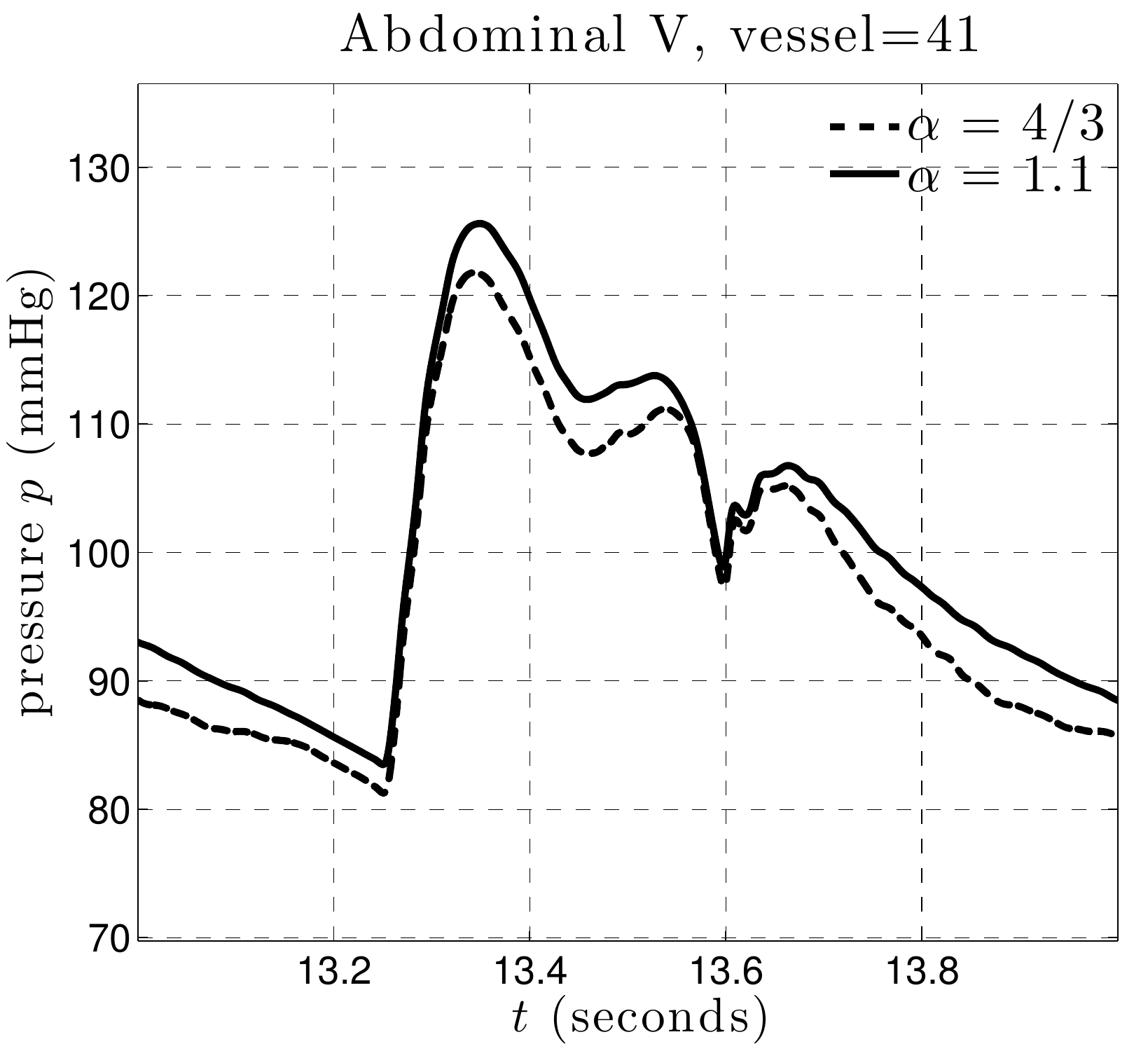} \\
\includegraphics[scale=0.35]{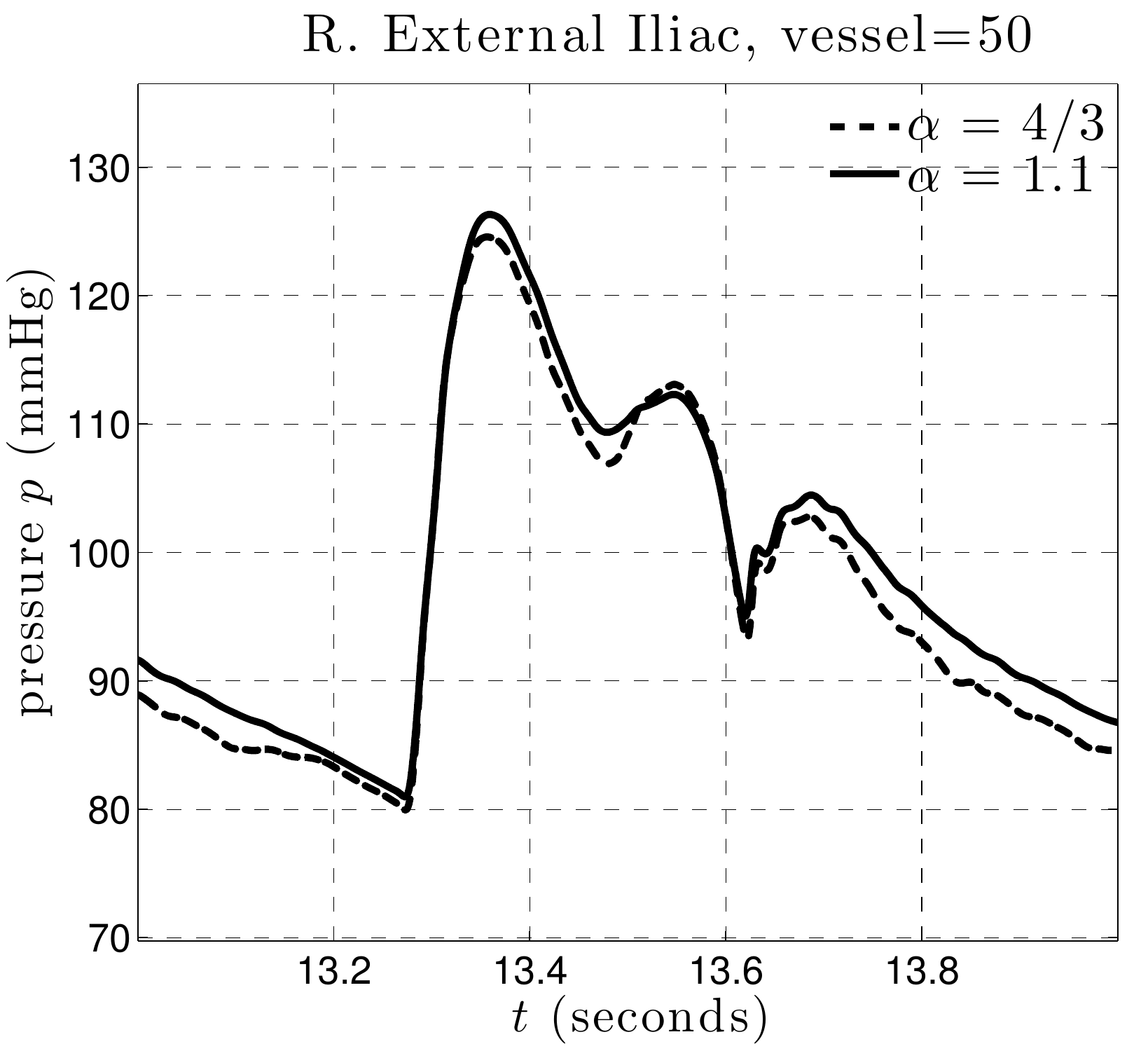} \includegraphics[scale=0.35]{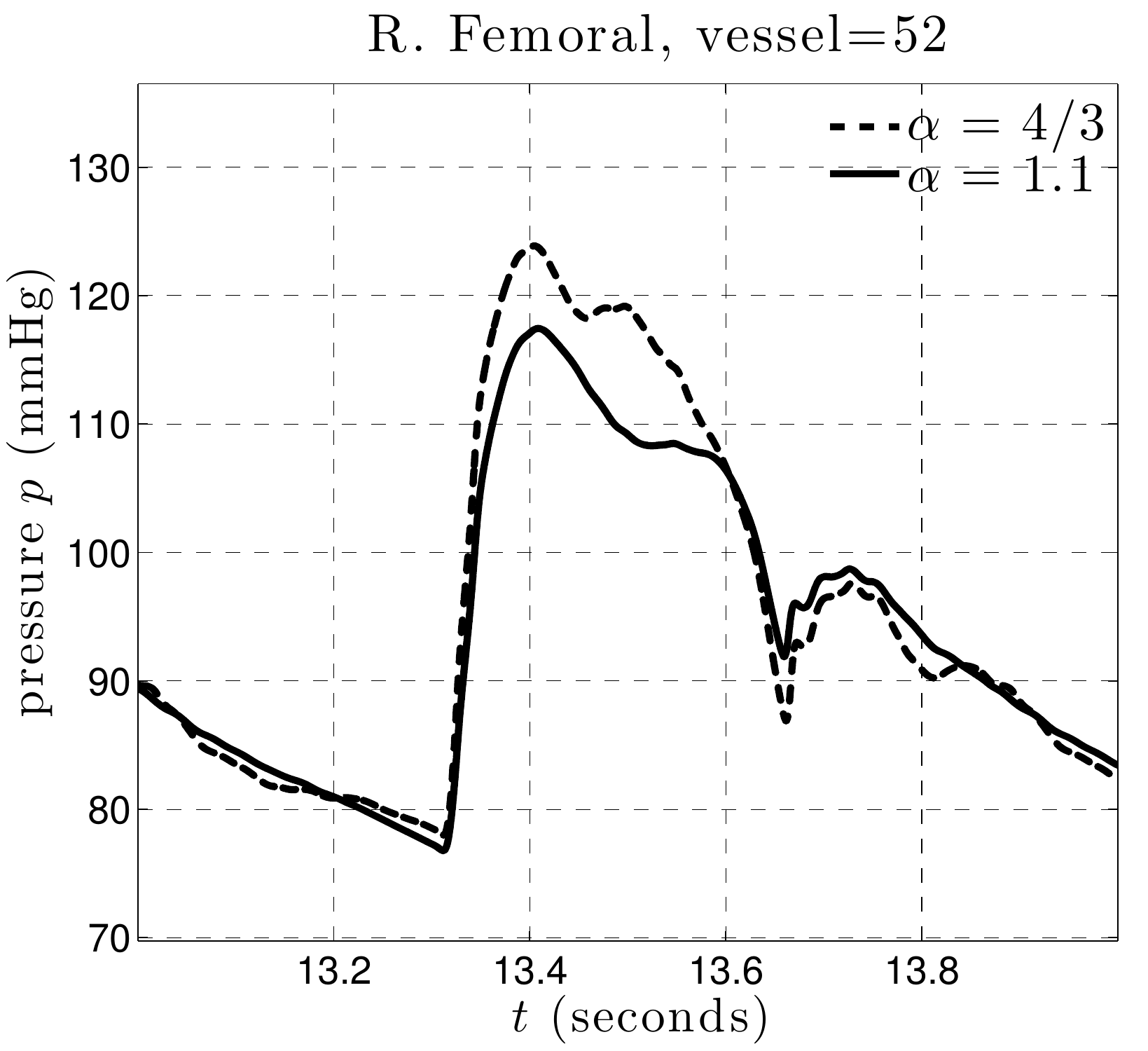} \includegraphics[scale=0.35]{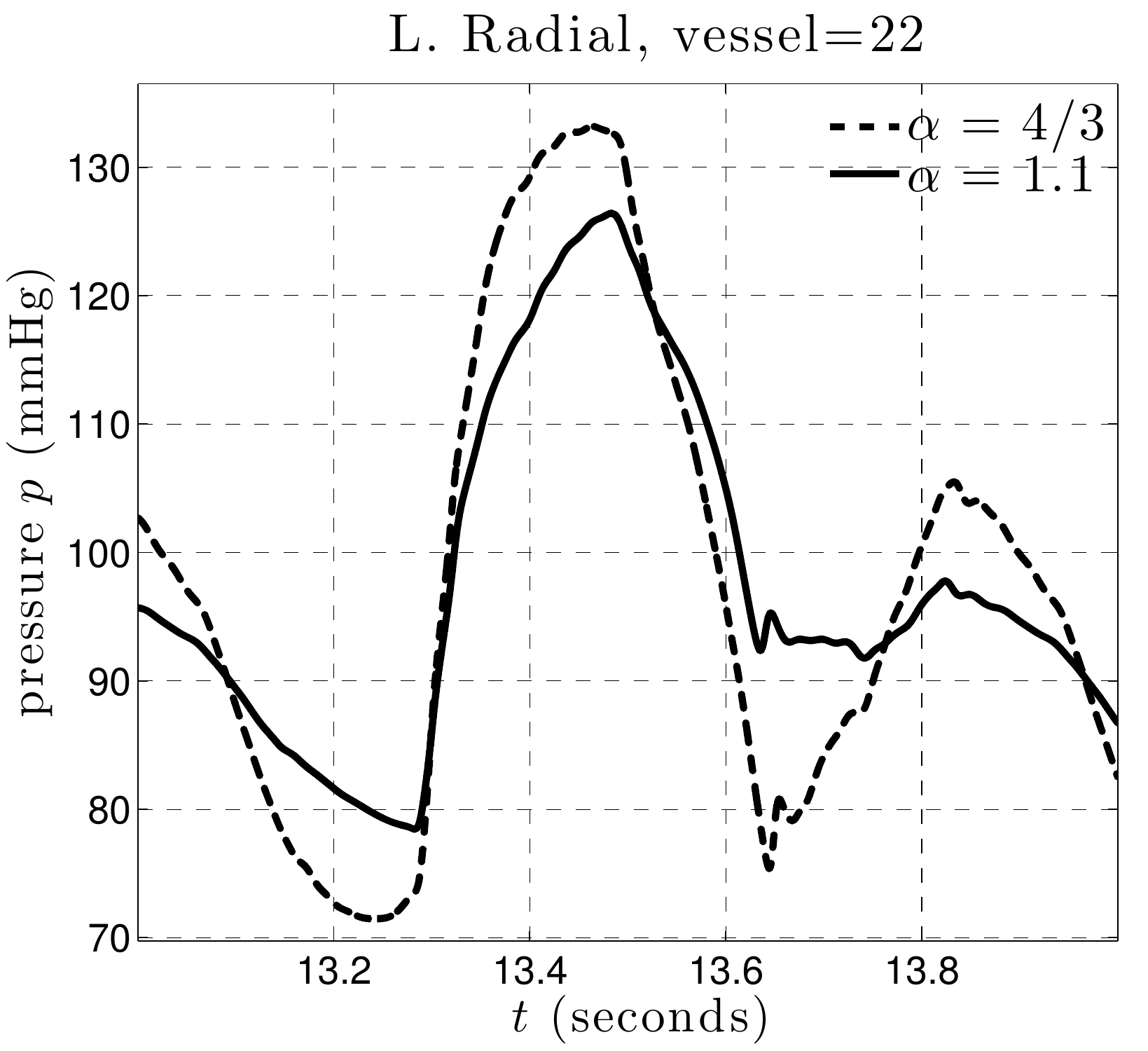} \\
\caption{A comparison of the pressure waveforms from the $(A,U)$ system with $\alpha = 1.1$ and $4/3$.}
\label{fig:55vespress2}
\end{center}
\end{figure}

\begin{figure}[!htb]
\begin{center}
\includegraphics[scale=0.35]{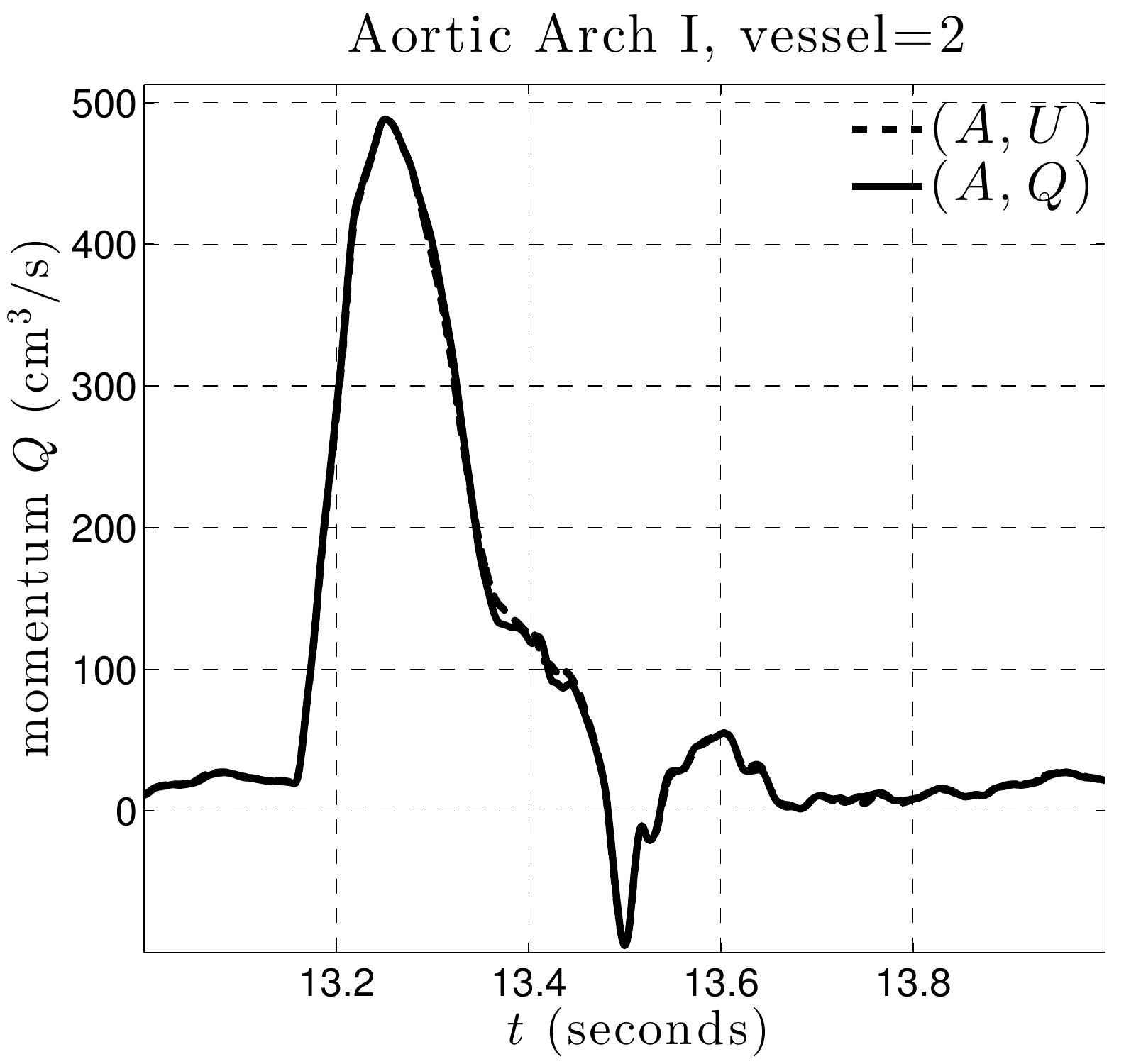} \includegraphics[scale=0.35]{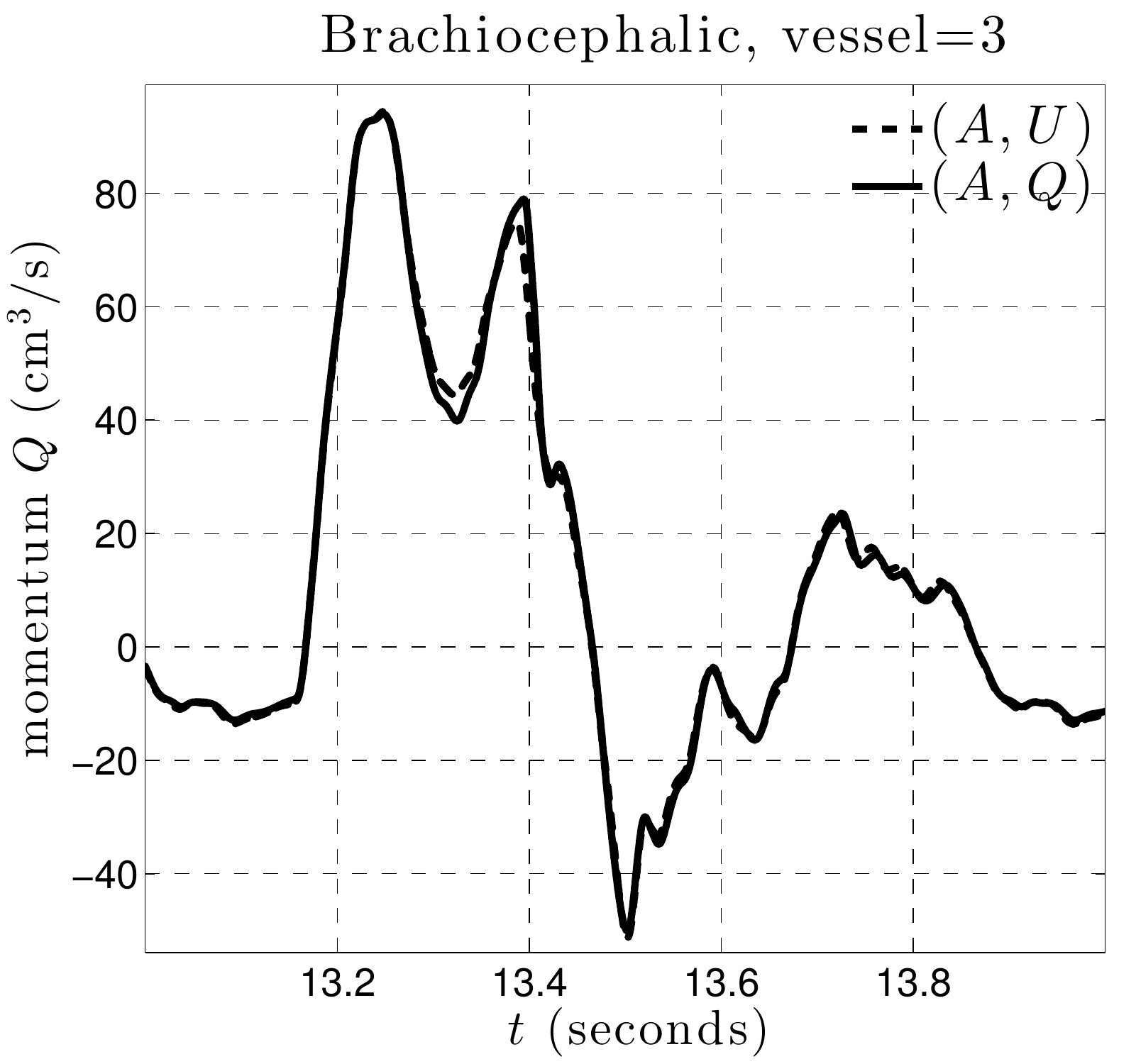} \includegraphics[scale=0.35]{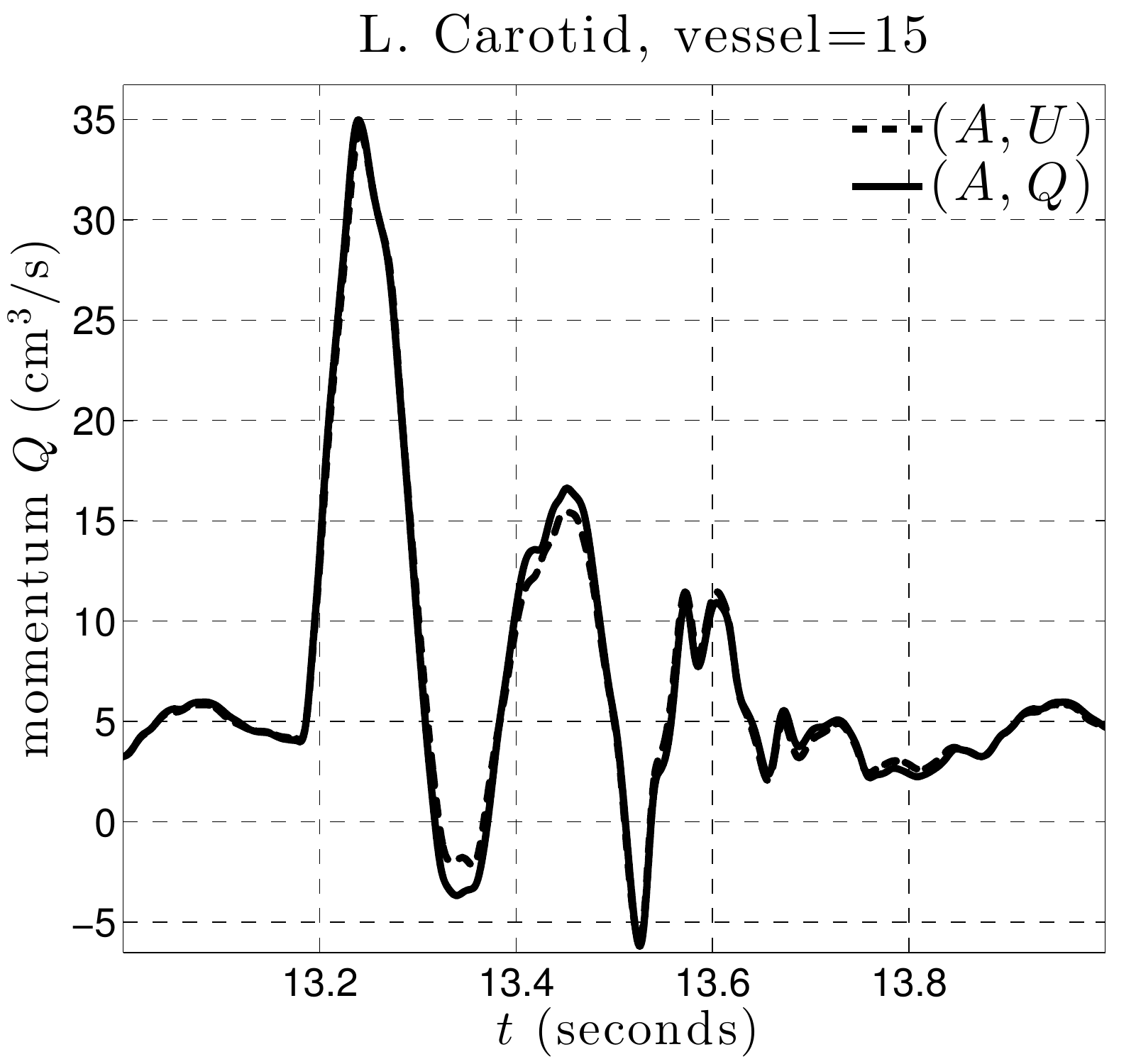} \\
\includegraphics[scale=0.35]{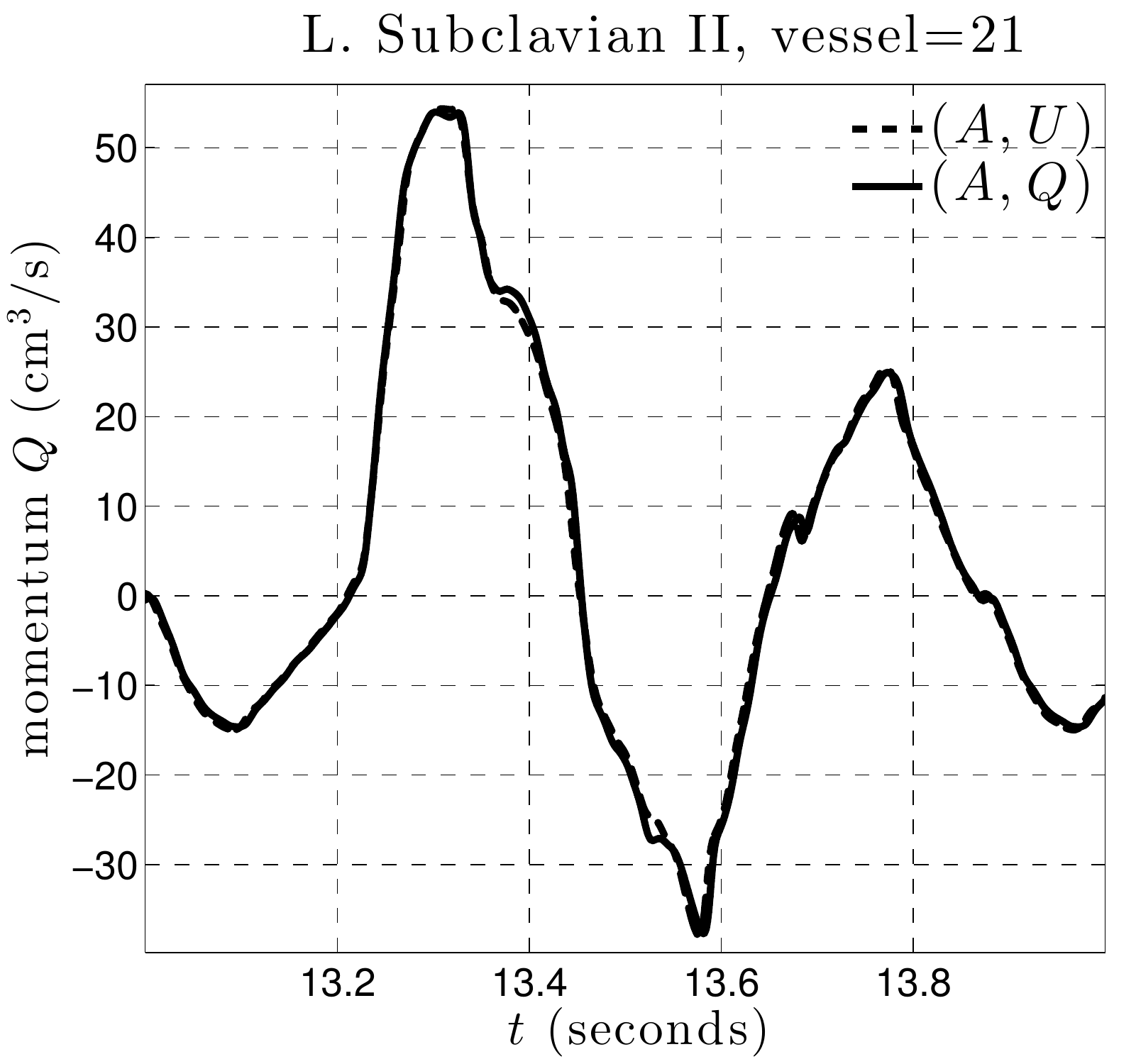} \includegraphics[scale=0.35]{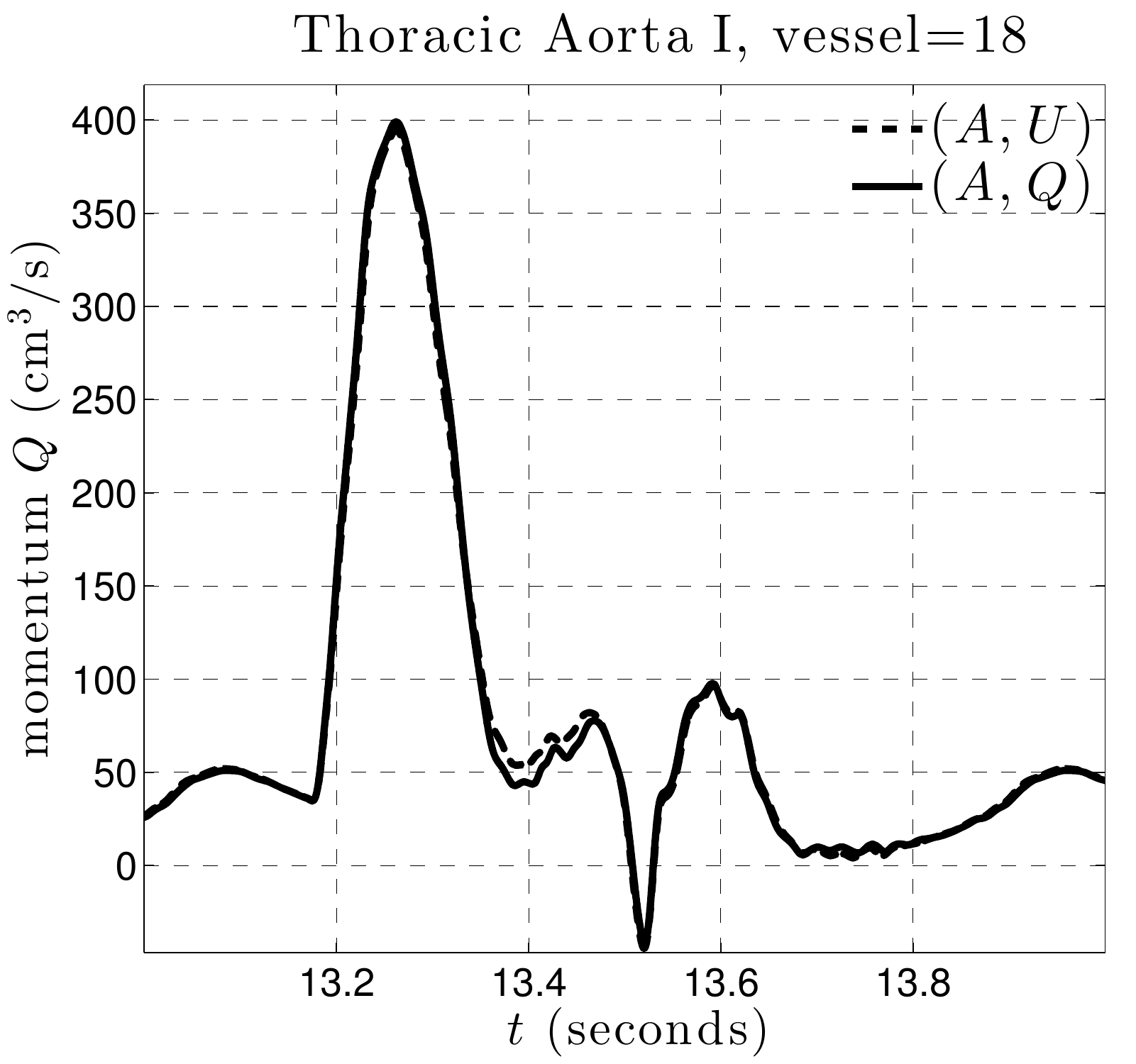} \includegraphics[scale=0.35]{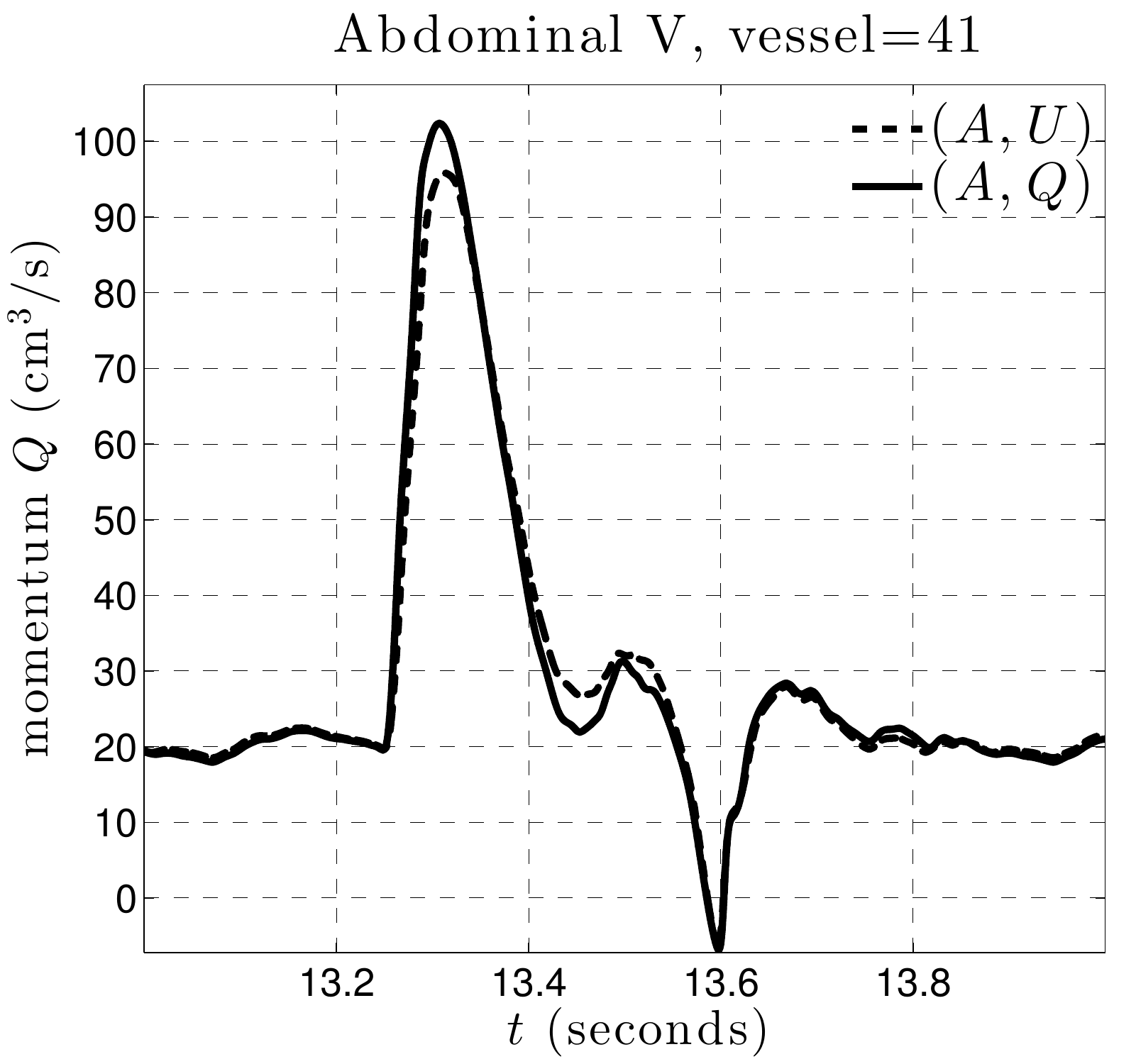} \\
\includegraphics[scale=0.35]{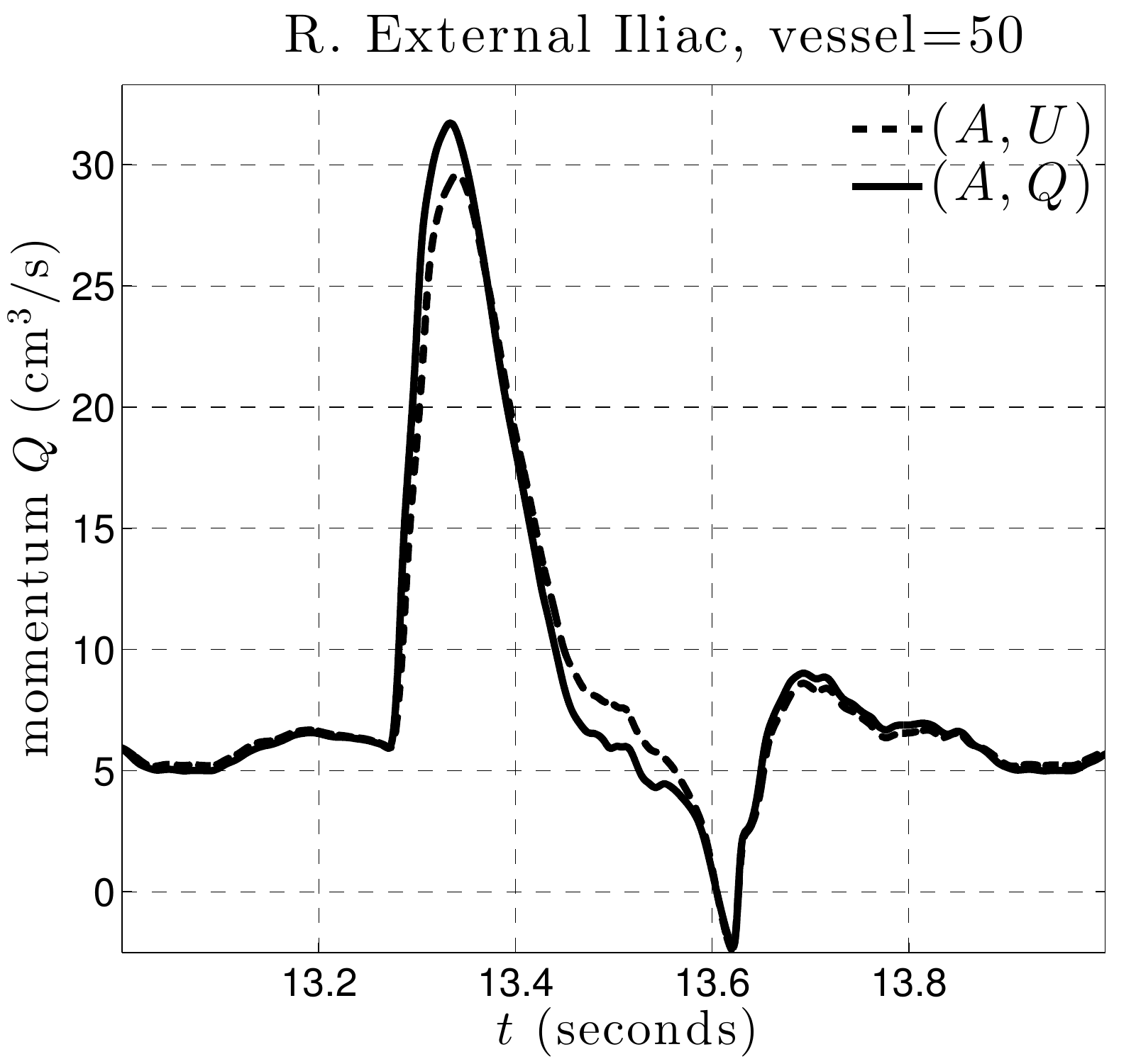} \includegraphics[scale=0.35]{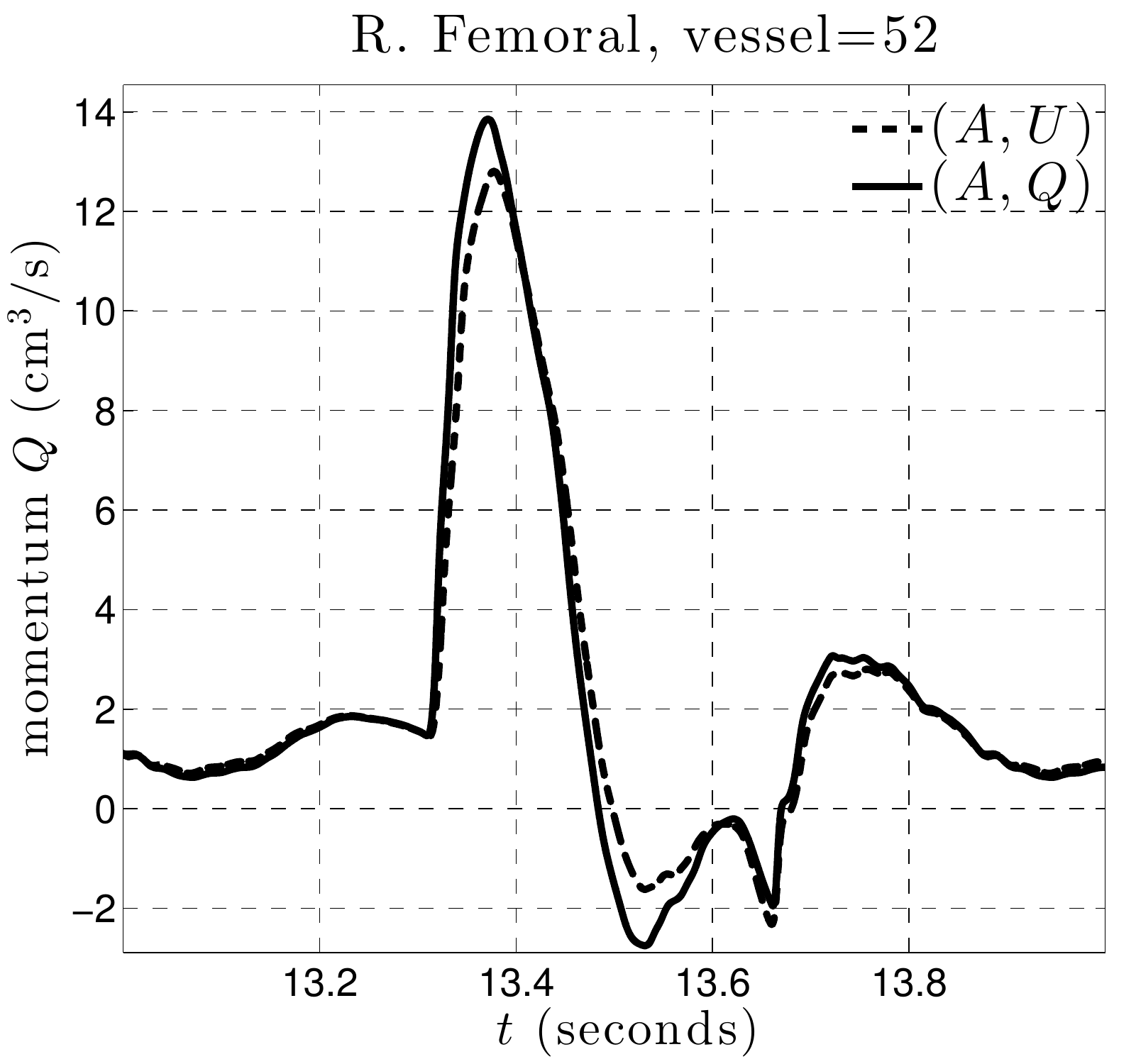} \includegraphics[scale=0.35]{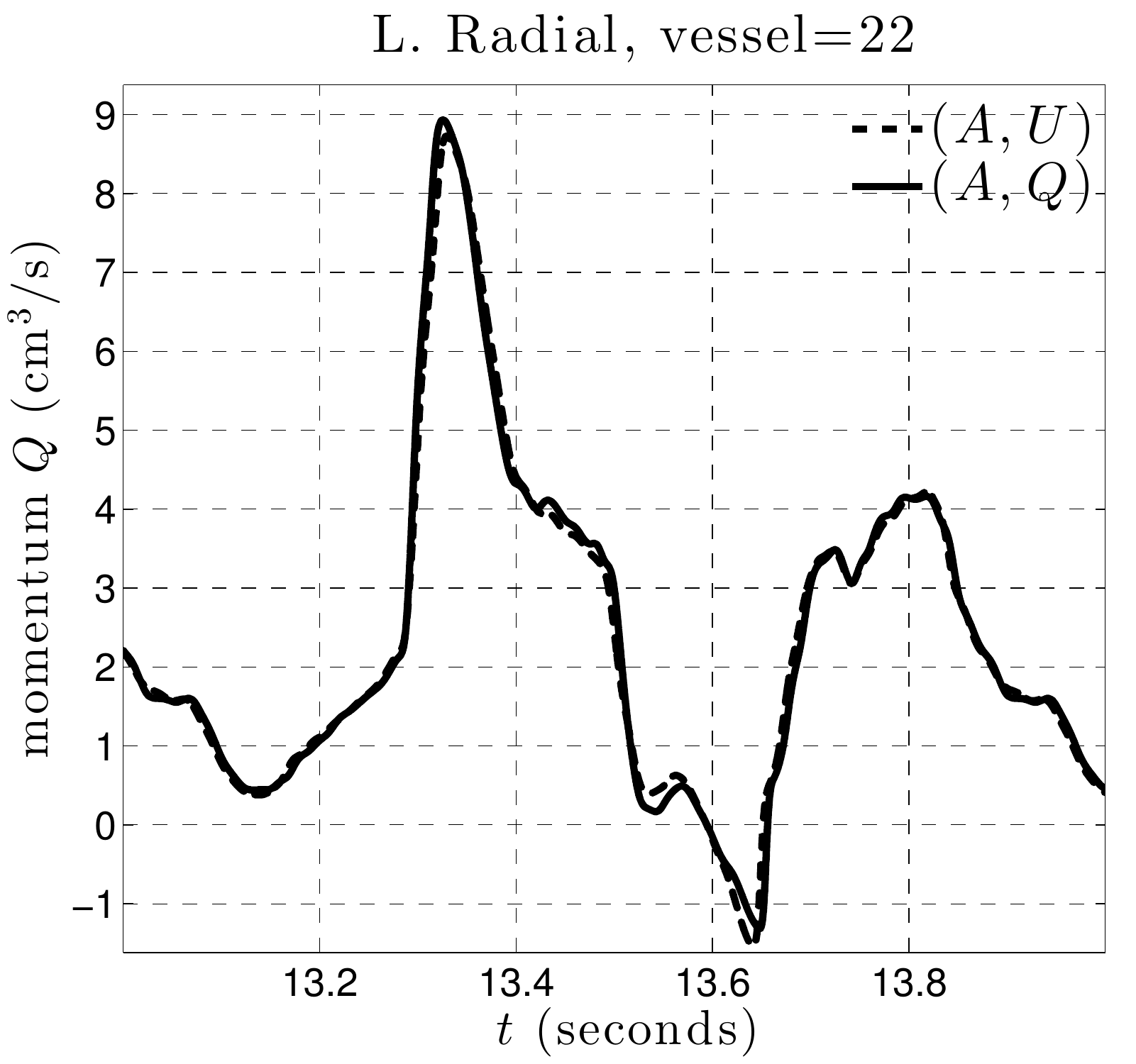} \\
\caption{A comparison of the momentum waveforms from the $(A,Q)$ and $(A,U)$ systems with $\alpha = 4/3$.}
\label{fig:55vesmom3}
\end{center}
\end{figure}

\begin{figure}[!htb]
\begin{center}
\includegraphics[scale=0.35]{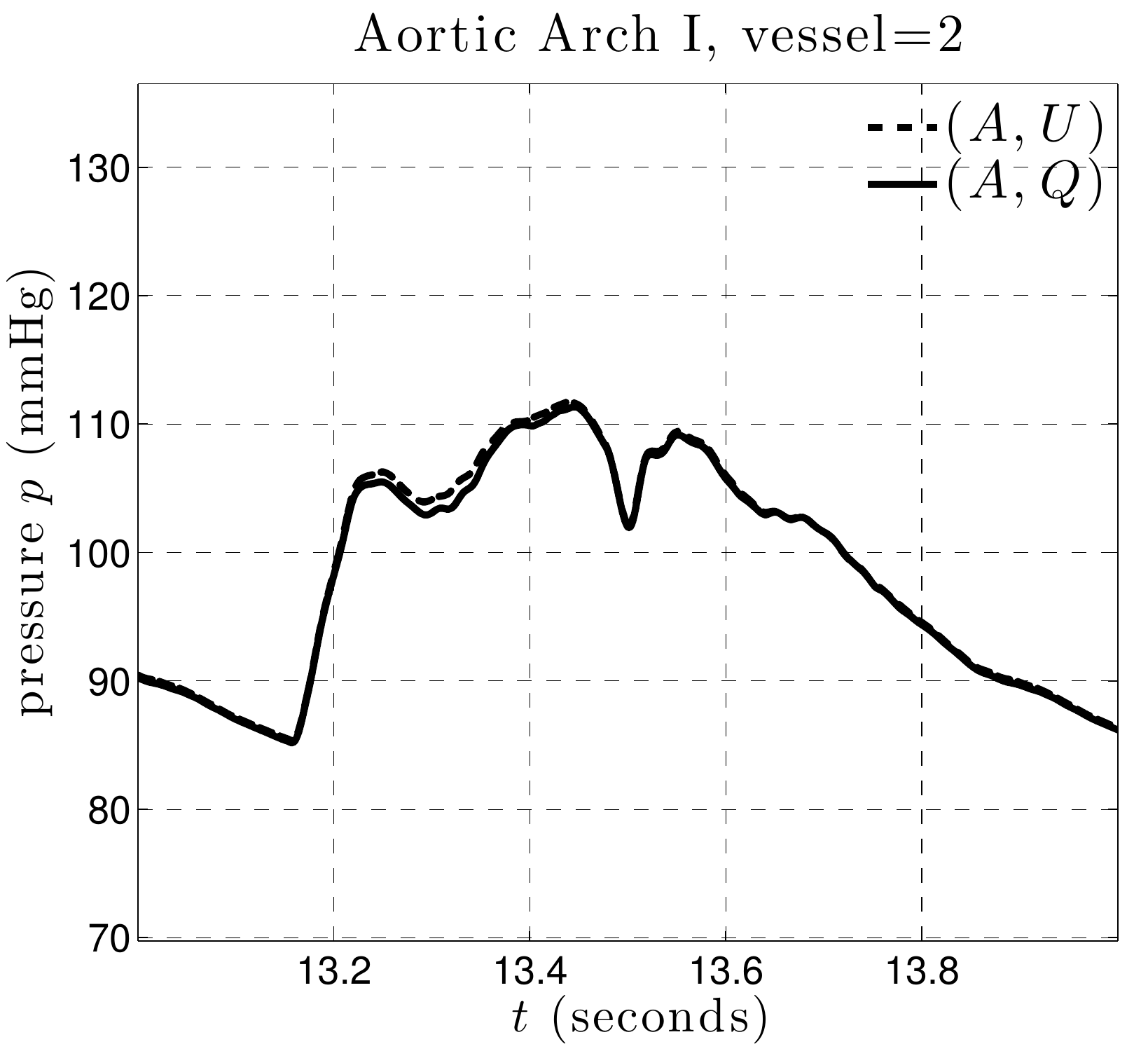} \includegraphics[scale=0.35]{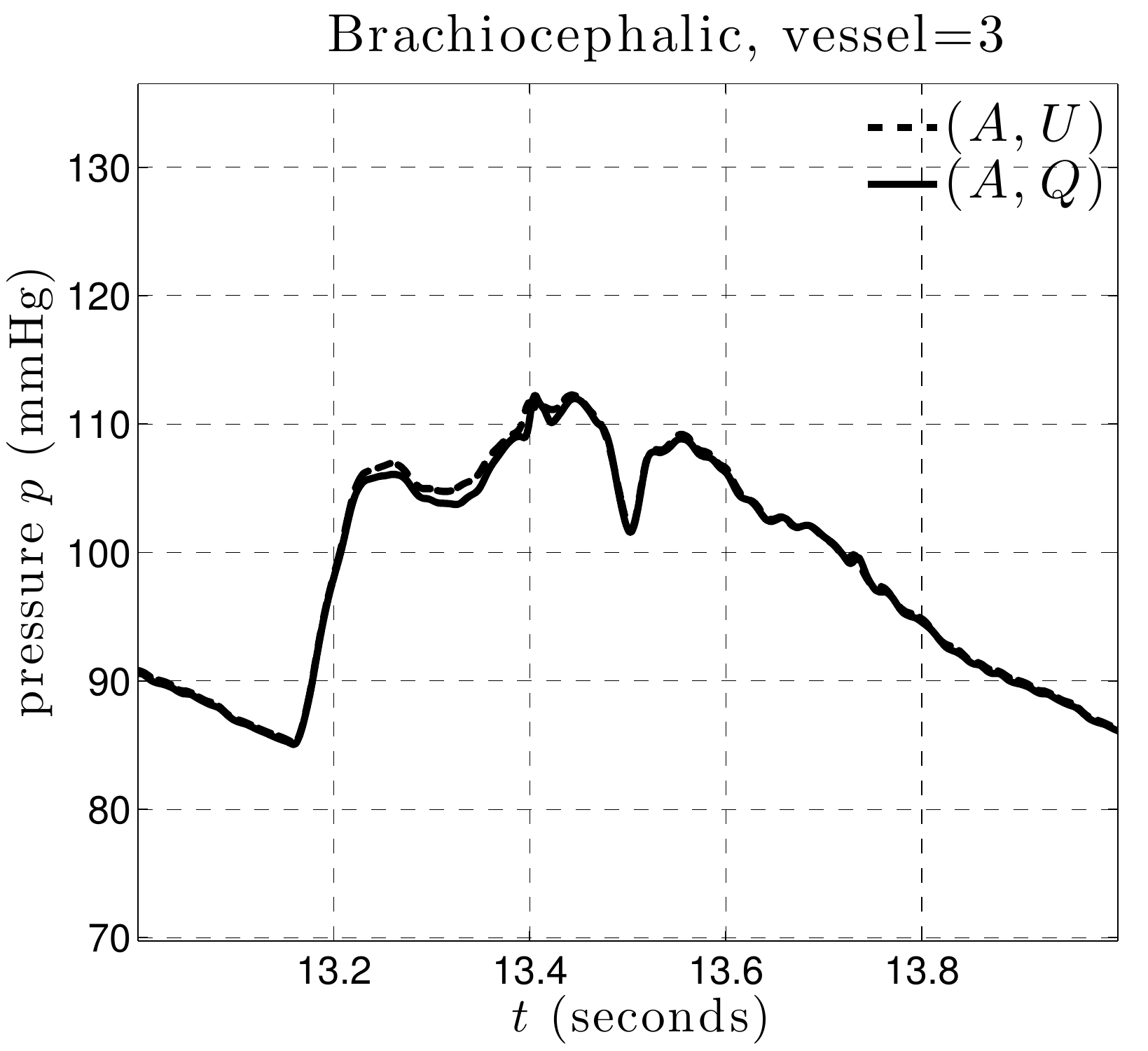} \includegraphics[scale=0.35]{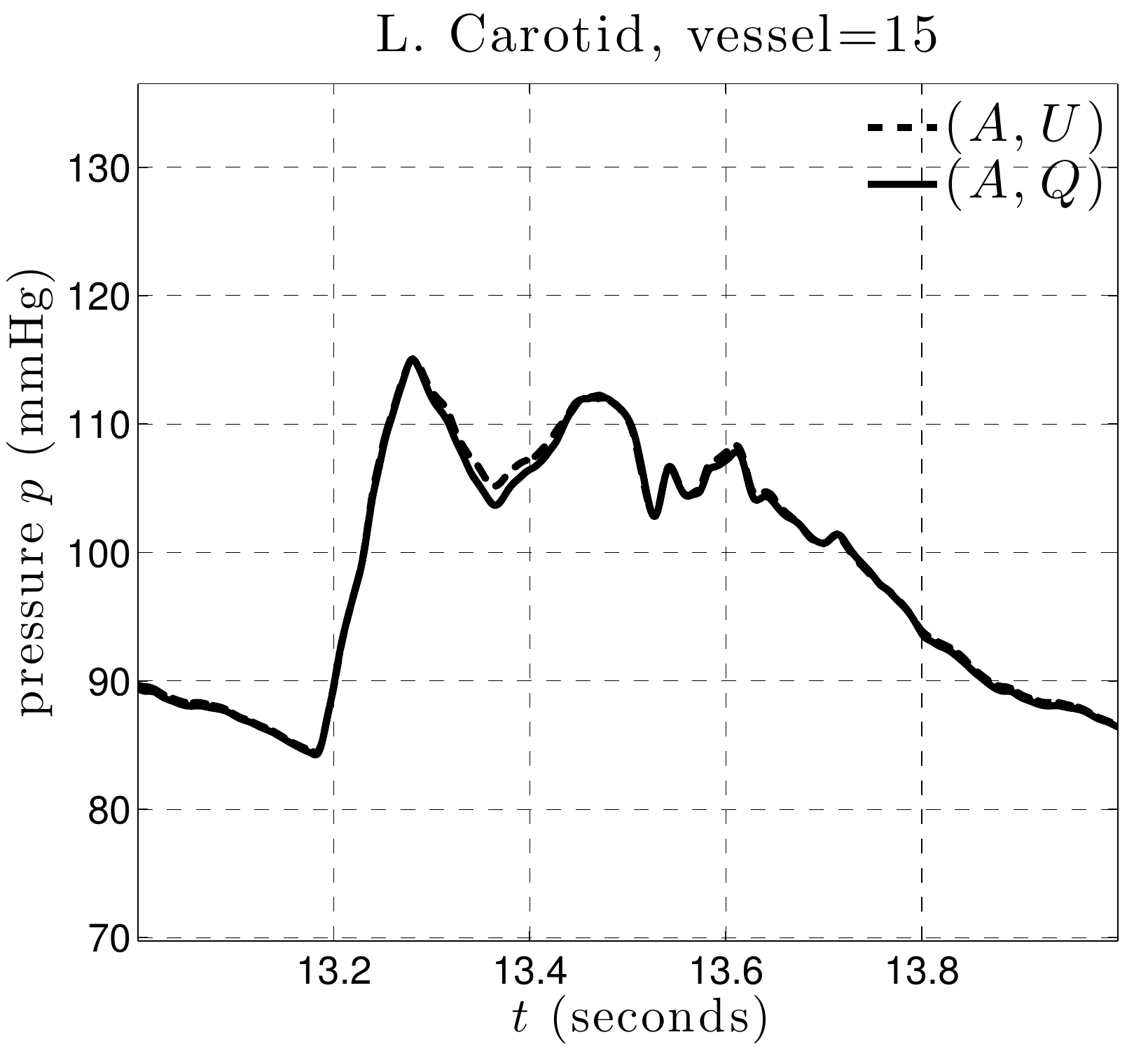} \\
\includegraphics[scale=0.35]{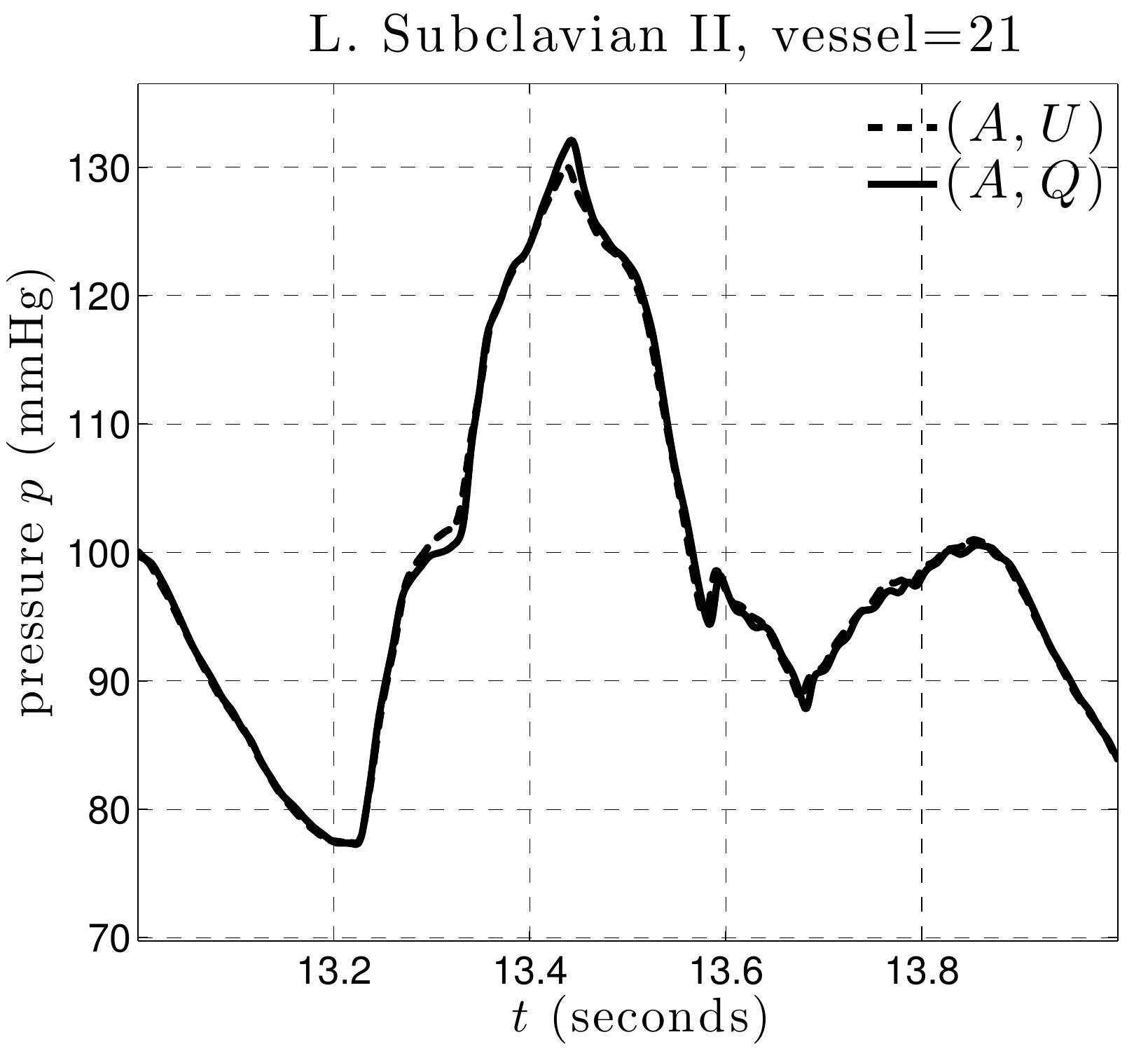} \includegraphics[scale=0.35]{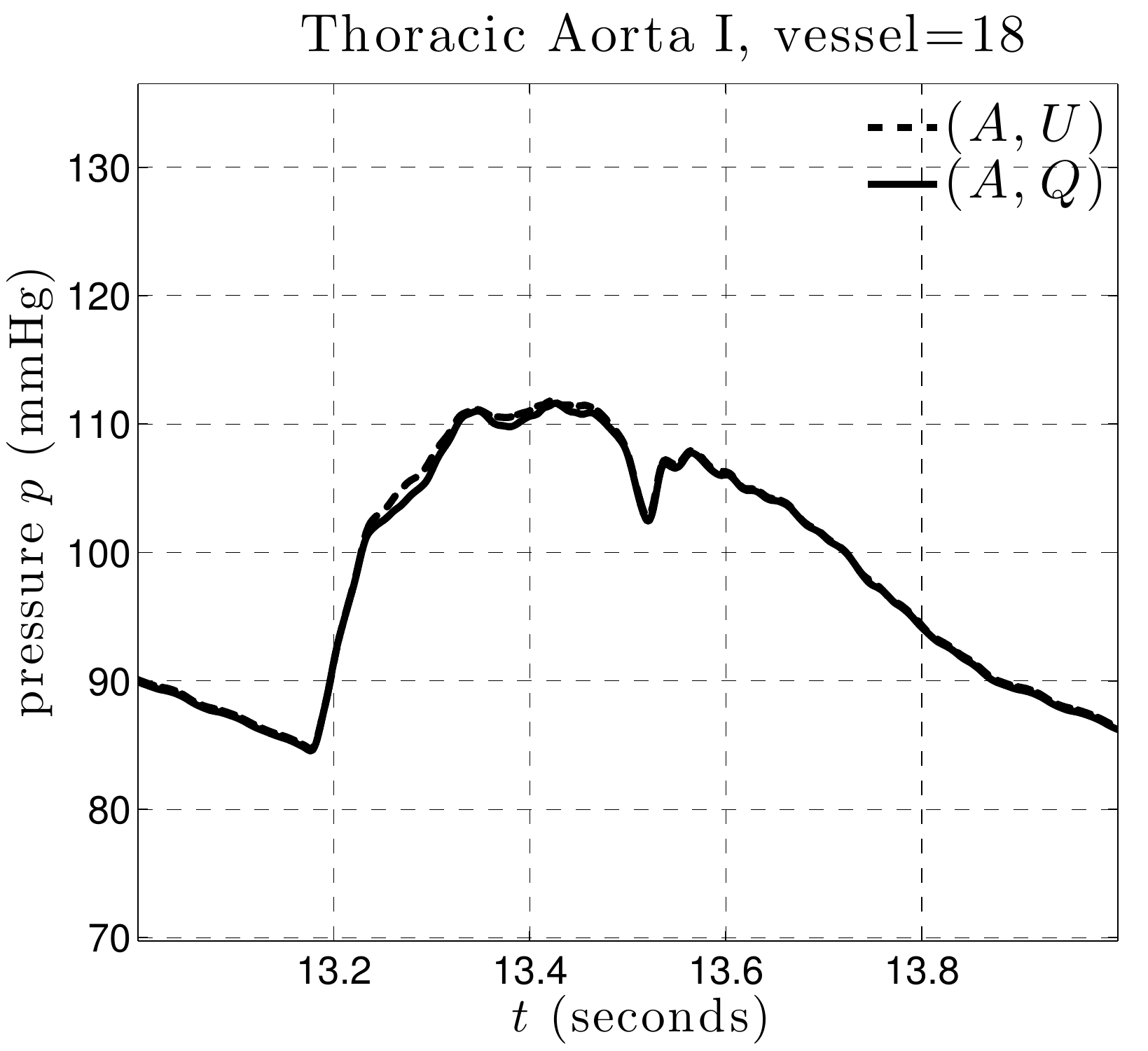} \includegraphics[scale=0.35]{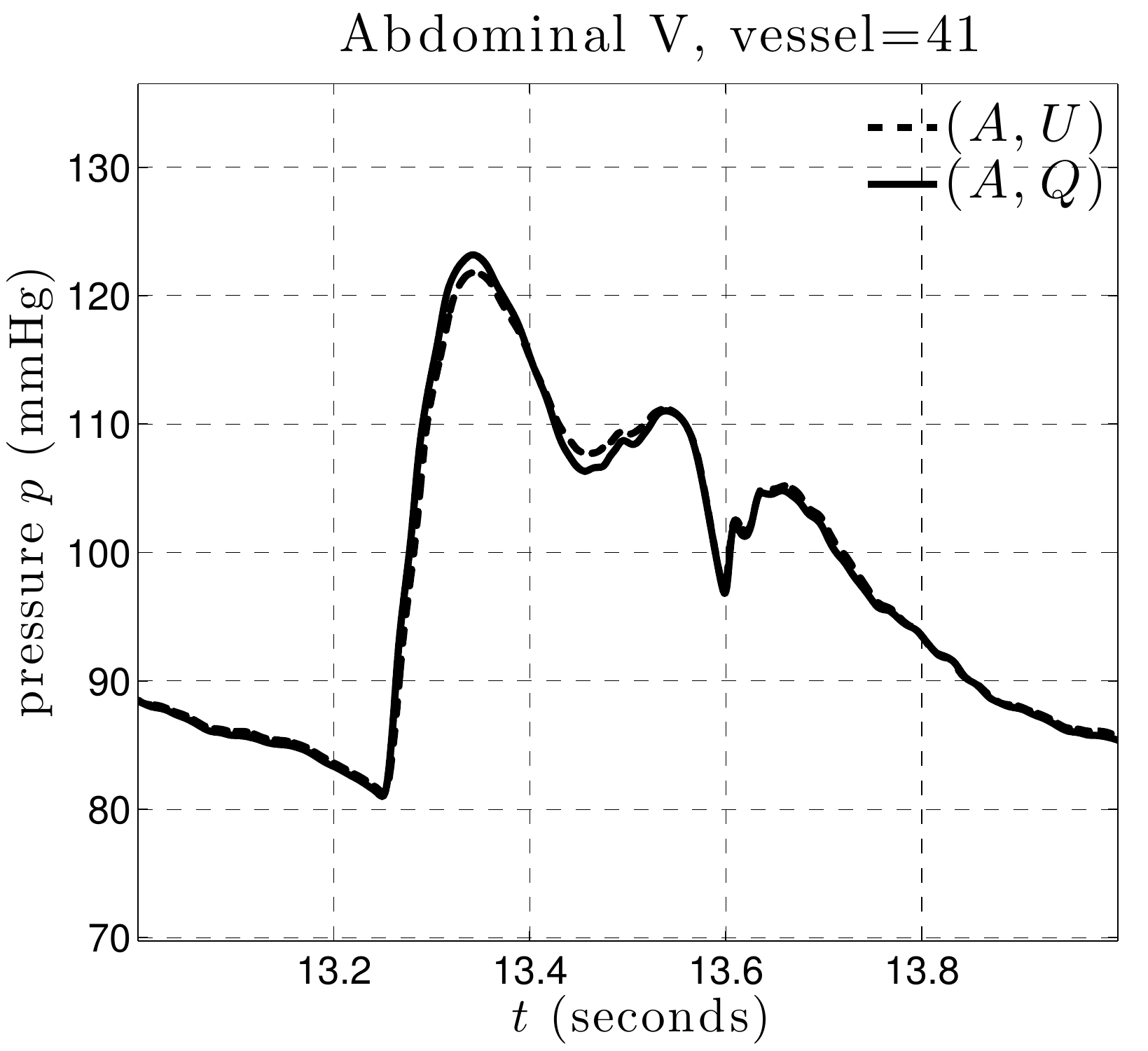} \\
\includegraphics[scale=0.35]{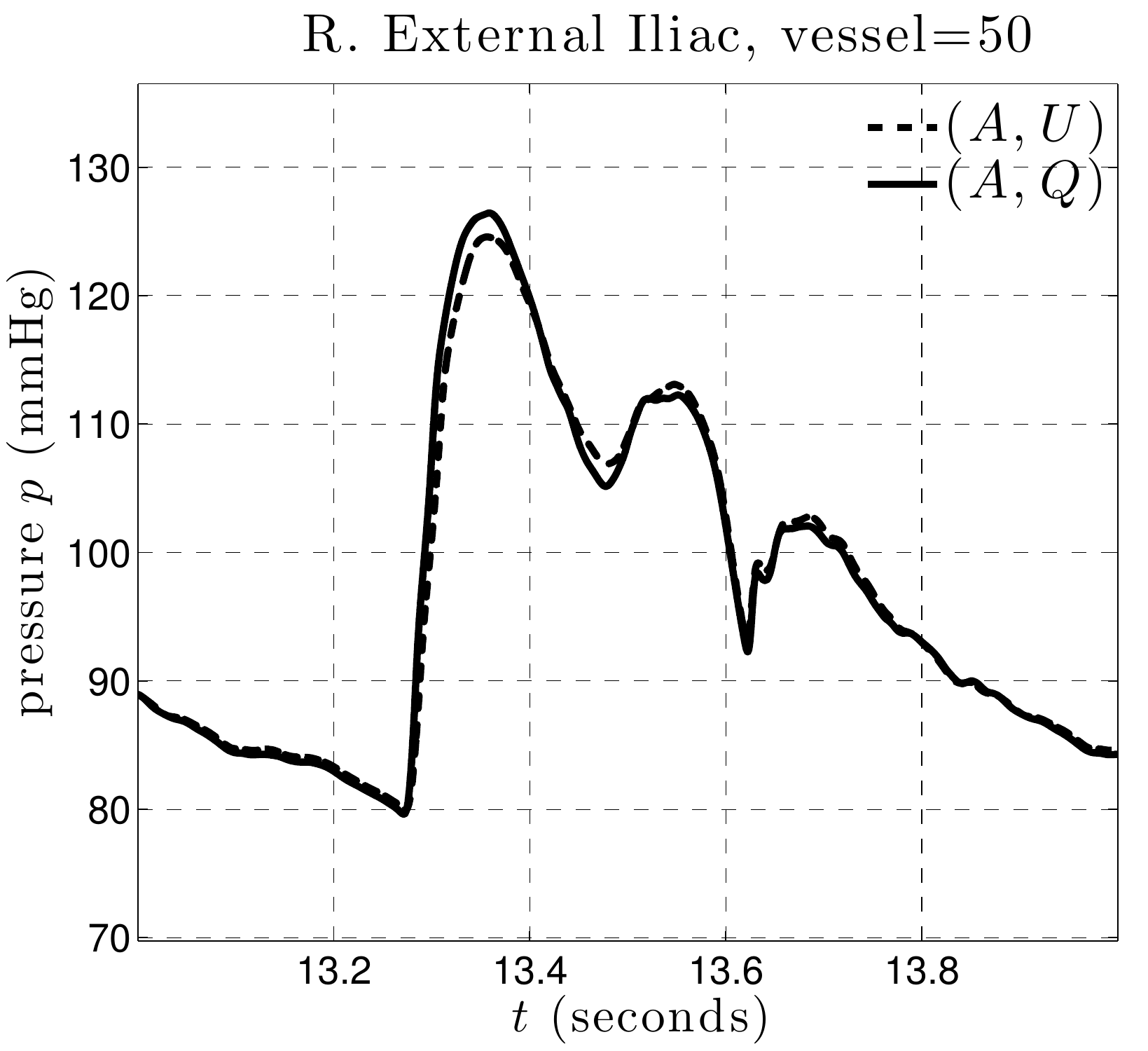} \includegraphics[scale=0.35]{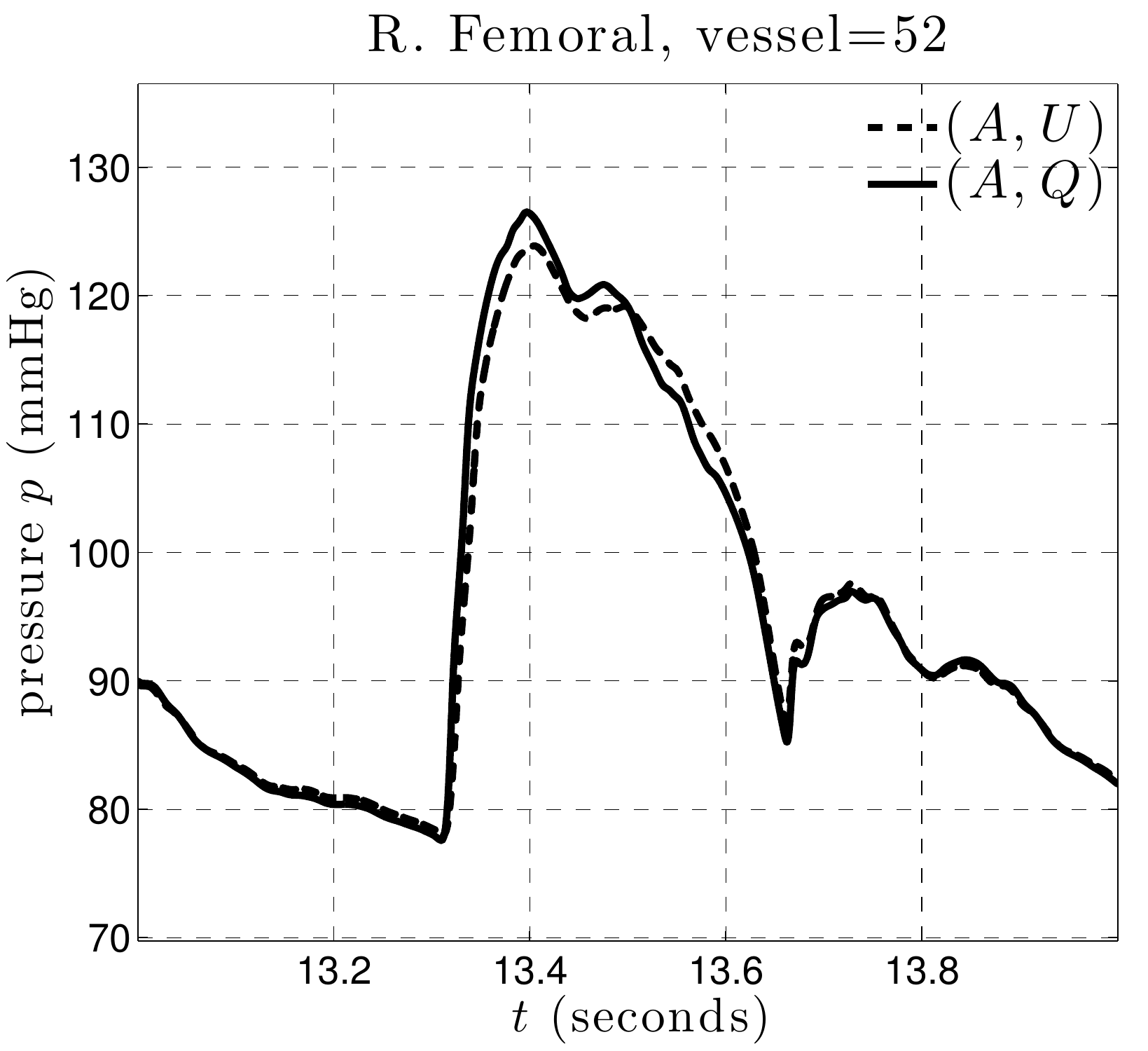} \includegraphics[scale=0.35]{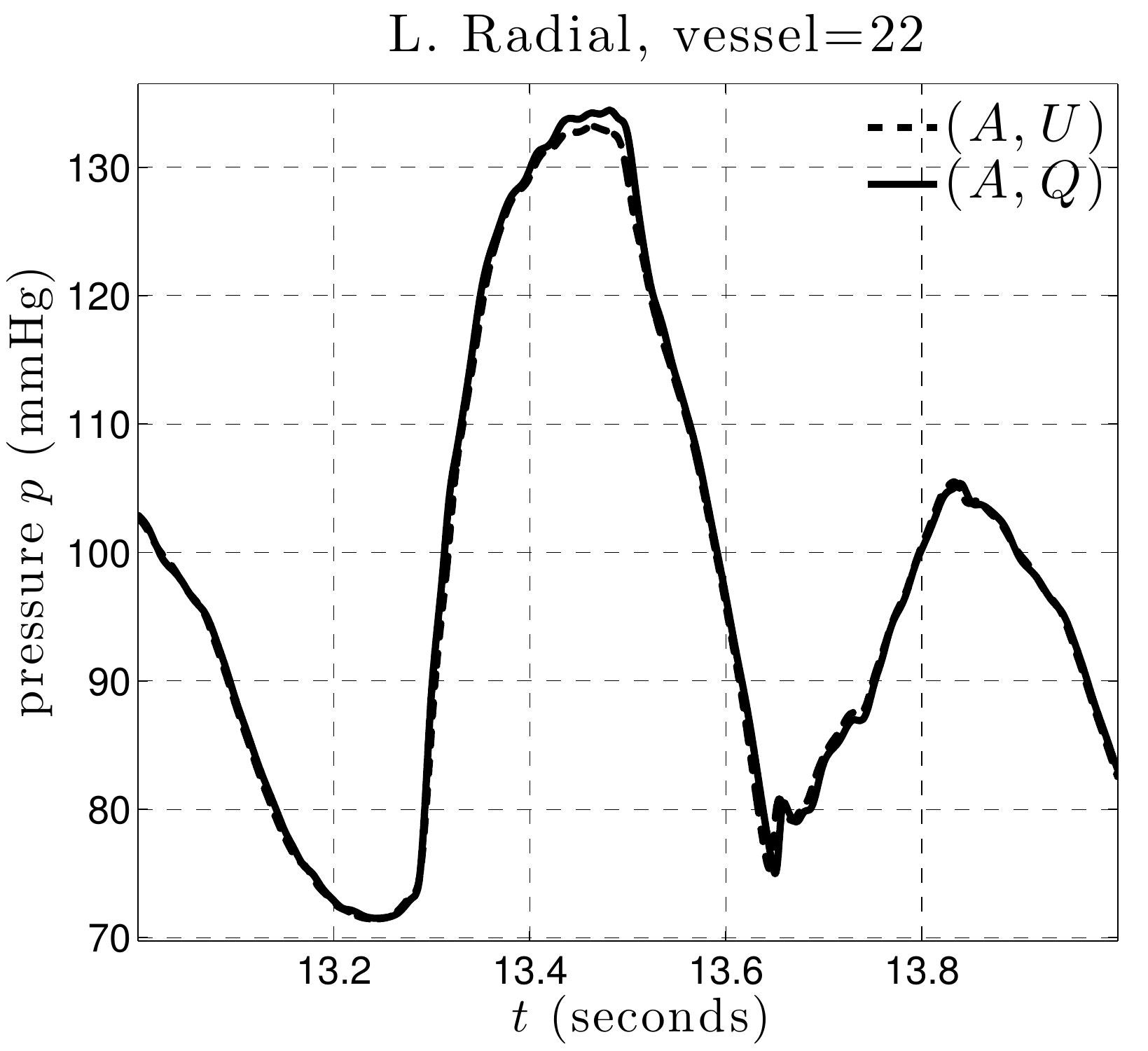} \\
\caption{A comparison of the pressure waveforms from the $(A,Q)$ and $(A,U)$ systems with $\alpha = 4/3$.}
\label{fig:55vespress3}
\end{center}
\end{figure}

\begin{figure}[!htb]
\begin{center}
\includegraphics[scale=0.35]{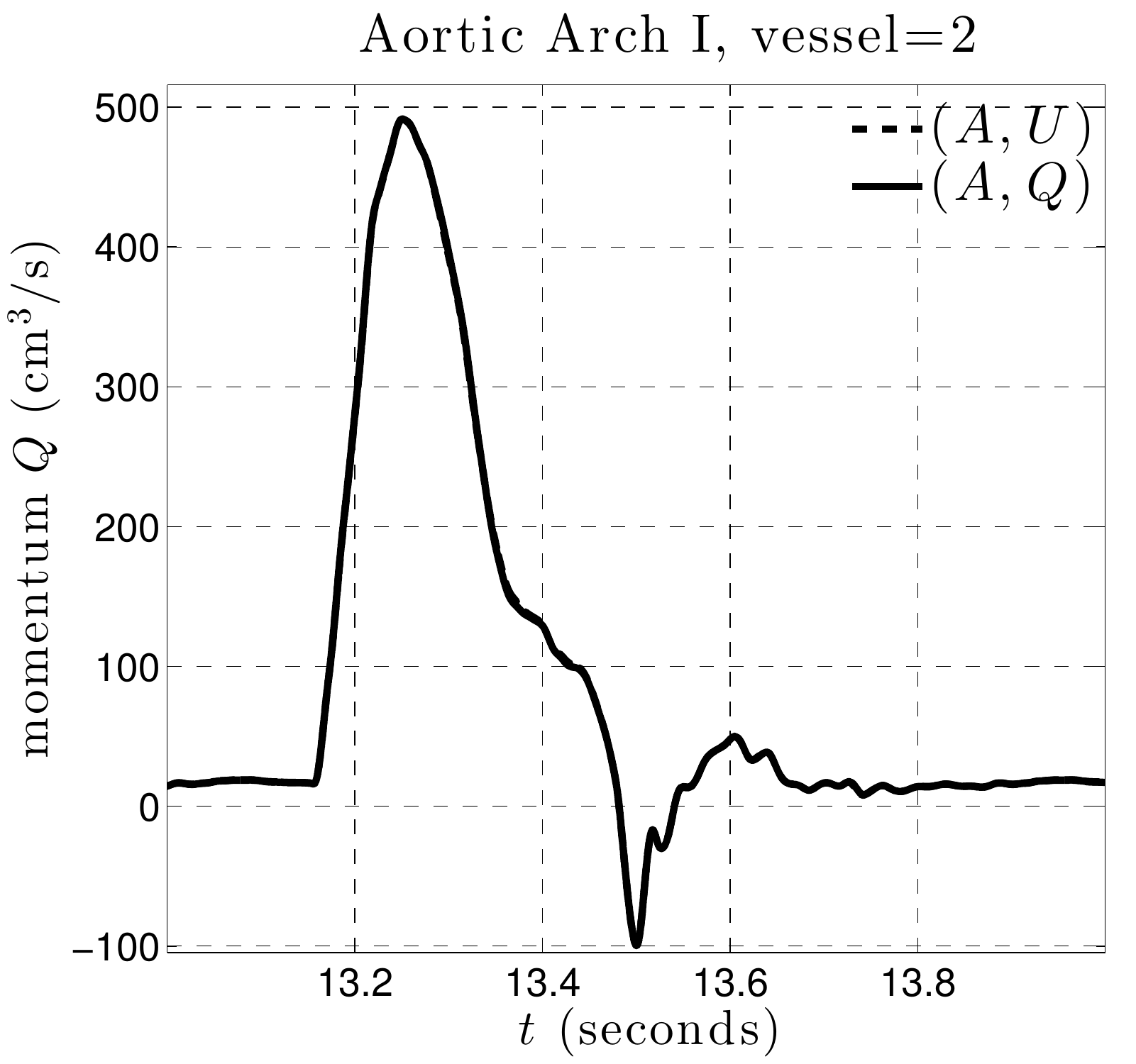} \includegraphics[scale=0.35]{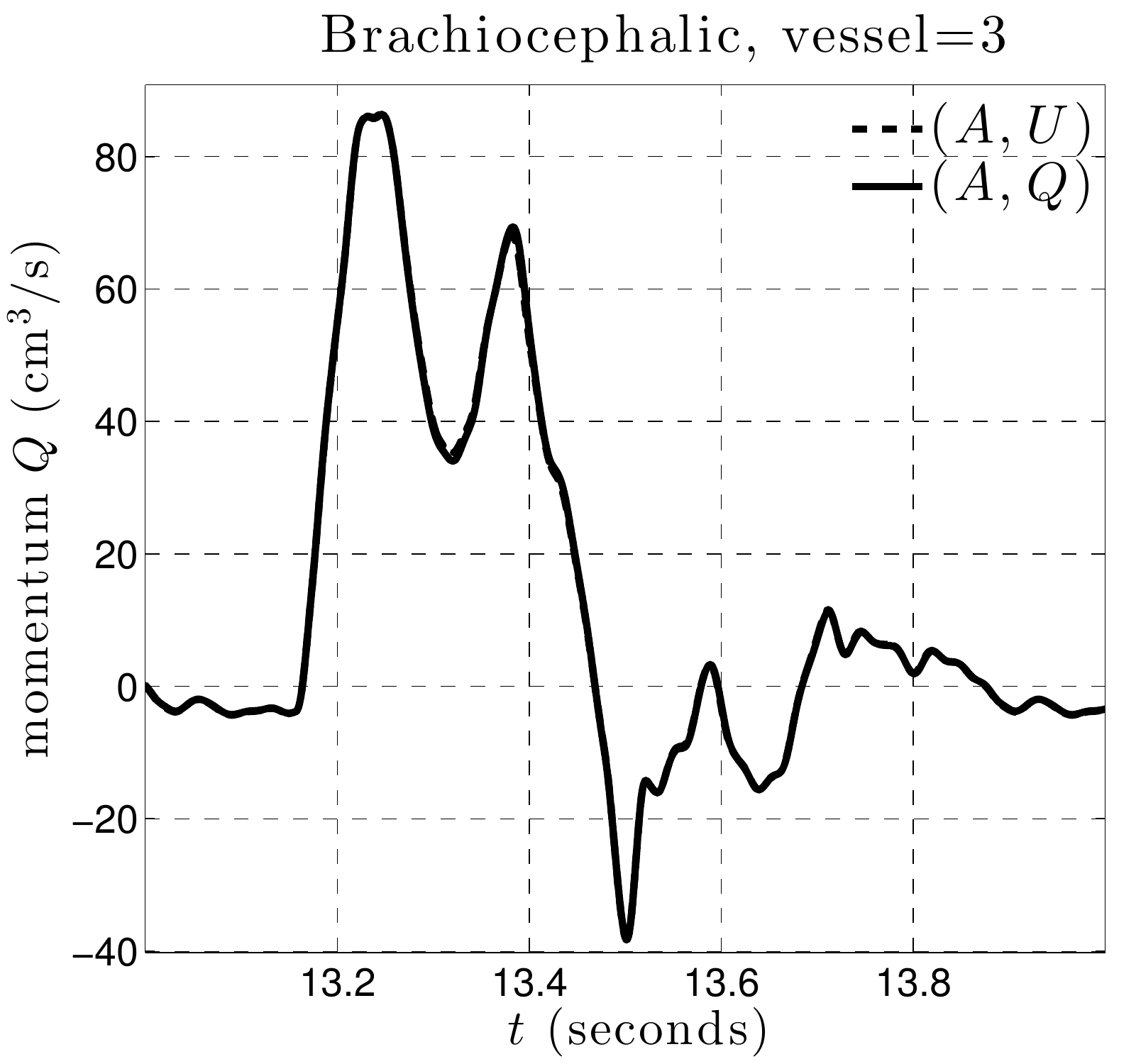} \includegraphics[scale=0.35]{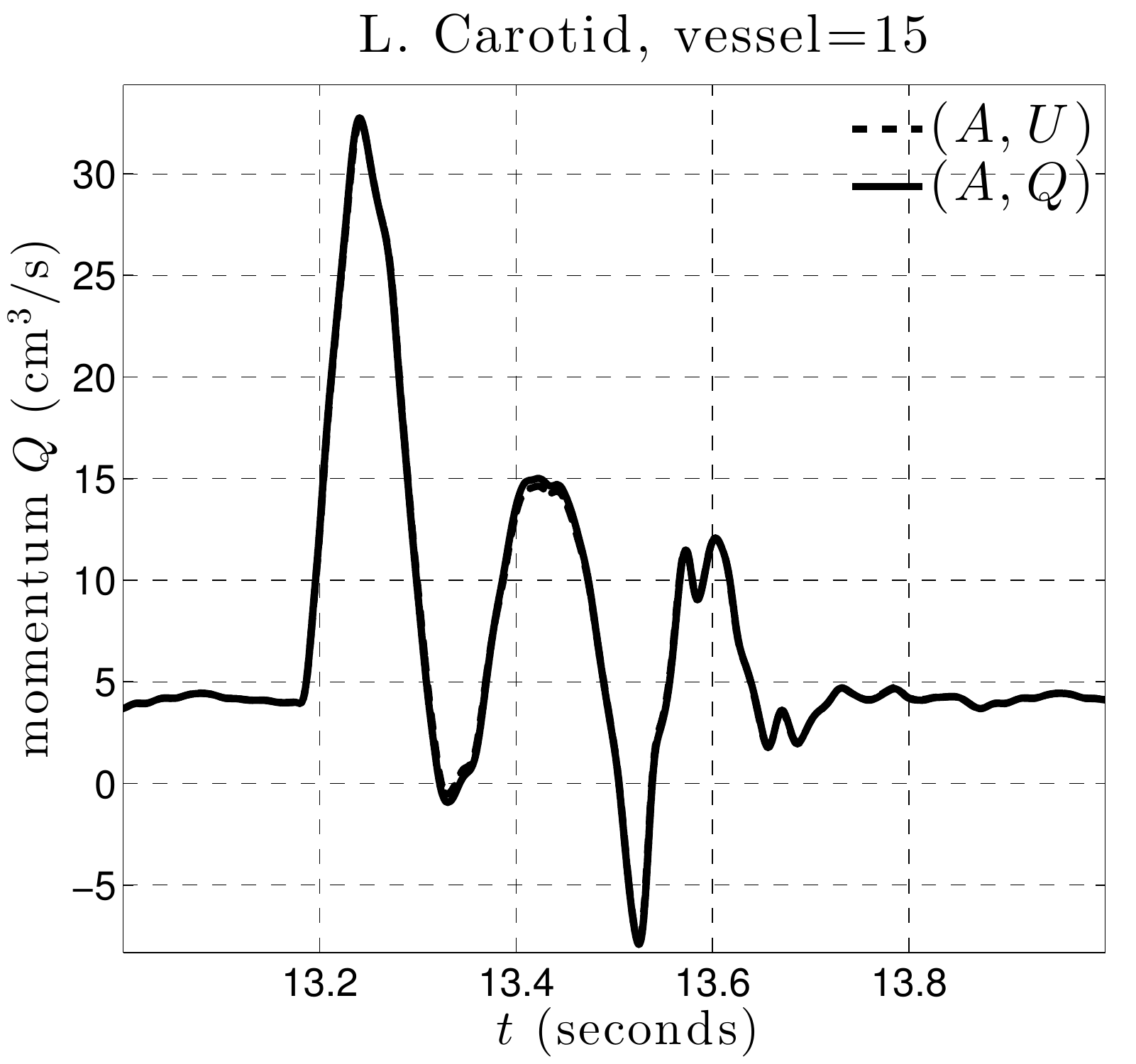} \\
\includegraphics[scale=0.35]{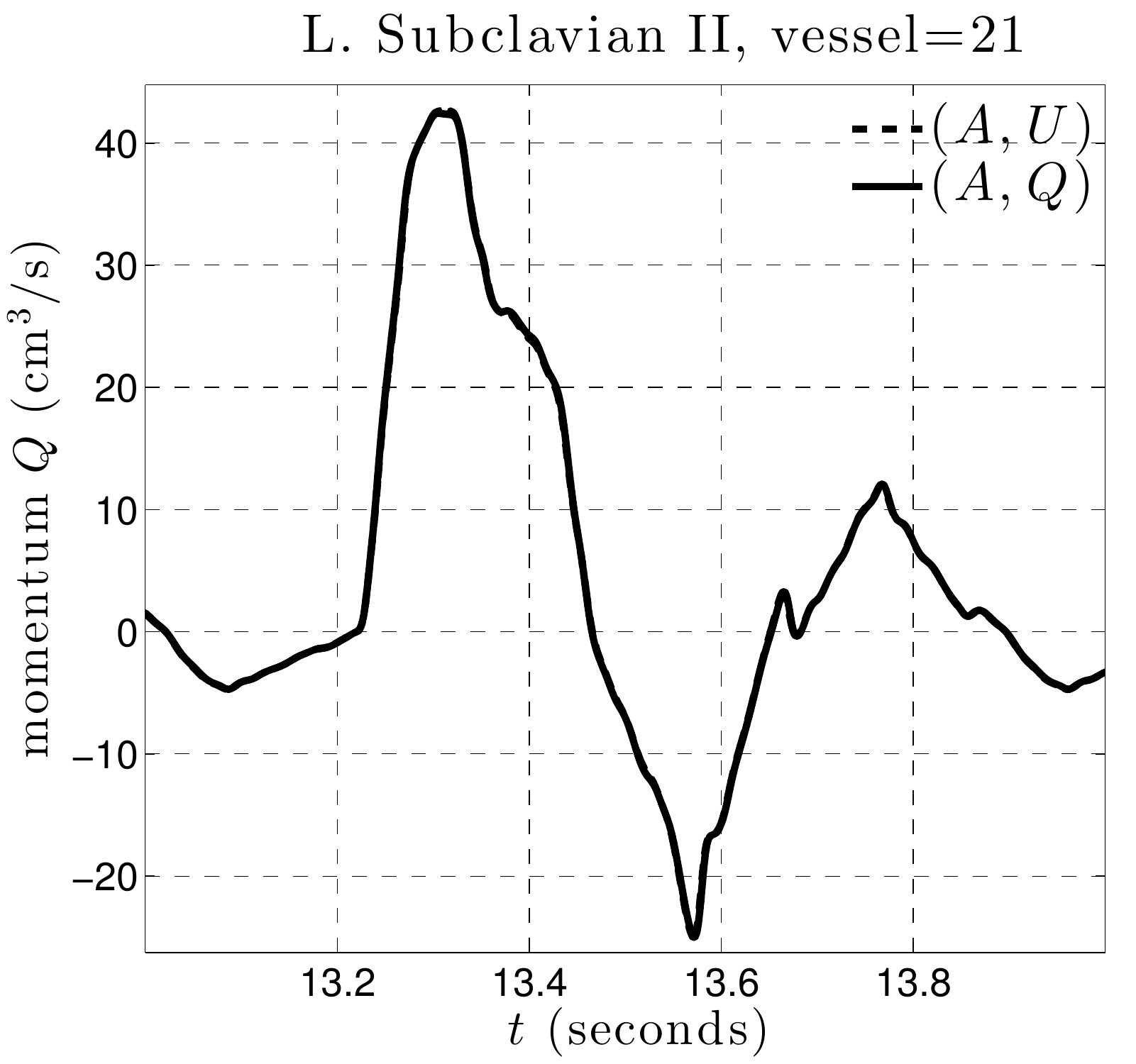} \includegraphics[scale=0.35]{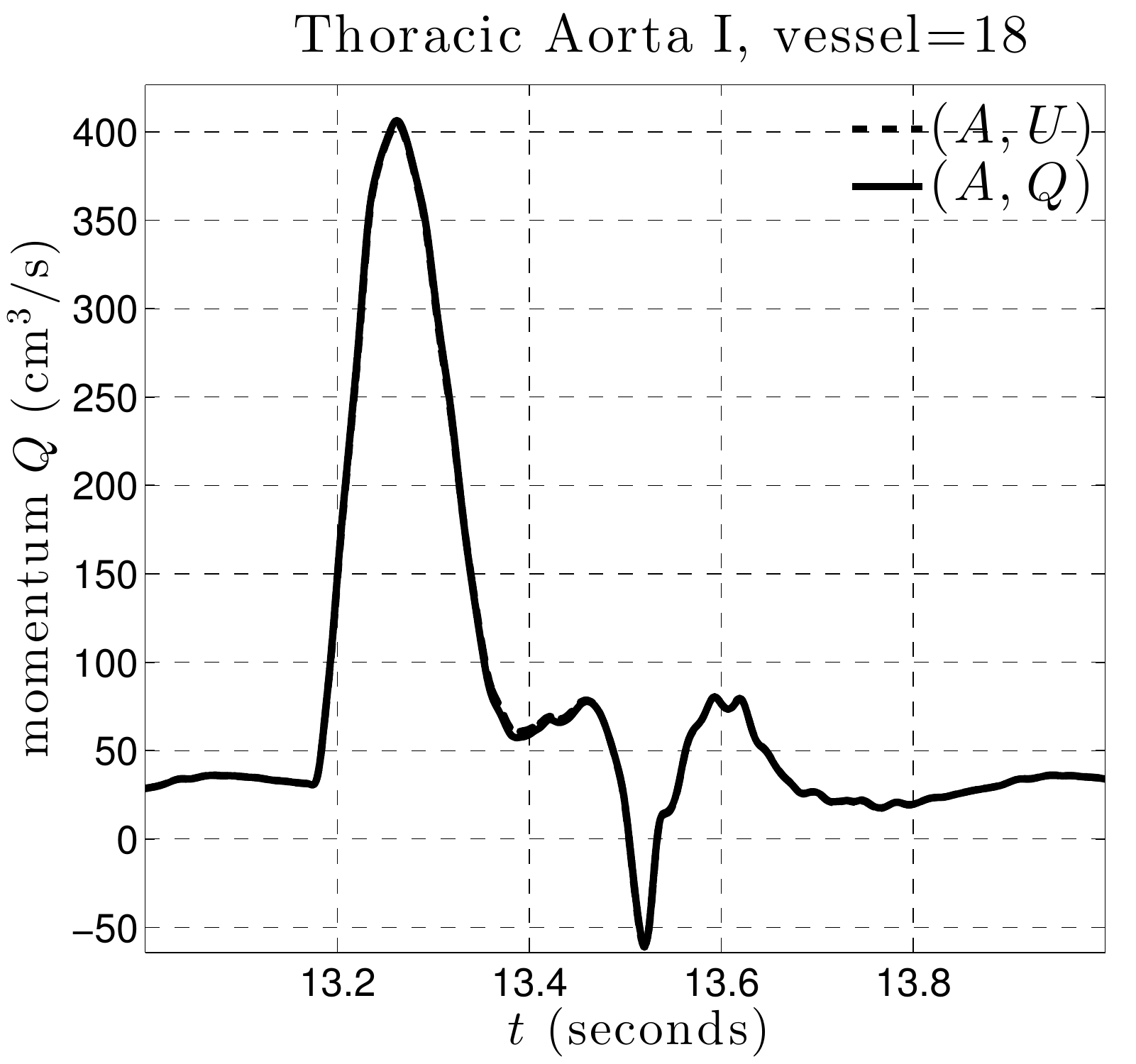} \includegraphics[scale=0.35]{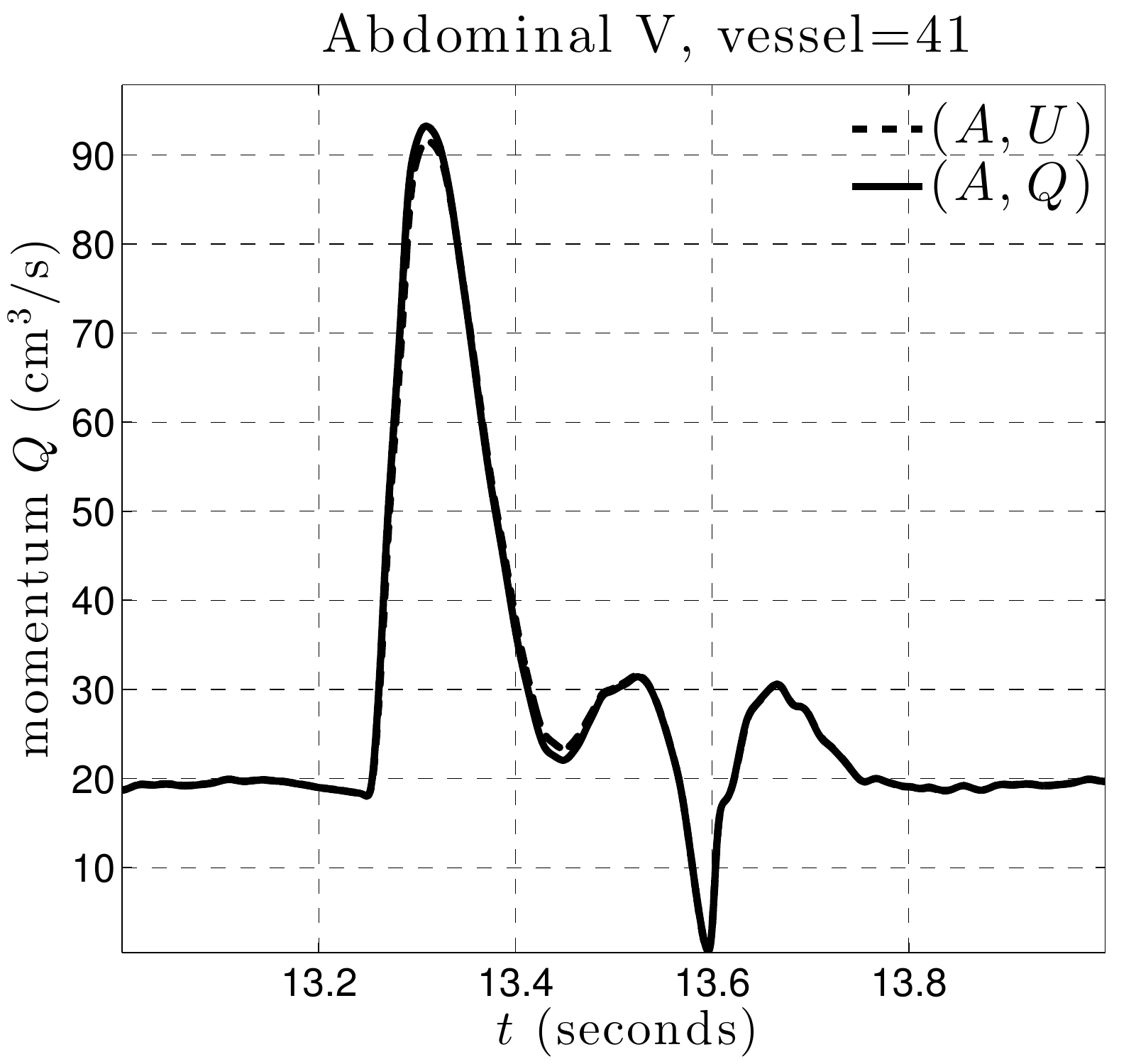} \\
\includegraphics[scale=0.35]{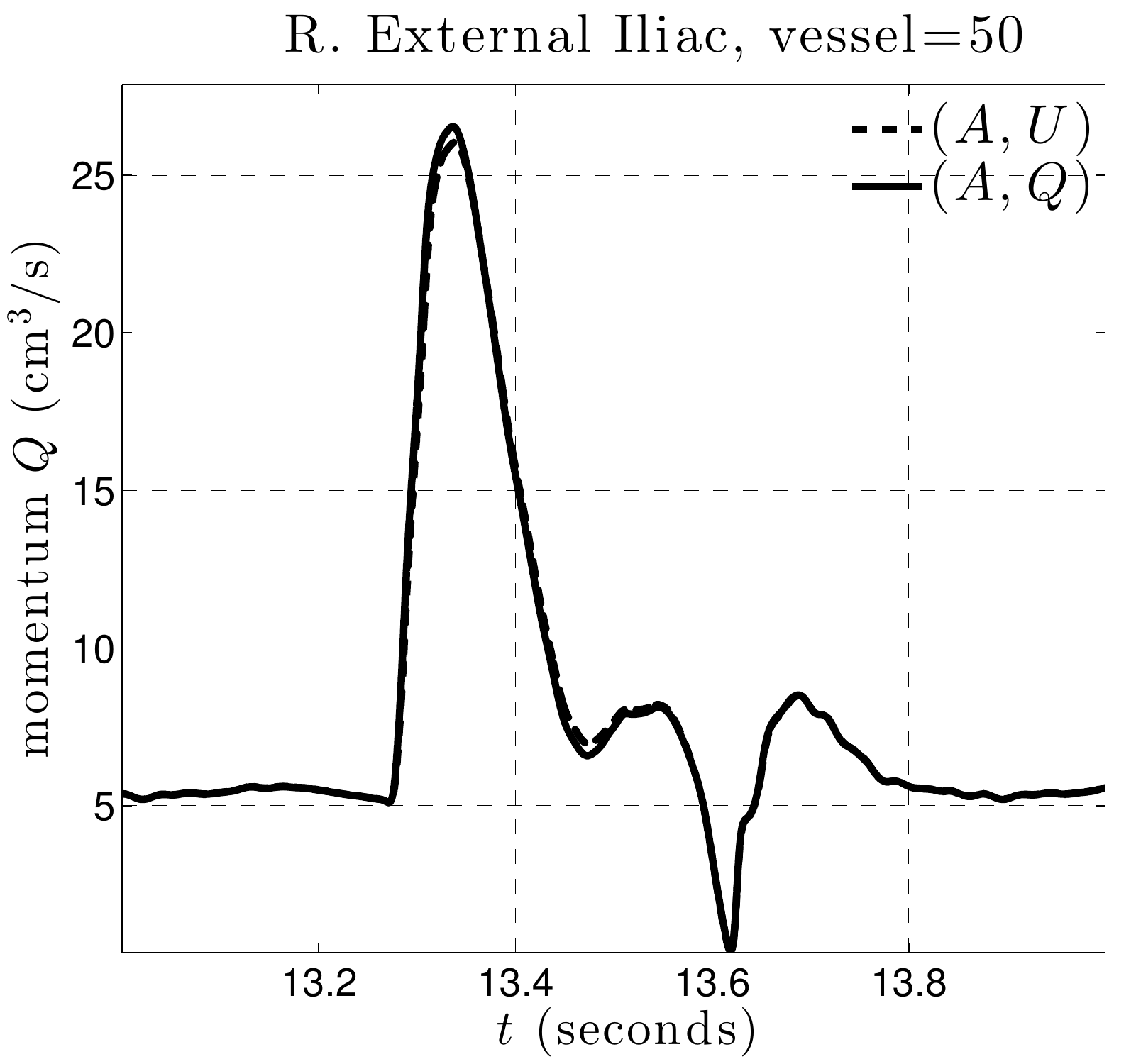} \includegraphics[scale=0.35]{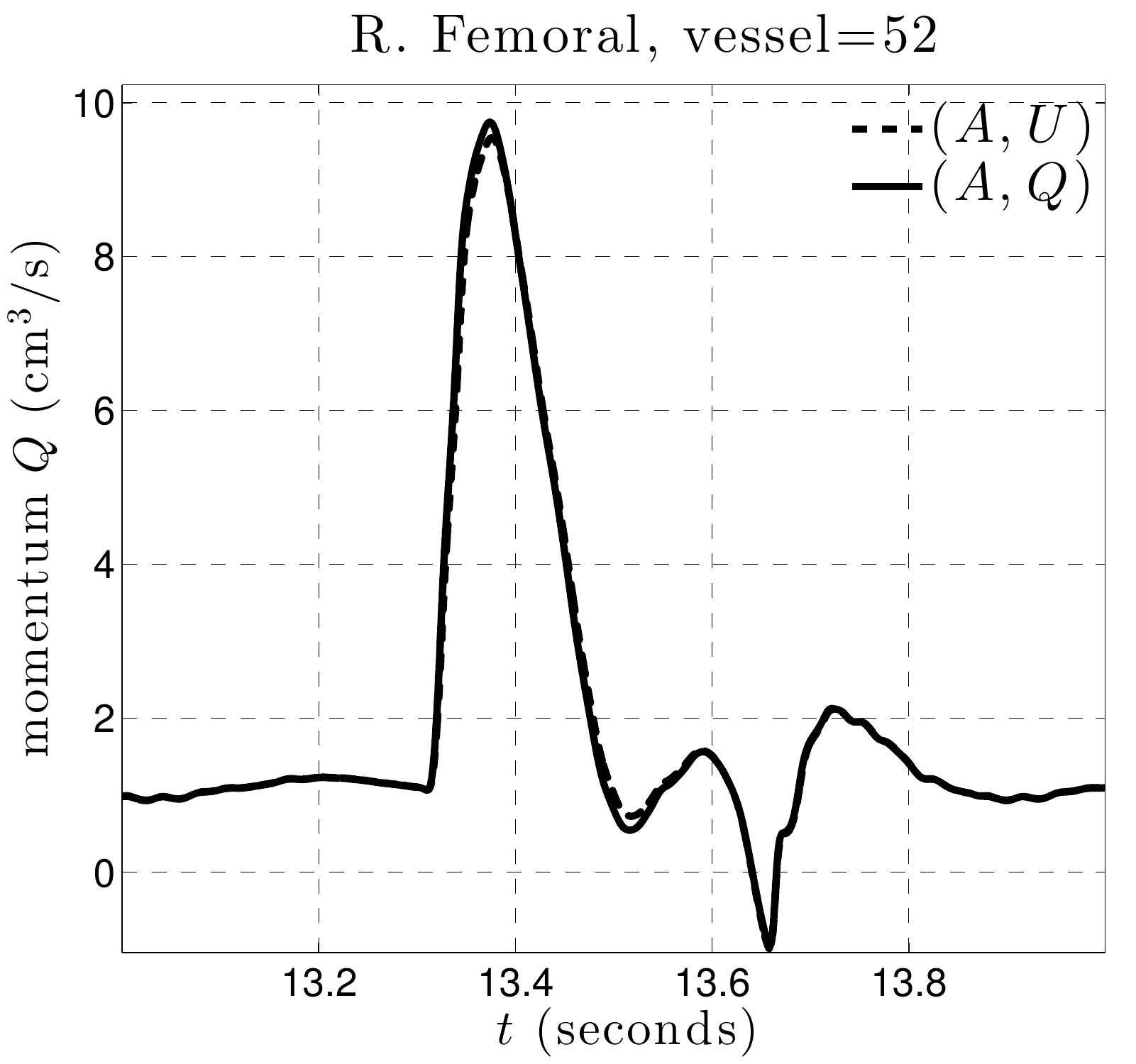} \includegraphics[scale=0.35]{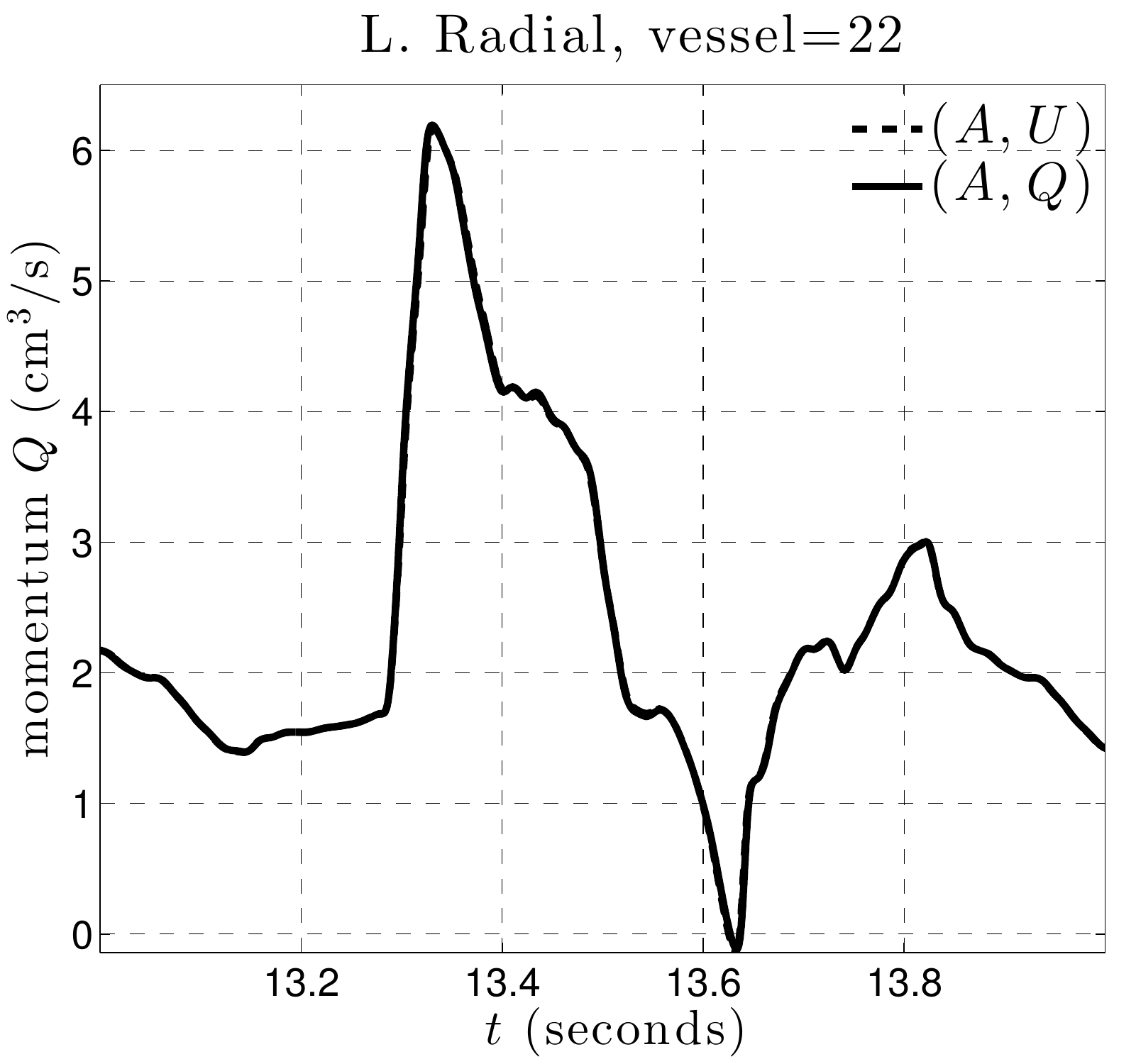} \\
\caption{A comparison of the momentum waveforms from the $(A,Q)$ and $(A,U)$ systems with $\alpha = 1.1$.}
\label{fig:55vesmom4}
\end{center}
\end{figure}

\begin{figure}[!htb]
\begin{center}
\includegraphics[scale=0.35]{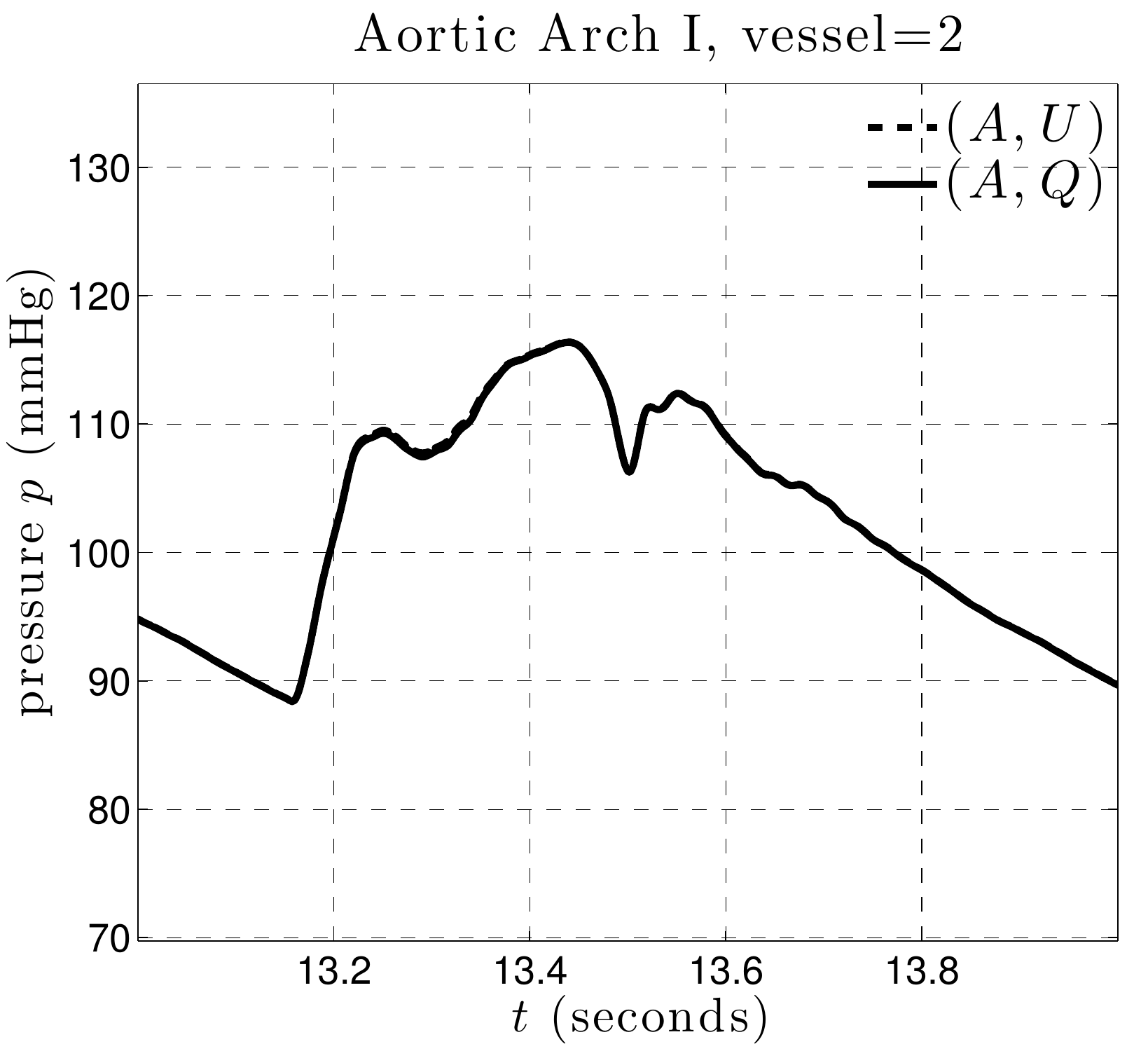} \includegraphics[scale=0.35]{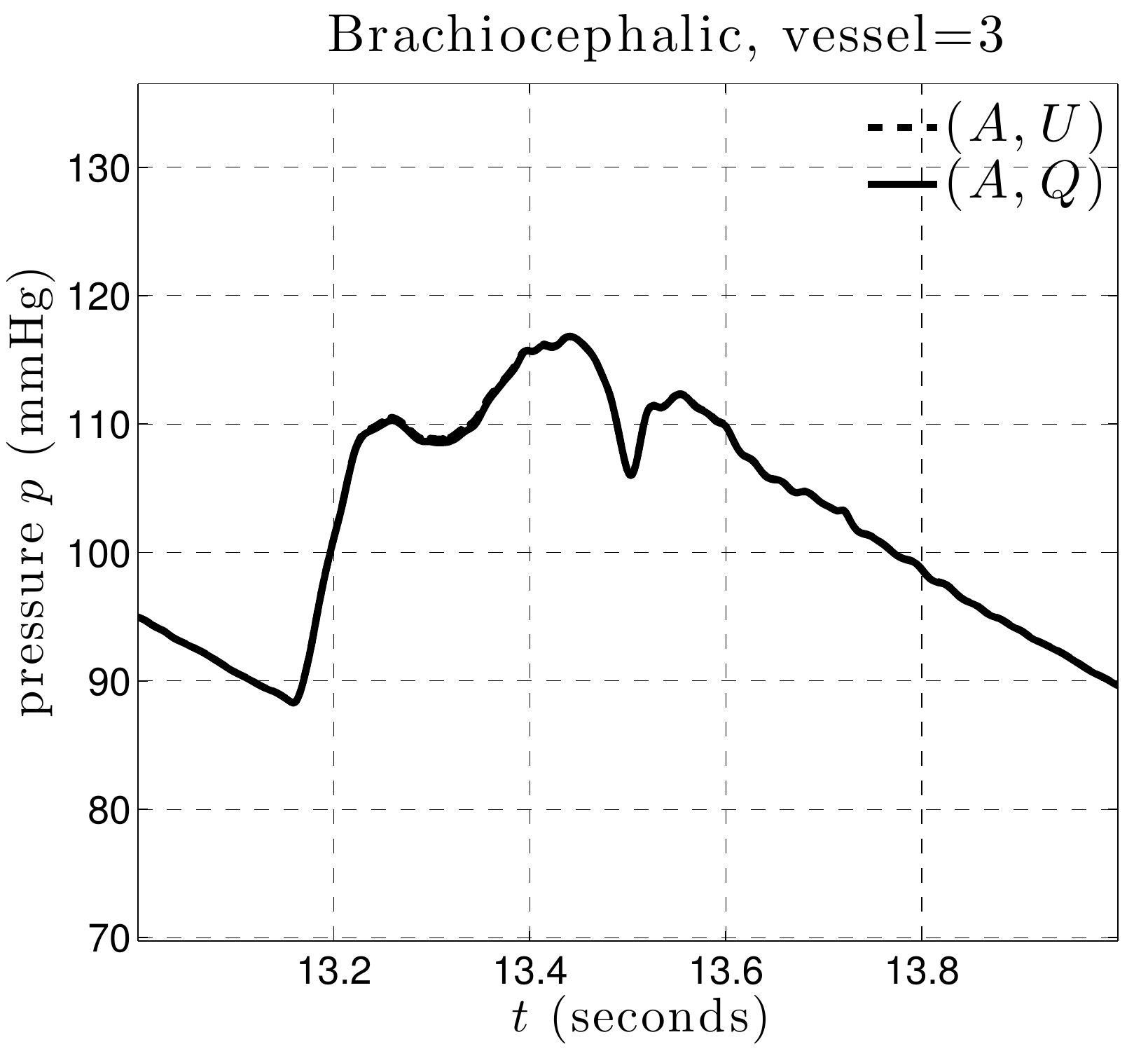} \includegraphics[scale=0.35]{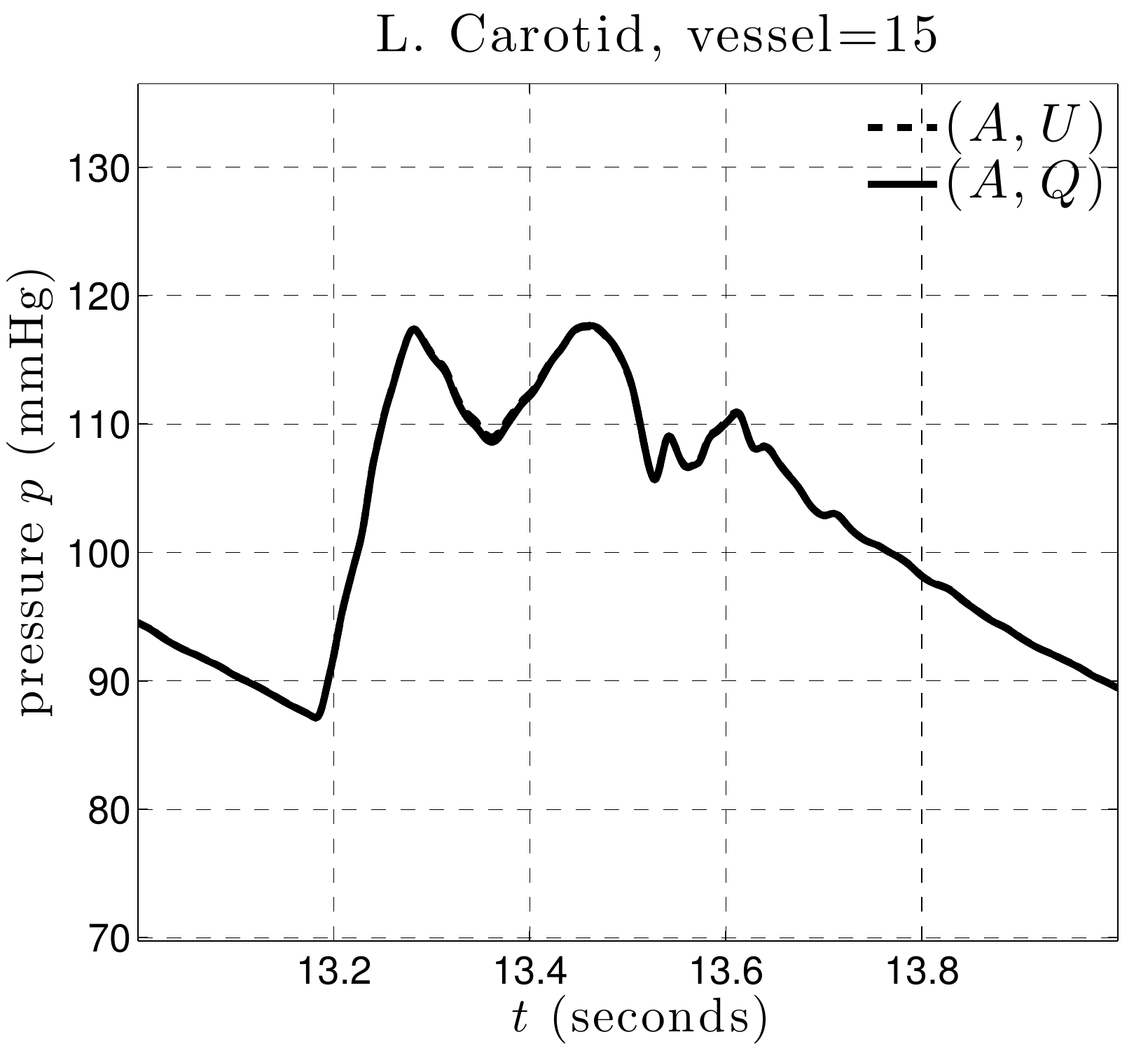} \\
\includegraphics[scale=0.35]{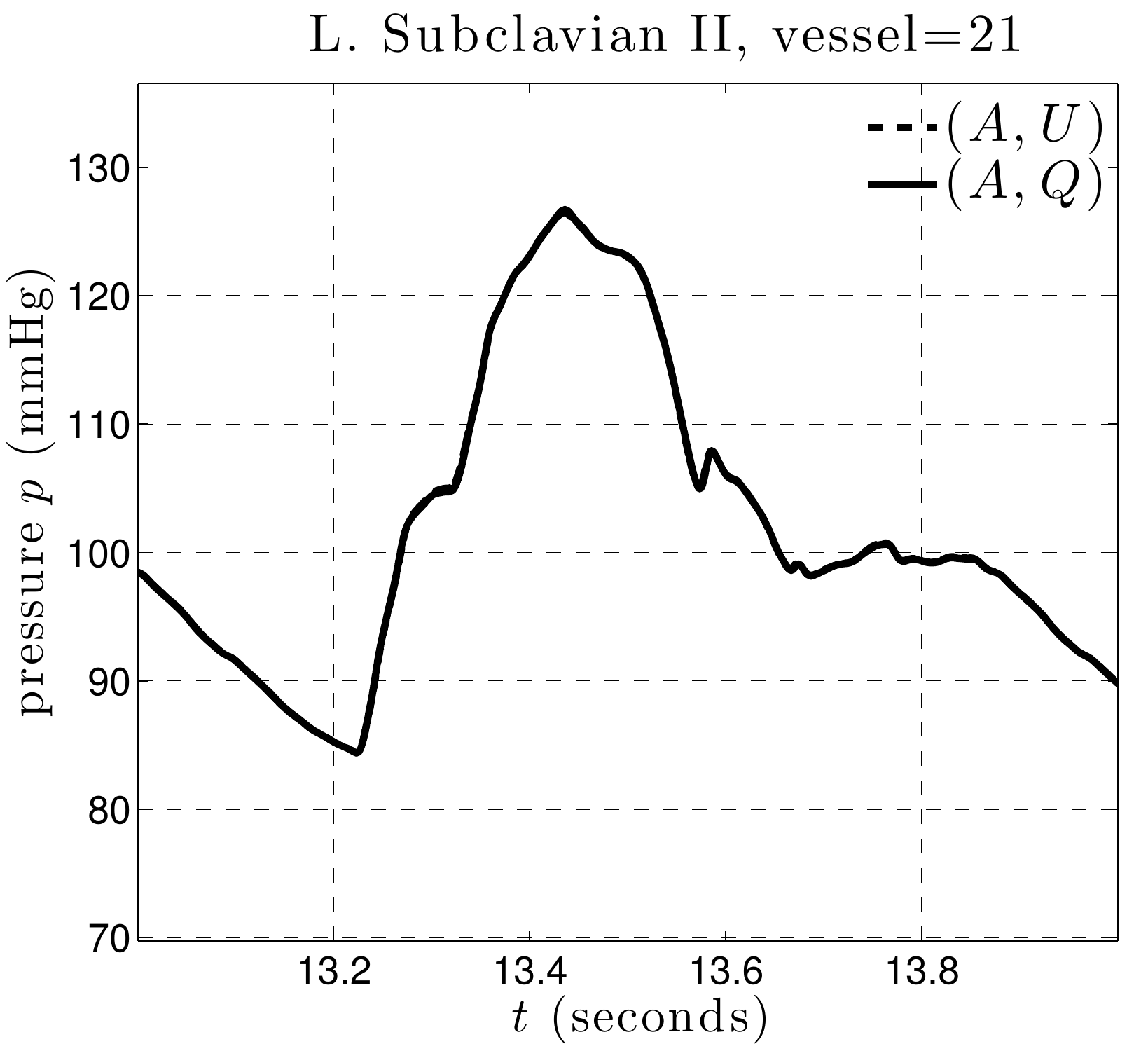} \includegraphics[scale=0.35]{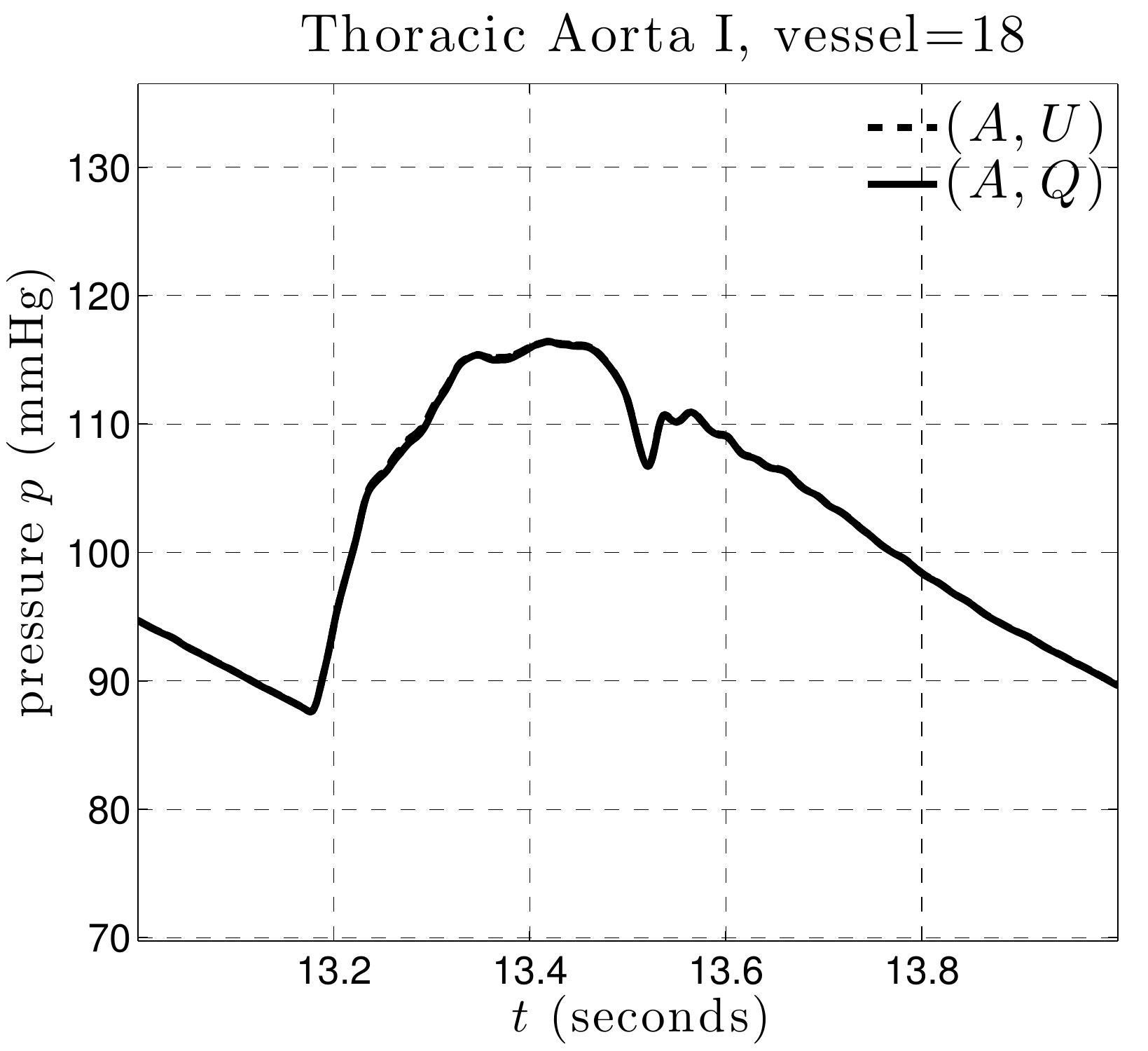} \includegraphics[scale=0.35]{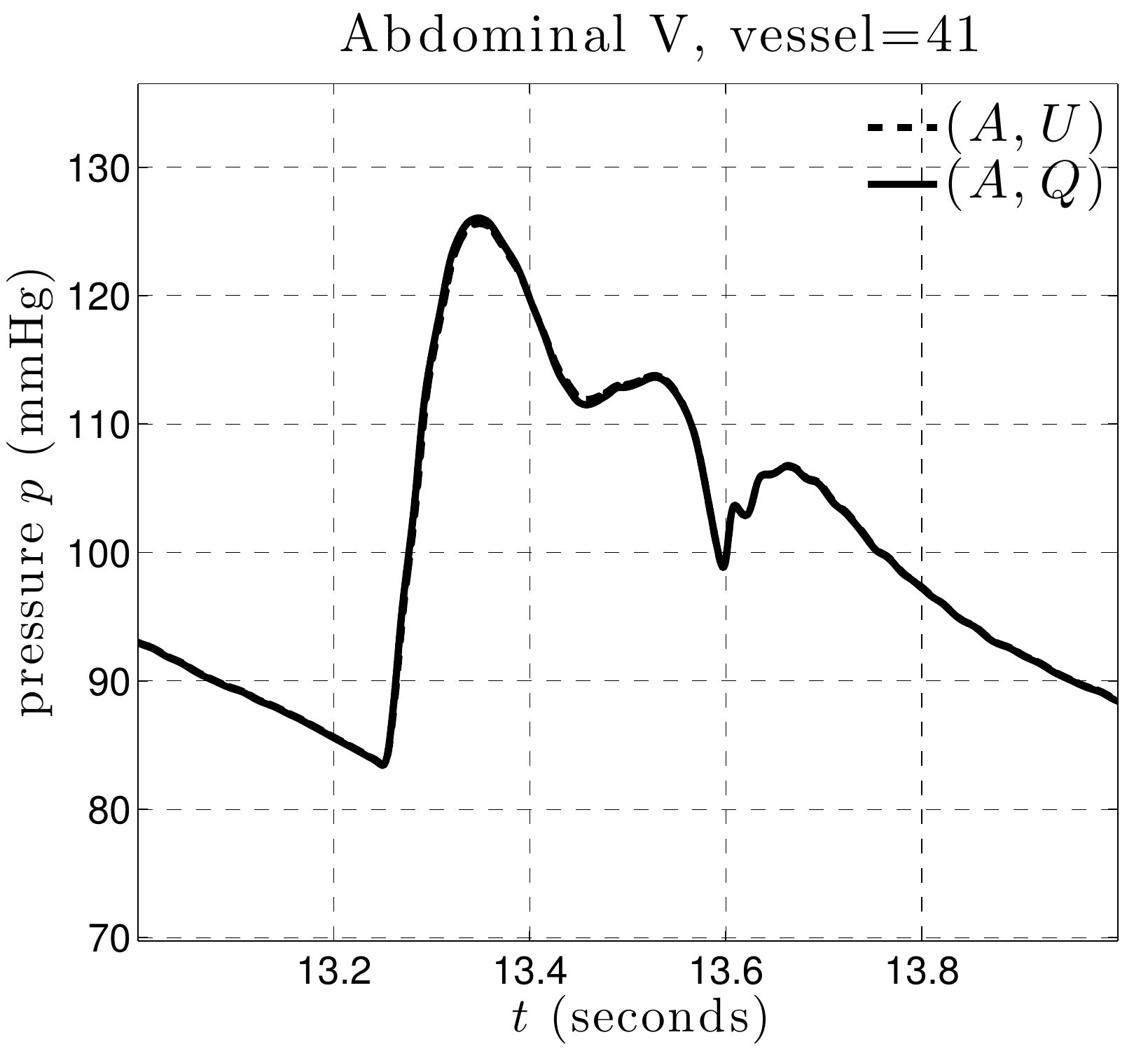} \\
\includegraphics[scale=0.35]{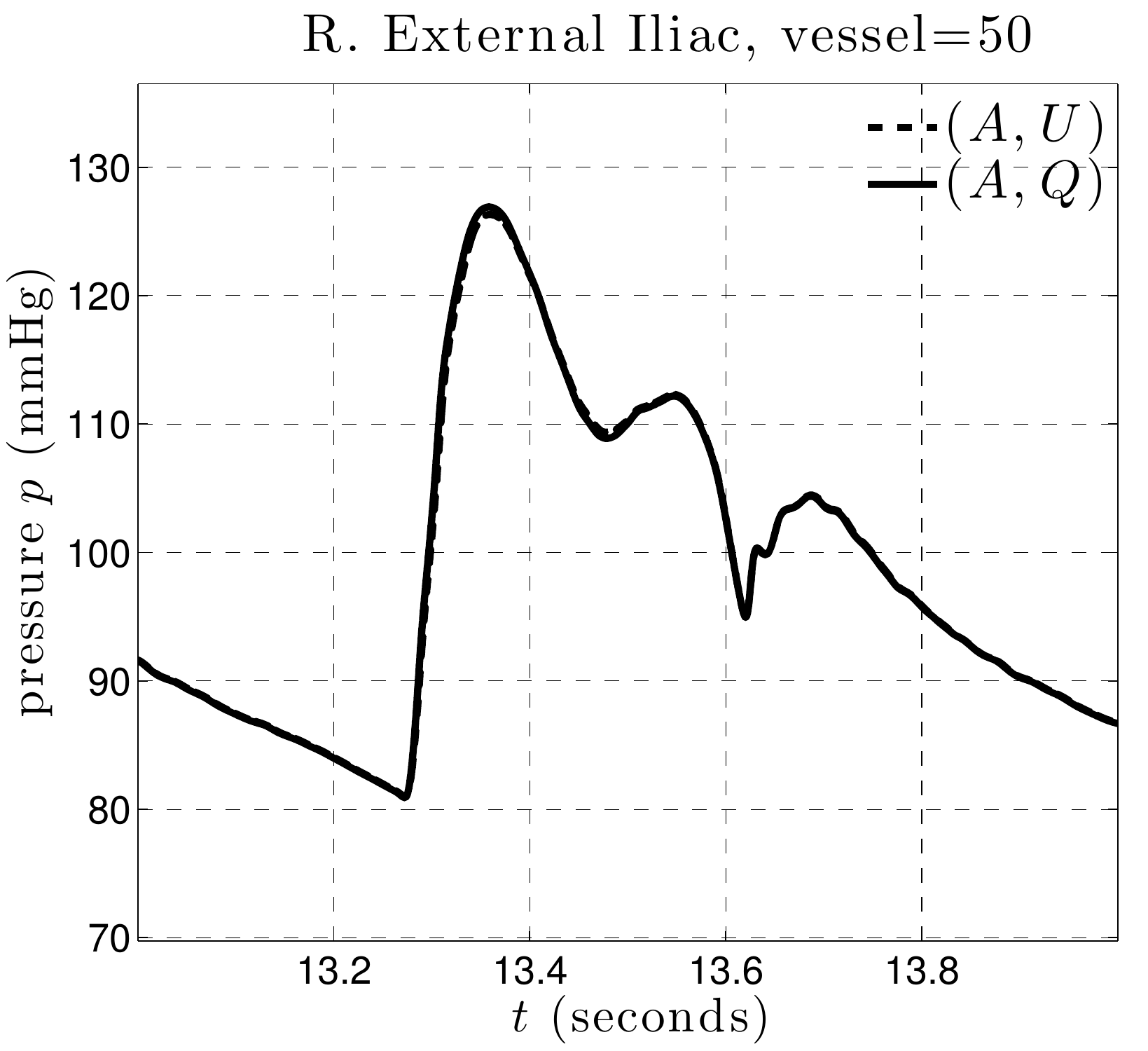} \includegraphics[scale=0.35]{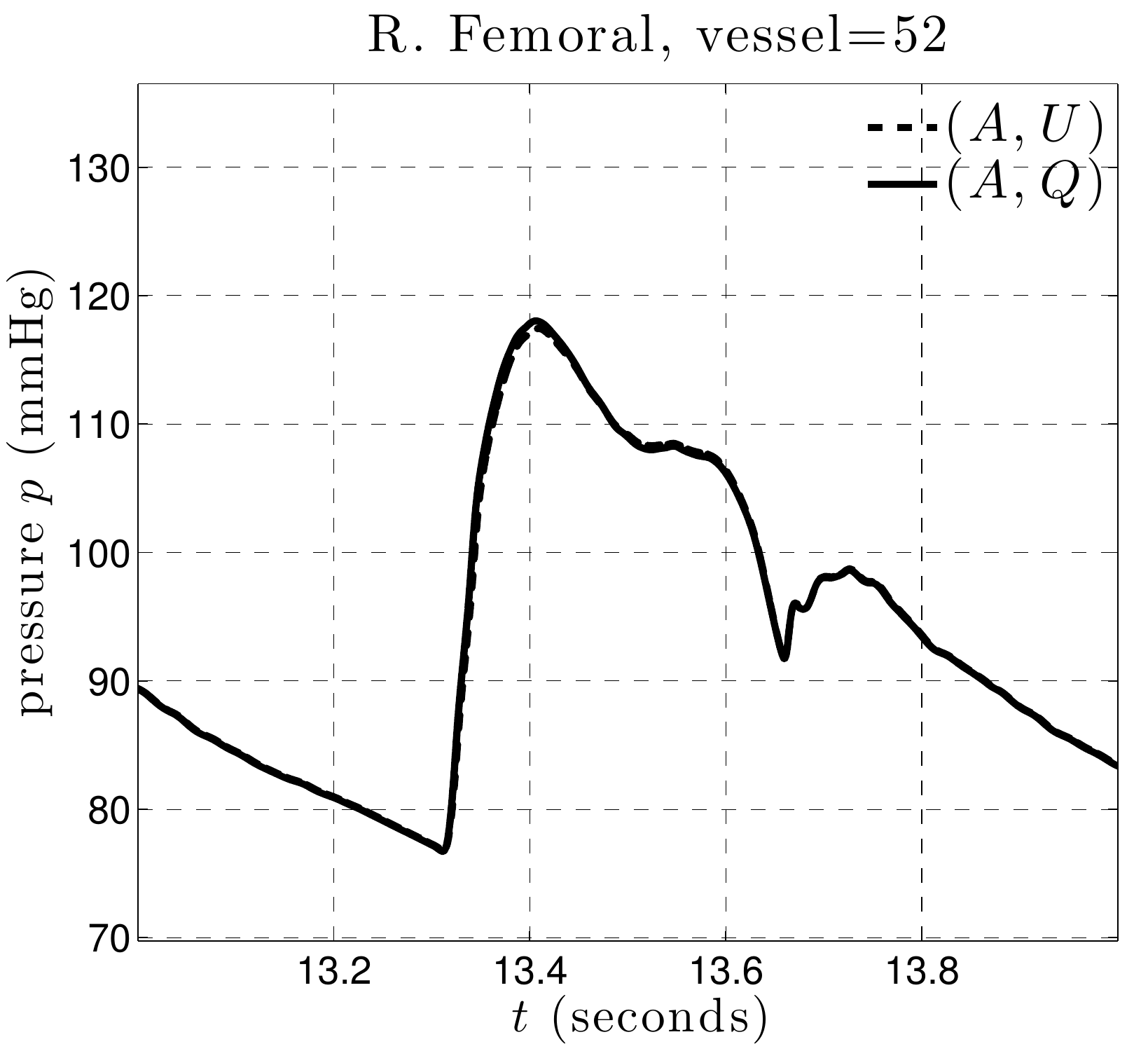} \includegraphics[scale=0.35]{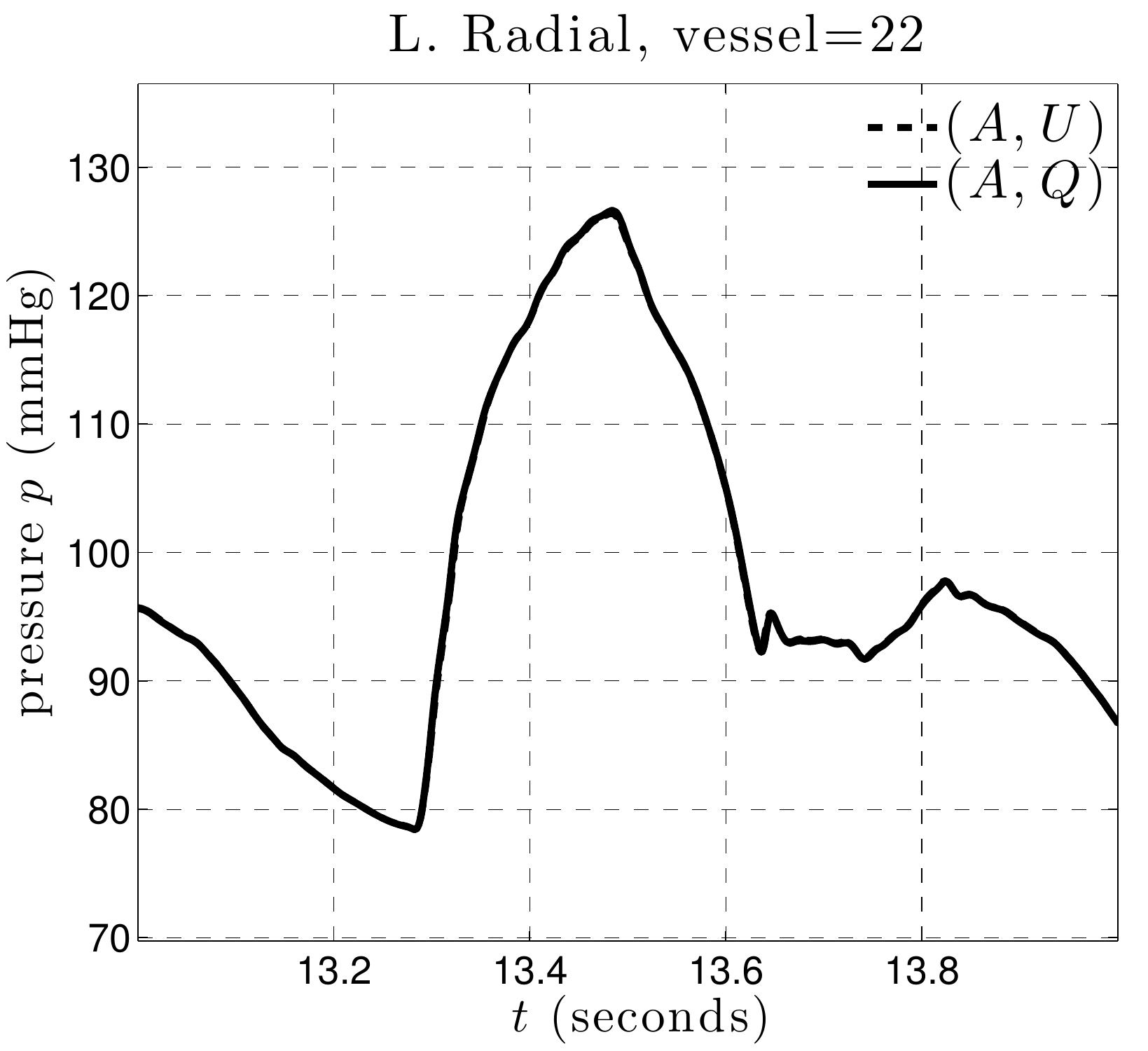} \\
\caption{A comparison of the pressure waveforms from the $(A,Q)$ and $(A,U)$ systems with $\alpha = 1.1$.}
\label{fig:55vespress4}
\end{center}
\end{figure}

Figures \ref{fig:55vesmom1} and \ref{fig:55vespress1} compare flow and pressure waveforms obtained throughout the network with the $(A,Q)$ system for $\alpha$ values $1.1$ and $4/3$.  In Figures \ref{fig:55vesmom2} and \ref{fig:55vespress2}, we show the same comparison for the $(A,U)$ system. Figures \ref{fig:55vesmom3} and \ref{fig:55vespress3} compare waveforms between the $(A,Q)$ and $(A,U)$ systems for $\alpha = 4/3$, and Figures \ref{fig:55vesmom4} and \ref{fig:55vespress4} show the same comparison but for $\alpha = 1.1$.  

We observe similar discrepancies in the waveforms for both the $(A,Q)$ and $(A,U)$ systems for different values of $\alpha$.   For the Poiseuille profile corresponding to $\alpha = 4/3$, the viscous term is smaller than for the flatter profile corresponding to $\alpha = 1.1$.  This difference yields waveforms with higher magnitude pressure gradients and oscillations for $\alpha = 4/3$; see for example the radial and subclavian arteries.  Further, note that most of the waveforms corresponding to $\alpha = 4/3$ exhibit a lower mean pressure, especially for the larger arteries.  This difference could be explained by the fact that a fluid with lower viscosity ($\alpha = 4/3$) moves more easily through a compliant cylinder and therefore renders a lower mean pressure. 

For a fixed value of $\alpha$, the $(A,Q)$ and $(A,U)$ systems generally produce waveforms with similar features and magnitudes.  As expected, the results agree relatively well for $\alpha = 1.1$, since the $(A,Q)$ and $(A,U)$ systems are equivalent for smooth solutions when $\alpha$ is set equal to one in the convective part of the $(A,Q)$ system.  However, when $\alpha = 4/3$, there are some discrepancies between these systems since they differ more in the convective term.

\begin{figure}[!htb]
\begin{center}
\includegraphics[scale=0.35]{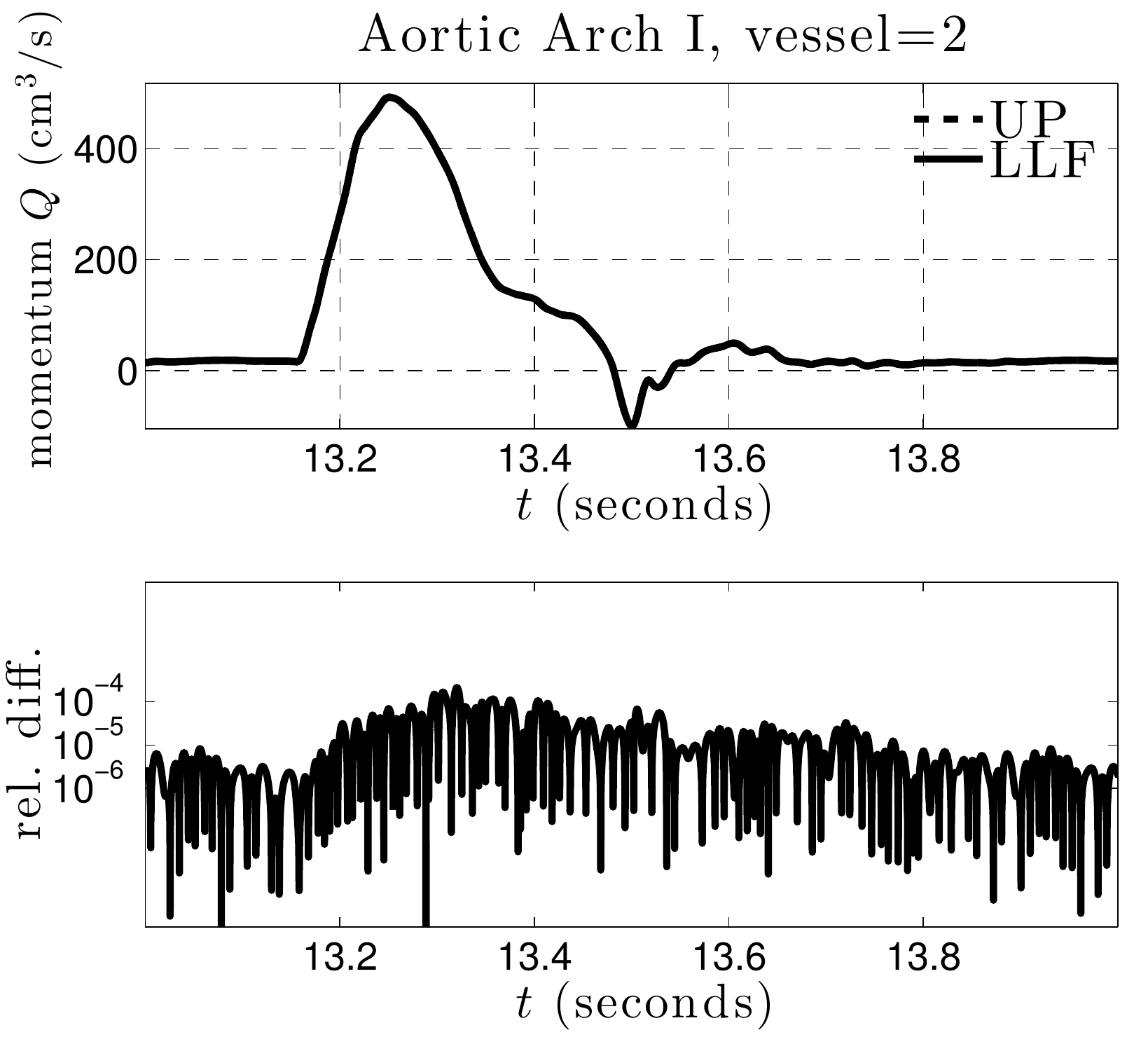} \includegraphics[scale=0.35]{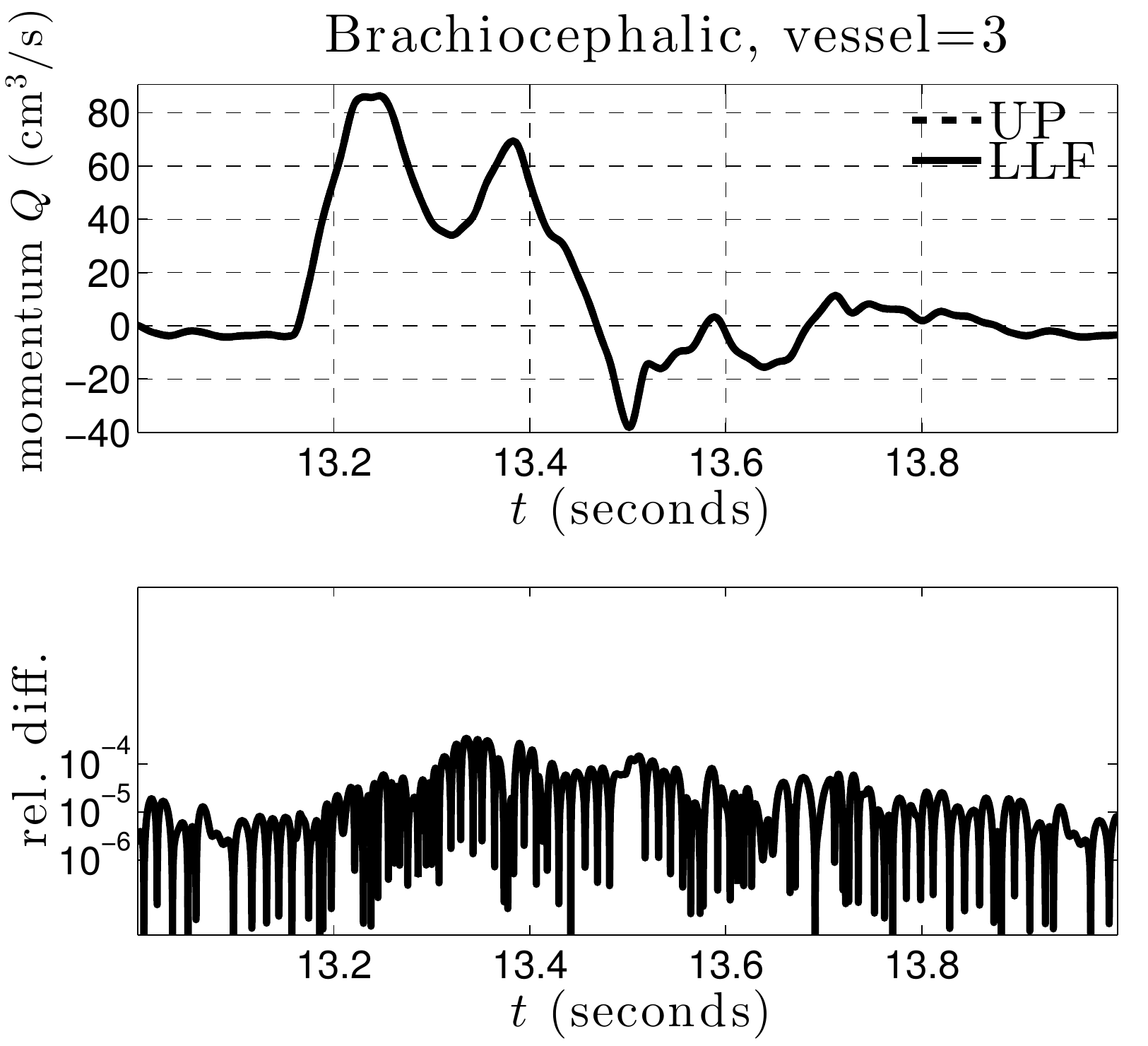} \includegraphics[scale=0.35]{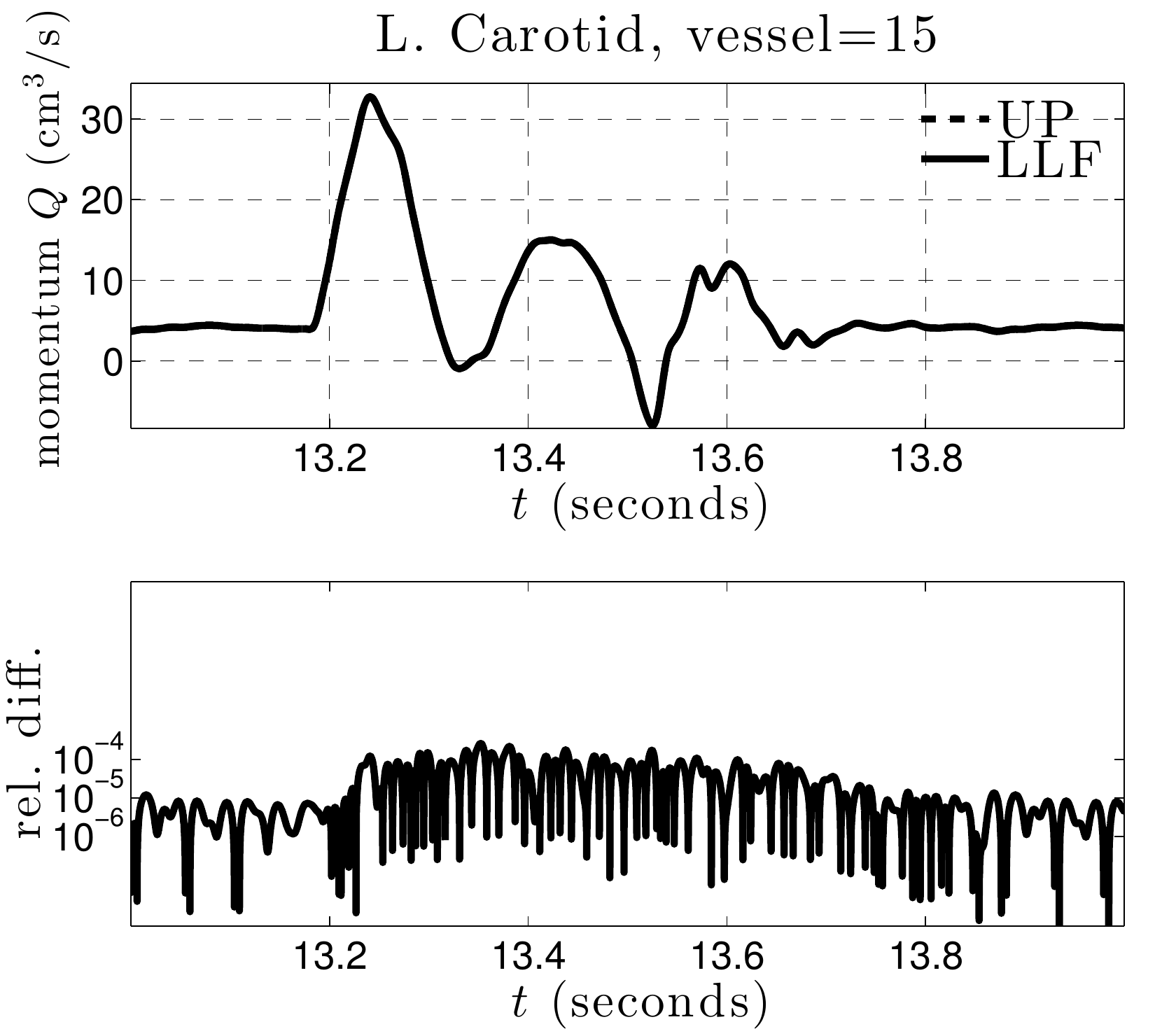} \\
\includegraphics[scale=0.35]{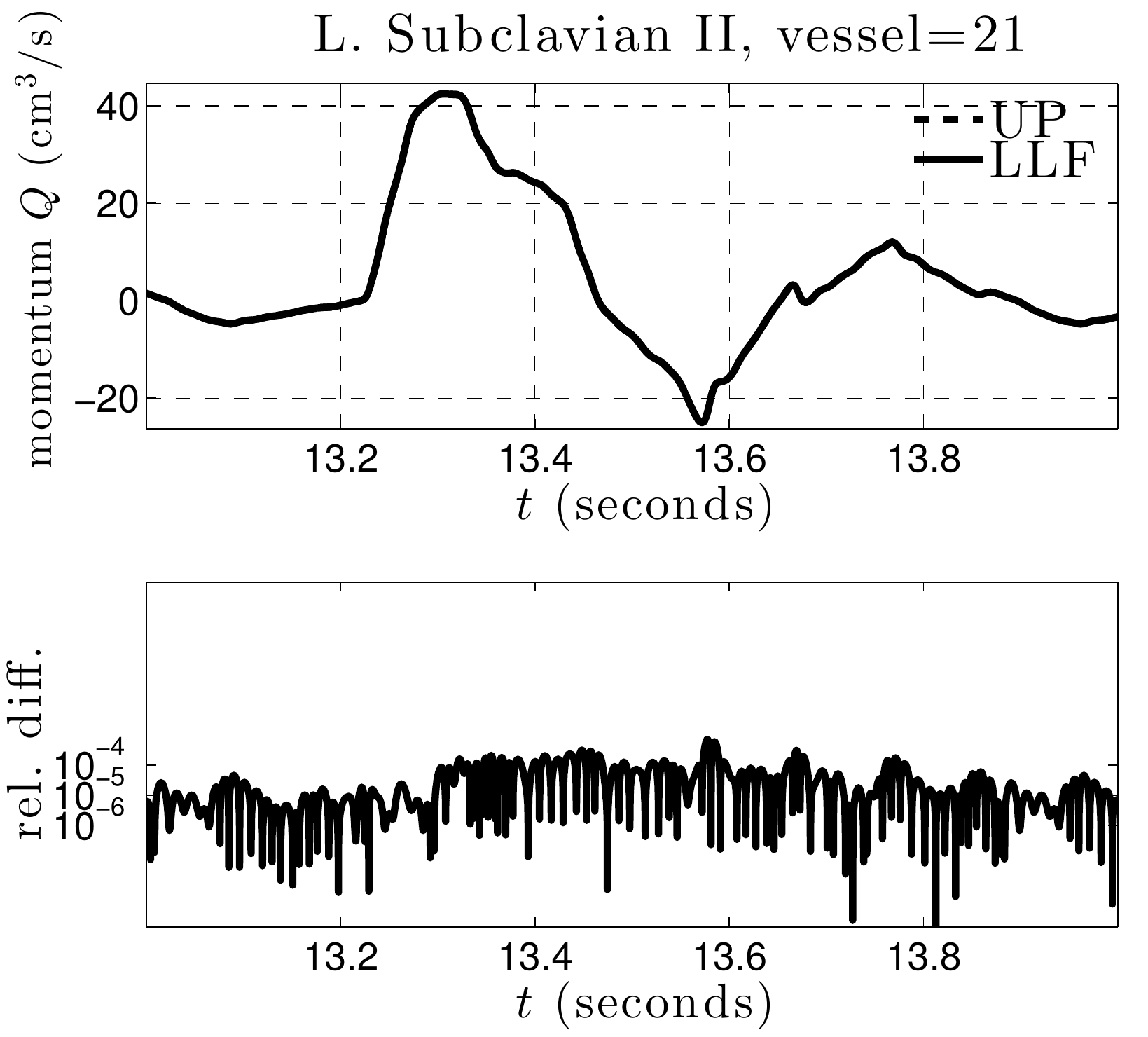} \includegraphics[scale=0.35]{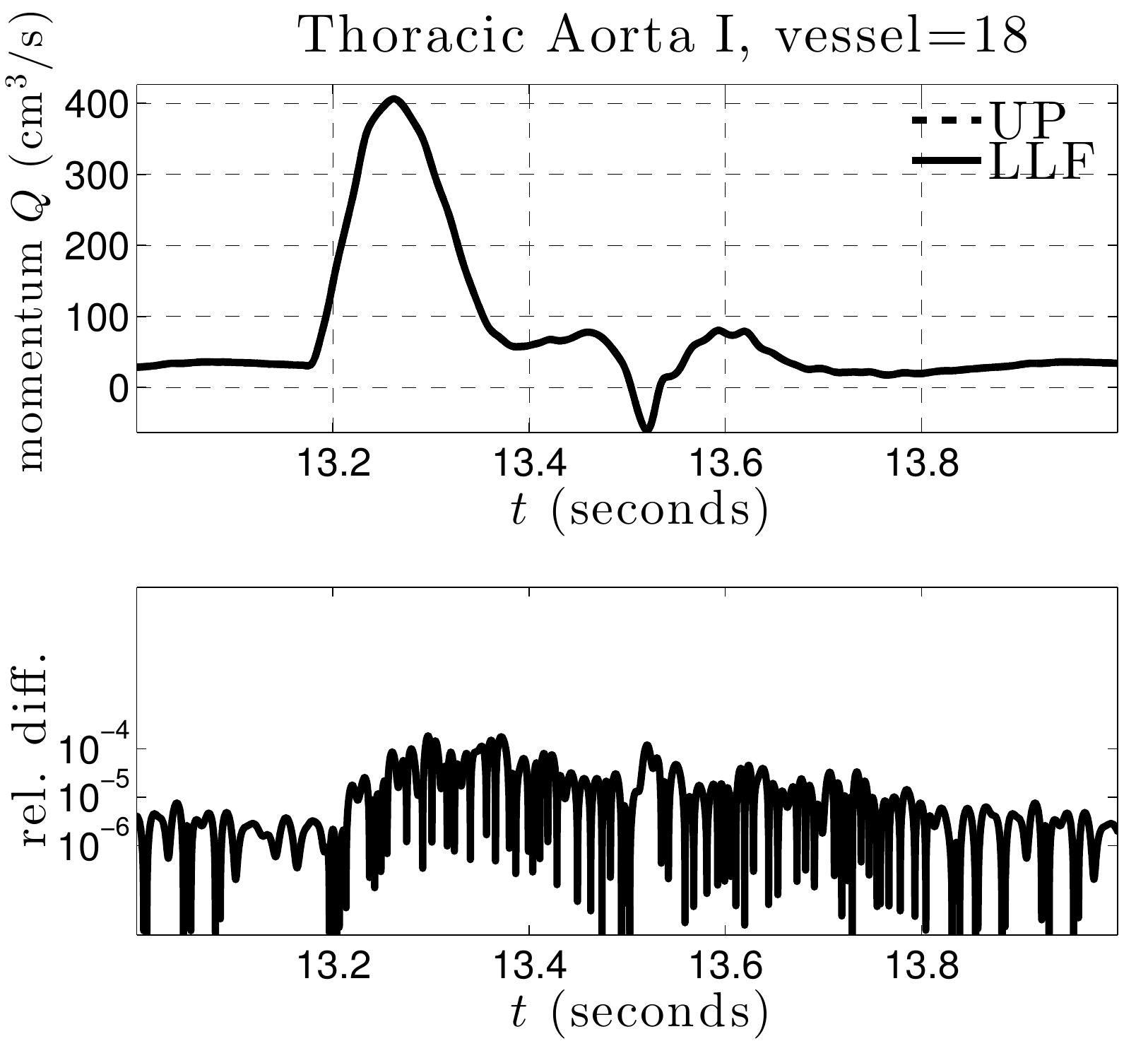} \includegraphics[scale=0.35]{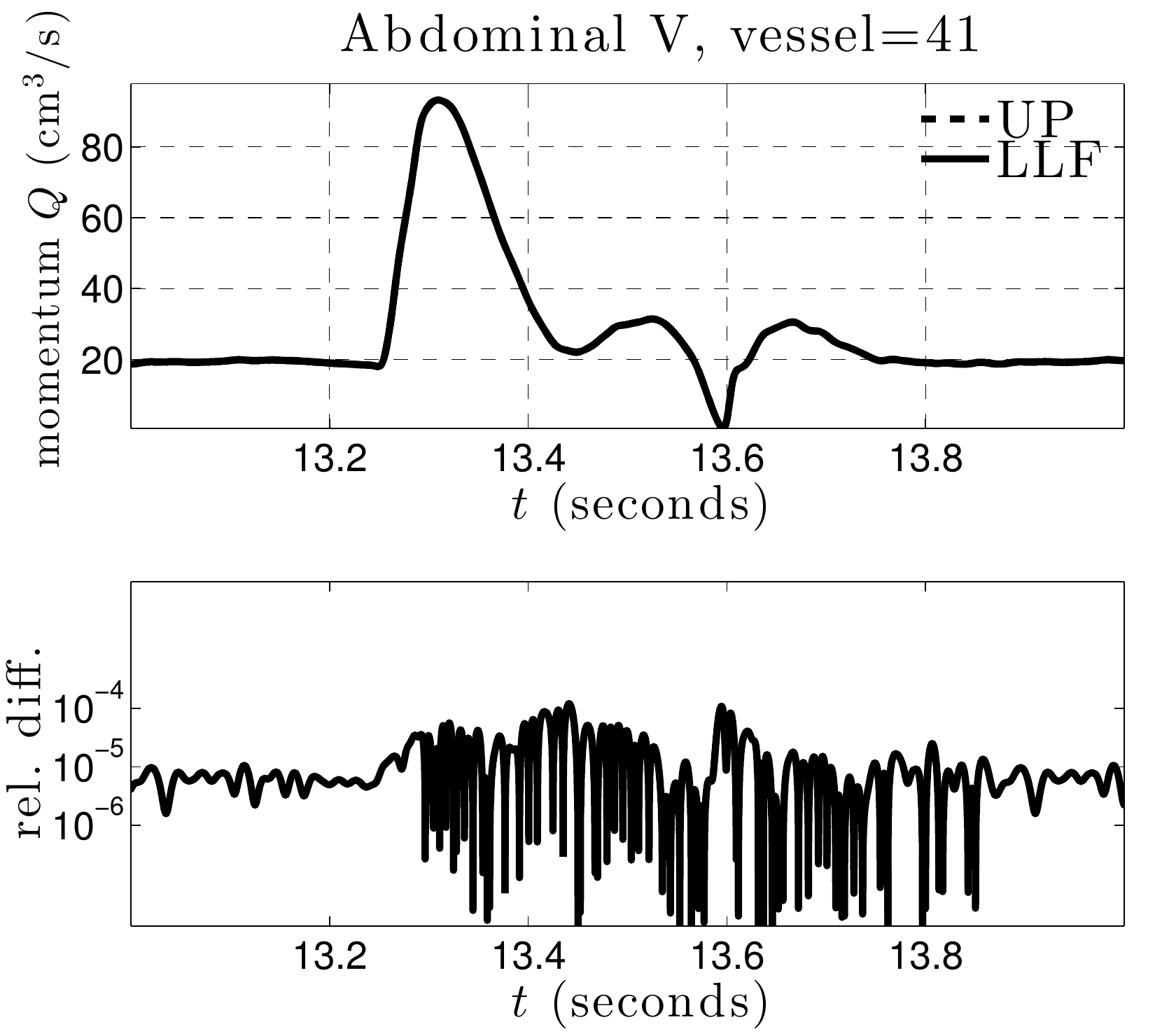} \\
\includegraphics[scale=0.35]{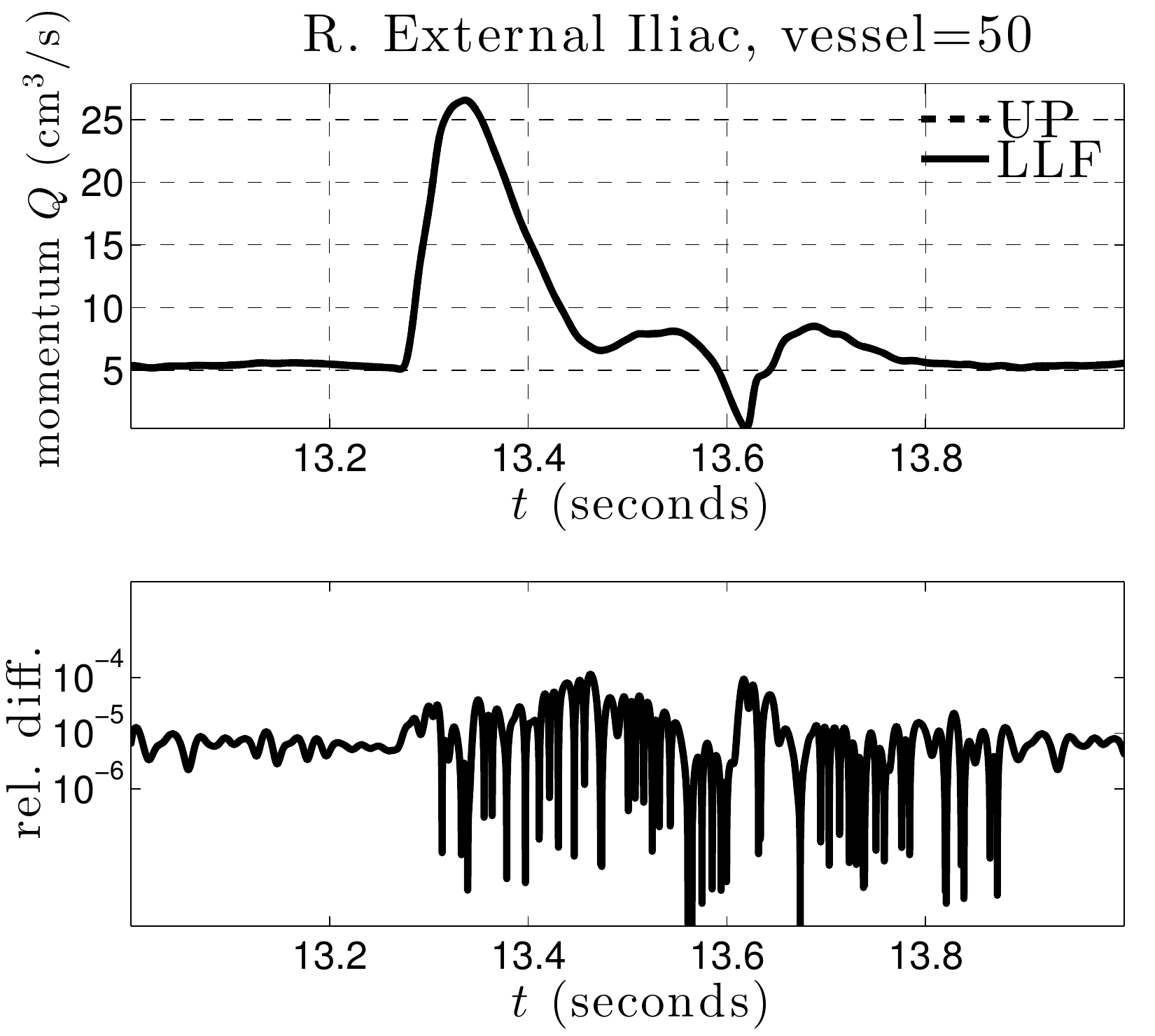} \includegraphics[scale=0.35]{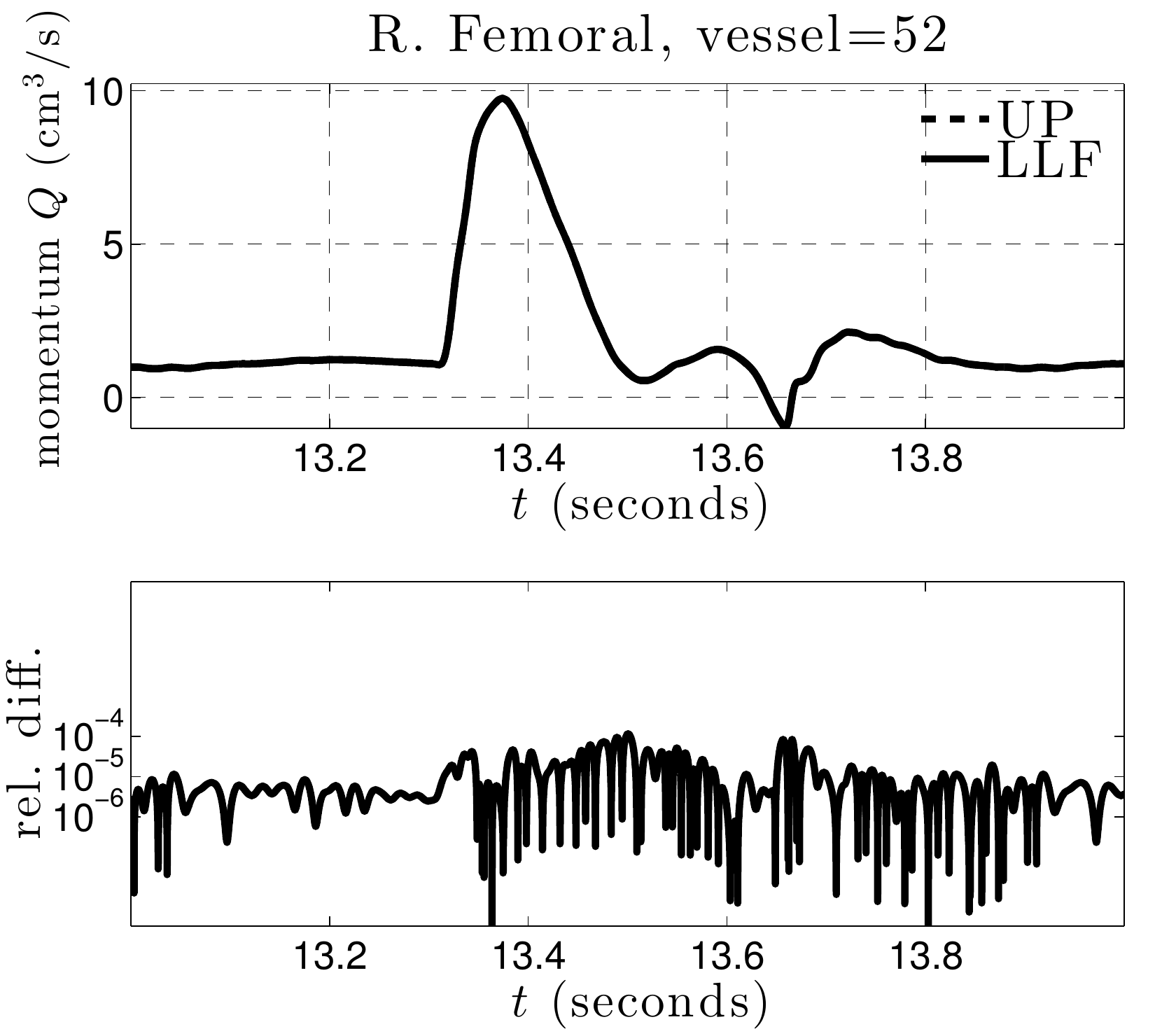} \includegraphics[scale=0.35]{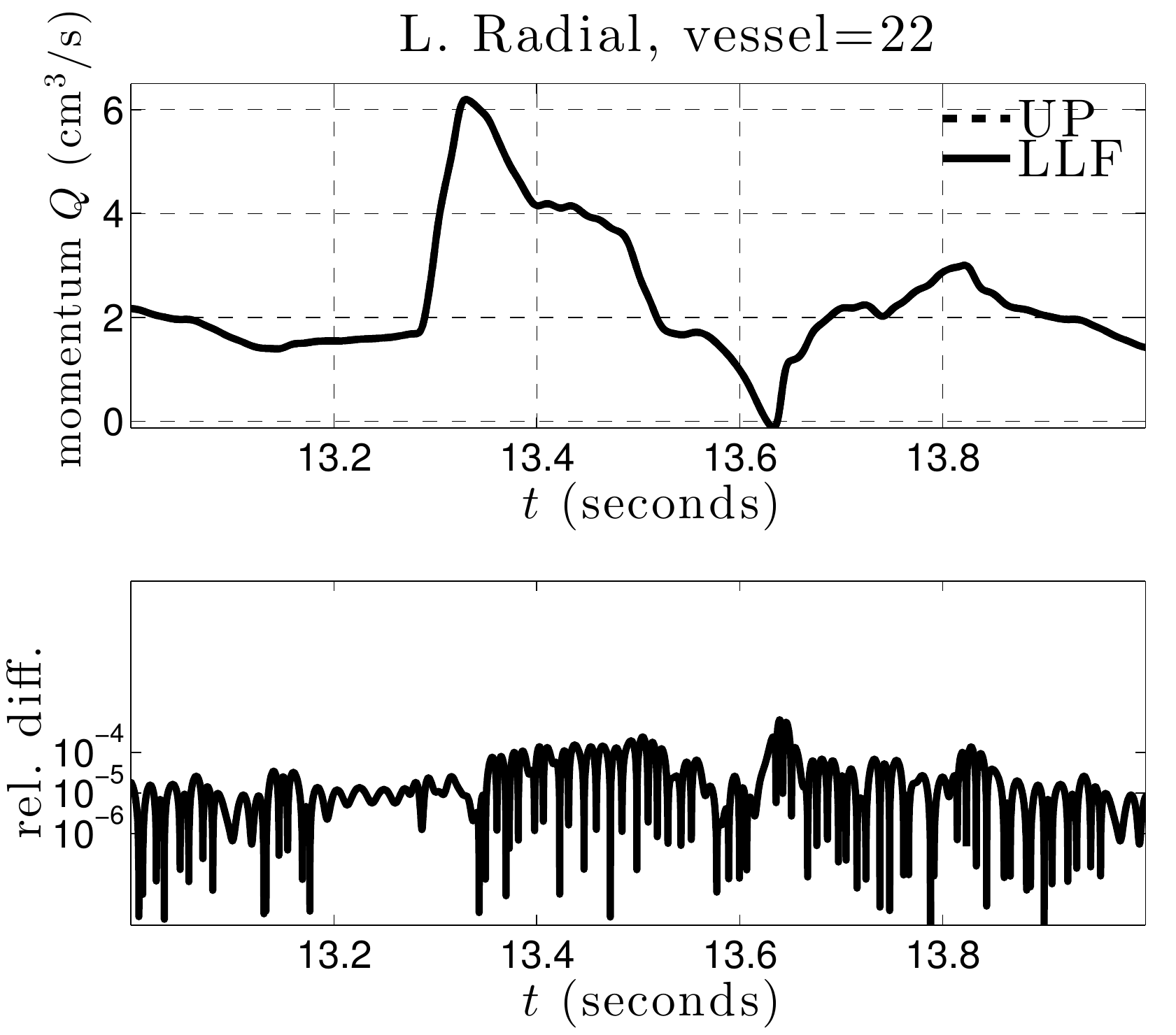} \\
\caption{A comparison of the momentum waveforms from the $(A,Q)$ system with $\alpha = 1.1$ for different choices of the numerical flux.}
\label{fig:55vesmom5}
\end{center}
\end{figure}

\begin{figure}[!htb]
\begin{center}
\includegraphics[scale=0.35]{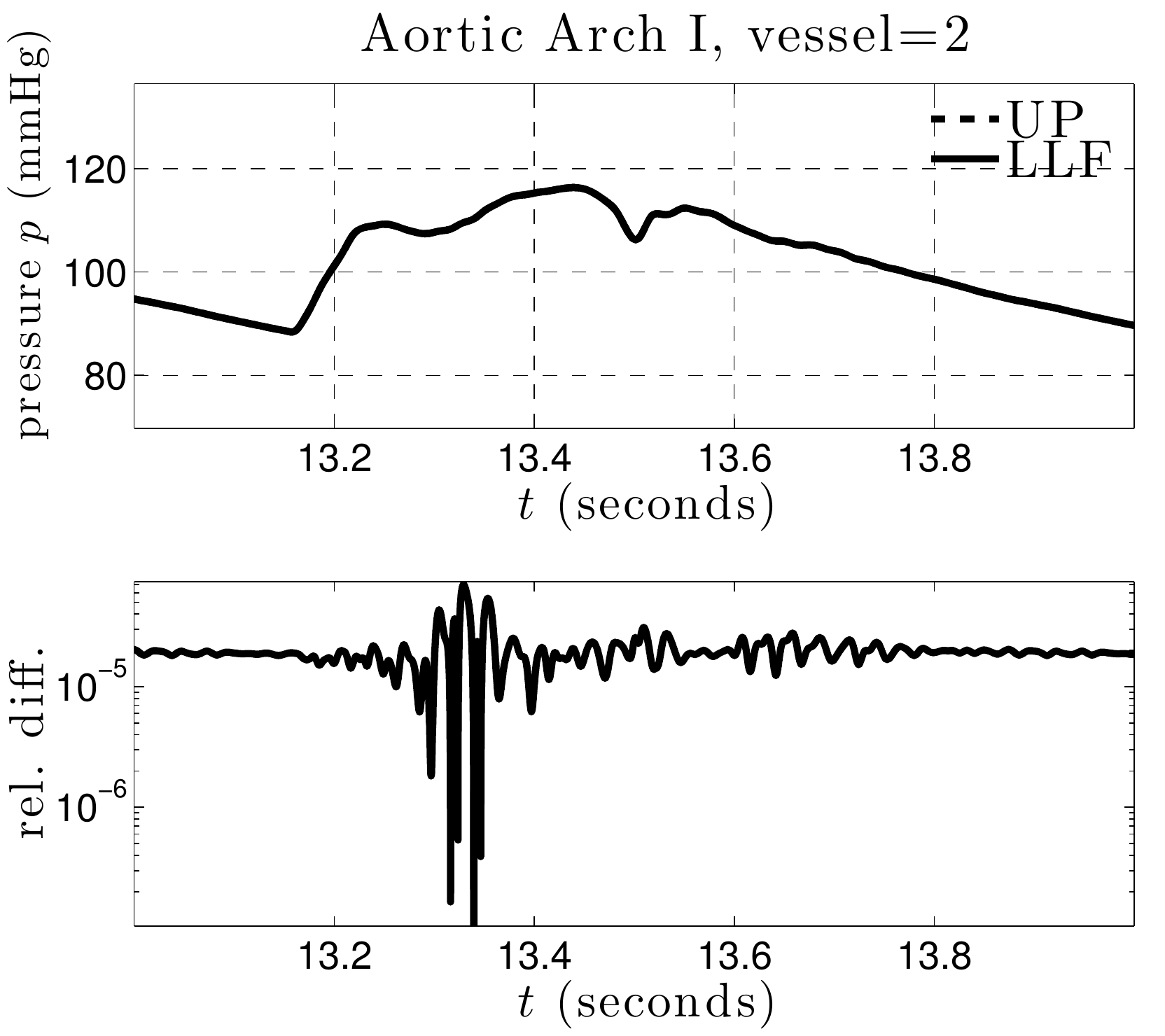} \includegraphics[scale=0.35]{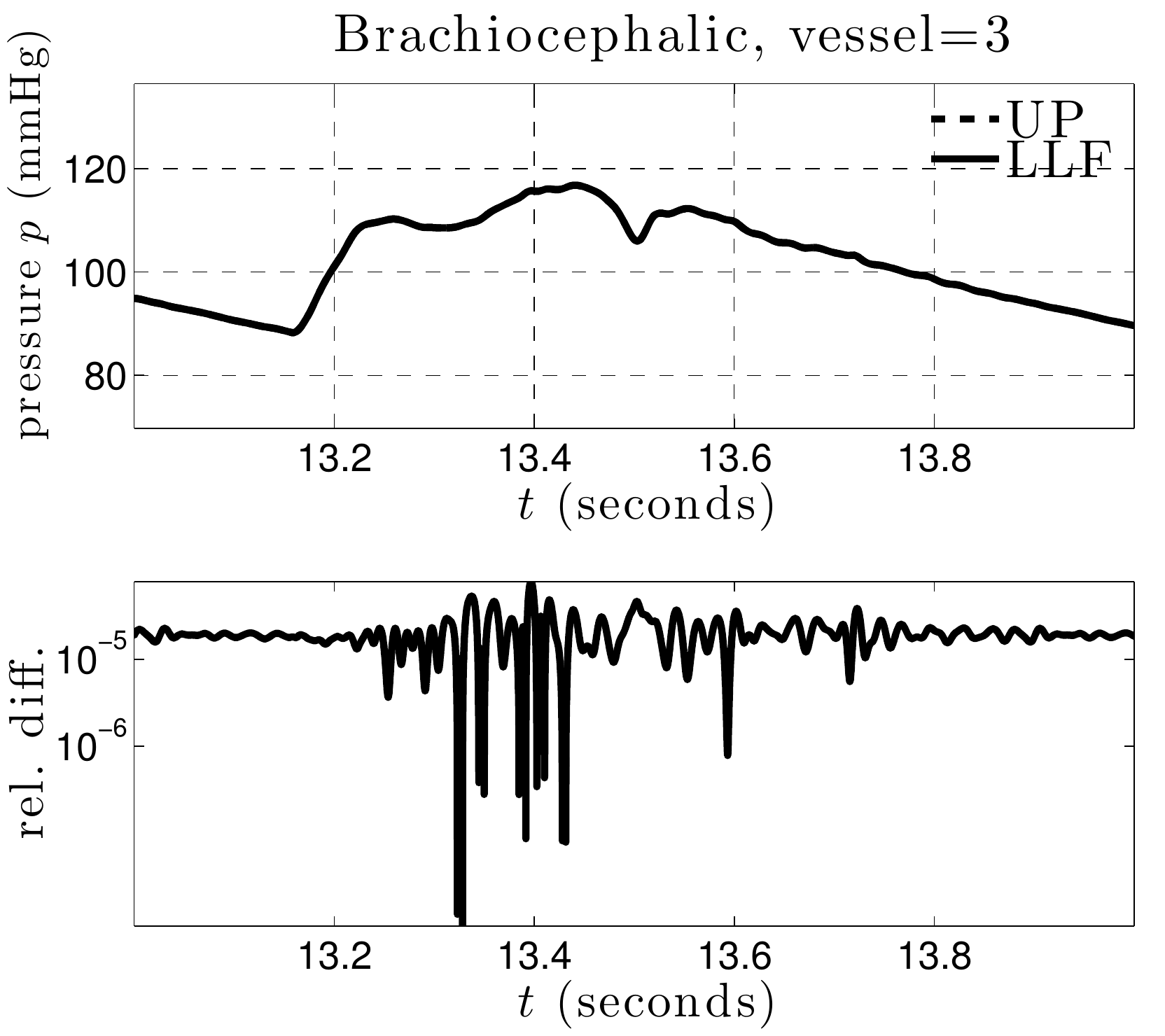} \includegraphics[scale=0.35]{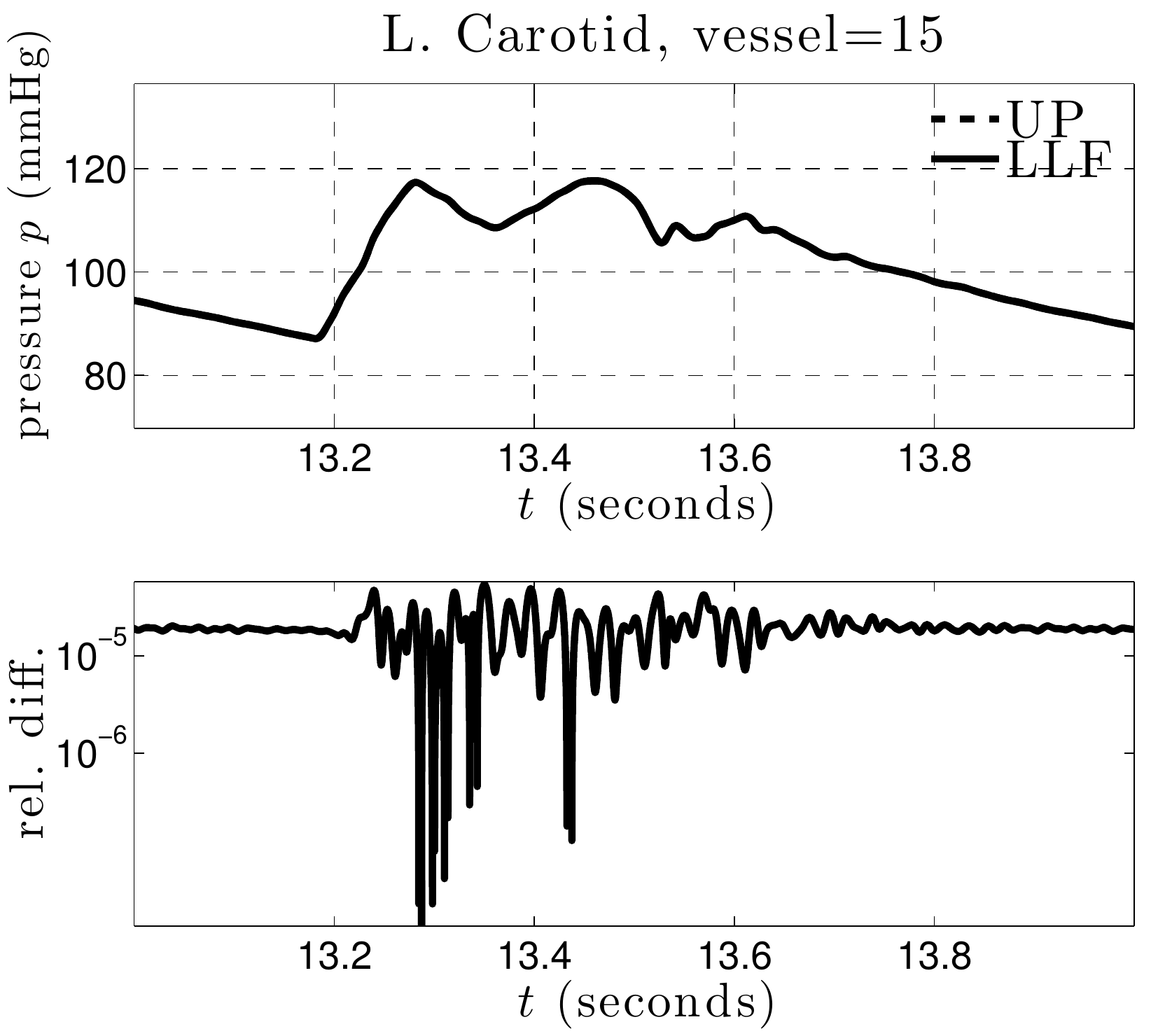} \\
\includegraphics[scale=0.35]{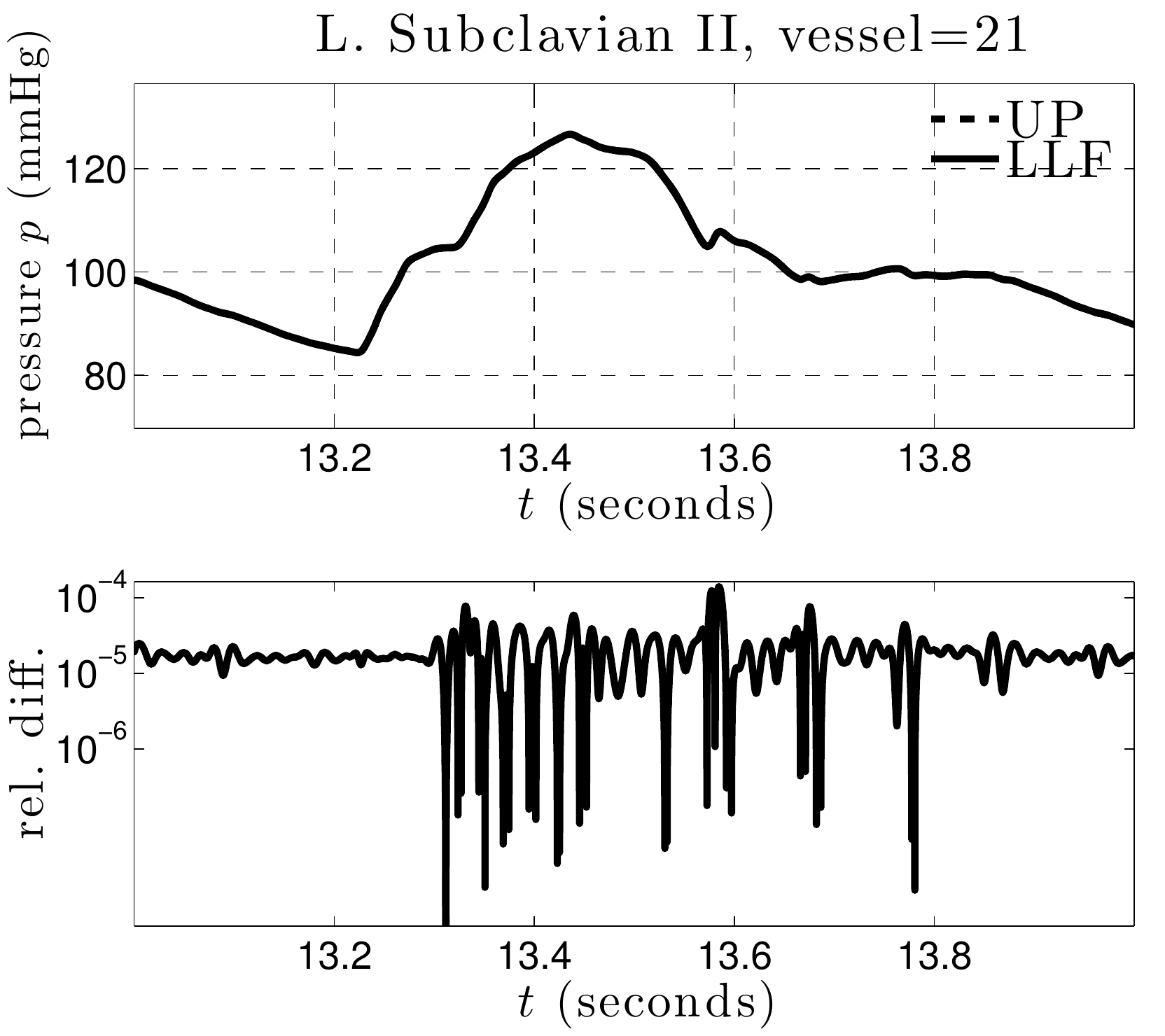} \includegraphics[scale=0.35]{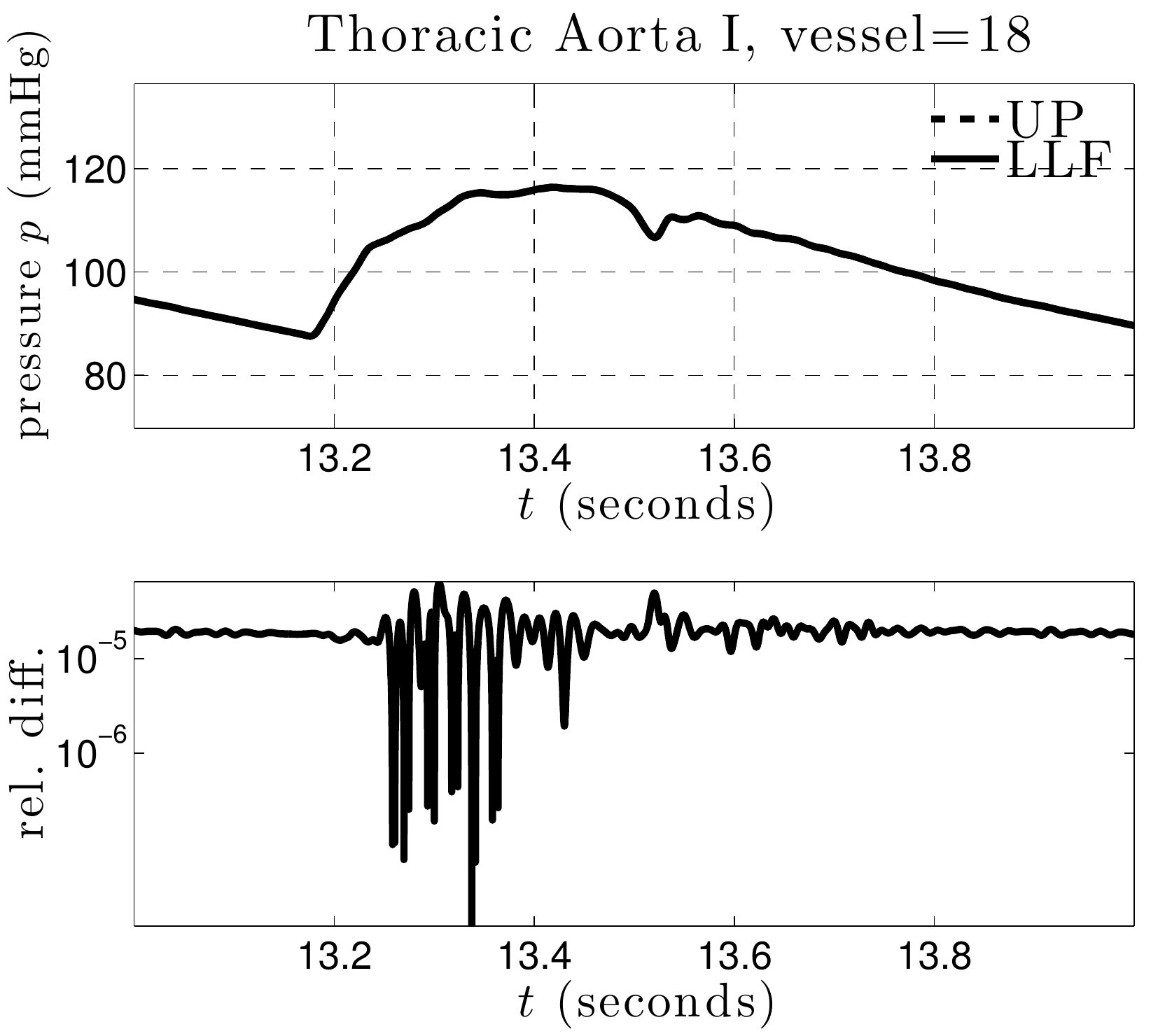} \includegraphics[scale=0.35]{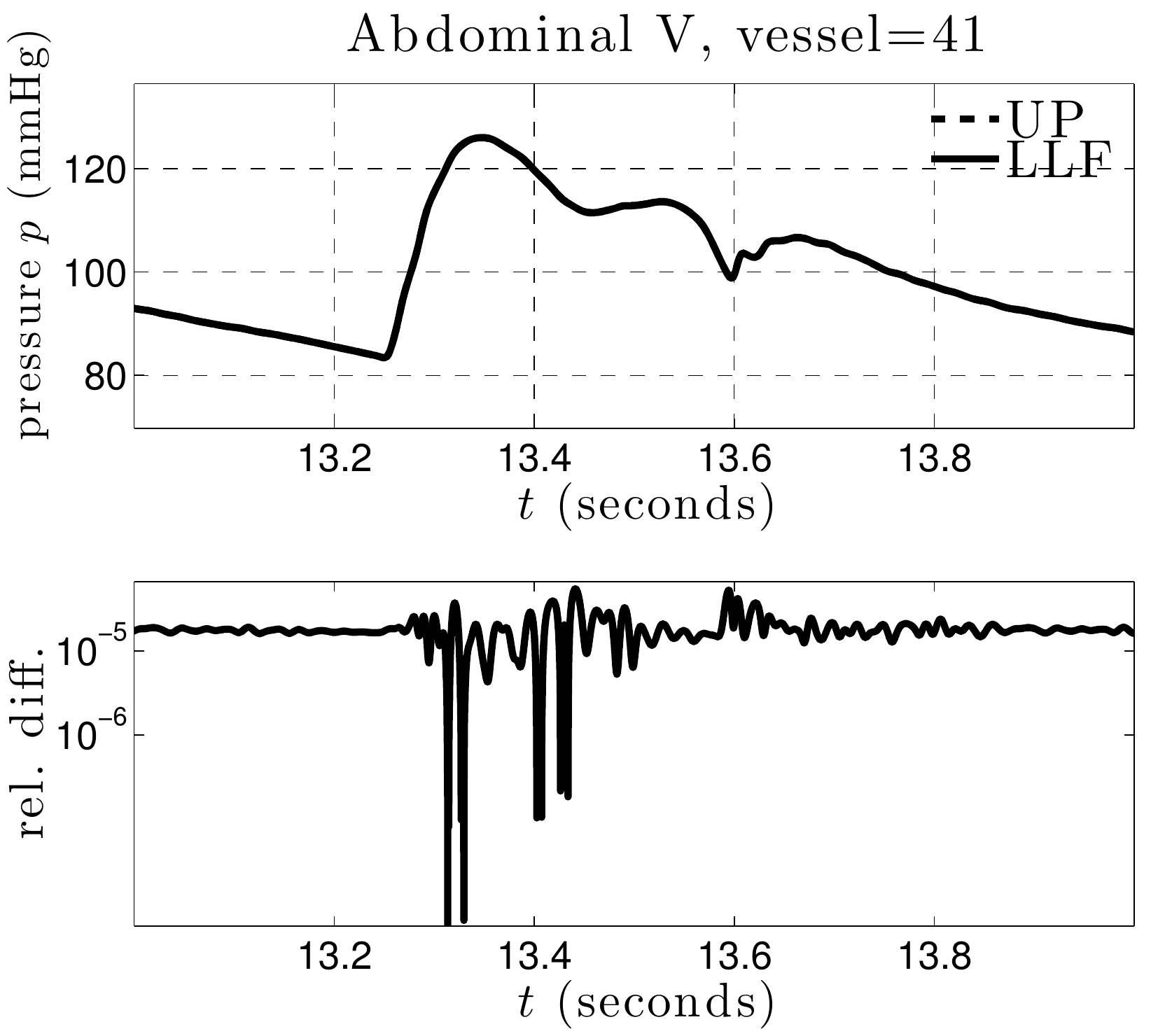} \\
\includegraphics[scale=0.35]{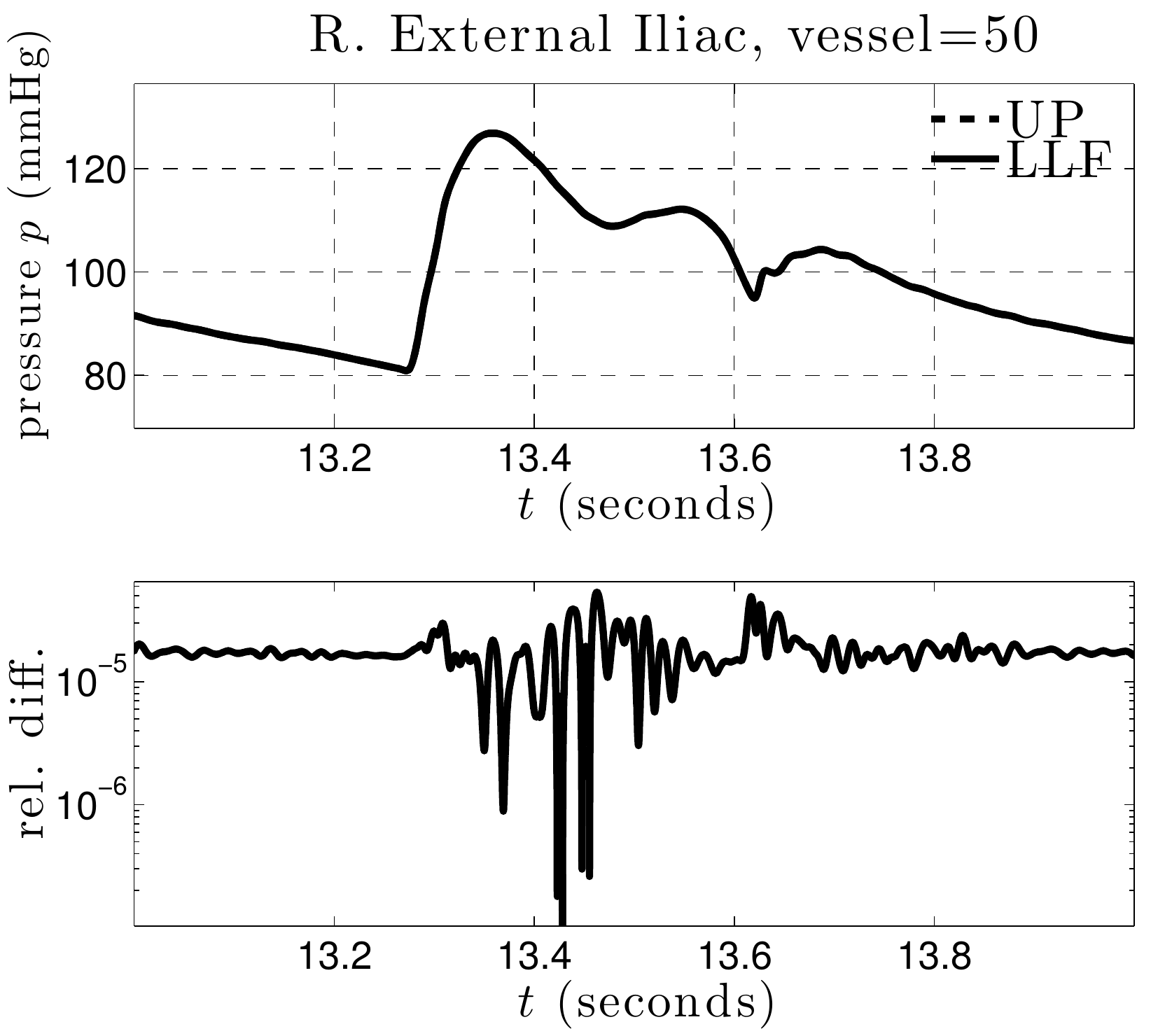} \includegraphics[scale=0.35]{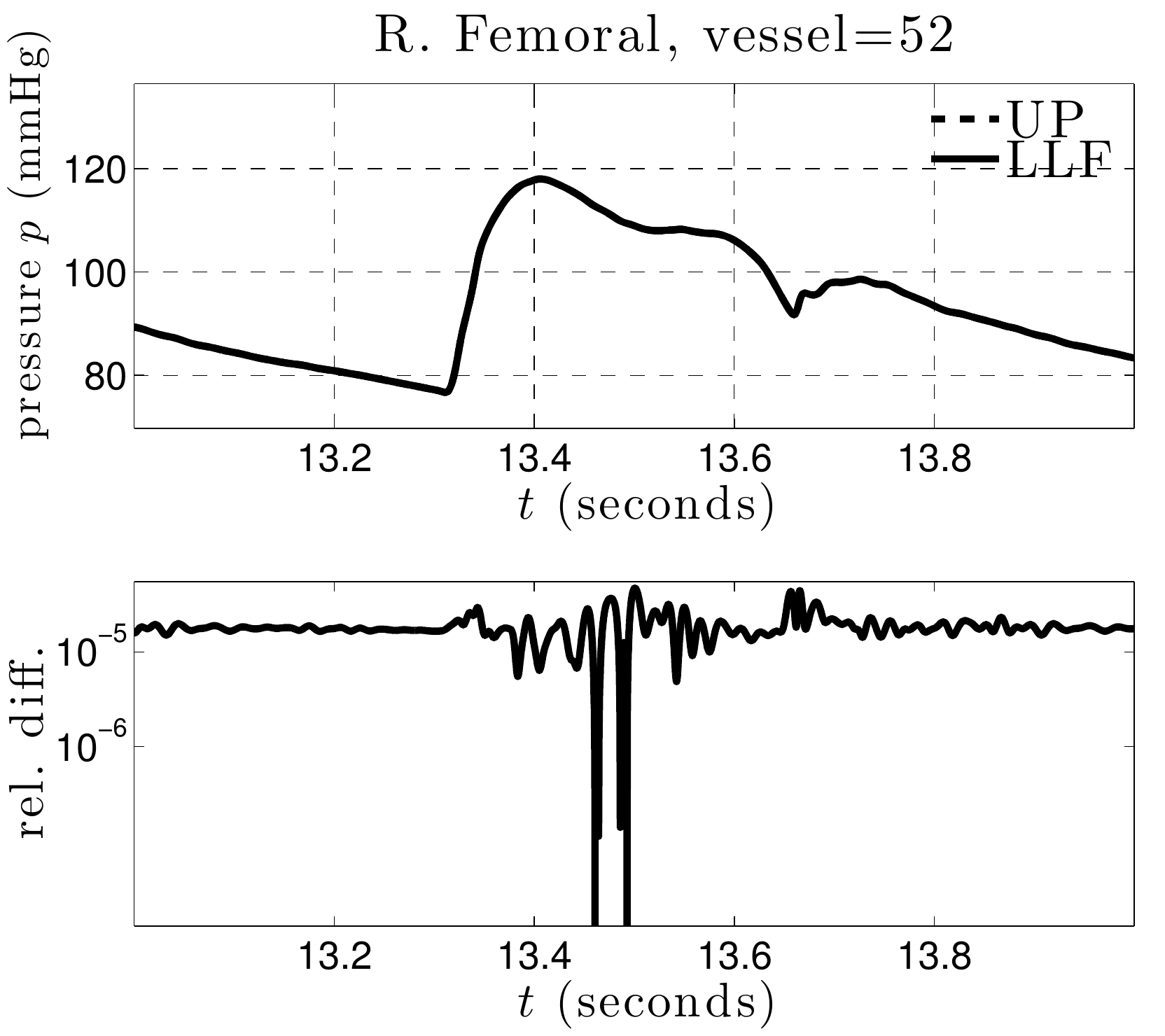} \includegraphics[scale=0.35]{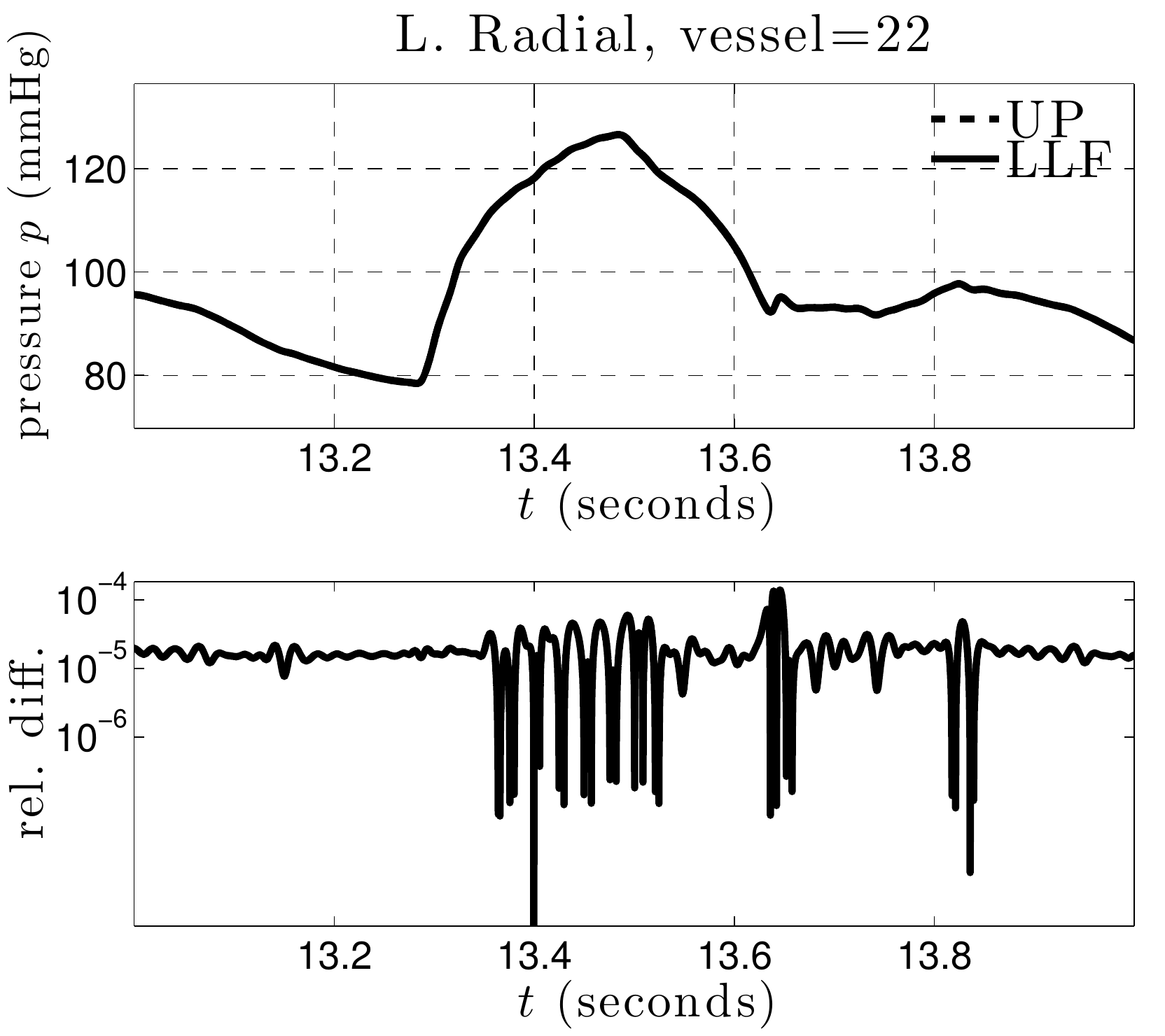} \\
\caption{A comparison of the pressure waveforms from the $(A,Q)$ system with $\alpha = 1.1$ for different choices of the numerical flux.}
\label{fig:55vespress5}
\end{center}
\end{figure}

Lastly, we compare results obtained with the local Lax--Friedrichs (LLF) and upwinding (UP) numerical fluxes in Figures \ref{fig:55vesmom5} and \ref{fig:55vespress5}.  The $(A,Q)$ system with $\alpha = 1.1$ is used; results are similar for the $(A,U)$ system and other values of $\alpha$.  The upper subfigure displays the waveforms from each numerical flux, and the lower subfigure displays the pointwise relative difference between the waveforms.  This difference is computed by normalizing by the maximum norm of the waveform produced by the LLF flux and is quite small ($\sim 10^{-5}$).

\clearpage
\subsection{Shock formation in the subclavian artery}

In this section, we are interested in comparing waveforms from different models in the realm of solutions with possible shocks.  Bloodflow waveforms are typically smooth in healthy individuals, but doctors speculate that a faulty heart produces flows that have the capacity to form sharp transitions within the body.  An important problem potentially leading to nonsmooth waveforms is aortic regurgitation, characterized by an aortic valve which leaks blood back into the left ventricle.  To compensate for the backflow of blood into the ventricle, the heart works harder and the pulse pressure (the difference between the maximum and minimum pressure) increases.  Interestingly, this physiological problem was important for early work on bloodflow modeling: experimental and clinical evidence, including ``pistol-shot'' sounds in the arteries of patients with aortic regurgitation, indicated the importance of nonlinear effects in reduced bloodflow equations (see e.g. \cite{Lan58, ANL71}).

\begin{figure}[!htb]
\begin{center}
\includegraphics[scale=0.4]{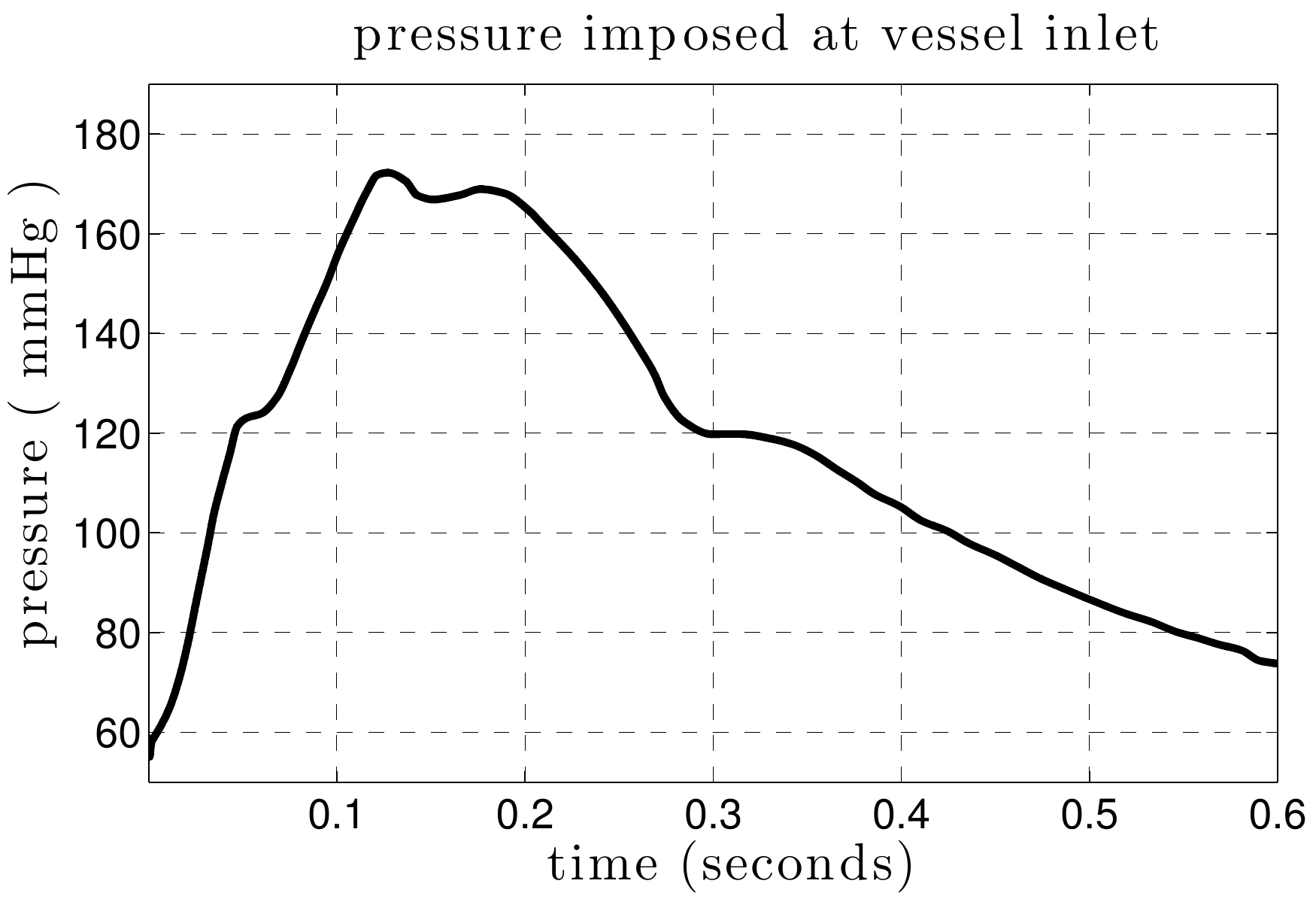}
\caption{Pressure imposed at inlet of the subclavian artery for the shock formation experiment, taken from reference \cite{Rem1956}. }
\label{fig:shock_inlet_data}
\end{center}
\end{figure}

Figure \ref{fig:shock_inlet_data} displays a pressure waveform measured in the subclavian artery of a patient with aortic regurgitation, taken from \cite{Rem1956}. Notice the pulse pressure is greater than 100 mmHg, while a typical healthy pulse pressure is 40 mmHg.

We use the waveform in Figure \ref{fig:shock_inlet_data} as the inlet boundary condition for the second subclavian vessel (``subclavian II'') from the fifty--vessel network given above (mechanical parameters for the vessel given in \cite{SFPF03}).  As with the fifty--five vessel network above, we set $p_0 = 75$ mmHg for these simulations.  The vessel and numerical parameters are given in Tables \ref{table:shock} and \ref{table:shockDG} respectively.  Further, since we expect the solution to develop sharp transitions and possibly shocks, we supplement the discontinuous Galerkin method with the minmod slope limiter \cite{CS_3}.

\begin{table}[!htb]
\begin{center}
\begin{tabular}{| c | c | c | c |}
\hline
{\bf name} & $\beta$ (g cm$^{-2}$ s$^{-2}$) & $A_0$ (cm$^2$) \\ \hline 
subclavian II & 466000 & 0.51 \\ \hline
\end{tabular}
\caption{Parameters for the vessel used in the shock formation experiment.}
\label{table:shock}
\end{center}
\end{table}

\begin{table}[!htb]
\begin{center}
\begin{tabular}{| c | c | c | c | c |}
\hline
$\Delta t$ (s) & $h$ (cm) & $k$ (polynomial degree) & numerical flux  & slope limiter? \\ \hline
$2\times10^{-5}$ & 0.25 & 1 & LLF & Yes, minmod \cite{CS_3} \\ \hline
\end{tabular}
\caption{Parameters of the numerical schemes for the shock formation experiment.}
\label{table:shockDG}
\end{center}
\end{table}

\begin{figure}[!htb]
\begin{center}
\includegraphics[scale=0.35]{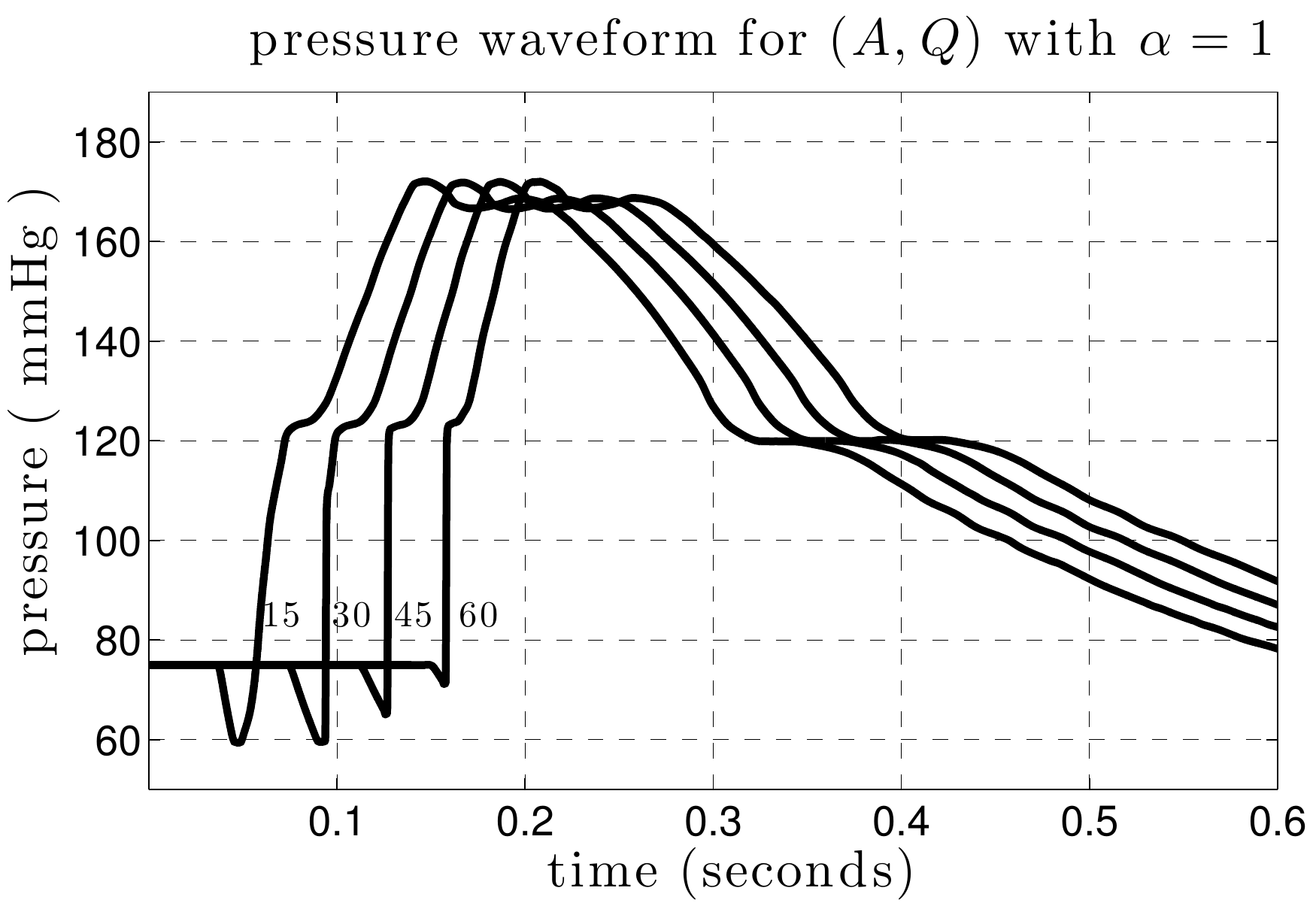}
\includegraphics[scale=0.35]{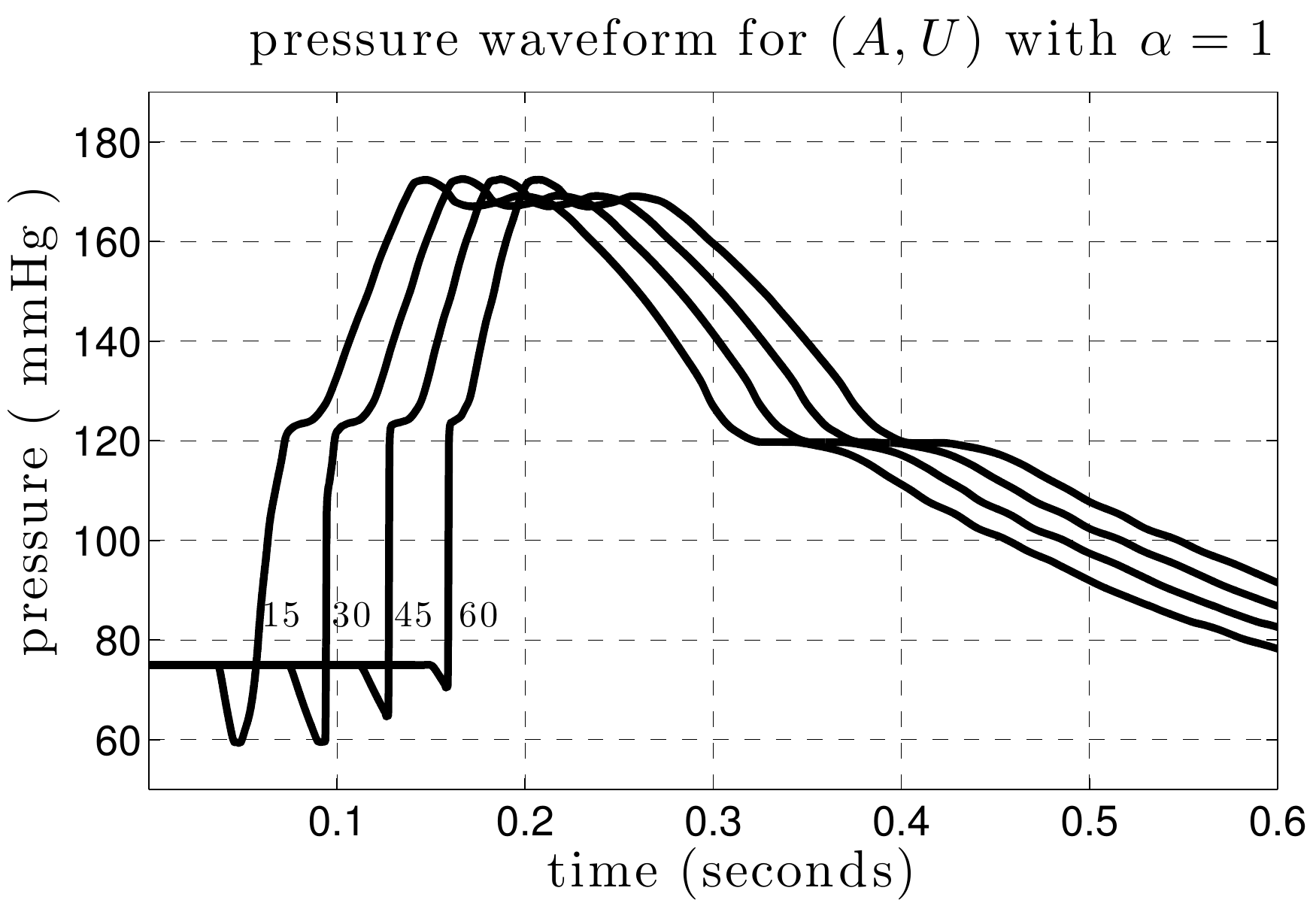} \\
\caption{Pressure waveforms measured at various distances from the vessel inlet.  The distance from the inlet is given in centimeters to the right of each waveform.}
\label{fig:shock1}
\end{center}
\end{figure}

In Figure \ref{fig:shock1}, we display the pressure waveforms measured at various distances from the vessel inlet for both the inviscid $(A,Q)$ and inviscid $(A,U)$ systems.  The results for both systems are similar.  Note the sharp transition from the minimum to maximum pressure at the beginning of each waveform; this transition increases in sharpness farther from the vessel inlet.  This feature indicates shock formation and also appears in clinical data \cite{Rem1956}.  Similar results are seen in the waveforms for  $\alpha = 1.1$ and $\alpha = 4/3$.

\begin{figure}[!htb]
\begin{center}
\includegraphics[scale=0.35]{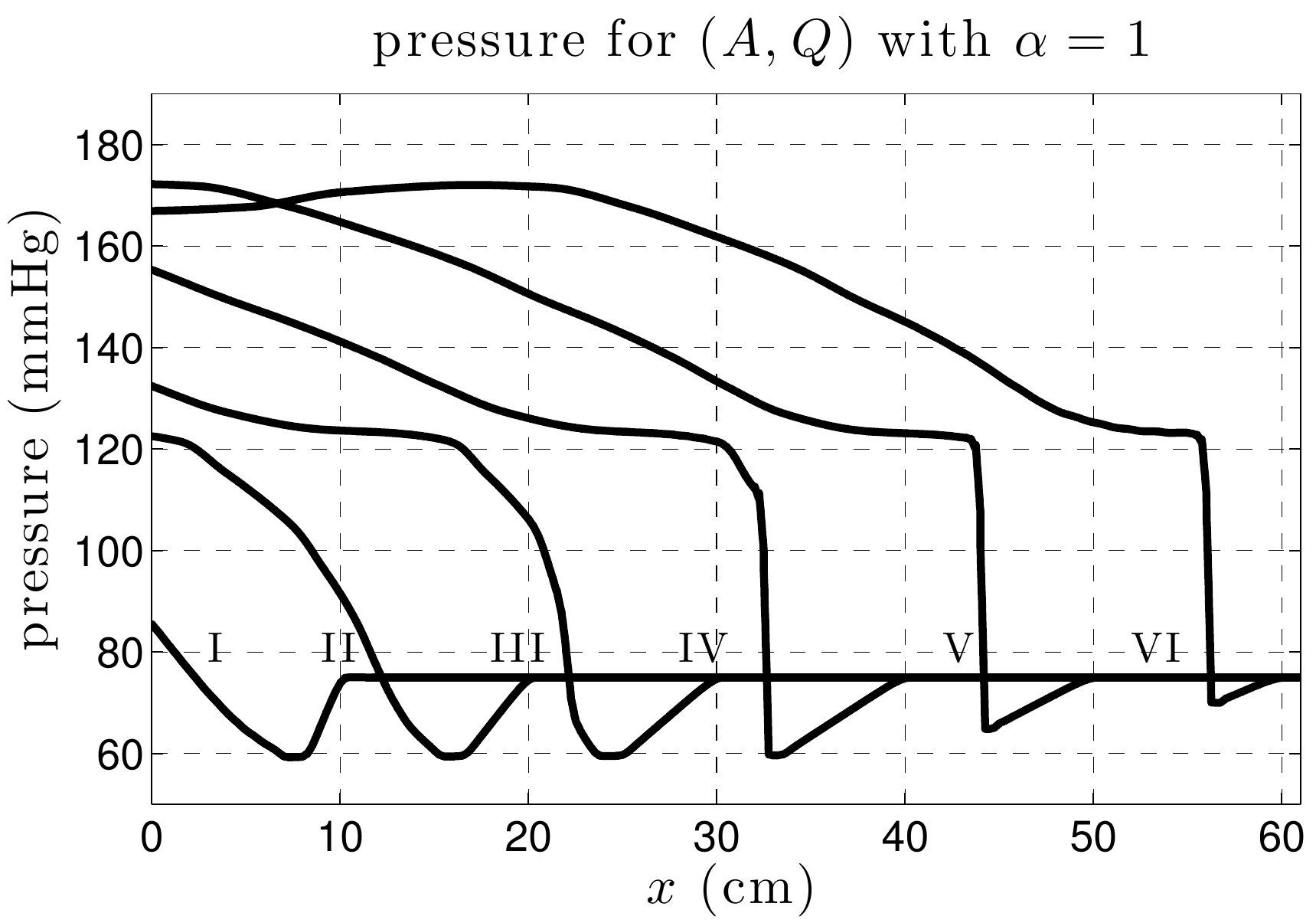}
\includegraphics[scale=0.35]{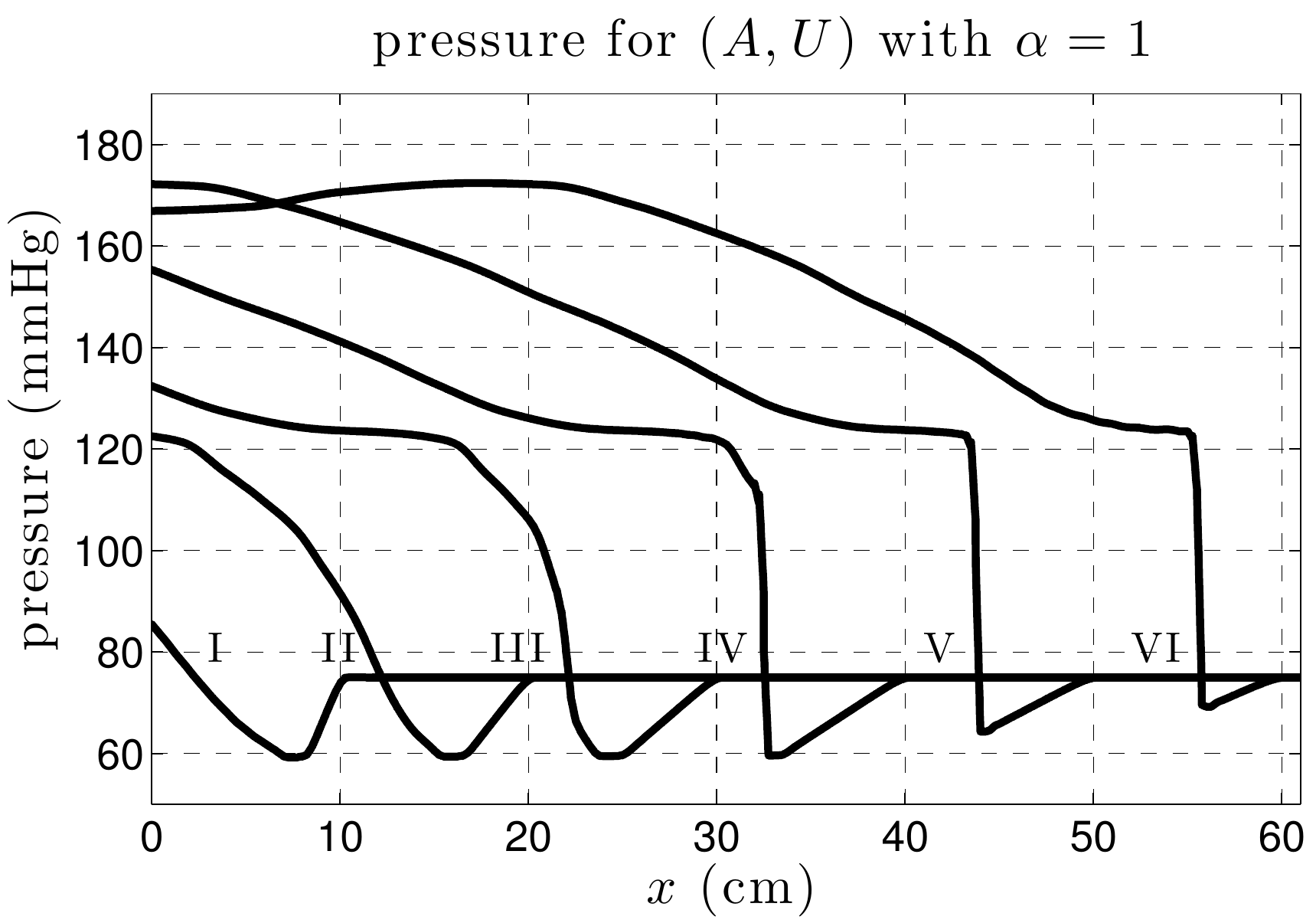} \\
\includegraphics[scale=0.35]{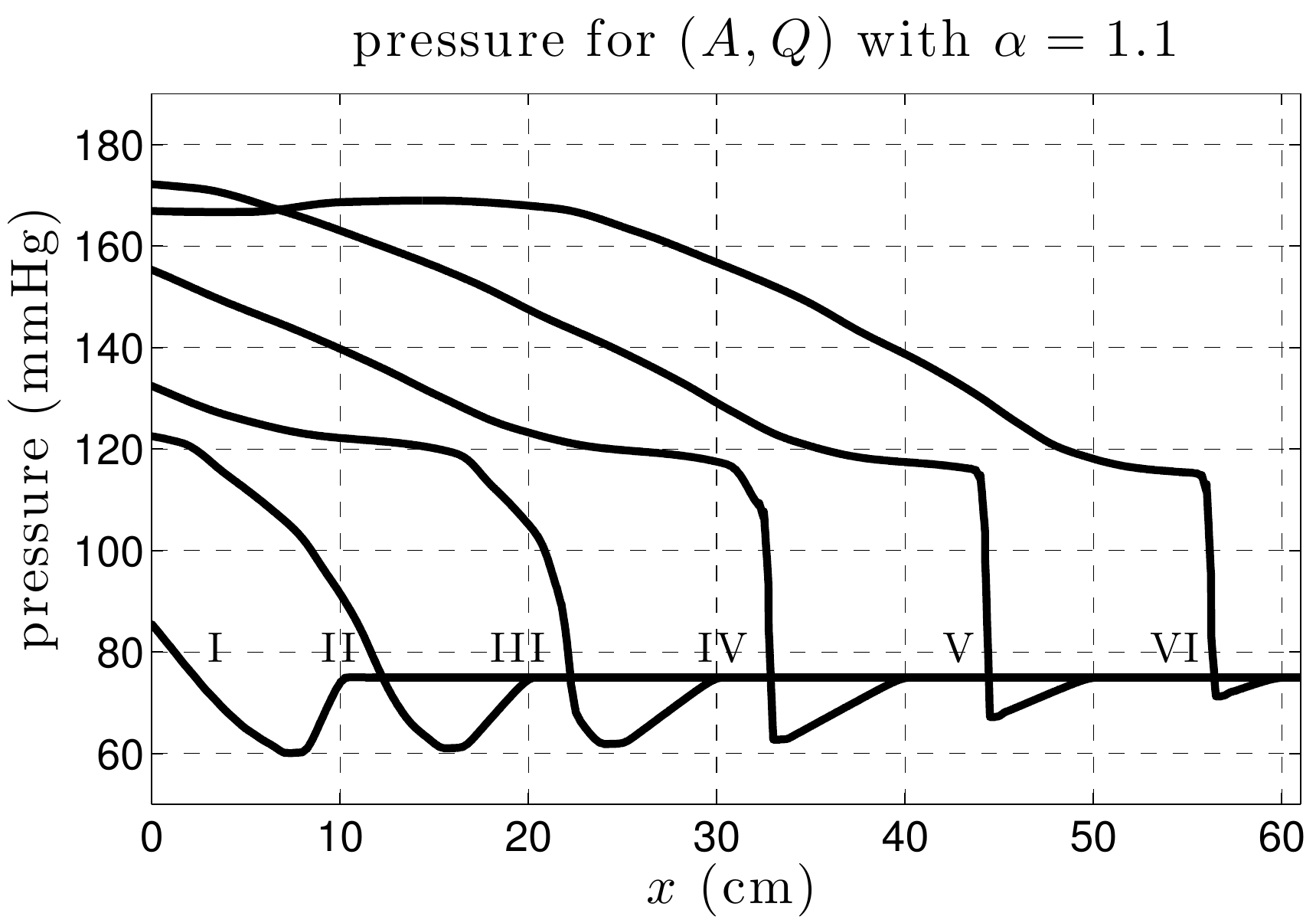}
\includegraphics[scale=0.35]{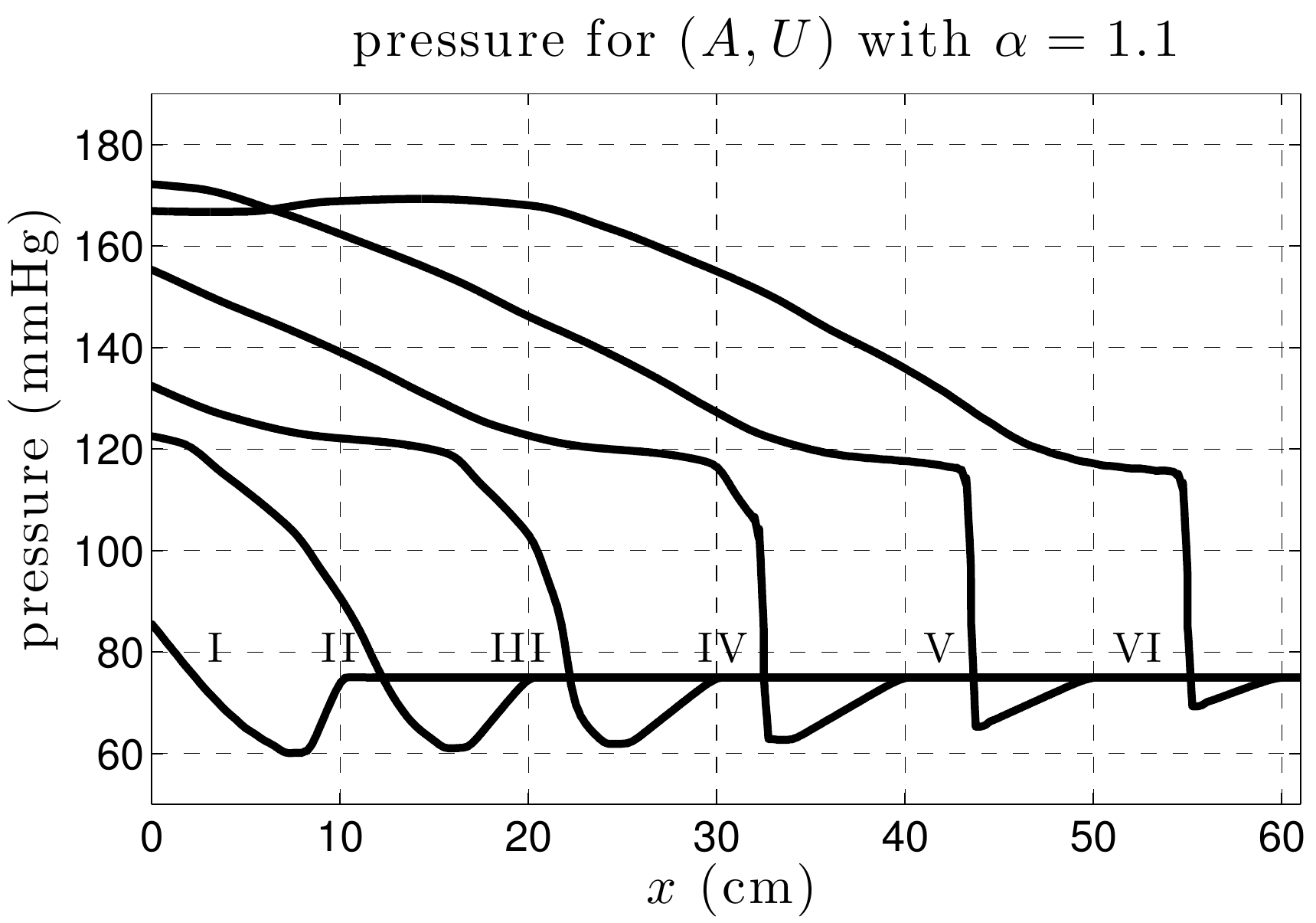} \\
\includegraphics[scale=0.35]{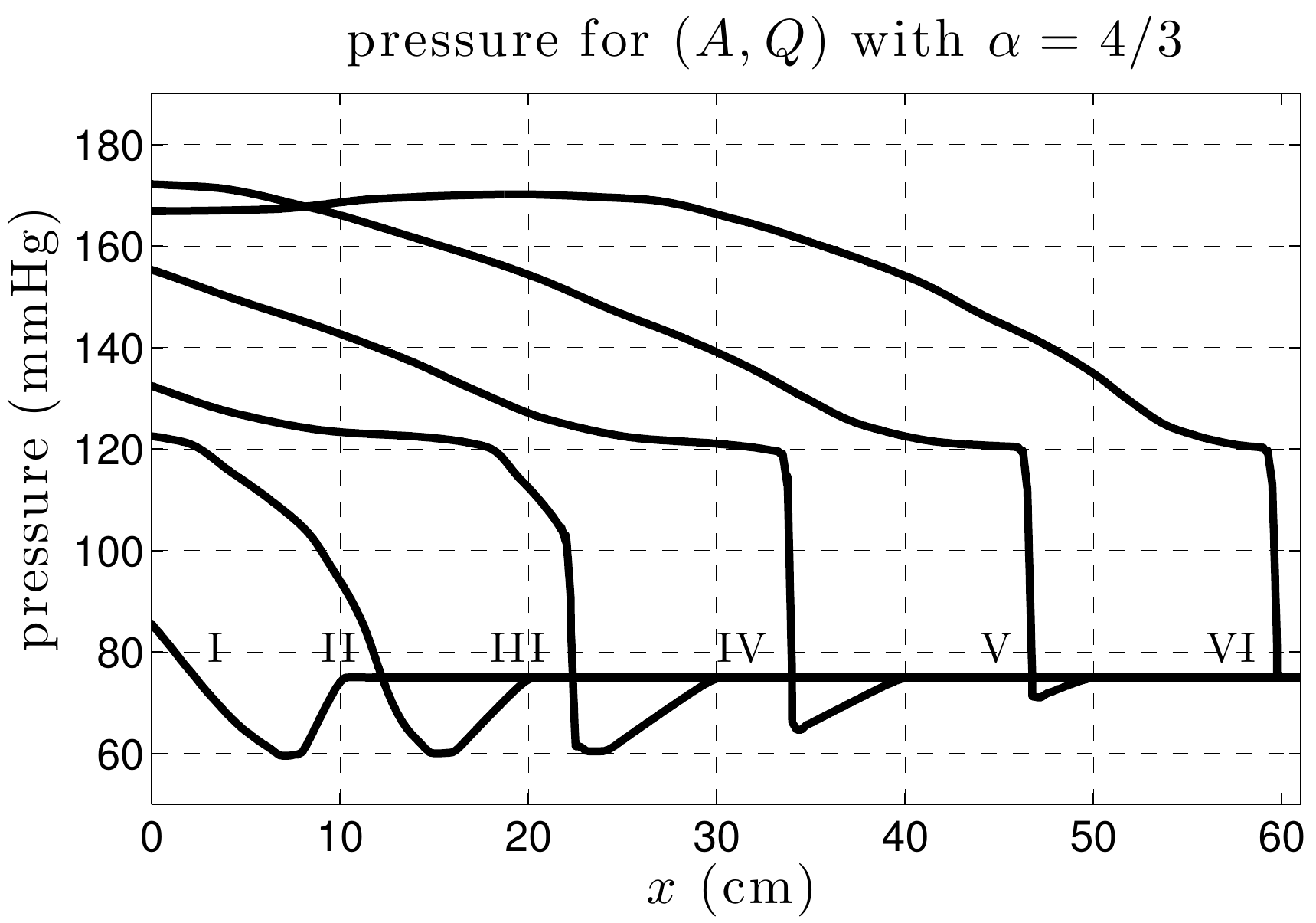}
\includegraphics[scale=0.35]{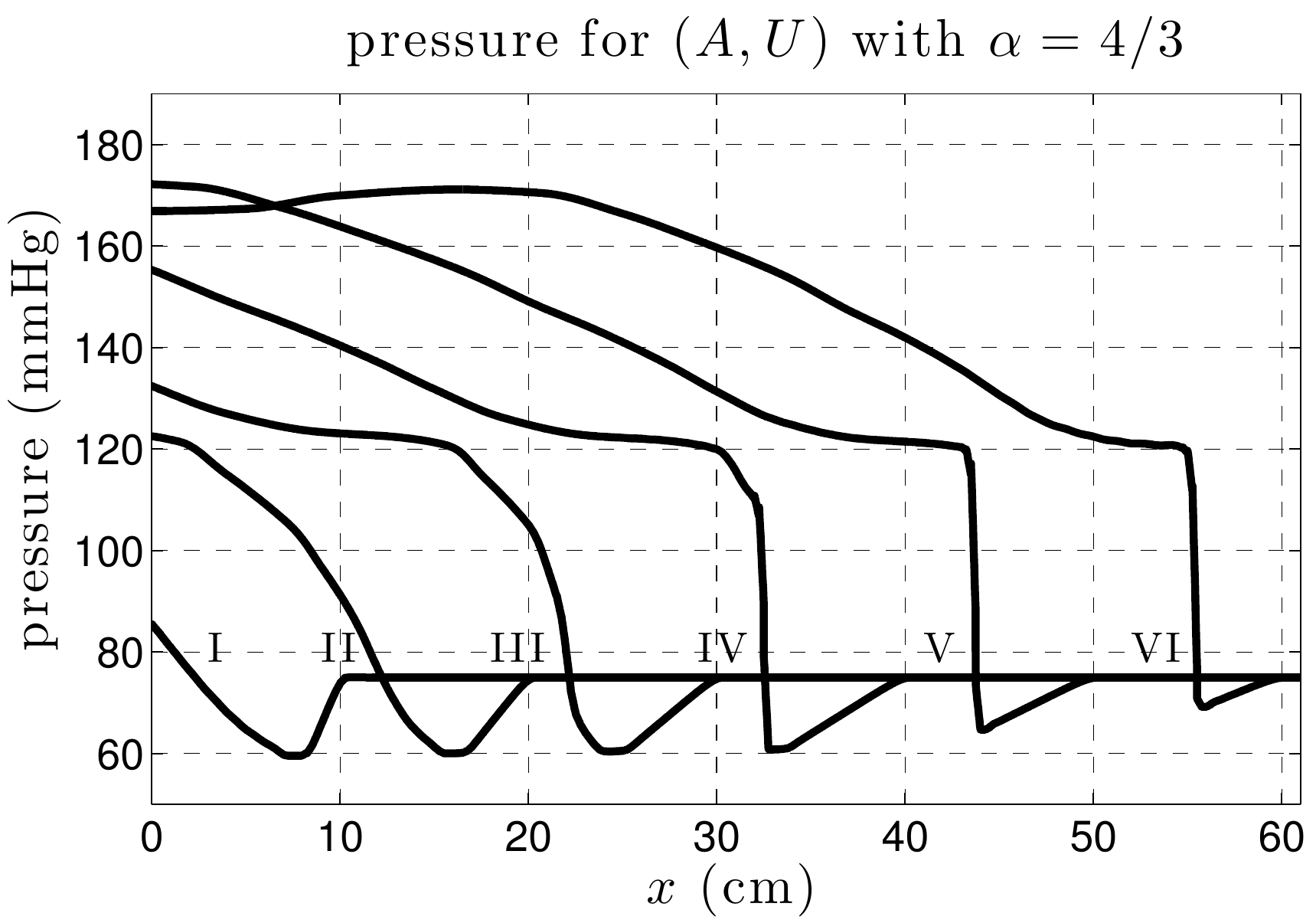} \\
\caption{The figures on the left are snapshots of pressure at uniformly spaced times: I(0.025 s), II(0.05 s), III(0.075 s), IV(0.1 s), V(0.125 s), and VI(0.15 s). Results for the $(A,Q)$ system are on the left and for the $(A,U)$ system are on the right.  Each panel corresponds to a different value of $\alpha$.}
\label{fig:shock2}
\end{center}
\end{figure}

In Figure \ref{fig:shock2}, we explore the formation of the shock.  The panels on the left correspond to the $(A,Q)$ system and the panels on the right correspond to the $(A,U)$ system.  Each panel displays six snapshots in time of pressure as a function of space.  First, we make the general comment that in all models, a sharp transition forms within the length of a typical arm; it is interesting that the nonlinearity in the model is able to capture the shock within this distance.  This modeling supports speculation that ``pistol--shot'' sounds in the body may indeed result from shock formation \cite{Rem1956, Lan58, ANL71}.   

Next, we see that for the $(A,Q)$ system, the pressure to the left of the shock on the curve VI is greater than 120 mmHg for $\alpha=1$, is equal to 120 mmHg for $\alpha=4/3$ and is smaller than 120 mmHg
for $\alpha=1.1$.  This behavior can be explained by an increase in the viscosity term as $\alpha$ varies from
from $1$ to $4/3$ to $1.1$. The pressure to the left of the shock on the curve VI for the $(A,U)$
system varies in the same way.
We also comment on the formation of the shock as seen in how curves I through V vary in shape.  Observe that the shock development for the $(A,Q)$ system is similar for the cases $\alpha=1$ and $\alpha=1.1$ but is different for the case $\alpha=4/3$.  For this value, the shock develops the fastest and appears fully developed in snapshot IV.  For the $(A,U)$ system, the shock formation is very similar for all values of $\alpha$.
%



\clearpage
\section{Conclusions}

This paper details a systematic comparison of two classes of reduced blood flow models, the $(A,Q)$  and $(A,U)$ systems.  Models within each class are further characterized by the Coriolis coefficient $\alpha$, which takes different values for different axial velocity profiles.
Discontinuous Galerkin in space with Runge--Kutta methods in time are used to discretize the system of conservation laws. 
Our results first show that the approximations of pressure, momentum and velocity did not depend on the particular choice of numerical flux.

In addition to validating our code with the results from \cite{SFPF03} and verifying theoretically justified convergence rates, we consider two separate experiments with different boundary conditions:  a fifty--five vessel network with flow imposed at the inlet of the ascending aorta and a single vessel with pressure from a patient with aortic regurgitation imposed at the inlet. In the former experiment, the solutions are smooth, and in the latter, solutions exhibit a shock.  In both cases we compare the $(A,Q)$ and $(A,U)$ models with varying values of $\alpha$ modeling inviscid flow ($\alpha = 1$) and viscous flow with either a Poiseuille profile ($\alpha = 4/3$) or a flatter profile ($\alpha = 1.1$).

For the fifty--five vessel network, we compare simple reflection terminal boundary conditions with three element windkessel boundary conditions.  As expected, the results with the windkessel boundary conditions do not exhibit high frequency oscillations like we see with the reflection boundary conditions.  Thus, we employ the windkessel boundary conditions when comparing the $(A,Q)$ and $(A,U)$ systems with different values of $\alpha$.

Our simulations reveal the selection of the Coriolis coefficient $\alpha$ does impact the solution; this effect is demonstrated in both the fifty--five vessel network and in the single vessel with shock formation.  The choice $\alpha = 4/3$ provides a smaller viscosity term than $\alpha = 1.1$ and produces solutions with higher pressure gradients.  In the case $\alpha = 1.1$, the $(A,U)$ system derived from a flat profile assumption compares reasonably well to the $(A,Q)$ system in all cases.  In light of these observations, the $(A,U)$ system with $\alpha = 1.1$ is a reasonable choice for modeling when the solutions are smooth, but generally we favor the use of the $(A,Q)$ system with $\alpha = 1.1$ since it describes the physically conserved variables.

In the shock formation experiments, the inviscid $(A,Q)$  and $(A,U)$ systems yield the same result.  In contrast, the models with $\alpha>1$ develop shocks in differing locations, although these differences are quite small for the $(A,U)$ system since $\alpha$ only appears in the viscous term. Knowledge of these discrepancies among models is important for physiological applications involving shock formation, like aortic regurgitation.

\clearpage 
\section{Acknowledgments}
Rivi\`ere and Puelz were funded in part by the grant NSF-DMS 1312391 and by a training fellowship from the Keck Center of the Gulf Coast Consortia, on the Training Program in Biomedical Informatics, National Library of Medicine (NLM) T15LM007093. \v{C}ani\'c was funded in part by the grants DMS-1262385, DMS-1311709, DMS-1318763, and NSF/NIGMS DMS-1263572.

\clearpage

\bibliography{dgc}

\begin{thebibliography}{10}
\expandafter\ifx\csname url\endcsname\relax
  \def\url#1{\texttt{#1}}\fi
\expandafter\ifx\csname urlprefix\endcsname\relax\def\urlprefix{URL }\fi
\expandafter\ifx\csname href\endcsname\relax
  \def\href#1#2{#2} \def\path#1{#1}\fi

\bibitem{Hughes74}
T.~J. Hughes, A study of the one-dimensional theory of arterial pulse
  propagation, Ph.D. thesis, University of California, Berkeley (1974).

\bibitem{Politi2016}
M.~T. Politi, A.~Ghigo, J.~M. Fern{\'a}ndez, I.~Khelifa, J.~Gaudric, J.~M.
  Fullana, P.-Y. Lagr{\'e}e, The dicrotic notch analyzed by a numerical model,
  Computers in {B}iology and {M}edicine 72 (2016) 54--64.
\newblock \href {http://dx.doi.org/doi:10.1016/j.compbiomed.2016.03.005}
  {\path{doi:doi:10.1016/j.compbiomed.2016.03.005}}.

\bibitem{FGNQ01}
L.~Formaggia, J.-F. Gerbeau, F.~Nobile, A.~Quarteroni, On the coupling of 3d
  and 1d {N}avier--{S}tokes equations for flow problems in compliant vessels,
  Computer {M}ethods in {A}pplied {M}echanics and {E}ngineering 191~(6) (2001)
  561--582.
\newblock \href {http://dx.doi.org/10.1016/S0045-7825(01)00302-4}
  {\path{doi:10.1016/S0045-7825(01)00302-4}}.

\bibitem{MN08}
J.~Mynard, P.~Nithiarasu, A \textsc{1D} arterial blood flow model incorporating
  ventricular pressure, aortic valve and regional coronary flow using the
  locally conservative \textsc{G}alerkin (\textsc{LCG}) method,
  {C}ommunications in {N}umerical {M}ethods in {E}ngineering 24~(5) (2008)
  367--417.
\newblock \href {http://dx.doi.org/10.1002/cnm.1117}
  {\path{doi:10.1002/cnm.1117}}.

\bibitem{SFPF03}
S.~Sherwin, L.~Formaggia, J.~Peiro, V.~Franke, Computational modelling of
  \textsc{1D} blood flow with variable mechanical properties and its
  application to the simulation of wave propagation in the human arterial
  system, International {J}ournal for {N}umerical {M}ethods in {F}luids
  43~(6-7) (2003) 673--700.
\newblock \href {http://dx.doi.org/10.1002/fld.543}
  {\path{doi:10.1002/fld.543}}.

\bibitem{XAF14}
N.~Xiao, J.~Alastruey, C.~Alberto~Figueroa, A systematic comparison between
  1-{D} and 3-{D} hemodynamics in compliant arterial models, International
  {J}ournal for {N}umerical {M}ethods in {B}iomedical {E}ngineering 30~(2)
  (2014) 204--231.
\newblock \href {http://dx.doi.org/10.1002/cnm.2598}
  {\path{doi:10.1002/cnm.2598}}.

\bibitem{OP00}
M.~S. Olufsen, C.~S. Peskin, W.~Y. Kim, E.~M. Pedersen, A.~Nadim, J.~Larsen,
  Numerical simulation and experimental validation of blood flow in arteries
  with structured-tree outflow conditions, Annals of {B}iomedical {E}ngineering
  28~(11) (2000) 1281--1299.
\newblock \href {http://dx.doi.org/10.1114/1.1326031}
  {\path{doi:10.1114/1.1326031}}.

\bibitem{HL73}
T.~J. Hughes, J.~Lubliner, On the one-dimensional theory of blood flow in the
  larger vessels, Mathematical {B}iosciences 18~(1) (1973) 161--170.
\newblock \href {http://dx.doi.org/doi:10.1016/0025-5564(73)90027-8}
  {\path{doi:doi:10.1016/0025-5564(73)90027-8}}.

\bibitem{BRV07}
D.~Bessems, M.~Rutten, F.~Van De~Vosse, A wave propagation model of blood flow
  in large vessels using an approximate velocity profile function, Journal of
  Fluid Mechanics 580 (2007) 145--168.

\bibitem{AP07}
K.~Azer, C.~S. Peskin, A one-dimensional model of blood flow in arteries with
  friction and convection based on the {W}omersley velocity profile,
  Cardiovascular Engineering 7~(2) (2007) 51--73.
\newblock \href {http://dx.doi.org/10.1007/s10558-007-9031-y}
  {\path{doi:10.1007/s10558-007-9031-y}}.

\bibitem{Coccarelli2015}
A.~Coccarelli, E.~Boileau, D.~Parthimos, P.~Nithiarasu, An advanced
  computational bioheat transfer model for a human body with an embedded
  systemic circulation, Biomechanics and {M}odeling in {M}echanobiology (2015)
  1--18\href {http://dx.doi.org/10.1007/s10237-015-0751-4}
  {\path{doi:10.1007/s10237-015-0751-4}}.

\bibitem{APRPR15}
S.~Acosta, C.~Puelz, B.~Rivi{\`e}re, D.~J. Penny, C.~G. Rusin, Numerical method
  of characteristics for one-dimensional blood flow, Journal of computational
  physics 294 (2015) 96--109.
\newblock \href {http://dx.doi.org/10.1016/j.jcp.2015.03.045}
  {\path{doi:10.1016/j.jcp.2015.03.045}}.

\bibitem{Sheng95}
C.~Sheng, S.~Sarwal, K.~Watts, A.~Marble, Computational simulation of blood
  flow in human systemic circulation incorporating an external force field,
  Medical and Biological Engineering and Computing 33~(1) (1995) 8--17.
\newblock \href {http://dx.doi.org/10.1007/BF02522938}
  {\path{doi:10.1007/BF02522938}}.

\bibitem{MWL09}
E.~Marchandise, M.~Willemet, V.~Lacroix, A numerical hemodynamic tool for
  predictive vascular surgery, Medical {E}ngineering and {P}hysics 31~(1)
  (2009) 131--144.
\newblock \href {http://dx.doi.org/10.1016/j.medengphy.2008.04.015}
  {\path{doi:10.1016/j.medengphy.2008.04.015}}.

\bibitem{Boileau15}
E.~Boileau, P.~Nithiarasu, P.~J. Blanco, L.~O. M{\"u}ller, F.~E. Fossan, L.~R.
  Hellevik, W.~P. Donders, W.~Huberts, M.~Willemet, J.~Alastruey, A benchmark
  study of numerical schemes for one-dimensional arterial blood flow modelling,
  International {J}ournal for {N}umerical {M}ethods in {B}iomedical
  {E}ngineering 31~(10).
\newblock \href {http://dx.doi.org/10.1002/cnm.2732}
  {\path{doi:10.1002/cnm.2732}}.

\bibitem{Muller2015}
L.~O. M{\"u}ller, P.~J. Blanco, S.~M. Watanabe, R.~A. Feij{\'o}o, A high-order
  local time stepping finite volume solver for one-dimensional blood flow
  simulations: application to the {ADAN} model, International {J}ournal for
  {N}umerical {M}ethods in {B}iomedical {E}ngineering\href
  {http://dx.doi.org/10.1002/cnm.2761} {\path{doi:10.1002/cnm.2761}}.

\bibitem{Zhang2015}
H.~Zhang, N.~Fujiwara, M.~Kobayashi, S.~Yamada, F.~Liang, S.~Takagi, M.~Oshima,
  Development of a numerical method for patient-specific cerebral circulation
  using 1d--0d simulation of the entire cardiovascular system with {SPECT}
  data, Annals of {B}iomedical {E}ngineering\href
  {http://dx.doi.org/10.1007/s10439-015-1544-8}
  {\path{doi:10.1007/s10439-015-1544-8}}.

\bibitem{Ster92}
N.~Stergiopulos, D.~Young, T.~Rogge, Computer simulation of arterial flow with
  applications to arterial and aortic stenoses, Journal of {B}iomechanics
  25~(12) (1992) 1477--1488.
\newblock \href {http://dx.doi.org/10.1016/0021-9290(92)90060-E}
  {\path{doi:10.1016/0021-9290(92)90060-E}}.

\bibitem{FNQ02}
L.~Formaggia, F.~Nobile, A.~Quarteroni, A one dimensional model for blood flow:
  application to vascular prosthesis, in: Mathematical Modeling and Numerical
  Simulation in Continuum Mechanics, Springer, 2002, pp. 137--153.
\newblock \href {http://dx.doi.org/10.1007/978-3-642-56288-4}
  {\path{doi:10.1007/978-3-642-56288-4}}.

\bibitem{Delestre13}
O.~Delestre, P.-Y. Lagr{\'e}e, A ``well--balanced'' finite volume scheme for
  blood flow simulation, International {J}ournal for {N}umerical {M}ethods in
  {F}luids 72~(2) (2013) 177--205.
\newblock \href {http://dx.doi.org/10.1002/fld.3736}
  {\path{doi:10.1002/fld.3736}}.

\bibitem{WFL14}
X.~Wang, J.-M. Fullana, P.-Y. Lagr{\'e}e, Verification and comparison of four
  numerical schemes for a 1d viscoelastic blood flow model, Computer {M}ethods
  in {B}iomechanics and {B}iomedical {E}ngineering 18~(15) (2015) 1704--1725.
\newblock \href {http://dx.doi.org/10.1080/10255842.2014.948428}
  {\path{doi:10.1080/10255842.2014.948428}}.

\bibitem{CK03}
S.~{\v{C}}ani{\'c}, E.~H. Kim, Mathematical analysis of the quasilinear effects
  in a hyperbolic model blood flow through compliant axi-symmetric vessels,
  Mathematical Methods in the Applied Sciences 26~(14) (2003) 1161--1186.
\newblock \href {http://dx.doi.org/10.1002/mma.407}
  {\path{doi:10.1002/mma.407}}.

\bibitem{FMQ05}
M.~{\'A}. Fern{\'a}ndez, V.~Milisic, A.~Quarteroni, Analysis of a geometrical
  multiscale blood flow model based on the coupling of {ODE}s and hyperbolic
  {PDE}s, Multiscale Modeling \& Simulation 4~(1) (2005) 215--236.
\newblock \href {http://dx.doi.org/10.1137/030602010}
  {\path{doi:10.1137/030602010}}.

\bibitem{ANL71}
M.~Anliker, R.~L. Rockwell, E.~Ogden, Nonlinear analysis of flow pulses and
  shock waves in arteries, Zeitschrift f{\"u}r angewandte Mathematik und Physik
  ZAMP 22~(2) (1971) 217--246.
\newblock \href {http://dx.doi.org/10.1007/BF01591407}
  {\path{doi:10.1007/BF01591407}}.

\bibitem{Rem1956}
J.~W. Remington, E.~H. Wood, Formation of peripheral pulse contour in man,
  Journal of applied physiology 9~(3) (1956) 433--442.

\bibitem{keener98}
J.~P. Keener, J.~Sneyd, Mathematical {P}hysiology, Vol.~1, Springer, 1998.

\bibitem{ZS06}
Q.~Zhang, C.-W. Shu, Error estimates to smooth solutions of {R}unge--{K}utta
  discontinuous {G}alerkin method for symmetrizable systems of conservation
  laws, SIAM Journal on Numerical Analysis 44~(4) (2006) 1703--1720.
\newblock \href {http://dx.doi.org/10.1137/040620382}
  {\path{doi:10.1137/040620382}}.

\bibitem{MGC07}
A.~Mikelic, G.~Guidoboni, S.~Canic, Fluid-structure interaction in a
  pre-stressed tube with thick elastic walls {I}: the stationary {S}tokes
  problem, Networks and Heterogeneous Media 2~(3) (2007) 397.
\newblock \href {http://dx.doi.org/10.3934/nhm.2007.2.397}
  {\path{doi:10.3934/nhm.2007.2.397}}.

\bibitem{CS_3}
B.~Cockburn, S.-Y. Lin, C.-W. Shu, {TVB} {R}unge-{K}utta local projection
  discontinuous {G}alerkin finite element method for conservation laws {III}:
  one-dimensional systems, Journal of Computational Physics 84~(1) (1989)
  90--113.
\newblock \href {http://dx.doi.org/10.1016/0021-9991(89)90183-6}
  {\path{doi:10.1016/0021-9991(89)90183-6}}.

\bibitem{Har83}
A.~Harten, On the symmetric form of systems of conservation laws with entropy,
  Journal of {C}omputational {P}hysics 49~(1) (1983) 151--164.
\newblock \href {http://dx.doi.org/10.1016/0021-9991(83)90118-3}
  {\path{doi:10.1016/0021-9991(83)90118-3}}.

\bibitem{APS12}
J.~Alastruey, K.~H. Parker, S.~J. Sherwin, Arterial pulse wave haemodynamics,
  in: 11th International Conference on Pressure Surges, Virtual PiE Led t/a BHR
  Group: Lisbon, Portugal, 2012, pp. 401--442.

\bibitem{murgo1980}
J.~P. Murgo, N.~Westerhof, J.~P. Giolma, S.~A. Altobelli, Aortic input
  impedance in normal man: relationship to pressure wave forms., Circulation
  62~(1) (1980) 105--116.
\newblock \href {http://dx.doi.org/10.1161/01.CIR.62.1.105}
  {\path{doi:10.1161/01.CIR.62.1.105}}.

\bibitem{Lan58}
R.~L. Lange, H.~H. Hecht, Genesis of pistol-shot and {K}orotkoff sounds,
  Circulation 18~(5) (1958) 975--978.
\newblock \href {http://dx.doi.org/10.1161/01.CIR.18.5.975}
  {\path{doi:10.1161/01.CIR.18.5.975}}.

\end{thebibliography}

\end{document}